\documentclass[twoside,11pt]{Latex/Classes/PhDthesisPSnPDF}
\pdfoutput=1
\setcitestyle{square}

\usepackage{siunitx}
\usepackage{amsmath}
\usepackage[textfont={bf,small}]{caption}
\usepackage{multirow}
\usepackage[final]{pdfpages}
\usepackage{lipsum}
\usepackage{mhchem}
\usepackage[percent]{overpic}
\usepackage{import}
\usepackage{subfig}
\usepackage{graphicx}
\usepackage{enumitem}
\usepackage{float}
\usepackage{minitoc}
\usepackage{bm}
\usepackage[toc,page]{appendix}

\DeclareCaptionLabelFormat{andtable}{#1~#2  \&  \tablename~\thetable}

\usepackage{silence}
\usepackage{rotfloat}

\usepackage{etoolbox}

\def\bary{\begin{array}}
\def\ear{\end{array}}
\def\beq{\begin{equation}}
\def\eeq{\end{equation}}

\newcommand{\ddi}[1]{\frac{\partial #1}{\partial x_i}}

\def\bve{\begin{verbatim}} \def\eve{\end{verbatim}}
\def\bit{\begin{itemize}} \def\eit{\end{itemize}}
\def\ben{\begin{enumerate}} \def\een{\end{enumerate}}
\def\bce{\begin{center}} \def\ece{\end{center}}
\def\bfl{\begin{flushleft}} \def\efl{\end{flushleft}}
\def\bfr{\begin{flushright}} \def\efr{\end{flushright}}
\def\bfi{\begin{figure}[!hbtp]} \def\efi{\end{figure}}
\def\bta{\begin{table}[!hbtp]} \def\etab{\end{table}}
\def\btb{\begin{tabular}} \def\etb{\end{tabular}}
\def\bdi{\begin{displaystyle}} \def\edi{\end{displaystyle}}
\def\bea{\begin{eqnarray}} \def\eea{\end{eqnarray}}

\def\ba{\begin{array}} \def\ea{\end{array}}
\def\be{\begin{equation}} \def\ee{\end{equation}}
\def\bea{\begin{eqnarray}} \def\eea{\end{eqnarray}}
\def\bc{\begin{center}} \def\ec{\end{center}}
\def\bi{\begin{itemize}} \def\ei{\end{itemize}}
\def\bfi{\begin{figure}[!hbtp]} \def\efi{\end{figure}}
\def\bse{\begin{subequations}} \def\ese{\end{subequations}}
\def\btb{\begin{tabular}} \def\etb{\end{tabular}}
\newdimen\boxfigwidth 
\def\bigbox{\begingroup
  \boxfigwidth=\hsize
  \advance\boxfigwidth by -2\fboxrule
  \advance\boxfigwidth by -2\fboxsep
  \setbox4=\vbox\bgroup\hsize\boxfigwidth
  \hrule height0pt width\boxfigwidth\smallskip%
  \linewidth=\boxfigwidth
}
\def\endbigbox{\smallskip\egroup\fbox{\box4}\endgroup}

  \author{YourName}
  \collegeordept{CollegeOrDept}
  \university{University}
\crest{} 
 
\degree{Philosophi\ae Doctor (PhD), DPhil,..}
\degreedate{year month}

\usepackage{array,makecell,booktabs}
\usepackage{multirow}
\usepackage[table,xcdraw]{xcolor}

\makeatletter
\patchcmd{\@makechapterhead}{\vspace*{50\p@}}{\vspace*{-40pt}}{}{}%
\patchcmd{\@makeschapterhead}{\vspace*{50\p@}}{\vspace*{-40pt}}{}{}%
\makeatother

\makeatletter
\newcommand{\extraPartText}[1]{\def\@extraPartText{#1}}
\pretocmd{\@endpart}{\vspace{5ex}\begingroup\centering\@extraPartText\par\endgroup\let\@extraPartText\relax}{}{}
\makeatother

\newenvironment{chapabstract}{\vspace{5mm}\rightskip1in\itshape}{\newline\rule{1.0\textwidth}{0.1mm}}

\usepackage{mathrsfs}
\usepackage{bm}
\usepackage{physics}
\usepackage{esint}
\DeclareMathAlphabet{\mathpzc}{OT1}{pzc}{m}{it}
\DeclareMathAlphabet{\mathscrbf}{OMS}{mdugm}{b}{n}
\DeclareMathAlphabet{\mathcalbf}{OMS}{cmsy}{b}{n}

\newcommand\T{\rule{0pt}{2.6ex}}       
\newcommand\B{\rule[-1.2ex]{0pt}{0pt}} 

\newcommand{\st}[1]{{#1}^{*}}

\begin{document}

\dominitoc
\setcounter{minitocdepth}{2}


\renewcommand\baselinestretch{1.2}
\baselineskip=18pt plus1pt

\includepdf[pages=-, offset=5 -5]{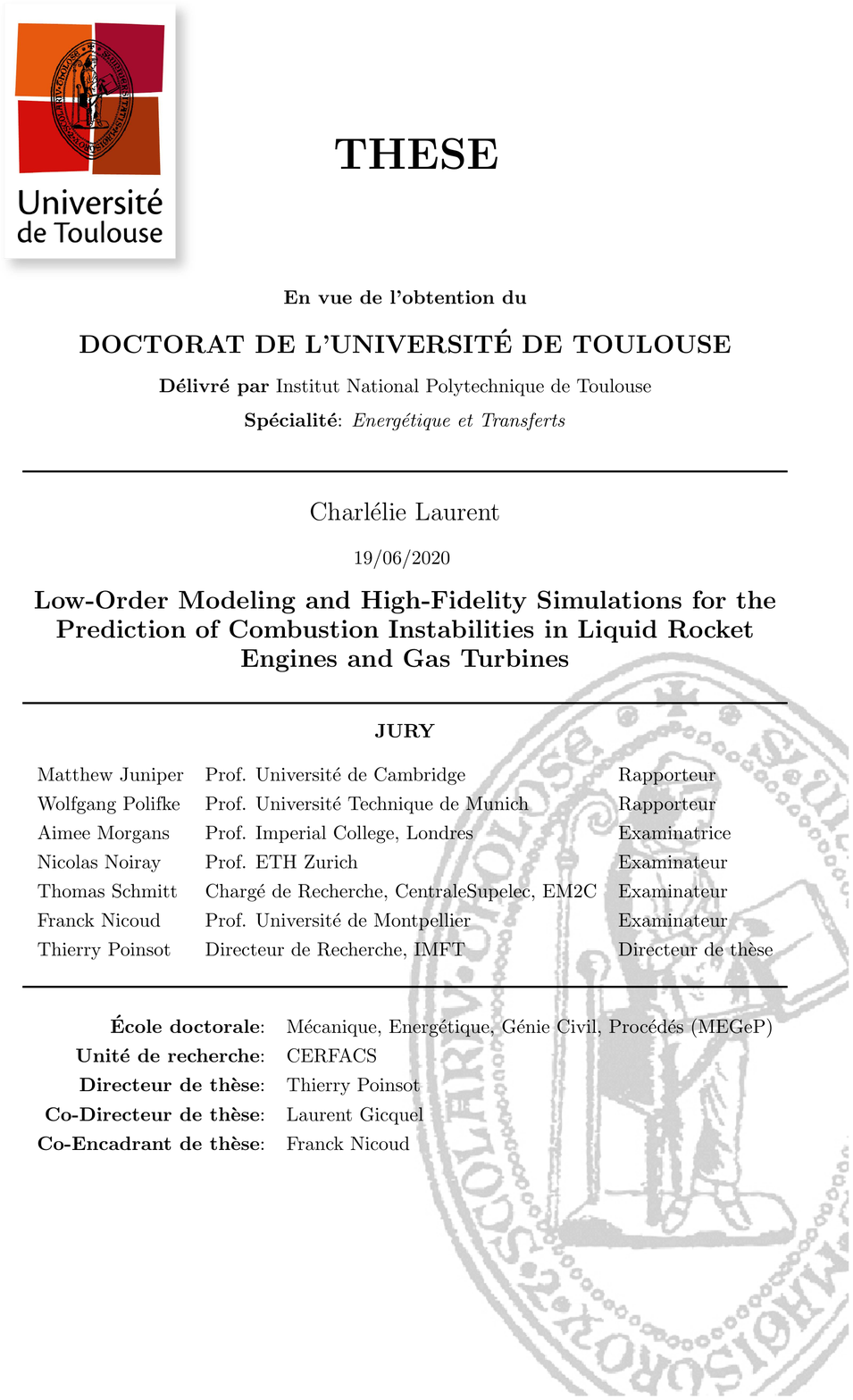}


%
%
%



\noindent \textsc{Abstract}\\

Over the last decades, combustion instabilities have been a major concern for a number of industrial projects, especially in the design of Liquid Rocket Engines (LREs) and gas turbines. Mitigating their effects requires a solid scientific understanding of the intricate interplay between flame dynamics and acoustic waves. During this PhD work, several directions were explored to provide a better comprehension of flame dynamics in cryogenic rocket engines, as well as more efficient and robust numerical methods for the prediction of thermoacoustic instabilities in complex combustors.

The first facet of this work consists in the resolution of unstable thermoacoustic modes in complex multi-injectors combustors, a task that often requires a number of simplifications to be computationally affordable. These necessary physics-based assumptions led to the growing popularity of acoustic Low-Order Models (LOMs), among which Galerkin expansion LOMs have shown a promising efficiency while retaining a satisfactory accuracy. Those are however limited to simple geometries that do not incorporate the complex features of industrial systems. A major part of this work therefore consists first in identifying the mathematical limitations of the classical Galerkin expansion, and then in designing a novel type of modal expansion, named a frame expansion, that does not suffer from the same restrictions. In particular, the frame expansion is able to accurately represent the acoustic velocity field, near non-rigid-wall boundaries of the combustor, a crucial ability that the Galerkin method lacks. In this work, the concept of surface modal expansion is also introduced to model topologically complex boundaries, such as multi-perforated liners encountered in gas turbines. These novel numerical methods are combined with the state-space formalism to build acoustic networks of complex systems. The resulting LOM framework is implemented in the code STORM (State-space Thermoacoustic low-ORder Model), which enables the low-order modeling of thermoacoustic instabilities in arbitrarily complex geometries.

The second ingredient in the prediction of thermoacoustic instabilities is the flame dynamics modeling. This work deals with this problem, in the specific case of a cryogenic coaxial jet-flame characteristic of a LRE. Flame dynamics driving phenomena are identified thanks to three-dimensional Large Eddy Simulations (LES) of the Mascotte experimental test rig where both reactants (CH4 and O2) are injected in transcritical conditions. A first simulation provides a detailed insight into the flame intrinsic dynamics. Several LES with harmonic modulation of the fuel inflow at various frequencies and amplitudes are performed in order to evaluate the flame response to acoustic oscillations and compute a Flame Transfer Function (FTF). The flame nonlinear response, including interactions between intrinsic and forced oscillations, is also investigated. Finally, the stabilization of this flame in the near-injector region, which is of primary importance to the overall flame dynamics, is investigated thanks to multi-physics two-dimensional Direct Numerical Simulations (DNS), where a conjugate heat transfer problem is resolved at the injector lip.

\vspace{1cm}

\noindent \textsc{R\'esum\'e}\\

Au cours des derni\`eres d\'ecennies, les instabilit\'es de combustion ont constitu\'e un important d\'efi pour de nombreux projets industriels, en particulier dans la conception de moteurs-fusées à ergols liquide et de turbines à gaz. L'atténuation de leurs effets n\'ecessite une solide compr\'ehension scientifique de l'interaction complexe entre la dynamique de flamme et les ondes acoustiques qu'elles impliquent. Au cours de cette thèse, plusieurs directions ont \'et\'e explor\'ees pour fournir une meilleure compr\'ehension de la dynamique des flammes dans les moteurs-fus\'ees cryog\'eniques, ainsi que des m\'ethodes num\'eriques plus efficaces et robustes pour la pr\'ediction des instabilit\'es thermoacoustiques dans les chambres de combustion à g\'eom\'etries complexes.

La premi\`ere facette de ce travail a consist\'e en la r\'esolution de modes thermoacoustiques dans les chambres de combustion complexes comportant à injecteurs multiples, une tâche qui n\'ecessite souvent des simplifications pour être abordable en termes de coût de calcul. Ces hypothèses physiques n\'ecessaires ont conduit à la popularit\'e croissante des mod\`eles bas-ordre acoustiques, parmi lesquels ceux utilisant l'expansion de Galerkin ont d\'emontr\'e une efficacit\'e prometteuse tout en conservant une pr\'ecision satisfaisante. Ceux-ci sont cependant limit\'es à des g\'eom\'etries simples qui n'int\`egrent pas les caract\'eristiques complexes des syst\`emes industriels. Une grande partie de ce travail a donc consist\'e tout d'abord à identifier clairement les limitations math\'ematiques de l'expansion classique de Galerkin, puis à concevoir un nouveau type d'expansion modale, appel\'e expansion sur frame, qui ne souffre pas des mêmes restrictions. En particulier, l'expansion sur frame est capable de repr\'esenter avec pr\'ecision le champ de vitesse acoustique pr\`es des parois de la chambre de combustion autres que des murs rigides, une capacit\'e cruciale qui manque à la m\'ethode Galerkin. Dans ce travail, le concept d'expansion modale de surface a \'egalement \'et\'e introduit pour mod\'eliser des fronti\`eres topologiquement complexes, comme les plaques multi-perfor\'ees rencontr\'ees dans les turbines à gaz. Ces nouvelles m\'ethodes num\'eriques ont \'et\'e combin\'ees avec le formalisme state-space pour construire des r\'eseaux acoustiques de syst\`emes complexes. Le mod\`ele obtenu a \'et\'e impl\'ement\'e dans le code STORM (State-space Thermoacoustic low-ORder Model), qui permet la mod\'elisation bas-ordre des instabilit\'es thermoacoustiques dans des g\'eom\'etries arbitrairement complexes.

Le deuxi\`eme ingr\'edient de la pr\'ediction des instabilit\'es thermoacoustiques est la mod\'elisation de la dynamique de flamme. Ce travail a trait\'e de ce point, dans le cas sp\'ecifique d'une flamme-jet cryog\'enique caract\'eristique d'un moteur-fus\'ee à ergols liquides. Les ph\'enom\`enes contrôlant la dynamique de flamme ont \'et\'e identifi\'es grâce à des Simulations aux Grandes \'echelles (SGE) du banc d'essai exp\'erimental Mascotte, où les deux r\'eactifs (CH4 et O2) sont inject\'es dans des conditions transcritiques. Une premi\`ere simulation donne un aperçu d\'etaill\'e de la dynamique intrins\`eque de la flamme. Plusieurs SGE avec modulation harmonique de l'injection de carburant, à diff\'erentes fr\'equences et amplitudes, ont \'et\'e effectu\'es afin d'\'evaluer la r\'eponse de la flamme aux oscillations acoustiques et de calculer une Fonction de Transfert de Flamme (FTF). La r\'eponse non-lin\'eaire de la flamme, notamment les interactions entre les oscillations intrins\`eques et forc\'ees, a \'egalement \'et\'e \'etudi\'ee. Enfin, la stabilisation de cette flamme dans la r\'egion proche de l'injecteur, qui est d'une importance primordiale sur la dynamique globale de la flamme, a \'et\'e \'etudi\'ee grâce à une simulation directe multi-physique, où un probl\`eme  coupl\'e de transfert de chaleur est r\'esolu au niveau de la l\`evre de l'injecteur.

\clearpage

\vspace*{\fill}
\begin{flushright}
\textit{}\\
\textit{}
\end{flushright}
\vspace*{\fill}




\begin{acknowledgements}      

First of all I would like to thank Thierry Poinsot who offered me the opportunity to do my PhD at CERFACS, Toulouse, France. The work environment in this lab approaches perfection, and I could not hope for something better. The second person I would like to thank is Franck Nicoud, who provided me an excellent scientific guidance throughout this PhD. He took the time and made the efforts to understand not only the \textit{big picture} of my work, but also the very intricate technical details, in order to assist me in the best possible manner. After three or so years working on scientific research, I realized that this technical ability is a very rare quality in the research community, and I am infinitely grateful to him for that. Working in the unique research environment that exist at CERFACS also implies that many people contributed directly or indirectly to my PhD work. In this matter, I would like to thank the CSG team who always made sure that our computational resources are working optimally; the administrative team for their constant kindness and helpfulness; the Coop team for their dedication to maintain and optimize the computational codes used at CERFACS, without which none of the work accomplished would have been possible. Obviously I would like to thank all PhD students at CERFACS with who I interacted during these 3.5 years: Fabien, Matthieu and Frederic undoubtedly stand out among them; their impact on my work and my everyday life as a PhD student is unquantifiable. I would also like to thank Lucien, Quentin M., Felix, Omar who were a constant source of jokes and laughter; Quentin Q. and Simon for our interesting discussions; Michael who helped me to understand the state-space theory in the early stage of my PhD; Abhijeet who had the patience to listen to my explanations for long hours. I am probably forgetting to mention others, but my gratitude to them is nonetheless profound and sincere. Finally, I cannot conclude this paragraph without mentioning people who deserve these acknowledgement the most: my family and friends, who supported me throughout these past years; nothing would have been possible without them and they deserve all the credit for this PhD thesis.

\end{acknowledgements}


\setcounter{secnumdepth}{3} 
\setcounter{tocdepth}{1}    
\tableofcontents            






%
%


\mainmatter




%


				

\chapter{Introduction} \label{chap:Introduction}
\minitoc

\section{Industrial context} \label{sec:Intro_industry_context}

\begin{figure}
\begin{center}
\includegraphics[width=144mm]{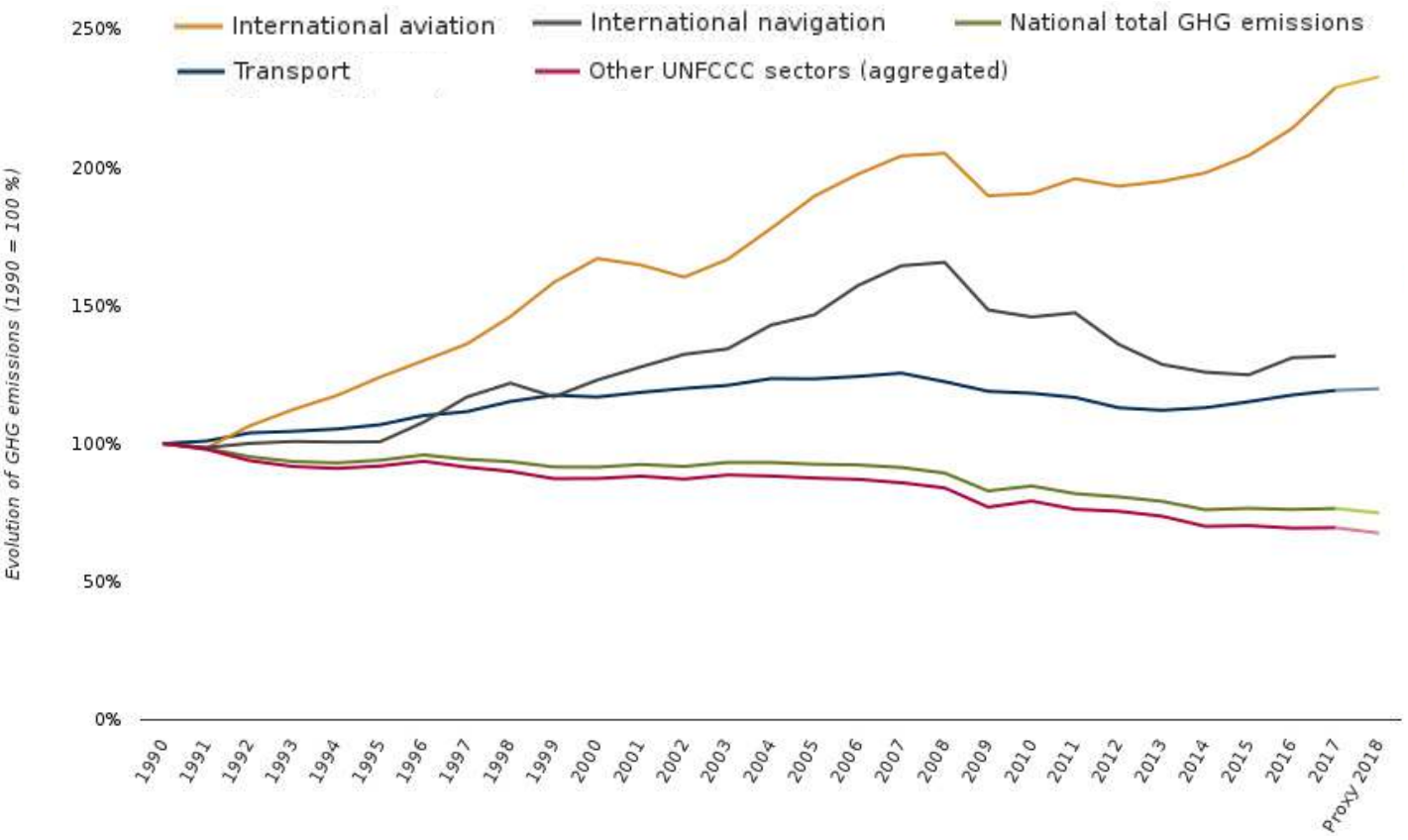}
\caption{Greenhouse gases emissions from transportation in the EU. Extracted from the European Environment Agency (EEA) database.}
\label{fig:graphs_emissions_transports}
\end{center}
\end{figure}

\begin{figure}
\begin{center}
\includegraphics[width=0.99\textwidth]{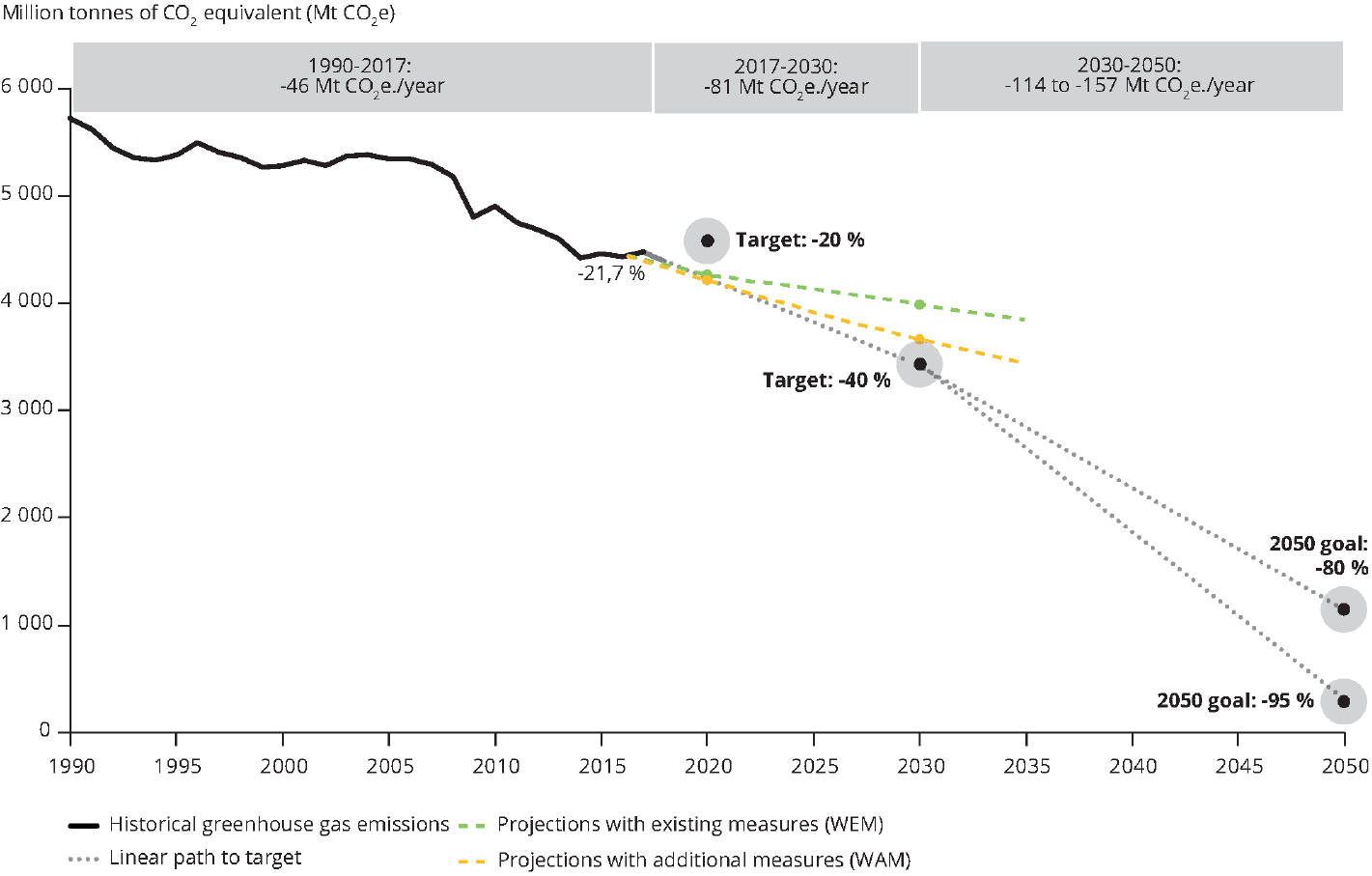}
\caption{Historical trends of total greenhouse gases emissions from EU countries, with projections and targets. Extracted from the EEA database. }
\label{fig:graphs_emissions_target}
\end{center}
\end{figure}

2020. The World is on the brink of a new revolution. The conflict opposing environmental activists to the ever powerful oil industry lobbying machine is raging. In this intense stand-off, where "\textit{Flygskam}" confronts climate change denial, tangible scientific facts weigh little. Yet, these two sides are neither willing, nor have the ability, to propose creative and constructive technical solutions to the existential issue that our world is facing. Air transportation has become the symbol of this opposition. On one hand it cannot be denied that commercial aviation is a substantial contributor to Greenhouse Gases (GHG) emissions responsible for global warming. Even worse: it is the only means of transportation that has seen its emission level soar by more than 100\% since 1990 (Fig.~\ref{fig:graphs_emissions_transports}), and this increase is expected to worsen over the next decades. 
In order to mitigate global warming before it reaches an irreversible point, the European Union fixed aggressive targets to decrease GHG emissions. It now appears that these targets are unlikely to be met, even if a number of additional regulations are enforced (see Fig.~\ref{fig:graphs_emissions_target}). 
If these reduction objectives are to be fulfilled, commercial aviation will have to drastically curb its negative impact. On the other hand, however, there is no silver bullet solution, as the dream of carbon-free, fully electric long-haul airplanes now seems more and more out of reach. Astonishingly, a 80 tons Airbus A320 would require 900 tons of state-of-the-art batteries to fly. These prohibitive electric energy limitations led the aviation industry to explore more reasonable directions, such as hybrid aircraft. Yet, the most significant improvements will have to be made on the combustion devices that generate the vast majority of the energy used by the turbofan engines propelling these airplanes. Thus, instead of becoming a collateral damage in the fight against global warming, combustion science is destined to be a key actor in the ongoing energy revolution, by assisting the development of cutting-edge technologies for a sustainable future.\par

\begin{figure}
\begin{center}
\includegraphics[width=144mm]{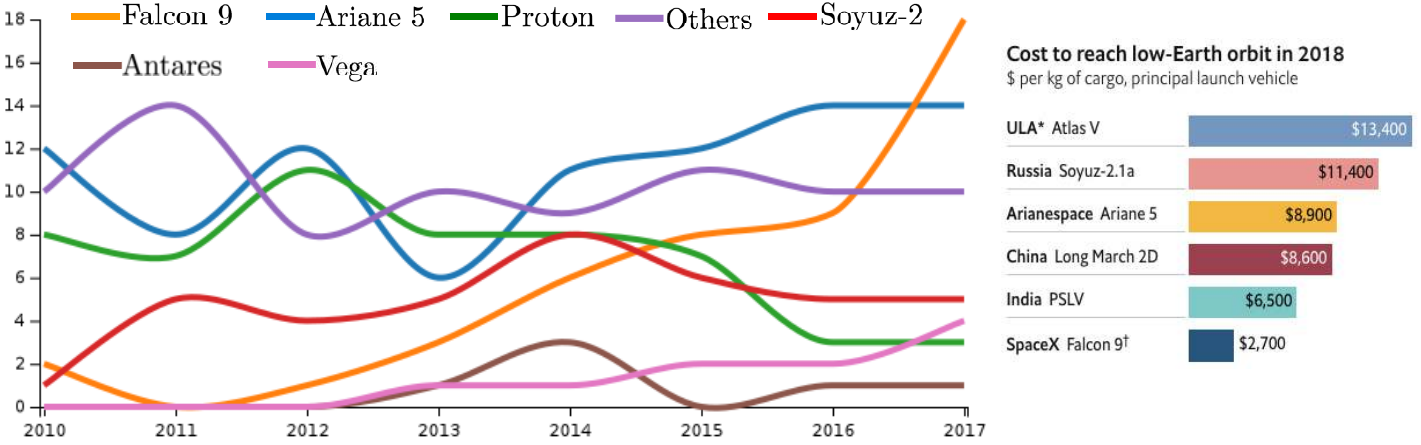}
\caption{Left: number of successful space launches per year, for major launcher models. Right: catalog price to send satellites to Low-Earth-Orbit (LOE). Extracted from \textit{The Economist}.}
\label{fig:graphs_space_market}
\end{center}
\end{figure}

On a completely opposite front, another technological revolution is taking place. The Cold War-area space race is reborn. Its actors are however different: the global space market is now dominated by dynamic private companies, rather than by the traditional governmental agencies. SpaceX is arguably the most striking example of this success, as it went from not achieving a single space launch in 2011, to becoming the world leader with 18 successful attempts in 2017 (see Fig.~\ref{fig:graphs_space_market}). The recipe for the achievement of this milsetone is simple: by designing the first reusable rocket launcher, dubbed the Falcon 9, SpaceX was able to cut its operation costs by a factor of 5 in comparison to its competitors, and thus to propose lower catalog prices to its customers.
Even more importantly, the Falcon 9 became the backbone of SpaceX expansion strategy. Its Merlin-1D engines, as well as the main components of its structure were used to assemble the Falcon Heavy, the most powerful rocket ever launched, and that may be used by SpaceX to send manned missions to Mars within the next few years. The versatile and inexpensive Falcon 9 is also the platform used by SpaceX to deploy its nano-satellites constellation Starlink, intended to bring low-latency broadband internet across the globe. These achievements did not go unnoticed in the space industry, as other private companies started to design their own reusable launchers, and others such as Blue Origin are already prepared to enter the space tourism market. Governmental agencies were also caught into the wake of these successes, and now display their will to develop their own nano-satellites constellations, or in the case of the Chinese National Space Administration, its own permanent lunar base. Perspective for combustion science are exciting: the development of a multitude of increasingly cheap and versatile space launchers inevitably require significant breakthroughs in the design of novel propulsion systems. On this front of the imminent energy revolution, combustion science does not need to help to extricate our world from a dead-end, but rather to push it beyond its limits.

\subsection{Gas turbines}

\begin{figure}
\begin{center}
\includegraphics[width=0.99\textwidth]{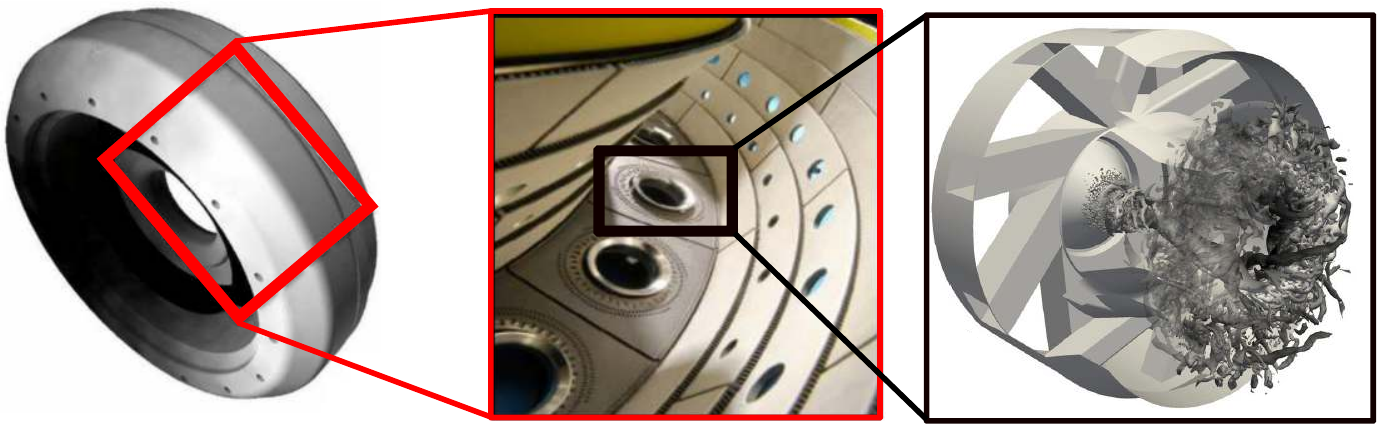}
\caption{Left: a typical annular combustion chamber from Safran Helicopter Engines. Middle: closeup view of a few sectors in a Rolls-Royce annular combustor. Right: an axial swirled-injector with a flow visualization, extracted from~\cite{brebion2017}.}
\label{fig:gas_turbine_design}
\end{center}
\end{figure}

Gas turbines, which are a ubiquitous combustion device for commercial aircraft propulsion, as well as for heavy-duty land-based power plants, have already been subjected to technical improvements intended to reduce their emissions. Modern designs include annular combustors divided into a number of sectors usually ranging from 10 to 30 (see Fig.~\ref{fig:gas_turbine_design}). Fuel and air are fed to each sector by a swirled-injector intended to enhance mixing and where a turbulent flame is stabilized.
Over the last decades, a common emission reduction strategy consisted in targeting lean combustion regimes~\cite{dunn2011}, for example with the RQL (Rich burn - Quick Mix - Lean burn) design. In this staged combustion process, a secondary air flow is injected downstream into the chamber thanks to a series of dilution holes perforated through the liner separating the hot primary flow from the cold casing. Further reducing GHG emission will require even more profound modifications to the existing gas turbine technologies. A promising direction consists in replacing conventional fossil fuels with renewable ones, such as biofuels. In this matter, Air France-KLM already started to operate a small number of flights where the conventional jet-fuel is mixed with up to 10\% of waste-derived biofuel. A more drastic approach, that is also scrutinized by gas turbines manufacturers, is the injection of a fraction of pure hydrogen into the combustion chamber in addition to the main hydrocarbon fuel, whether it is a fossil or a renewable one. More than simply replacing a conventional fuel,  strategically injecting the pure hydrogen at specific locations can be used to unlock its unique combustion properties in order to control the flame regime, structure, and stabilization mechanisms. These multi-fuel combustion systems, where a large variety of biofuels blends can be combined with pure hydrogen, open the way to an infinity of new perspectives in the transition to cleaner combustion.\par
A broad use of these alternative fuels will, however, necessitate to profoundly rethink existing gas turbines designs. Annular combustion chambers have long been known to be subjected to azimuthal thermoacoustic instabilities~\cite{Crighton:1992,Candel:1992,Poinsot:2005,Lieuwen:2005a,Poinsot:2017}, a phenomenon resulting from an intricate interplay between flames fluctuations and the acoustic waves they emits. These instabilities, first identified by Lord Rayleigh~\cite{rayleigh1878} and Rijke~\cite{Rijke:1859a} in the 19\textsuperscript{th} century, have since then been responsible for performance loss, or even irreversible damages, in gas turbines (see Fig.~\ref{fig:instability_damage}, left). The lean combustion regime, highly desired for its favorable emissions characteristics, is especially well-known to promote the apparition of thermoacoustic instabilities. The design of novel gas turbines optimized for bio and multi-fuel lean combustion, will therefore inevitably require a thorough comprehension and modeling of thermoacoustic instabilities in these complex systems.

\subsection{Liquid Rocket Engines}

\begin{figure}
\begin{center}
\includegraphics[width=0.99\textwidth]{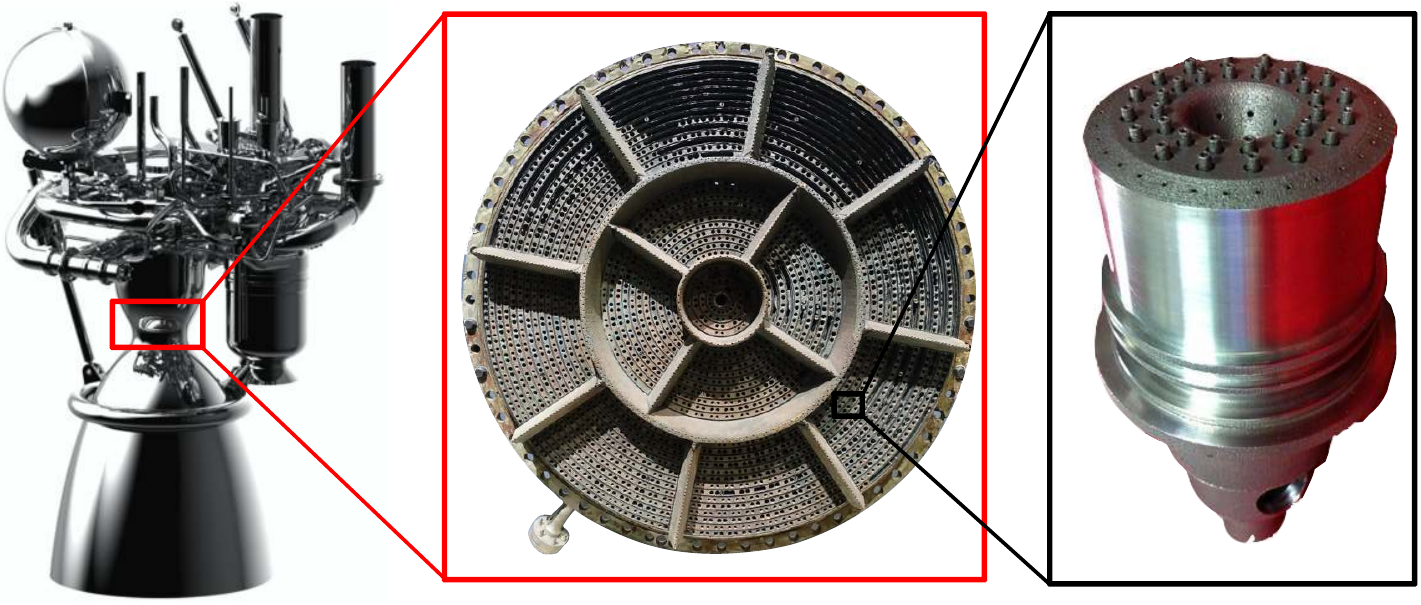}
\caption{Left: Prometheus, a reusable, re-ignitable, LO\textsubscript{2}/LCH\textsubscript{4} rocket engine, currently under development by ArianeGroup (ESA picture: \url{http://www.esa.int/Enabling_Support/Space_Transportation/Prometheus_to_power_future_launchers}). Middle: injector plate from the F-1 LRE propelling the Saturn V during the Apollo missions. Right: a single multi-nozzles injector, 3D-printed at NASA's Marshall Space Flight Center in 2013 for the Space Launch System (\url{http://www.flickr.com/photos/nasamarshall}).}
\label{fig:lre_design}
\end{center}
\end{figure}

On the opposite side of the spectrum, space propulsion combustion is also undergoing a radical evolution. The technology push initiated by SpaceX lead the almost entirety of the space industry to accelerate their development of reusable launchers. Modern launchers are usually propelled by Liquid Rocket Engines (LRE) fed in fuel and liquid cryogenic oxygen through an injection plate comprising hundreds of injectors (see Fig.~\ref{fig:lre_design}), that is itself alimented by a powerful turbo-pump.
Injectors often comprise coaxial round jets of fuel and oxidizer, even though other slightly different designs, such as multi-nozzles injectors, have also been used. The specificity of LRE combustion is the high pressure in the combustion chamber, usually around or above 100 bar, and the huge temperature difference existing between cryogenic reactants at -200 ${}^{\circ}$C, and the flames at 3500 ${}^{\circ}$C. These extreme thermodynamic conditions impose severe loads to the engine components, which make the development of a reusable launcher a challenge where controlling the combustion process is of vital importance. Reusability is not the only target of rocket manufacturers: engine re-ignitability and ability to operate at variable thrust levels (from 30\% to 110\% of its nominal value), are also highly desirable features intended to give a greater maneuverability to future space launchers. In order to meet these specifications, a large number of LRE manufacturers decided to shift from the classical liquid oxygen-hydrogen combustion (LO\textsubscript{2}/H\textsubscript{2}) to liquid oxygen-methane combustion (LO\textsubscript{2}/CH\textsubscript{4}). In this matter, SpaceX already started the tests of their future LO\textsubscript{2}/CH\textsubscript{4} Raptor engine, and the Euopean Space Agency with its prime contractor ArianeGroup are well underway in the development of the Prometheus LRE. Methane combustion and thermodynamic properties strongly differ from that of hydrogen or other previously used rocket propellants, and once again this new fuel and the possibilities it offers will ultimately lead to a wide variety of combustion regimes that are little-studied and mostly unknown.\par

\begin{figure}
\begin{center}
\includegraphics[width=0.99\textwidth]{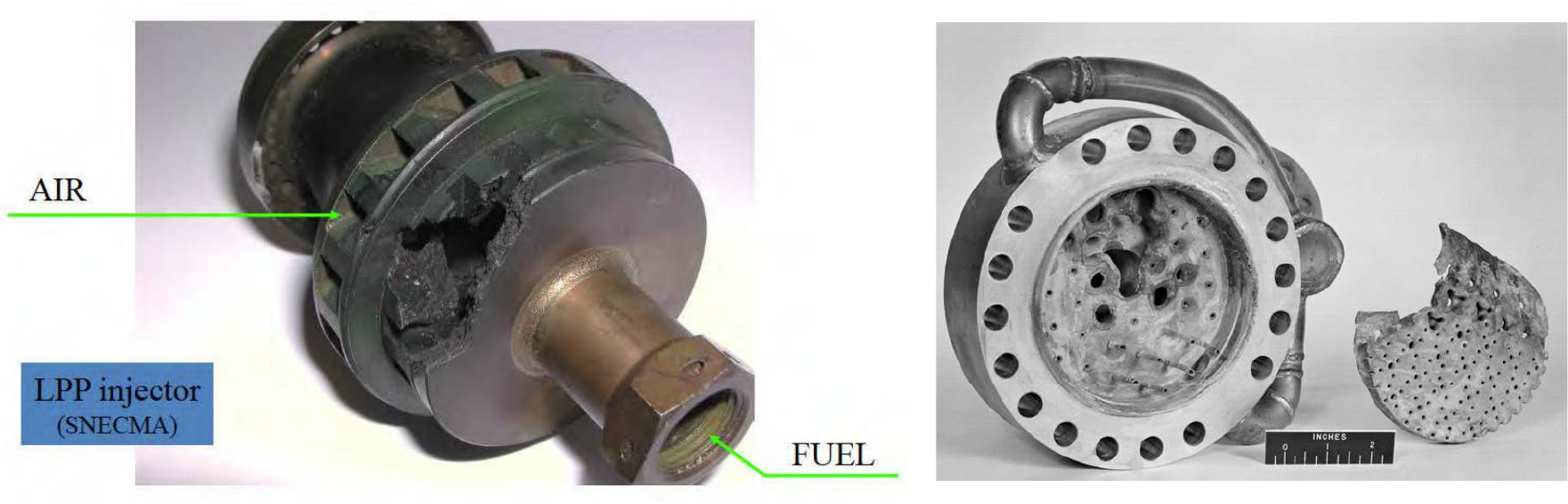}
\caption{Left: an injector of a Safran Aircraft Engines Lean Premixed Prevaporized combustor damaged by a thermoacoustic instability. Right: a NASA LRE combustion chamber destroyed by a flashback due to a thermoacoustic instability during the early days of the Apollo program. Extracted from~\cite{poinsot1987vortex}.}
\label{fig:instability_damage}
\end{center}
\end{figure}

In a similar fashion to the development of new cleaner gas turbines, the design of reusable more flexible LREs will inevitably face the problem of themoacoustic instabilities. This difficulty is even more present in the domain of rocket engines, where thermoacoustic instabilities have historically been known to plague the progress of numerous industrial projects since the early days of the space race. The most famous of them is arguably the F-1 engine propelling the Saturn V used during the 60s and 70s Apollo missions. In the early phase of its development, combustion instabilities observed were so strong that they would lead to the spectacular destruction of the combustion chamber (see Fig.~\ref{fig:instability_damage}, right).
To mitigate them, Rocketdyne and NASA engineers carried out a trial and error test campaign, where the main design parameter was the spatial layout of the injectors on the injection plate. This strategy led to the destruction of numerous prototypes and a billions dollars over budget, but eventually produced one of the most powerful LRE ever built. NASA and other LRE industry actors gained a valuable experience throughout the development of subsequent projects, leading to an improved and well-documented understanding of thermoacoustic instabilities~\cite{crocco1956,harrje1972,oefelein1993,Culick:2006}. The design of novel reusable LO\textsubscript{2}/CH\textsubscript{4} LREs will need on one hand to take full advantage of this existing knowledge, and on the other to complement it with the study of combustion instabilities in conditions that heretofore are largely unexplored.

\section{Combustion instabilities mechanisms} \label{sec:Intro_instabilities_mechanisms}

\begin{figure}
\begin{center}
\includegraphics[width=0.90\textwidth]{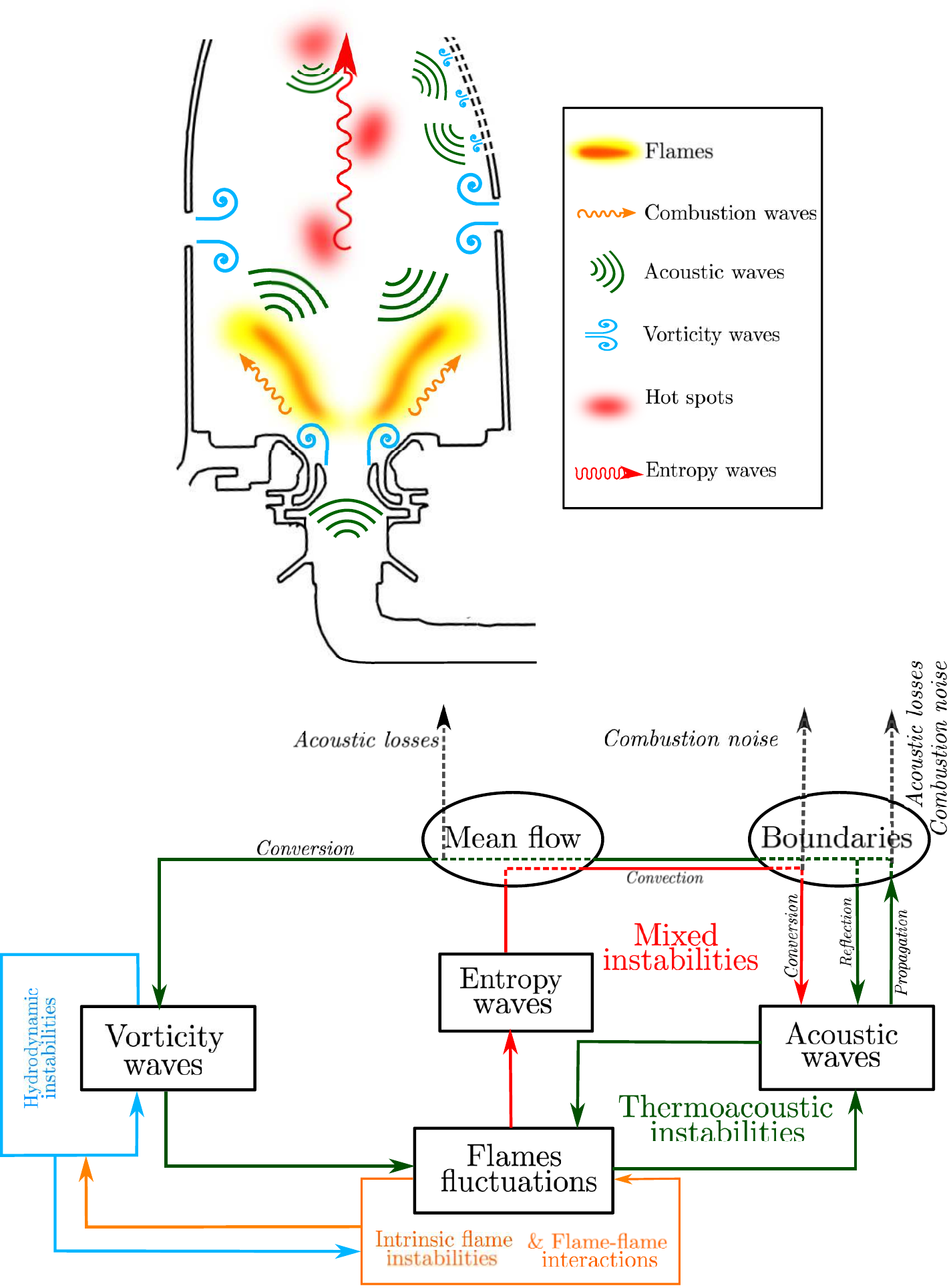}
\caption{Top: schematic of a gas turbine combustor and the major types of fluctuations that may occur. Bottom: diagram detailing the interactions that may take place between the different sources of fluctuations and the components of the combustor. The green arrows represent the closed-loop coupling leading to a thermoacoustic instability.}
\label{fig:instability_illustration}
\end{center}
\end{figure}

Thermoacoustic instabilities in gas turbines and LREs are governed by common mechanisms that are detailed in Fig.~\ref{fig:instability_illustration}. Those are extensively detailed in a number of comprehensive reviews~\cite{Lieuwen:2003,Lieuwen:2005,Ducruix:2003,Ducruix:2005,OConnor:2015,Poinsot:2017}.
Numerous sources of oscillations may exist in an industrial combustion chamber. The flames are not only subjected to chaotic turbulent oscillations, but can also be disturbed by coherent vortices due to an hydrodynamic instability, by perturbations of the mixture composition, through interactions with neighboring flames, or can even experience their own self-sustained intrinsic oscillations. When a flame is disrupted, it creates an unsteady heat-release, that in turn produces an expansion of gases resulting in the formation of an acoustic wave. Acoustic waves propagate both downstream and upstream in the combustor, interacting on the way with the mean flow and with existing vorticity waves, until they reach a boundary of the chamber. Depending on the geometry of the boundary, acoustic waves are then reflected back towards the flames or undergo a hydrodynamic interaction through which they are converted into vorticity waves. Both contributions can then further disrupt the  flames, thus resulting in the formation of even more acoustic waves, and so forth. This closed loop coupling between flame fluctuations and acoustic waves, with a possible intermediate conversion into vortical waves, is the keystone mechanism of a thermoacoustic instability. The progressive build-up of acoustic waves and heat-release oscillations produces increasingly intense fluctuations, that may in the worst case irreversibly disrupt the combustion process or damage the combustor structure.\par

A secondary mechanism may trigger the apparition of an instability: as they oscillate, flames also produce temperature inhomogeneities called hot spots, that are convected downstream by the mean flow. As they reach the combustor outlet, that can for example consist of a turbine or a chocked nozzle, these entropy waves are  partially converted into acoustic waves that propagates back upstream in the combustion chamber. Those can additionally perturb the flames, thus creating another closed-loop coupling mechanism, called a mixed instability~\cite{Marble:1977,Abouseif:1984,Keller:1985b,Goh:2013,Motheau:2014}.

\subsection{Acoustics}

The first ingredient in the unstable coupling mechanism described above is the acoustics, and they therefore require to be correctly understood and modeled. Even though this point may seem relatively simple in comparison to the multitude of other phenomena involved, the complexity of an industrial combustor is a challenge that greatly complicates this task. The following points are particularly crucial in the accurate description of acoustic waves in combustion chambers:
\begin{itemize}
\item{The formation of acoustic waves is due to flames oscillations, and an equation governing the acoustics dynamics with the heat-release fluctuations as a source term should therefore be derived.}
\item{The mean temperature, density, and sound speed fields in a combustor are highly inhomogeneous, due to the large difference between cold reactants and hot combustion products. The propagation of acoustic waves in these conditions is largely impacted by these spatial distributions. Additionally, in LRE conditions the heat-capacity ratio field is also inhomogeneous, which further complicates this problem.}
\item{Most importantly, a combustor is a confined domain and acoustics are in this case strongly dependent on its geometry: acoustic waves propagate freely in the fluid volume, but are affected by the numerous reflections on its walls. As a result, they self-organize into coherent patterns corresponding to the combustor eigenmodes. The geometrical complexity of the combustion chamber, that can for instance include dozens of injectors, is therefore a major difficulty in the modeling of thermoacoustic instabilities.}
\item{Industrial combustion devices include complex elements, such as multi-perforated liners intended to cool down the chamber walls. When acoustic waves reflect on those, they are subjected to elaborate hydrodynamic interactions that can be the cause of acoustic losses. Those should be accounted for to correctly predict the unstable nature of a combustor.}
\item{Finally, during a LRE combustion instability it is not uncommon that the acoustic pressure oscillations reach 40\% of the base chamber pressure. Acoustic waves then enter a nonlinear regime that affects their propagation and may result in the formation of weak shocks~\cite{Culick:1994}.}
\end{itemize}

\subsection{Flame Dynamics}

The second ingredient in the thermoacoustic instability closed-loop mechanism is the response of the flame to acoustic perturbations, also called flame dynamics. Since a flame is strongly dependent on both the nature of the injector and the burner where it is stabilized, modeling the flame dynamics must often consider the generation of vortical waves that may also participate in the flame perturbation. Flame dynamics is arguably the most challenging problem in the modeling of thermoacoustic instabilities, as it must account for rich physics such as multi-components fuel chemical kinetics, and chemistry-turbulence, flame-vortex, and flame-wall interactions. Simplifying assumptions are often necessary. The most common of them is to consider that the flame behaves as a Linear Time Invariant (LTI) system, in which case its response can be embedded into a Flame Transfer Function (FTF). The first heuristic FTF, introduced by Crocco~\cite{Crocco:1952} in the 50s to model LRE flame dynamics, simply consists in assuming that a flame responds to an incoming acoustic perturbation with a relative intensity $n$ and a time-delay $\tau$. Over the last decades, as the Flame Transfer Function became a ubiquitous tool to predict thermoacoustic instabilities, many extensions of this model were proposed to account for nonlinear effects~\cite{Noiray:2012}, wall-heat losses~\cite{Kedia:2011}, or swirled injector interactions~\cite{Palies:2011c}.

\section{The role of numerical simulation} \label{sec:Intro_simulation_role}

As in many fields of engineering, numerical simulation plays an increasingly prevalent role in the prediction of thermoacoustic instabilities. It can not only be used to replace costly test campaigns carried out by rocket engine and gas turbine manufacturers, but it can also advantageously complement those by providing data that remain inaccessible even to the most advanced diagnostics. It must nonetheless be noted that the characteristic timescale of thermoacoustic oscillations can be as large as $10^{-2}$~s, and that the chemical timescales existing in a flame can be as low as $10^{-10}$~s. Similarly, acoustic wavelengths in industrial combustors are of the order of $50$~cm, while the flame thickness can be below $0.1$~mm. These extreme separations of scales are a daunting challenge that led to the development of an array of numerical tools to predict thermoacoustic instabilities. These methods, that vary in fidelity and cost, are listed below.
\begin{itemize}
\item{\textbf{Direct Numerical Simulation:} Direct Numerical Simulation (DNS) is the most accurate, but also the most costly method to perform combustion simulations. Due to this prohibitive cost, DNS is relatively little used for the prediction of thermoacoustic instabilities in complex configurations. It is however useful to gain a detailed insight into small-scale fundamental problems, such as single laminar flame configurations~\cite{Silva:2015}, or the effect of flame-wall interaction on flame dynamics~\cite{Kaiser:2019}.}
\item{\textbf{Large Eddy Simulation:} Large Eddy Simulation (LES) has already proved to be an invaluable tool for the computation of a wide variety of combustion phenomena~\cite{Pitsch:2006,Poinsot:2005}, due to its ability to capture unsteady fluctuations at a cost significantly lower than that of DNS. It was applied to the simulation of limit-cycle thermoacoustic instabilities in full-scale industrial systems, such as a helicopter engine gas turbine~\cite{Wolf:2012} or a 42-injectors LO\textsubscript{2}/LH\textsubscript{2} rocket engine~\cite{Urbano:2017}. This approach however necessitates considerable computational resources. A more pragmatic strategy consists in isolating a single or a few injectors from an industrial system: LES of this simplified configuration are then performed to specifically study the flame dynamics problem. This approach was notably employed to compute the FTFs of swirl-stabilized flames~\cite{Gicquel:2012a,Chong:2010,Merk:2019}.}
\item{\textbf{Low-Order Model:} Both DNS and LES necessitate considerable ressources to simulate thermoacoustic instabilities, which makes them impractical to perform tasks requiring a large number of repeated computations. This limitation motivates the design of Low-Order Models (LOMs) to predict instabilities at a fraction of the cost of LES. A large range of methods are available to reduce the order of a thermoacoustic problem: they usually consist in simplifying the resolution of the acoustic fields, while the flame dynamics are embedded into a FTF previously obtained through LES, experimental measurements, or theoretical derivations~\cite{Schuller:2003}. LOMs were successfuly applied in computationally intensive tasks such as Uncertainty Quantification (UQ) in a model annular combustor~\cite{Bauerheim:2016a}, or adjoint optimization~\cite{Magri:2019}. However, as existing thermoacoustic LOMs rely on a set of simplifying assumptions, they suffer from severe limitations regarding the complexity of the combustor, and are therefore only applicable to model problems.}
\end{itemize}

\section{PhD objectives and thesis outline} \label{sec:Intro_outline}

The goal of this PhD thesis is to improve state-of-the-art capabilities in the prediction of thermoacoustic instabilities in both gas turbines and Liquid Rocket Engines. Two distinct directions are explored, with on one hand the introduction of a novel class of numerical methods that enable the low-order modeling of combustion instabilities in these complex systems, and on the other hand the investigation of flame dynamics in conditions that are currently little studied. These two distinct aspects are detailed in two independent parts that are organized as follows.

\begin{itemize}
\item{\textbf{Part I:} First, a detailed overview of low-order modeling in thermoacoustics is provided. Existing methods, with their strengths and weaknesses are presented and compared on simple test cases. The concept of acoustic network is introduced and formalized thanks to the state-space framework. The second chapter presents one of the major contributions of this work:  a generalization of the classical Galerkin expansion, a method widely used in thermoacoustic LOMs, is proposed. This novel expansion, called the \textit{frame modal expansion}, has the potential to represent any type of boundary conditions, whereas the classical Galerkin expansion is limited to either rigid-wall or pressure-release boundaries. This advantageous characteristic can not only be used to include impedance boundary conditions in LOMs, but also to build more elaborate acoustic networks. Comprehensive comparisons between these two types of modal expansions are presented on a series of examples, with an emphasis on their respective accuracy and convergence properties. It is in particular shown that the solutions obtained through the Galerkin expansion are affected by a Gibbs phenomenon near non-rigid-wall boundaries, while the frame expansion succesfully mitigates these oscillations. A third chapter introduces the concept of \textit{surface modal expansion}, used to model topologically complex boundaries in thermoacoustic LOMs. This approach, combined with the frame modal expansion, is applied to resolve thermoacoustic instabilities in geometries comprising complex impedance, such as curved multi-perforated liners or chocked nozzles. These low-order methods, along with an entire library of acoustic elements, have been compiled to give birth to the LOM code STORM (STate-space thermOacoustic Reduced-order Model).}

\item{\textbf{Part II:} The second part of this thesis presents a series of high-fidelity simulations intended to study the dynamics of a cryogenic LO\textsubscript{2}/LCH\textsubscript{4} coaxial jet-flame in the Mascotte test rig~\cite{vingert2000}, representative of a Liquid Rocket Engine. This combustion regime, dubbed \textit{doubly-transcritical}, is characteristic of future methane-fueled reusable LREs and is largely unexplored in the literature. To the knowledge of the author, this work is one of the first and most complete attempts to characterize the dynamics of a LRE flame in such complex conditions. This task constituted a computational challenge that could only be achieved thanks to significant resources awarded through PRACE, GENCI, and CNES. A first chapter recalls the key challenges of transcritical and supercritical combustion. It also introduces the numerical methods used in the real-gas version of the solver AVBP employed to perform the simulations. Then, a first important contribution of this PhD is presented: it consists in the derivation of a kinetic scheme for CH\textsubscript{4} combustion in LRE conditions. A second chapter describes a first set of LES that were performed to evaluate the flame linear response to small-amplitude acoustic oscillations at the fuel inlet. An axial Flame Transfer Function is computed over a wide frequency range spanning from 1000~Hz to 17000~Hz, and preferential response regions are evidenced. It also shows that the flame response is directly governed by the dynamics of annular vortex rings convected in the annular methane jet. A theoretical analysis is then carried out to identify the major physical contributions driving the heat-release fluctuations. In a third chapter, the nonlinear flame dynamics are investigated thanks to another set of LES at a larger acoustic forcing amplitude. A wide variety of nonlinear phenomena are considered and thoroughly evaluated: for instance, the interaction between forced and intrinsic flame dynamics, the saturation of the flame response, the apparition of higher harmonics, the nonlinear vortex convection, and so forth. Finally, a last chapter questions an important assumption made in the previous simulations. The lip of the coaxial injector was indeed assumed adiabatic, which is highly disputable given its thinness and its proximity to both the hot combustion products and the cryogenic propellants. In this matter, two-dimensional DNS of the near injector region were performed: a first one with adiabatic boundaries, and a second one where a conjugate heat-transfer problem is resolved by coupling the flow solver to a heat-conduction solver within the lip wall. This latter DNS evidenced a distinctively different flame root stabilization mechanism, characterized by coherent 450~Hz oscillations affecting both the flame root and the temperature field within the lip.}
\end{itemize}
  
Finally, the work accomplished during this PhD is expected to bring more robust methods for low-order modeling of thermoacoustic instabilities in realistic industrial combustors. In particular, the LOM code STORM may be used by future generations of PhD students, and by research engineers at CERFACS industrial partners. On another hand, the set of high-fidelity simulations that were performed will bring a new insight into the challenging problem of flame dynamics in LRE conditions, which is an invaluable asset to anticipate the apparition of thermoacoustic instabilities during development of new space launchers.

\extraPartText{ \hfill \\[5ex]
\textbf{This part led to the following publications:} \\[8mm]
\textit{\textbf{1. Laurent, C.}, Bauerheim, M., Poinsot, T., \& Nicoud, F. (2019). A novel modal expansion method for low-order modeling of thermoacoustic instabilities in complex geometries.} \textbf{Combustion and Flame}, 206, 334-348. \\[5mm]
\textit{\textbf{2. Laurent, C.}, Badhe, A., Nicoud, F. (2020). Including complex boundary conditions in low-order modeling of thermoacoustic instabilities.} (Submitted to \textbf{Combustion and Flame})  }
\part{Low-Order Modeling of thermoacoustic instabilities in complex geometries}\label{part: LOM}
\newcommand{\td}{(\vec{x},t)}
\newcommand{\fd}{(\vec{x},\omega)}

\chapter{Introduction to low-order modeling in thermoacoustics} \label{chap:intro_LOM}
\minitoc				

\begin{chapabstract}
This chapter provides an introduction to low-order modeling for the prediction of thermoacoustic instabilities. It starts with a brief reminder of the acoustic theory relevant to combustion systems, including a derivation of the Helmholtz equation. The principles used to reduce the order of a thermoacoustic problem are then presented. An emphasis is placed on the state-space formalism, a framework widely used to build acoustic networks of complex systems; existing LOMs found in the literature are then listed. This chapter concludes with a few observations on the Galerkin expansion method, a classical approach used to build thermoacoustic LOMs, that was selected in this work for its potential to deal with complex geometries.
\end{chapabstract}

\section{The Helmholtz equation} \label{sec:intro_lom_helmholtz}

As previously stated, a fundamental process in thermoacoustic instabilities is the propagation of acoustic waves in the combustor. Although acoustics are fully contained in the classical Navier-Stokes equations, they can also be described by a much simpler set of equations. This simplification is the first necessary step in the derivation of thermoacoustic LOMs. It is achieved thanks to a number of hypotheses, starting with:
\begin{itemize}
\item{\textbf{H1}: Molar weight and heat capacity of all species are equal.}
\item{\textbf{H2}: Viscous diffusion of heat and momentum are neglected.}
\item{\textbf{H3}: The fluid behaves as an ideal gas.}
\end{itemize}
\textbf{H1} eases the derivation without significantly affecting its outcome; it can also be relaxed by defining an equivalent fictive species with physical properties representative of that of the mixture. \textbf{H2} essentially discards any form of acoustic dissipation due to hydrodynamic interactions (\textit{e.g.} conversion into vorticty waves). Acoustic losses can however be reintegrated in later steps through lumped models. Note that \textbf{H3} is not applicable in rocket engine configurations where real-gas effects may be of primary importance. A common approximation then consists in assuming that all real-gas effects on acoustics can be embedded into the mean fields of density, sound speed, and adiabatic factor~\cite{Urbano:2016,Urbano:2017}, without modifying the governing equations. This point is however disputable, and is still the subject of current research\cite{Migliorino:2020}. Under these conditions, the flow variables are governed by the Euler equations:
\begin{align}
\label{eq:euler_def}
& \pdv{\rho}{t} + \vec{u} . \vec{\nabla} \rho = - \rho \nabla. \vec{u} \\
& \rho \pdv{\vec{u}}{t} + \vec{u} . \vec{\nabla} \vec{u} = - \vec{\nabla} p \\
& \pdv{s}{t} + \vec{u} . \vec{\nabla} s = \dfrac{r \omega_T}{p}
\end{align}
with the perfect-gas law $p = \rho r T$, and the entropy $s$ defined from its reference value $s_{ref}$:
\begin{align}
\label{eq:entropy_definition}
s = s_{ref} + \int_{T_{ref}}^{T} \dfrac{C_p (T')}{T'} \ dT' - r \ln \left( \dfrac{p}{p_{ref}} \right)
\end{align}
Flow variables are then decomposed as $f (\vec{x},t) = f_0 (\vec{x}) + f' (\vec{x},t) $, where $f_0$ is the mean field, and $f'$ is the coherent fluctuating part. It is worth noticing that acoustic-turbulence interactions are not considered here. In this case, it would be necessary to introduce a triple decomposition $f = f_0 + f' + f''$, with $f''$ the non-coherent turbulent oscillations. Other hypotheses are necessary to continue the derivation:
\begin{itemize}
\item{\textbf{H4}: For any flow variable $f$, the fluctuating part $f'$ is assumed very small in comparison to the mean value $f_0$: $f'/f_0 = \varepsilon \ll 1$}
\item{\textbf{H5}: The Mach number $M$ is assumed small.}
\end{itemize}
Once again, \textbf{H4} is not valid during LRE instabilities where $p'$ can be as large as 40\% of $p_0$. These strong pressure fluctuations can result in nonlinear phenomena that were discussed by Culick \textit{et al.}~\cite{Culick:1976,Culick:1994} and Yang~\cite{Yang:1987}. \textbf{H4} is however verified at the onset of an instability, that is before the pressure fluctuations reach a large-amplitude limit-cycle. \textbf{H5} is an important assumption that usually does not strongly affect the pure acoustic modes of a	chamber, but that is unsuitable to account for mixed entropic-acoustic modes~\cite{Nicoud:2009,Nicoud:2007,Motheau:2013b,Motheau:2014,Chen:2016}. The following sections in this thesis are based on this zero-Mach number assumption, and would require important adaptation to be able to capture the mixed-instabilities that may exist. The decomposition $f = f_0 + f'$ is then introduced in Eq~\eqref{eq:euler_def}, \textbf{H4} is used to linearize the equations and only retain the terms of order $\varepsilon$, and the convection terms $ \vec{u}_0 . \vec{\nabla (.)}$ are neglected thanks to \textbf{H5}. Pressure, velocity, and entropy fluctuations  are then governed by:
\begin{align}
\label{eq:euler_linearized}
& \pdv{\rho'}{t} + \rho_0 \nabla . \vec{u}' + \vec{u}' . \vec{\nabla} \rho_0 = 0 \\
& \rho_0 \pdv{\vec{u}'}{t} + \vec{\nabla} p' =0 \\
& \pdv{s'}{t} + \vec{u}' . \vec{\nabla} s_0 = \dfrac{r \omega_T'}{p_0} \\
& \dfrac{p'}{p_0} - \dfrac{\rho'}{\rho_0} - \dfrac{T'}{T_0} = 0 \ , \ s' = C_p \dfrac{T'}{T_0} - r \dfrac{p'}{p_0}
\end{align}
Combining these relations leads to the following equation that governs the propagation of linear acoustic waves in the fluid:
\begin{align}
\label{eq:dalembert_wave}
\nabla . \left( \dfrac{1}{\rho_0} \vec{\nabla} p' \right) - \dfrac{1}{\gamma p_0} \pdv[2]{p'}{t} = - \dfrac{\gamma-1}{\gamma p_0} \pdv{\omega_T'}{t}
\end{align}
Note that no assumption were made regarding the spatial dependence of the mean fields $\rho_0 (\vec{x})$, $c_0 (\vec{x})$ and $\gamma (\vec{x})$. When the fluid is a perfect gas at relatively low $p_0$, $\gamma$ can be considered uniform such that the spatial derivative becomes $\nabla . (c_0^2 \vec{\nabla} p')$, but in high pressure rocket engine configurations $\gamma (\vec{x})$ significantly varies between the cryogenic reactants and the hot combustion products, which requires to retain its spatial distribution. Such considerable variations of $\gamma (\vec{x})$ can also be due to fuel droplets sprays, usually encountered in aeronautical propulsion systems. Acoustic losses distributed over the volume can be included Eq.~\eqref{eq:dalembert_wave}, for instance by adding a term $\alpha \partial p' / \partial t$ in the left-hand side, $\alpha$ being a loss coefficient.\par

When dealing with linear acoustics it is convenient to introduce $\hat{p} (\vec{x}, \omega)$, the frequency-domain counterpart of $p' (\vec{x},t)$. These quantities are related through the Fourier transform and its inverse:
\begin{align}
\label{eq:Fourier_transform_definition}
\hat{p}(\vec{x},\omega) = \int_{- \infty}^{+ \infty} e^{- j \omega t} p(\vec{x},t)  dt \ \ , \ \ p(\vec{x},t) = \int_{- \infty}^{+ \infty} e^{j \omega t} \hat{p}(\vec{x},\omega) d \omega
\end{align}
Applying the Fourier transform to Eq.~\eqref{eq:dalembert_wave} yields the frequency-domain Helmholtz equation:
\begin{align}
\label{eq:helmholtz_definition} 
\nabla . \left( \dfrac{1}{\rho_0} \vec{\nabla} \hat{p} \right) + \dfrac{\omega^2}{\gamma p_0} \hat{p} = - j \omega \dfrac{\gamma-1}{\gamma p_0} \hat{\omega}_T
\end{align}
It is also worth presenting a simplified version of the Helmholtz equation, valid for uniform mean fields $c_0$, $\rho_0$, and $\gamma$:
\begin{align}
\label{eq:helmholtz_uniform_definition}
c_0^2 \nabla^2 \hat{p} \fd + \omega^2 \hat{p} \fd = - j \omega (\gamma-1) \hat{\omega}_T
\end{align}
In Eq.~\eqref{eq:helmholtz_definition} and~\eqref{eq:helmholtz_uniform_definition}, the flame dynamics embedded into the right-hand side source term are clearly dissociated from the acoustic propagation. In the following chapters, for the sake of simplicity most derivations will be based on Eq.~\eqref{eq:helmholtz_uniform_definition}, but can be easily adapted to the nonuniform case of Eq.~\eqref{eq:helmholtz_definition}. Appropriate adjustments will be specified when necessary.\par

The Helmholtz equation is completed with boundary conditions that can be sorted in three categories:
\begin{itemize}
\item{The rigid-wall boundary $\hat{\vec{u}} . \vec{n}_s = 0$, where $\vec{n}_s$ is the surface normal vector pointing outwards, is a homogeneous Neumann boundary condition for $\hat{p}$ (it can also be written $\vec{\nabla} \hat{p} . \vec{n}_s = 0$).} 
\item{The atmosphere opening, or pressure release, $\hat{p}=0$ is a homogeneous Dirichlet condition.}
\item{The complex-valued impedance boundary condition $\hat{p} + Z (j \omega) \rho_0 c_0 \hat{\vec{u}} . \vec{n}_s = 0$ is more generic and covers a wide variety of cases, such as an inlet linked to a compressor, an outlet connected to a turbine, or a multi-perforated liner. The frequency-dependent impedance $Z (j \omega)$ can for instance model acoustic losses due to hydrodynamic interactions occuring at an aperature through wall~\cite{howe1979,DongYang:2017}.}
\end{itemize}


\section{Low-order modeling strategies} \label{sec:intro_lom_strategies}

Low-order modeling resides in two basic ideas: (1) the number of Degrees of Freedom (DoF) should be reduced as much as possible to permit fast computations, and (2) the model should be flexible and highly modular, in the sense that it should allow for the straightforward modification of most geometrical or physical parameters. Fast and modular LOMs have been promisingly applied to intensive tasks demanding a large number of repeated resolutions, such as Monte Carlo Uncertainty Quantification~\cite{bauerheim2016,avdonin2018}, or passive control through adjoint geometrical optimization~\cite{aguilar2018}. Note also that usual thermoacoustic LOMs are physics-based rather than data-based, and therefore rely on a set of simplifying physical assumptions. The two key aspects in the resolution of thermoacoustic instabilities are the ability of the method to account for complex geometries that are encountered in industrial combustors, and the accurate representation of the flame dynamics. Equation~\eqref{eq:helmholtz_definition} evidences a clear separation between acoustics and flame dynamics, which therefore suggests to apply low-order modeling principles to these two difficulties separately thanks to a \textit{divide and conquer} strategy.

\paragraph{Low-order flame dynamics} \mbox{}\\[-4mm]

Full-order flame dynamics modeling of multiple burners located in a combustor requires the costly resolution of the reactive Navier-Stokes equations on meshes comprising $O(10^{7})$ to $O(10^{9})$ DoF. Reducing the order of the flame dynamics problem is therefore primordial to build an efficient thermoacoustic LOM. The simplest approach to do so consists in assuming that the flames behave as Linear Time-Invariant (LTI) systems, in which case their response can be modeled through a transfer function called a Flame Transfer Function (FTF). This strategy relies on Crocco's seminal ideas~\cite{Crocco:1952,Crocco:1956} postulating that the heat-release fluctuations at a time $t$ is proportional to the velocity fluctuation in the injector at a time $t- \tau$, where the proportionality factor $n$ is called the gain and $\tau$ is the time-delay. The heat-release source term in Eq.~\eqref{eq:helmholtz_definition} then writes:
\begin{align}
\label{eq:FTF_first_definition}
\hat{\omega}_T (\vec{x}, \omega) = n(\vec{x},\omega) e^{- j \omega \tau (\vec{x},\omega)} \ \hat{\vec{u}}_{ref} (\vec{x}_{ref}, \omega) . \vec{e}_{ref}
\end{align}
where $\vec{x}_{ref}$ is the location of the reference point and $\vec{e}_{ref}$ a unitary vector indicating a reference direction. A large number of FTF models are available in the literature: they can be obtained through theoretical derivation for laminar premixed flames~\cite{Cuquel:2013,Fleifil:1996,Schuller:2003}, laminar diffusion flames~\cite{Magina2013,Magina2016}, or turbulent swirled premixed flames~\cite{Palies:2011c,Candel:2014}. They can also be measured experimentally~\cite{Merk:2019,Mejia:2018,Gatti:2019}, or computed thanks to full-order numerical simulations such as LES or DNS~\cite{Hermeth:2014,Merk:2019}. Once an FTF is determined for a single flame in a simplified configuration, it can be implemented into Eq.~\eqref{eq:helmholtz_definition} to resolve thermoacoustic modes in any multi-burner system (if flame-flame interactions are neglected). An FTF is however only applicable to small-amplitude fluctuations appearing at the onset of an instability, and cannot account for the rich nonlinear flame dynamics, such as the saturation phenomena responsible for the establishment of a limit-cycle. Simple nonlinear flame response can be modeled thanks to an extension of the FTF, called a Flame Describing Function (FDF)~\cite{Noiray:2008}. Analytical FDF are rare due to the mathematical difficluty that they involve~\cite{Preetham:2008}, but several methods exist to formulate an FDF from available experimental data~\cite{Ghirardo:2015b}, or to extend the FDF formalism to more complex nonlinear behaviors~\cite{Orchini:2016,Haeringer:2019}.\par

Although some generalizations are possible, FTF and FDF remain limited to relatively simple situations, where the fluctuations are purely harmonic signals, and are therefore not adapted to deal with fast transients, or chaotic regimes. A more advanced class of flame dynamics LOMs is based on a simplification of the combustion governing equations, for instance through the resolution of a level-set equation, also called G-equation. This approach was extensively used to model two-dimensional laminar premixed flames~\cite{kashinath2014,Orchini:2015b,orchini2015,Ren:2018}, and was later extended to partially-premixed flames~\cite{Semlitsch:2017}. An analogous approach for two-dimensional diffusion flames consists in solving a single mixture fraction equation instead of the full Navier-Stoke equations~\cite{Balasubramanian:2008,Illingworth2013,Magri:2014}. For instance, Orchini \textit{et al.}~\cite{Orchini:2015b} reduced the DoF to $O(10^{4})$ for a single premixed flame modeled through a G-equation, and Balasubramanian \textit{et al.}~\cite{Balasubramanian:2008} used $O(10^{3})$ DoF for a single diffusion flame modeled thanks to a mixture fraction transport equation. In a different fashion, the recent work of Avdonin et al.~\cite{Avdonin:2019} implemented a linearized reactive flow solver that resolves a large part of the flame and flow dynamics while drastically reducing the complexity of the combustion governing equation. Note also that these physics-based low-order models can be combined with data assimilation methods to enhance their fidelity~\cite{Yu:2019}. These different classes of flow solvers resolving simplified versions of the flame dynamics governing equations, can be directly embedded into an acoustic solver to compute the source term of Eq.~\eqref{eq:dalembert_wave} in the time-domain or Eq.~\eqref{eq:helmholtz_definition} in the frequency-domain.\par

Although low-order flame dynamics modeling is a promising direction to build efficient thermoacoustic LOMs, this work does not focus on this point. It is rather interested in dealing with the second major difficulty in the prediction of thermoacoustic instabilities, that is the modeling of acoustics in geometrically complex systems. Thus, all the LOMs derived in this work only use the FTF or FDF formalism to account for the flame dynamics. It is however possible to couple the acoustic LOMs introduced in this work to the more advanced flame dynamics models presented above.

\paragraph{Geometrical simplification} \mbox{}\\[-4mm]

The resolution of the acoustics thanks to Eq.~\eqref{eq:dalembert_wave} or Eq.~\eqref{eq:helmholtz_definition} strongly depends on the combustor complexity. However, not every geometrical detail of the system has first order effects on the acoustic eigenmodes, and some features may be simplified. For instance, removing a swirler and replacing it with a straight tube of equivalent length and section area is a simple method to reduce the number of DoF, since the vanes of a swirled-injector usually require a large number of mesh cells to be correctly discretized. This relatively low-sensitivity of the acoustics with respect to small geometrical details can also be exploited to avoid directly discretizing apertures on a multi-perforated liner. Those can instead be represented thanks to a homogenized impedance boundary condition~\cite{howe1979,gullaud2012}.\par

Geometrical simplification can also consist in reducing the spatial dimension of some combustor components. For instance, injectors are often long and narrow ducts where only planar acoustic waves propagate. They can therefore be considered as one-dimensional, which implies that a single spatial dimension needs to be discretized, instead of 3 in the actual system. This principle can also be applied to thin annular domains that can be considered as one-dimensional (if only azimuthal modes are targeted), or two-dimensional (if longitudinal modes are also targeted). This dimensionality reduction underlines that the ability of a LOM order model to combine heterogeneous elements of different nature and dimensions (0D, 1D, 2D, 3D) in a same system is desirable. This principle is the basis of the acoustic network concept presented in Sec.~\ref{sec:intro_lom_network_state_space}.

\paragraph{Numerical discretization methods} \mbox{}\\[-4mm]

Utilizing an appropriate numerical discretization method to solve Eq.~\eqref{eq:dalembert_wave} or Eq.~\eqref{eq:helmholtz_definition} may result in a significant lower number of DoF. A brute force Finite Element of Finite Difference method discretization is likely to produce many unnecessary DoF. For instance, in~\cite{Semlitsch:2017} a tracking method is designed to resolve the G-equation-based flame dynamics LOM, by selectively adding discretization points near the flame front. A comparable approach is used in~\cite{sayadi2014}, where a specific high-order numerical scheme is used in the vicinity of a flat flame to capture the large acoustic velocity gradient it induces. One of the most remarkable discretization method used in thermoacoustics LOMs is the decomposition of the acoustic pressure and velocity onto a set of known acoustic eigenmodes, called Galerkin modes. These modes are solutions of the \textit{homogeneous} Helmholtz equation, and their spatial structures are therefore close to that of the problem under consideration, such that a small number of them is usually sufficient for an accurate resolution. This type of \textit{spectral discretization} thus often results in fewer DoF than spatial discretization approaches.

\section{Acoustic network and state-space formalism} \label{sec:intro_lom_network_state_space}

\begin{figure}
\begin{center}
\includegraphics[width=0.90\textwidth]{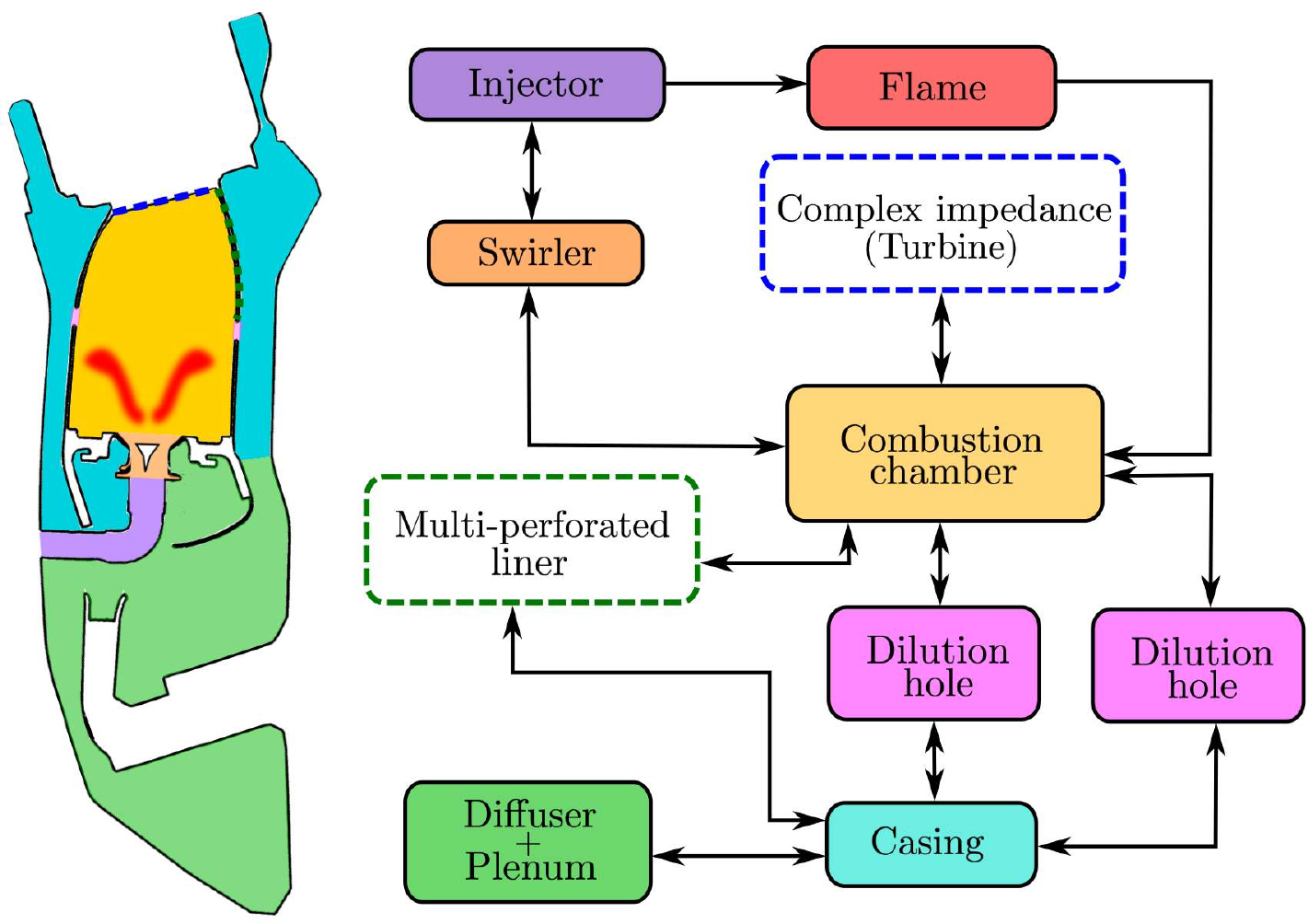}%
\caption{Schematic of a gas turbine combustor split into an acoustic network. Arrows indicate coupling between connected subsystems. Each subsystem in the network can be modeled separately.}
\label{fig:acoustic_network_intro}
\end{center}
\end{figure}

The \textit{divide and conquer} principle can be applied to a complex combustor in order to split it into a collection of smaller subsystems, where the resolution of the thermoacoustic equations is easier. This collection of subsystems is called an acoustic network, and is exemplified in Fig~\ref{fig:acoustic_network_intro}.
Acoustic networks not only simplify the resolution of the thermoacoustic problem, but also yield two key characteristsics of LOMs: (1) they are modular, since any element in the network can be modified or added without affecting the others, and (2) they are heterogeneous, in the sense that they can contain elements of different nature and dimensions. The second point is primordial to take advantage of the dimensionality reduction previously mentioned.\par

In order to resolve the acoustic flow in the whole combustor, individual network elements need to be connected together, or in other words, coupling relations between subsystems must be enforced. An elegant formulation to connect subsystems is to use a \textit{state-space} approach. This method, already popular in the thermoacoustic community~\cite{bothien2007,schuermans2003,schuermans2003_th,emmert2016}, is adopted in this work with some adaptations. Implementation details are given below; further developments relative to state-space representations can be found in control theory textbooks~\cite{friedland2012}. For any physical system described by a set of coordinates $\mathbf{X}(t)$ in a phase-space, we call linear state-space realization of this system a set of equations under the form:
\begin{align}
\label{eq:definition_statespace}
\left\{
\begin{aligned}
& \dot{\mathbf{X}} (t) = \mathbf{A} \ \mathbf{X} (t) + \mathbf{B} \ \mathbf{U}(t) \\[5pt]
& \mathbf{Y} (t) = \mathbf{C} \ \mathbf{X}(t) + \mathbf{D}\ \mathbf{U}(t)
\end{aligned} \right.
\end{align}
where $\mathbf{X}(t)$ is the coordinates vector in the phase-space, also called state vector, $\mathbf{A}$ is the dynamics matrix, $\mathbf{B}$ is the input matrix, $\mathbf{U}(t)$ the input vector, $ \mathbf{Y}(t)$ the output vector, $\mathbf{C}$ the output matrix, and $\mathbf{D}$ is the action, or feedthrough matrix. The first equation of the state-space representation governs the dynamical evolution of the state vector under the forcing exercised by the input vector. The second equation defines a way to compute any desired outputs from the knowledge of the state vector and the forcing term. Note that the output $ \mathbf{Y}(t)$ depends on the state $\mathbf{X}(t)$, but the reverse is not true: $\mathbf{X}(t)$ evolves independently of the selected output $ \mathbf{Y}(t)$.\par

\begin{figure}[h!]
\centering
\includegraphics[width=0.95\textwidth]{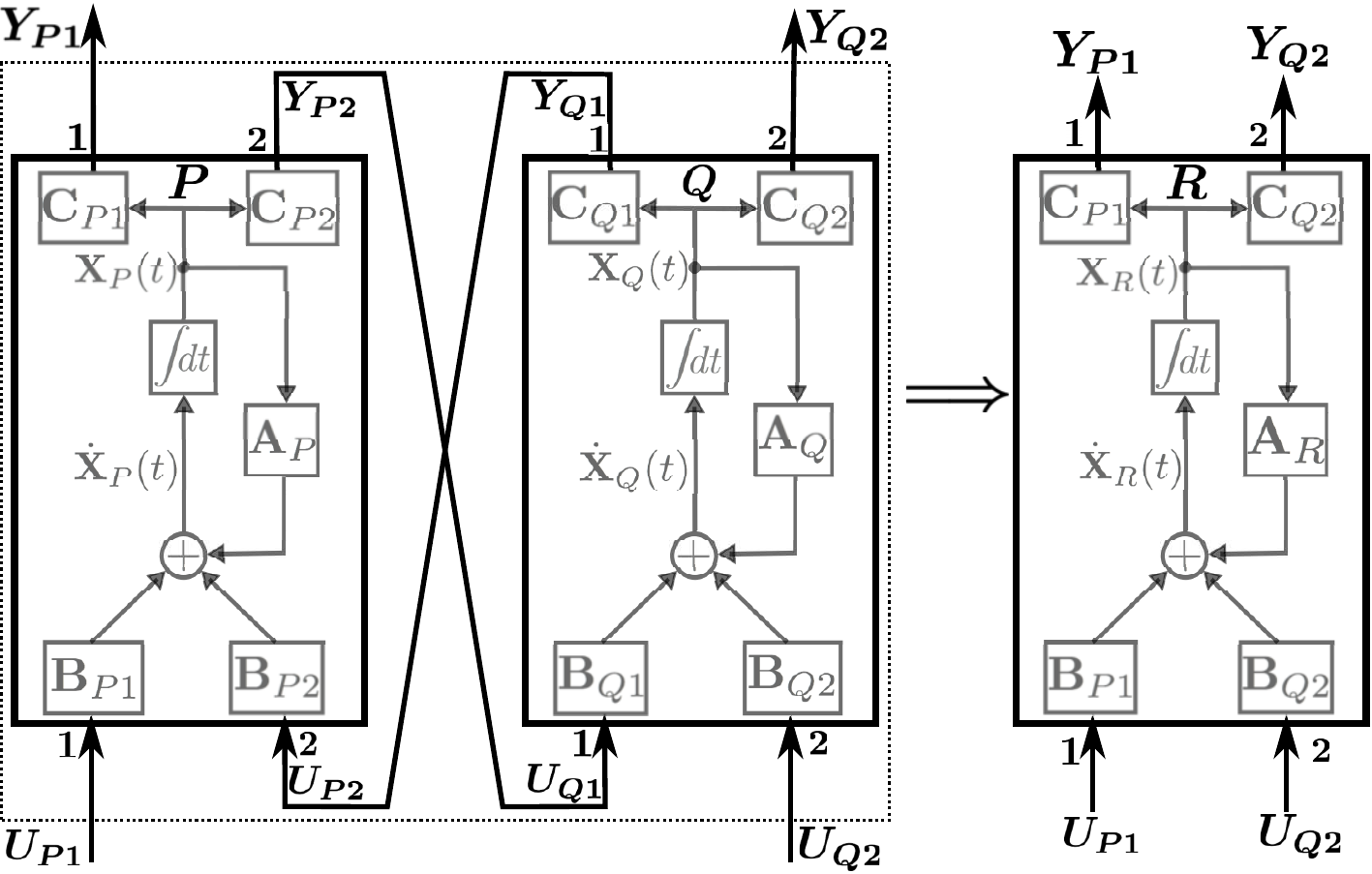}
\caption{Connection of two MIMO systems $\bm{P}$ and $\bm{Q}$. Subscripts $\bm{P}$ (resp.~$\bm{Q}$) designate state-space representation matrices and vectors for the subsystem P (resp.~$\bm{Q}$). Inputs and outputs ports 2 in $\bm{P}$ are those connected to the subsystem $\bm{Q}$, while inputs and outputs ports 1 in $\bm{P}$ are connected to other systems in the network. Conversely, inputs and outputs ports 1 in $\bm{Q}$ are those connected to the subsystem $\bm{P}$, while inputs and outputs ports 2 in $\bm{Q}$ are connected to other systems in the network}
\label{fig:redheffer}
\end{figure}

The state-space formalism, through the \textit{Redheffer star-product}~\cite{redheffer1960}, provides a direct way to connect two systems represented by their state-space realizations, by relating their respective inputs and outputs. Let $P$ and $Q$ be two Multiple-Input-Multiple-Output (MIMO) systems with their respective state-space representations as shown in Fig.~\ref{fig:redheffer}. For clarity, and because it is the case for most of the state-space realizations considered presently, all the feedthrough matrices $\mathbf{D}$ are zero.
Let us also introduce the state-space representation of $P$ with the block-structure:
\begin{align}
\label{eq:redheffer_P_1}
\dot{\mathbf{X}}_{P}(t) = \mathbf{A}_{P} \mathbf{X}_{P}(t) +
 \underbrace{\begin{pmatrix}
\mathbf{B}_{P1} &  \mathbf{B}_{P2}
\end{pmatrix}}_{\mathbf{B}_{P}}
\underbrace{\begin{pmatrix}
\mathbf{U}_{P1}(t) \\
\mathbf{U}_{P2}(t)
\end{pmatrix}}_{\mathbf{U}_{P}(t)}
\end{align}
\begin{align}
\label{eq:redheffer_P_2}
\mathbf{Y}_{P}(t) =
\begin{pmatrix}
\mathbf{Y}_{P1}(t) \\
\mathbf{Y}_{P2}(t)
\end{pmatrix} =
\underbrace{ \begin{pmatrix}
\mathbf{C}_{P1} \\
\mathbf{C}_{P2}
\end{pmatrix}}_{\mathbf{C}_{P}}
 \mathbf{X}_{P}(t) 
\end{align}
where indices $1$ and $2$ in the input and output matrices refer to the respective inputs and outputs ports of $P$. Analogous notations hold for the state-space $Q$. Then, forming the Redheffer star-product of $P$ and $Q$ as indicated in Fig.~\ref{fig:redheffer} yields the block-structure for $R$:
\begin{align}
\label{eq:redheffer_R_1}
\underbrace{
\begin{pmatrix}
\dot{\mathbf{X}}_{P}(t) \\
\dot{\mathbf{X}}_{Q}(t)
\end{pmatrix}}_{\dot{\mathbf{X}}_{R}(t)} =
\underbrace{
\begin{pmatrix}
\mathbf{A}_{P}  & \mathbf{B}_{P2}  \mathbf{C}_{Q1} \\
\mathbf{B}_{Q1} \mathbf{C}_{P2} & \mathbf{A}_{Q}
\end{pmatrix}}_{\mathbf{A}_{R}}
 \begin{pmatrix}
\mathbf{X}_{P}(t) \\
\mathbf{X}_{Q}(t)
\end{pmatrix} + \underbrace{\begin{pmatrix}
\mathbf{B}_{P1}  \\
\mathbf{B}_{Q2} 
\end{pmatrix}}_{\mathbf{B}_{R}}
\underbrace{
\begin{pmatrix}
\mathbf{U}_{P1}(t) \\
\mathbf{U}_{Q2}(t)
\end{pmatrix}}_{\mathbf{U}_{R}(t)}
\end{align}
\begin{align}
\label{eq:redheffer_R_2}
\underbrace{
\begin{pmatrix}
\mathbf{Y}_{P1}(t) \\
\mathbf{Y}_{Q2}(t)
\end{pmatrix}}_{\mathbf{Y}_{R}(t)} =
\underbrace{\begin{pmatrix}
\mathbf{C}_{P1} \\
\mathbf{C}_{Q2} 
\end{pmatrix}}_{\mathbf{C}_{R}}
 \begin{pmatrix}
\mathbf{X}_{P}(t) \\
\mathbf{X}_{Q}(t)
\end{pmatrix}
\end{align}
In Eq.~\eqref{eq:redheffer_R_1}, extra-diagonal blocks are coupling terms between subsystems $P$ and $Q$: more precisely, $\mathbf{B}_{P2}  \mathbf{C}_{Q1}$ represents the effect from $Q$ onto the dynamics of $P$, and conversely $\mathbf{B}_{Q1} \mathbf{C}_{P2}$ represents the influence of $P$ onto the dynamics of $Q$. Equivalent results can also be obtained through an analogous procedure named state-space interconnect~\cite{emmert2016}. After iteratively applying the Redheffer star-product to connect together state-space representations of every subsystems, the full state-space of the whole combustor is obtained as:
\begin{equation}
\label{eq:statespace_full}
\dot{\mathbf{X}}^f (t) = \mathbf{A}^f \ \mathbf{X}^f (t) + \mathbf{B}^f \ \mathbf{U}^f(t)
\end{equation}
where the input vector $\mathbf{U}^f$ and matrix $\mathbf{B}^f$ may represent an external forcing or a self-excited instability. Two approaches are then possible:
\begin{itemize}
\item{\textbf{Time-domain resolution:} Eq.~\eqref{eq:statespace_full} can be integrated over time to obtain the temporal evolution of the acoustic flow under the external forcing $\mathbf{U}^f(t)$}
\item{\textbf{Frequency-domain resolution:} the complex eigenvalues and eigenvectors of the dynamics matrix $\mathbf{A}^f$ can be solved for, yielding the global acoustic eigenfrequencies and eigenmodes of the whole domain. If $\lambda_n = 2 \pi \sigma_{n} + j 2 \pi f_{n}$ is the n\textsuperscript{th} complex eigenvalue of the matrix $\mathbf{A}^f$, then $f_n$ is the eigen-frequency of the n\textsuperscript{th} acoustic mode of the whole geometry. In the absence of acoustic losses, volume sources or complex boundary impedances, $\sigma_n$ is zero. Conversely, if the system comprises acoustic sources, then $2 \pi \sigma_n,$ is the growth-rate of the n\textsuperscript{th} acoustic mode of the whole geometry: $\sigma_n>0$ (resp.~$\sigma_n<0$) implies that the mode is unstable (resp.~stable). The mode shape can also be reconstructed from the modal components contained in the eigenvector $\mathbf{v}_n$ associated to the eigenvalue $\lambda_n$.}
\end{itemize}
\par

The challenge now consists in deriving a state-space representation as that of Eq.~\eqref{eq:definition_statespace} for every type of subsystems in the combustor acoustic network. This can be achieved in a number of different fashions and will be the object of the next chapters. The state-space realizations derived in this work are gathered in Appendix~\ref{Appendix:A}.

\section{Distinct classes of thermoacoustic LOMs} \label{sec:intro_lom_classes}

The modeling principles described in Sec.~\ref{sec:intro_lom_strategies} can be applied in various extents to derive thermo-acoustic LOMs. Low-order models existing in the literature can be classified into five main categories, some of which are based on the state-space formalism and the acoustic network decomposition presented above.


\subsection{Finite Element Helmholtz solvers}

One of the key aspects for the resolution of thermoacoustic eigenmodes is the ability of the method to accurately account for complex geometries that are encountered in industrial combustors. The most straightforward approach is the direct discretization of the Helmholtz equation that is then solved thanks to a Finite Element Method (FEM) solver. State-of-the-art FEM Helmholtz solvers are able to solve for thermaoustic eigenmodes in elaborate geometries comprising active flames and complex impedance boundary conditions~\cite{nicoud2007,camporeale2011,mensah2016,krebs2001}, such as multi-perforated liners~\cite{gullaud2012,andreini2011,giusti2013} or other dissipative elements~\cite{ni2018}. The FDF formalism can also be incorporated to capture nonlinear limit-cycle behaviors~\cite{silva2013,laera2017}.\par

However, frequency-domain FEM Helmholtz solvers remain a costly alternative to full-order methods. They usually resolve a nonlinear eigenvalue problem of the form:
\begin{align}
\label{eq:nonlinear_eigenvalue_problem}
\mathcalbf{A} \mathbf{V} = \mathcalbf{B} (j \omega) \mathbf{V}
\end{align} 
where $\mathcalbf{A}$ is a constant-valued matrix resulting from the FEM discretization, $\mathcalbf{B} (j \omega)$ is a matrix containing the $\omega$-nonlinearity arising for instance from the FTF model, and $\mathbf{V}$ are the eigenmodes being sought for. Since Eq.~\eqref{eq:nonlinear_eigenvalue_problem} is nonlinear, it is usually resolved thanks to costly iterative fixed-point algorithms~\cite{nicoud2007}, that do not offer any guarantee of convergence or of capturing all the eigenmodes. Some improved algorithms were proposed in recent works~\cite{Buschmann:2020,Mensah:2019thesis}, but this point is still considered a serious limitation for FEM solvers. In addition, FEM Helmholtz solvers often result in a large number of DoF (approximately $O(10^{5})$ to $O(10^6)$), synonymous of a considerable computational cost, and only permit little modularity, as any change in the geometrical parameters requires a new geometry and mesh generation. Because of these reasons, FEM Helmholtz solvers are sometimes not referred as actual LOMs, and are rather seen as intermediate-order methods. It is however also worth mentioning the slightly different approach undertaken by Hummel \textit{et al.}~\cite{hummel2016}, who designed a two-steps strategy where FEM simulations are combined with a data reduction technique, to formulate a model where the number of DoF are reduced to $O(10^{2})$.

\subsection{Riemann invariants based LOMs}

Another class of LOM is a wave-based 1D network approach, where the acoustic pressure and velocity are written in function of the Riemann invariants $A^{+}$ and $A^{-}$. This method was first successfully used to predict the stability of purely annular gas turbine combustors in the 90s~\cite{keller1995} and early 2000s~\cite{polifke2001AIAA}. Among others, the LOTAN tool~\cite{Dowling2003_lotan} was for instance designed to resolve in the frequency domain linearly unstable thermoacoustic modes in simplified configurations. More recently and in a similar fashion, the open-source LOM solver Oscilos~\cite{li2014} developed at Imperial College, London, was used to perform for example time-domain simulations of thermoacoustic limit cycles in longitudinal combustors~\cite{li2015,han2015}, whose operation conditions may approach those of industrial systems~\cite{xia2019}. Wave-based low order modeling was also generalized to more complex cases, including azimuthal modes in configurations comprising an annular combustion chamber linked to an annular plenum through multiple burners. This procedure allowed Bauerheim \textit{et al.} to conduct a series of studies based on a  network decomposition of an idealized annular combustor~\cite{bauerheim2014,bauerheim2015,bauerheim2016} to capture its azimuthal modes. This approach, where 1D elements are combined into a multi-dimensional acoustic network, is sometimes referred as 1.5D. It is also worth underlining that the wave-based resolution of the linearized Euler equations permits to naturally include mean flow effects, such as the convection of entropy and vorticity waves, in addition to the acoustics.\par

Wave-based decomposition of the acoustic variables in the frequency-domain leads to a scalar dispersion relation $F(j \omega) = 0$ in the complex plane. For simple systems, its resolution is fast and can be achieved thanks to: (1) iterative optimization algorithms requiring an initial guess, such as the gradient descent, or (2) algorithms based on Cauchy's argument principle~\cite{brebion2017} that do not require an initial guess and guarantee to identify all solutions within a given window. For more complex systems, the dispersion relation becomes strongly nonlinear, and the second class of algorithms is more difficult to utilize: iterative local optimization must then be used, and initial guesses must be provided. Even though wave-based LOMs are the most adequate to deal with networks of longitudinal elements where acoustic waves can be assumed as planar, they also suffer strict limitations: they are indeed unable to capture non-planar modes in complex geometries, and are therefore limited to idealized annular combustors at best. 

\subsection{Galerkin expansion based LOMs}

A large category of LOMs relies on Galerkin modal expansions to express the acoustic pressure field as a combination of known acoustic modes. Modal expansion was first introduced and formalized in an acoustic context by Morse and Ingard in their influential book \textit{Theoretical Acoustics}~\cite{morse1968} dated from 1968. In the field of thermoacoustics, Zinn\textit{ et al.}~\cite{zinn1971} and Culick~\cite{culick1976,culick1988,culick1995,culick2006} were among the first to use it to study combustion instabilities in LREs. Similarly to the wave-based approach, multiple studies utilizing modal expansions are dealing with the Rijke tube~\cite{rijke1859}: for example by Balasubramanian \textit{et al.}~\cite{Balasubramanian:2008}, Juniper~\cite{juniper2011}, and Waugh \textit{et al.}~\cite{waugh2011}. Idealized annular configurations were also examined thanks to pressure modal expansion: Noiray \textit{et al.}~\cite{Noiray2011,Noiray2013a,Noiray2013b} and Ghirardo \textit{et al.}~\cite{ghirardo2016} conducted a series of theroretical studies in such geometries. More complex modal expansion-based networks were developed for multi-burners chamber-plenum geometries, by Stow \textit{et al.}~\cite{stow2009}, Schuermans \textit{et al.}~\cite{schuermans2003,schuermans2003_th}, and Belluci \textit{et al.}~\cite{bellucci2004}. Their strategy is to perform modal expansions for the pressure in the chamber/plenum and to assume acoustically compact burners that can be lumped and represented by simple transfer matrices. Unlike wave-based low-order modeling, this method is not limited to planar acoustic waves, and can resolve both azimuthal and longitudinal chamber modes. Even though their approach does not rely on an acoustic network decomposition, Bethke \textit{et al.}~\cite{bethke2005} showed that arbitrarily complex geometries can be incorporated in a thermoacoustic LOM by expanding the pressure onto a set of basis functions computed in a preliminary step thanks to a FEM Helmholtz solver.\par

Although they appear more general than wave-based LOMs, modal expansion-based LOMs are also subjected to strict limitations, which mainly resides in the choice of the modal basis employed to expand the acoustic pressure. This point is one of the main object of this work, and is discussed in more details later.

\subsection{Hybrid waves-modal expansions LOMs}

The fourth class of thermoacoustic LOM is a mixed method, that takes advantage of a heterogeneous network decomposition to combine both Riemann invariants and modal expansions. The former are used to account for longitudinal propagation, while the latter are used to account for multi-dimensional geometries. This mixed method was employed by Evesque \textit{et al.}~\cite{evesque2002} to model azimuthal modes in multi-burners annular chamber/plenum configurations. In more recent works~\cite{orchini2018,dong2018} the mixed strategy was further developed to investigate nonlinear spinning/standing limit cycles in the MICCA annular combustor. More precisely, in~\cite{orchini2018} the planar acoustic field in the ducted burners is represented by Riemann invariants, while it is represented through modal expansions in the chamber and in the plenum. The LOTAN tool~\cite{stow2001,stow2009} also has this ability to combine the advantages of Riemann invariants and Galerkin expansion to resolve longitudinal and mixed modes in idealized annular combustors. Once again, a key benefit of this hybrid approach lies in the possibility to make use of the Riemann invariants to naturally combine acoustic, entropy, and vorticity waves in a same formulation.

\subsection{Direct discretization LOMs}

Finally, the last class of LOMs consists in those based on a direct spatial discretization. This method is relatively close to the FEM Helmholtz solvers, except that it can combine in a same system 1D, 2D, and 3D spatial discretization. For example, Sayadi \textit{et al.}~\cite{sayadi2014} made use of several finite difference schemes to build a dynamical system representation of a one-dimensional thermoacoustic system comprising a volumetric heat source localized within the domain. A more generic direct discretization LOM is the taX low-order model developed at Technical University of Munich~\cite{emmert2016}. Its modularity lies in the ability to combine in a same thermoacoustic network one-dimensional elements discretized by finite difference, geometrically complex elements discretized through FEM, and other types of elements such as FTFs and scattering matrices. Note that thanks to the state-space formalism, the taX LOM can express $\omega$-nonlinearities as higher-order state-space realizations that are \textit{linear with} $\omega$. Unlike the FEM Helmholtz solvers, the resulting eigenvalue problem is therefore linear and can be resolved with classical eigenvalue algorithms. Since this LOM is built upon direct discretization of linearized partial differential equations, it can also potentially incorporate richer physics, including mean flow effects or acoustic-vortex interactions.\par

However, the price to pay for this high modularity is a large number of DoF: for instance, in the taX LOM, about $O(10^5)$ DoF were needed to obtain acoustic eigenmodes of an annular combustor comprising 12 ducted injectors (but no active flame).

\section{Conclusion}  \label{sec:intro_lom_conclusion}  

A rapid examination of the different methods presented above suggests that Galerkin expansions produce LOMs with relatively low numbers of DoF, that also have the ability to account for complex geometries, through the preliminary construction of the expansion basis thanks to a FEM Helmholtz solver. Consequently, since this work particularly targets the geometrical complexity inherent to realistic combustion systems, the Galerkin method is selected to continue our developments.\par

Interestingly, most Galerkin expansion LOMs use the same type of modal basis, namely the basis composed of the \textit{rigid-wall cavity modes}, or in other words acoustic eigenmodes satisfying homogeneous Neumann boundary conditions (\textit{i.e.} zero normal velocity) over the entire boundaries of the domain (and without internal volume sources). This work focuses on the nature of the acoustic eigenmodes basis used to decompose the pressure, and the convergence properties resulting from this expansion. Although the rigid-wall modal basis presents the huge advantage of being orthogonal, many actual systems obviously comprise frontiers with far more complex boundary conditions than just a homogeneous Neumann condition (for example an inlet or an outlet where the impedance has a finite value). The use of such basis then appears paradoxical: \textit{how is it possible that a solution expressed as a rigid-wall modes series converges towards a solution satisfying a non rigid-wall boundary condition?} This singularity in the pressure modal expansion was already noticed by Morse and Ingard~\cite{morse1968}, but they did not study its impact on the convergence of the whole method. Later, Culick~\cite{culick2006} provided more explanations about this singularity: the modal expansion does not converge uniformly over the domain, but only in the less restrictive sense of the Hilbert norm (L\textsubscript{2} norm); as a result, even though each individual term of the expansion does not satisfy the appropriate boundary condition, the infinite sum of these terms may satisfy it. In other words, "\textit{the limit of the sum is not equal to the sum of the limit}" in the neighborhood of the boundary. An interesting examination regarding this singularity is provided in a recent work by Ghirardo \textit{et al.}~\cite{ghirardo2018}, where a projection onto a mode satisfying a specific non rigid-wall inlet boundary condition was discussed. The convergence issue arising from rigid-wall modal expansion is even more problematic in the case of acoustic LOMs where the geometry is decomposed into a network of subsystems that need to be coupled together at their boundaries (see Fig.~\ref{fig:acoustic_network_intro}). For each individual subdomain, these coupling boundaries are not rigid-wall but each term of the basis corresponds to a rigid-wall: the convergence singularity may arise at each one of the coupling interfaces. Although Culick~\cite{culick2006} proposed an explanation to this singularity based on a \textit{local} justification, only very few studies deal with \textit{global} effects, such as for example convergence speed of the eigenfrequencies.\par
Thus, the main object of the next chapters consists in: (1) identifying the mathematical limitations of the classical Galerkin modal expansion, and (2) proposing a proper formulation for the modeling of non-rigid-wall boundaries, a primordial point in the construction of elaborate acoustic networks representing complex geometries.

				
\chapter{A novel modal expansion method for the low-order modeling of thermoacoustic instabilities} \label{chap:FRAME}
\minitoc				

\begin{chapabstract}
This chapter introduces the core contribution of this work to low-order modeling of thermoacoustic instabilities. It starts by recalling the classical Galerkin expansion, where acoustic variables are expanded onto on orthogonal basis of known acoustic eigenmodes verifying rigid-wall boundary conditions over the frontiers of the domain under consideration. It then presents a novel type of modal expansion, called a \textit{frame} expansion: acoustic variables are now expanded onto an \textit{over-complete} family of eigenmodes constructed by gathering the rigid-wall and the pressure-release orthogonal bases. Both types of modal expansions are then implemented in a state-space-based LOM. Their respective convergence properties are assessed on a series of one-dimensional test cases: it is in particular evidenced that the rigid-wall Galerkin expansion results in a Gibbs phenomenon affecting the acoustic velocity representation near non-rigid-wall boundaries, which deteriorates its precision. The frame modal expansion successfully mitigates these Gibbs oscillations and leads to significantly greater accuracy and convergence speed. The LOM modularity and its ability to handle complex geometries are then illustrated by considering a configuration featuring an annular chamber, an annular plenum, as well as multiple burners. However, the use of an over-complete frame comes at a price, namely the apparition of non-physical \textit{spurious} eigenmodes. A number of strategies that are implemented to identify and attenuate these spurious components are discussed. Finally, this chapter concludes with a brief presentation of the LOM code STORM (STate-space thermOacoustic Reduced-order Model) that is built upon the frame modal expansion.
\end{chapabstract}

\section{The classical Galerkin modal expansion} \label{sec:galerkin_expansion}

\begin{figure}[h!]
\centering
\includegraphics[width=0.80\textwidth]{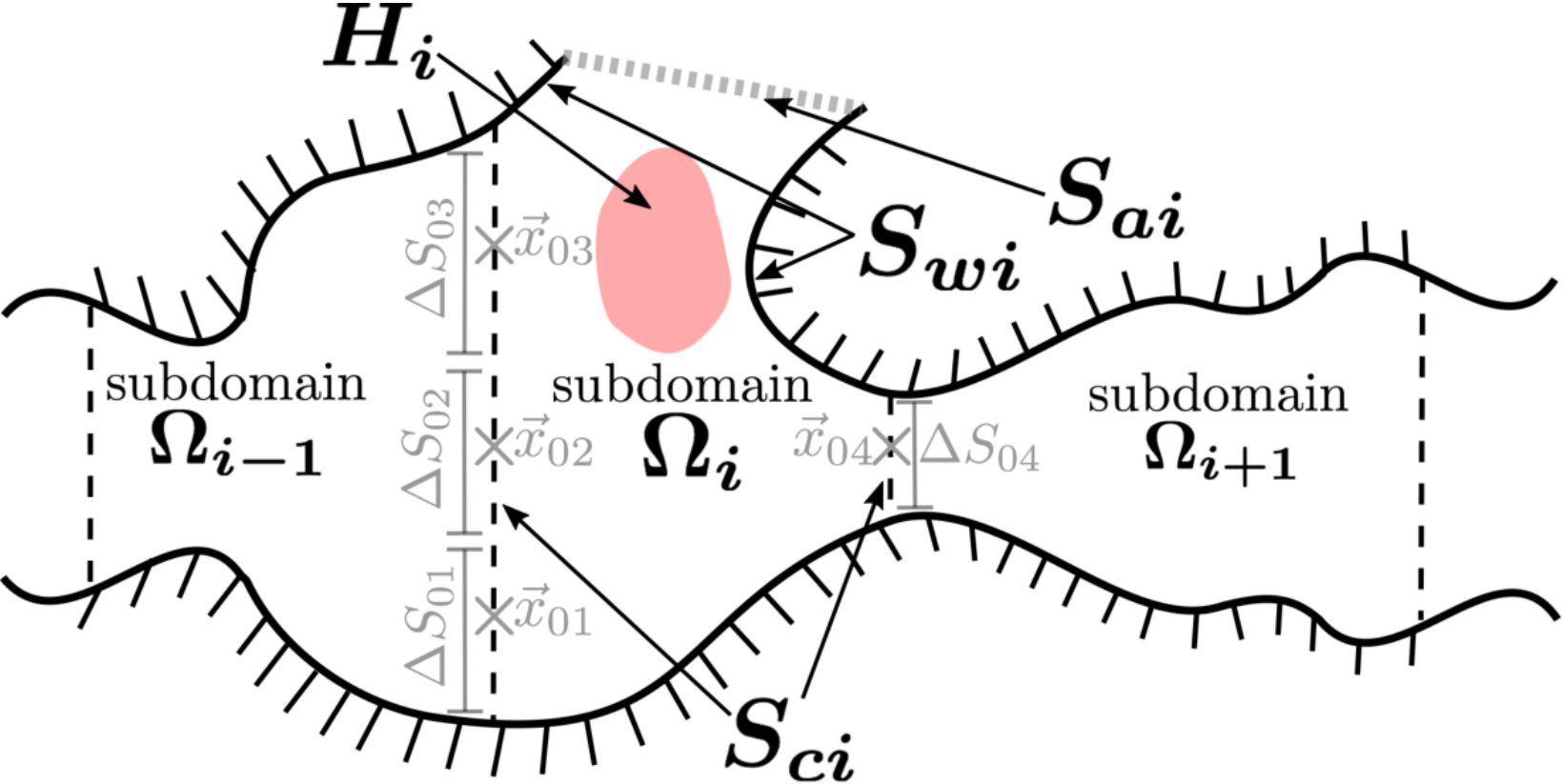}
\caption{Schematic representation of an acoustic network. Hashed lines show rigid walls. Gray dotted lines are opening to the atmosphere. The geometry is split into distinct subsystems, that are connected together. Those can include volume subdomains (\textit{e.g.} $\bm{\Omega_i}$), active flames (\textit{e.g.} $\bm{H_i}$), and complex boundaries (\textit{e.g.} $\bm{S_{ci}}$). The subdomain $\bm{\Omega_i}$ is delimited by its rigid wall boundary $\bm{S_{wi}}$, its boundary opened to the atmosphere $\bm{S_{ai}}$, and its connection boundaries $\bm{S_{ci}}$ that are to be linked to other subsystems in the acoustic network. The interface $\bm{S_{ci}}$ is split into several surface elements $\bm{S_{0j}}$ of area $\bm{\Delta S_{0j}}$ located at the connection points $\bm{\vec{x}_{0j}}$. The subdomain  $\bm{\Omega_i}$ also contains $\bm{M_H}$ volume heat sources $\bm{H_i^{(l)}}$ (here $\bm{M_H = 1}$).}
\label{fig:illustration_rom}
\end{figure}

As previously mentioned, acoustic LOM networks for complex configurations usually start with the splitting of the system into a set of smaller subsystems. Let us consider a subdomain $\Omega_i$ as in Fig.~\ref{fig:illustration_rom}, defined as a bounded domain delimited by $\partial \Omega_i = S_{wi} \cup S_{ai} \cup S_{ci}$. 
For the sake of simplicity, in the following the sound speed field is assumed uniform and the baseline flow to be at rest. Note however that these hypotheses are not necessary and could be omitted. The frequency-domain acoustic pressure $\hat{p}(\vec{x},\omega)$ in the subdomain $\Omega_i$ is then solution of the following Helmholtz equation:
\begin{align}
\label{eq:press_ss0}
\left\{ \begin{aligned}
&c_0^2 \nabla^2 \hat{p} \fd - j \alpha \omega \hat{p} \fd + \omega^2 \hat{p} \fd = \hat{h} \fd \ \  \mathrm{ for } \ \ \vec{x} \in \Omega_i \\
&\nabla_s \hat{p} = 0 \ \ \mathrm{ for } \ \ \vec{x}_s \in S_{wi} \ \ , \ \ \hat{p} = 0 \ \ \mathrm{ for } \ \ \vec{x}_s \in S_{ai} \\
& \nabla_s \hat{p} =  \hat{f} (\vec{x}_s,\omega) \ \ \mathrm{ for } \ \ \vec{x}_s \in S_{ci} \\
\end{aligned}\right.
\end{align}
where the notation $\nabla_s \hat{p}  = \vec{\nabla} \hat{p} (\vec{x}_s).\vec{n}_s$ is introduced, $\vec{x}_s$ being a geometrical point belonging to a boundary of the flow domain and $\vec{n}_s$ the normal unity vector pointing outward. In Eq.~\eqref{eq:press_ss0},  $\alpha$ is an acoustic loss coefficient, the term $\hat{h}(\vec{x},\omega)$ is a volume acoustic source, while $\hat{f} (\vec{x},\omega)$ is a surface forcing term imposed on the connection boundary $S_{ci}$: it is an external input to the subdomain $\Omega_i$ exerted by adjacent subsystems. The pressure verifies rigid-wall boundary condition on $S_{wi}$ and is null on $S_{ai}$. The volumic source $\hat{h}(\vec{x},\omega)$ represents fluctuations of heat release, and it may be written as follows:
\begin{align}
\label{eq:hr_definition}
\hat{h} (\vec{x},\omega) = - j \omega \left( \gamma - 1 \right) \hat{\omega}_T(\vec{x}, \omega) =  - j \omega \left( \gamma - 1 \right) \sum_{l=1}^{M_H} \mathcal{H}_i^{(l)} (\vec{x}) \hat{Q}_l(\omega)
\end{align} 
where $\gamma$ is the heat capacity ratio and $\hat{\omega}_T(\vec{x}, \omega)$ is the local fluctuating heat release rate resulting from flame dynamics. In Eq.\eqref{eq:hr_definition}, $\hat{\omega}_T(\vec{x}, \omega)$ is decomposed into the contributions from $M_H$ independent heat sources $H_i^{(l)}$ contained in $\Omega_i$, characterized by their respective global heat release rates $\hat{Q}_l(\omega)$ and spatial volume densities $\mathcal{H}_i^{(l)} (\vec{x})$ representing the flame shape (the integral of $\mathcal{H}_i^{(l)} (\vec{x})$ over $\Omega_i$ is unity). In the following, only one flame in the subdomain $\Omega_i$ is considered for conciseness (\textit{i.e.} $M_H = 1$, and the superscripts ${}^{(l)}$ are dropped), but the reasoning can be extended without difficulty to any number of distinct and independent flames located in a same subdomain $\Omega_i$.\par

Galerkin expansion usually makes use of the inner product defined for any functions $f (\vec{x})$ and $g(\vec{x})$ as:
\begin{align}
\label{eq:inner_prod}
\langle f , g \rangle = \iiint_{\Omega_i} f(\vec{x}) g(\vec{x}) \ d^3 \vec{x}
\end{align}
The associated $L^2$ norm is noted $||f||_2 =  \langle f , f \rangle^{1/2}$. The second Green's identity, which is used multiple times throughout this derivation is also recalled:
\begin{align}
\label{eq:second_green_id}
\begin{aligned}
& \iiint_{\Omega} \left( f (\vec{x}) \nabla^2 g (\vec{x})  -  g (\vec{x}) \nabla^2 f (\vec{x}) \right) d^3 \vec{x} = \\
& \oiint_{\partial \Omega} \left(   f (\vec{x}_s) \nabla_s g (\vec{x}_s) - g (\vec{x}_s) \nabla_s f (\vec{x}_s) \right)   d^2 \vec{x}_s
\end{aligned}
\end{align}

Modal expansion based LOMs follow the seminal ideas of Morse~\cite{morse1946,morse1968}. First, on each subdomain $\Omega_i$ the pressure is decomposed onto a family $\left( \psi_n (\vec{x}) \right)_{n \geqslant 1}$ formed of known acoustic eigenmodes of $\Omega_i$. This expansion writes $\hat{p}(\vec{x},\omega) = \sum_{n} \hat{\gamma}_n (\omega) \psi_n (\vec{x})$. The purpose is then to derive a set of governing equations for the modal amplitudes. The set $\left( \psi_n (\vec{x}) \right)_{n \geqslant 1}$ is classically chosen as the rigid-wall eigenmodes of the subsystem $\Omega_i$ (without volume sources and acoustic damping). These eigenmodes satisfy rigid-wall conditions (\textit{i.e.} zero normal velocity) over $S_{wi}$, but also over the connection boundary $S_{ci}$. In the presence of boundaries that are known to be opened to the atmosphere ($S_{ai}$ in Fig.~\ref{fig:illustration_rom}), the eigenmodes basis can be chosen to satisfy the appropriate condition on $S_{ai}$ (\textit{i.e.} zero pressure), without further difficulty, since the expansion basis is still orthogonal. The set  $\left( \psi_n (\vec{x}) \right)_{n \geqslant 1}$ is solution of the following eigenvalue problem: 
\begin{align}
\label{eq:modes_ss0}
\left\{ \begin{aligned}
&c_0^2 \nabla^2 \psi_n + \omega_n^2 \psi_n  = 0 \ \  \mathrm{ for } \ \ \vec{x} \in \Omega_i \\
&\nabla_s \psi_n = 0 \ \ \mathrm{ for } \ \ \vec{x} \in S_{wi} \ \ , \ \ \psi_n = 0 \ \ \mathrm{ for } \ \ \vec{x} \in S_{ai} \\
&\nabla_s \psi_n = 0 \ \ \mathrm{ for } \ \ \vec{x} \in S_{ci}
\end{aligned}\right.
\end{align}
where $\omega_n$ is the eigen-pulsation of the n\textsuperscript{th} eigenmode.  By making use of the second Green's identity (Eq.~\eqref{eq:second_green_id}), it can be shown that the set $\left( \psi_n (\vec{x}) \right)_{n \geqslant 1}$ defined by Eq.~\eqref{eq:modes_ss0} is indeed an orthogonal basis, that is $\langle \psi_n,\psi_m \rangle=0$ for any $n \neq m$.\par

A solution to the pressure Helmholtz equation (Eq.~\eqref{eq:press_ss0}) is sought by making use of the Green's function $\hat{G} (\vec{x},\omega | \vec{x}_0)$, where $\vec{x}_0$ designates the location of a source point in $\Omega_i$. The Green's function has the advantage of recasting the problem into a set of boundary integral equations, which facilitates the decomposition of complex geometries into networks of simpler subsystems, as two adjacent subdomains are coupled together through their respective boundary source terms. The equation governing the Green's function is:
\begin{align}
\label{eq:F_green_ss0}
\left\{ \begin{aligned}
&c_0^2 \nabla^2 \hat{G} (\vec{x},\omega|\vec{x}_0) - j  \omega \alpha  \hat{G} (\vec{x},\omega|\vec{x}_0) + \omega^2 \hat{G} (\vec{x},\omega|\vec{x}_0) = \delta(\vec{x}-\vec{x}_0) \  \mathrm{ for } \ \vec{x} \in \Omega_i \\
&\nabla_s \hat{G} = 0 \ \ \mathrm{ for } \ \ \vec{x} \in S_{wi} \ \ , \ \ \hat{G} = 0 \ \ \mathrm{ for } \ \ \vec{x} \in S_{ai} \\
&\nabla_s \hat{G} = 0 \ \ \mathrm{ for } \ \ \vec{x} \in S_{ci}
\end{aligned}\right.
\end{align}
where $\delta(\vec{x}-\vec{x}_0)$ is the Dirac delta function. The Green's function is  chosen such that it satisfies the same homogeneous boundary conditions as $\hat{p}$ over the rigid-wall frontier $S_{wi}$ and the boundary opened to the atmosphere $S_{ai}$. Note however that unlike the acoustic pressure, the Green's function verifies a \textit{homogeneous} Neumann boundary condition ($u_s=0$) on the connection surface $S_{ci}$. The decomposition of the Green's function is sought under the form:
\begin{align}
\label{eq:F_decomp_G_ss0}
\hat{G}(\vec{x},\omega | \vec{x}_0) = \sum_{n = 0}^{\infty} \hat{\eta}_n(\omega | \vec{x}_0) \psi_n (\vec{x})
\end{align}
Let us inject the modal decomposition of Eq.~\eqref{eq:F_decomp_G_ss0} into the Green's function equation Eq.~\eqref{eq:F_green_ss0} and use the fact that $\psi_n$ is an eigenmode. Then, forming the inner product with $\psi_m$, and using both the orthogonality of the modes and the properties of the Dirac function yields: 
\begin{align}
\label{eq:G_decomp_fin_ss0}
\hat{G}(\vec{x},\omega | \vec{x}_0) = \sum_{n=0}^{\infty} \dfrac{  \psi_n (\vec{x}_0) \psi_n (\vec{x})}{\Lambda_{n}(\omega^2 - j \omega \alpha - \omega_n^2)}
\end{align}
where $\Lambda_{n} = ||\psi_n||_2^2 $. Thus, the Green's function is now known through its modal expansion onto the rigid-wall eigenmodes basis. Equation~\eqref{eq:press_ss0} and Eq.~\eqref{eq:F_green_ss0} are then evaluated in $\vec{x}_0$, multiplied by $\hat{G}(\vec{x}_0, \omega | \vec{x})$ and $\hat{p} (\vec{x}_0,\omega)$ respectively, and integrated over the volume $\Omega_i$ with respect to $\vec{x}_0$. Finally, using the Dirac properties, the second Green's identity (Eq.~\eqref{eq:second_green_id}), and the reciprocity property of the Green's function $\hat{G}(\vec{x_0},\omega | \vec{x}) = \hat{G}(\vec{x},\omega | \vec{x}_0)$, the following general expression relating the acoustic pressure to its Green's function is obtained:
\begin{align}
\label{eq:Green_to_press}
\begin{aligned}
&\hat{p}(\vec{x},\omega) =  \iiint_{\Omega_i} \hat{G}(\vec{x}, \omega | \vec{x}_0) \hat{h} (\vec{x}_0,\omega) d^3 \vec{x}_0  \ + \\
 & \oiint_{\partial \Omega_i} c_0^2 \left( \hat{p}(\vec{x}_{0s},\omega) \nabla_{0,s} \hat{G}(\vec{x}, \omega | \vec{x}_{0s}) \ - \ \hat{G}(\vec{x}, \omega | \vec{x}_{0s}) \nabla_{0,s}   \hat{p}(\vec{x}_{0s},\omega ) \right) \ d^2 \vec{x}_{0s}
\end{aligned}
\end{align}
where the notation $\nabla_{0,s} () = \vec{\nabla}_{0}().\vec{n}_s $ is used. Note that here the fluxes $\nabla_{0,s}$ in the surface integral are taken with respect to the source location $\vec{x}_{0s}$. Using the boundary conditions verified by the pressure (Eq.~\eqref{eq:press_ss0}) and its Green's function (Eq.~\eqref{eq:F_green_ss0}), this equation simplifies to:
\begin{align}
\label{eq:press_integrated_ss0}
\hat{p}(\vec{x},\omega) = \iiint_{\Omega_i} \hat{G}(\vec{x}, \omega | \vec{x}_0) \hat{h} (\vec{x}_0,\omega) d^3 \vec{x}_0 - \iint_{S_{ci}} c_0^2  \hat{G}(\vec{x}, \omega | \vec{x}_{0s})  \hat{f}(\vec{x}_{0s},\omega) \ d^2 \vec{x}_{0s}
\end{align}
In this expression, the vector $\vec{x}_{0}$ (resp.~$\vec{x}_{0s}$) refers to the location of volume sources (resp.~boundary forcing).\par

The next step of the derivation requires to evaluate the surface integral in Eq.~\eqref{eq:press_integrated_ss0}. The most straightforward way is to decompose the surface $S_{ci}$ into $M_S$ elements $S_{0j}$ of surface area $\Delta S_{0j}$, that connects $\Omega_i$ with the adjacent subdomains $\Omega_j$ at the boundary points $\vec{x}_{0j}$. Note that the subdomains $\Omega_j$ are not necessarily distinct, since there may exist several connection points between $\Omega_i$ and a same neighbor (\textit{i.e.} we can have $\vec{x}_{0j1} \neq \vec{x}_{0j2}$, but $\Omega_{j1} = \Omega_{j2}$). It is assumed that these surface elements are small enough such that acoustic variables can be considered uniform over each one of them. An example of such surface discretization is given in Fig.~\ref{fig:illustration_rom}. The surface integral in Eq.~\eqref{eq:press_integrated_ss0} can then be expressed through piece-wise approximations, which yields the pressure field given by:
\begin{align}
\label{eq:press_sum_Sj_ss0}
\begin{aligned}
\hat{p}(\vec{x},\omega) =  & - \sum_{j=1}^{M_S} c_0^2 \Delta S_{0j}\hat{G}(\vec{x}, \omega | \vec{x}_{0j}) \hat{f}(\vec{x}_{0j},\omega) \\
& - j \omega \left( \gamma - 1 \right) \hat{Q}(\omega)  \iiint_{\Omega_i} \hat{G}(\vec{x}, \omega | \vec{x}_0) \mathcal{H}_i (\vec{x}_0) d^3 \vec{x}_0
\end{aligned}
\end{align}
Note that the number and the locations of the surface elements $S_{0j}$ necessary to achieve an accurate approximation of the surface integral is highly disputable. This particular point is out of the scope of this chapter, and is the subject of a more general and robust methodology presented in Chap.~\ref{chap:complex_boundaries}. In Eq.~\eqref{eq:press_sum_Sj_ss0}, the surface forcing $\hat{f}(\vec{x}_{0j},\omega)$ is related to the normal acoustic velocity $\hat{u}_s^{\Omega_j} (\vec{x}_{0j},\omega)$ in the adjacent subsystem $\Omega_j$ as $\hat{f}(\vec{x}_{0j}) = - \rho_0 j \omega \hat{u}_s^{\Omega_j} (\vec{x}_{0j})$. The negative sign comes from the fact the acoustic velocity $\hat{u}_s^{\Omega_j} (\vec{x}_{0j})$ in $\Omega_j$ is computed with respect to the outer normal of $\Omega_i$, which is pointing in the opposite direction of the normal of $\Omega_j$. This yields:
\begin{align}
\label{eq:press_sum_Sj_ss0_2}
\begin{aligned}
\hat{p}(\vec{x},\omega) =  & \sum_{j=1}^{M_S} \rho_0 c_0^2 \Delta S_{0j}\hat{G}(\vec{x}, \omega | \vec{x}_{0j}) j \omega \hat{u}_s^{\Omega_j}(\vec{x}_{0j}) \\
& - j \omega \left( \gamma - 1 \right) \hat{Q}(\omega)  \iiint_{\Omega_i} \hat{G}(\vec{x}, \omega | \vec{x}_0) \mathcal{H}_i (\vec{x}_0) d^3 \vec{x}_0
\end{aligned}
\end{align}
Using the Green's function modal decomposition of Eq.~\eqref{eq:G_decomp_fin_ss0} into Eq.~\eqref{eq:press_sum_Sj_ss0_2} gives:
\begin{align}
\label{eq:press_sum_Sj_ss0_bis}
\begin{aligned}
\hat{p}(\vec{x},\omega) = & \sum_{j=1}^{M_S} \sum_{n=0}^{\infty} \dfrac{\rho_0 c_0^2 \Delta S_{0j}  j \omega \hat{u}_s^{\Omega_j} (\vec{x}_{0j}) \psi_n (\vec{x}_{0j}) \psi_n (\vec{x})}{\Lambda_{n}(\omega^2 - j \omega \alpha - \omega_n^2)} \\
& - \sum_{n=0}^{\infty}  \dfrac{ j \omega \left( \gamma - 1 \right) \mathcal{H}_{i,n} \hat{Q}(\omega)  \psi_n (\vec{x})}{\Lambda_{n}(\omega^2 - j \omega \alpha - \omega_n^2)}
\end{aligned}
\end{align}
where $\mathcal{H}_{i,n} = \langle \mathcal{H}_i, \psi_n \rangle$ is the projection of the flame shape $\mathcal{H}_i (\vec{x}_0)$ onto $\psi_n (\vec{x}_0)$. Equating Eq.~\eqref{eq:press_sum_Sj_ss0_bis} to the decomposition of the pressure field on the basis $(\psi_n)_{n \geqslant 0}$, $\hat{p}(\vec{x},\omega) = \sum_{n} \hat{\gamma}_n (\omega) \psi_n (\vec{x})$, it comes that the $\hat{\gamma}_n$ coefficients are such that:
\begin{align}
\label{eq:ss0_calculation_press1_anex}
\begin{aligned}
(\omega^2 - j \omega \alpha - \omega_n^2) \left( \dfrac{\hat{\gamma}_n(\omega)}{j \omega}  \right) = & \sum_{j=1}^{M_S} \dfrac{\rho_0 c_0^2 \Delta S_{0j} \psi_n (\vec{x}_{0j})}{\Lambda_n} \hat{u}_s^{\Omega_j} (\vec{x}_{0j} ) \\
& -   \dfrac{ \left( \gamma - 1 \right) \mathcal{H}_{i,n} }{\Lambda_{n}}  \hat{Q}(\omega)
\end{aligned}
\end{align}
Recasting the equation into the time-domain, and introducing $\hat{\Gamma}_n = \hat{\gamma}_n/(j \omega)$ (\textit{i.e.} $\dot{\Gamma}_n(t) = \gamma_n(t)$) finally leads to the dynamical system governing the evolution of the acoustic pressure in the subdomain $\Omega_i$:
{\small  
\setlength{\abovedisplayskip}{6pt}
\setlength{\belowdisplayskip}{\abovedisplayskip}
\setlength{\abovedisplayshortskip}{0pt}
\setlength{\belowdisplayshortskip}{3pt}
\begin{align}
\label{eq:press_final_equation_original}
\left\{ \begin{aligned}
& p \td = \sum_{n=1}^{\infty} \dot{\Gamma}_n(t) \psi_n (\vec{x}) \\
& \ddot{\Gamma}_n(t) = - \alpha \dot{\Gamma}_n (t) - \omega_n^2 \Gamma_n (t) - \sum_{j=1}^{M_S} \dfrac{\rho_0 c_0^2  \Delta S_{0j} \psi_n (\vec{x}_{0j})}{\Lambda_n} \ u_s^{\Omega_j} (\vec{x}_{0j},t) + \dfrac{ \left( \gamma - 1 \right) \mathcal{H}_{i,n} }{\Lambda_{n}} Q(t)
\end{aligned} \right.
\end{align}}%

This dynamical system governs the temporal evolution of the pressure field in the subdomain $\Omega_i$, under the normal velocity forcing $u_s^{\Omega_j} (\vec{x}_{0j},t)$ imposed by adjacent subsystems $\Omega_j$, and under the volume forcing $Q(t)$ imposed by fluctuating flames contained within $\Omega_i$. This set of equations was used for example in~\cite{schuermans2003,stow2009}, where the infinite series was truncated up to a finite order $N$. It is also worth noting that the acoustic velocity can be calculated from the knowledge of the modal amplitudes $\Gamma_n(t)$ as $\vec{u}(x,t) = -\sum_{n} \Gamma_n (t) \vec{\nabla} \psi_n (\vec{x}) / \rho_0$. The state-space approach~\ref{sec:intro_lom_network_state_space} can then be used to couple together the subsystems defining the whole thermoacoustic system of interest.\par
Finally, since the acoustic pressure is a linear combination of the modal basis vectors $\psi_n$, it necessarily verifies the same boundary conditions, in particular $\nabla_s p = 0$, \textit{viz.}~$\vec{u}.\vec{n}_s = 0$ on $S_{ci}$. Since the acoustic velocity should not be zero over the boundaries of the (arbitrarily chosen) sub-domain $\Omega_i$, this may result in a singularity in the representation of the acoustic velocity field. The impact of this singularity on the convergence properties of the method is discussed in Sec.~\ref{sec:frame_convergence}. The following section proposes a mathematical reformulation of the pressure modal expansion to mitigate this undesirable feature.


\section{The over-complete frame modal expansion} \label{sec:frame_expansion}

The purpose is now to introduce a modal expansion of the acoustic pressure that would allow for satisfying any boundary condition on $S_{ci}$ (and not $\nabla_s p = 0$ only). In this matter, it is necessary to retain in the modal expansion an additional degree of freedom, such that both acoustic pressure and normal acoustic velocity at the connection boundary $S_{ci}$ remain \textit{a priori} undetermined. Let us then introduce two distinct families $(\xi_m)_{m \geqslant 1}$ and $(\zeta_k)_{k \geqslant 1}$ of acoustic eigenmodes of the subdomain $\Omega_i$, characterized by the two following eigenproblems:
\begin{align}
\label{eq:eigenproblem_revisited_1}
\left\{ \begin{aligned}
&c_0^2 \nabla^2 \xi_m + \omega_m^2 \xi_m  = 0 \ \  \mathrm{ for } \ \ \vec{x} \in \Omega_i \\
& \nabla_s \xi_m = 0 \ \ \mathrm{ for } \ \ \vec{x} \in S_{wi} \mathrm{ , } \ \ \xi_m = 0 \ \ \mathrm{ for } \ \ \vec{x} \in S_{ai}  \\
& \nabla_s \xi_m = 0 \ \ \mathrm{ for } \ \ \vec{x} \in S_{ci}  \\
\end{aligned}\right.
\end{align}
\begin{align}
\label{eq:eigenproblem_revisited_2}
\left\{ \begin{aligned}
&c_0^2 \nabla^2 \zeta_k + \omega_k^2 \zeta_k  = 0 \ \  \mathrm{ for } \ \ \vec{x} \in \Omega_i \\
&\nabla_s \zeta_k = 0 \ \ \mathrm{ for } \ \ \vec{x} \in S_{wi} \mathrm{ , } \ \ \zeta_k = 0 \ \ \mathrm{ for } \ \ \vec{x} \in S_{ai}  \\
& \zeta_k = 0 \ \ \mathrm{ for } \ \ \vec{x} \in S_{ci}  \\
\end{aligned}\right.
\end{align}
Both eigenmodes families $(\xi_m)_{m \geqslant 1}$ and $(\zeta_k)_{k \geqslant 1}$ verify the same rigid-wall (resp.~open) boundary conditions on $S_{wi}$  (resp.~$S_{ai}$). The eigenmodes family $(\xi_m)_{m \geqslant 1}$ is similar to the orthogonal basis $(\psi_n)_{n \geqslant 1}$ used in Sec.~\ref{sec:galerkin_expansion}. Conversely, the eigenmodes family $(\zeta_k)_{k \geqslant 1}$ differs since it verifies pressure-release boundary conditions on $S_{ci}$.\par

Consider now the eigenmodes family $(\phi_n)_{n \geqslant 1}$ formed as the concatenation of $(\xi_m)_{m \geqslant 1}$ and $(\zeta_k)_{k \geqslant 1}$: $(\phi_n)_{n \geqslant 1} = (\xi_m)_{m \geqslant 1} \cup (\zeta_k)_{k \geqslant 1} $. The eigenpulsations associated to the eigenmodes $\phi_n$ are noted $(\omega_n)_{n \geqslant 1} = (\omega_m)_{m \geqslant 1} \cup (\omega_k)_{k \geqslant 1} $. As a concatenation of two orthogonal bases, $(\phi_n)_{n \geqslant 1}$ is not a basis but is instead an \textit{over-complete} set of eigenmodes, also called a \textit{frame}~\cite{daubechies1986,adcock2019}. The concept of frame was first introduced in the context of nonharmonic Fourier analysis~\cite{Duffin:1952}, and later used to build the theory of wavelet analysis~\cite{daubechies1986,Daubechies:1992}, which found great sucess in signal and image processing. Frames started to gain popularity in numerical analysis and Partial Differential Equations resolution only very recently. Adcock and Huybrechs~\cite{adcock2019,Adcock:2018b} give a comprehensive review of these applications. Those include for instance the Fourier embedding~\cite{Huybrechs:2010}, or Fourier extension, which approximates functions in complex geometries thanks to a frame built from the classical Fourier basis of a cube encompassing the domain of interest. This embedding approach was applied to CFD, with the development of immersed boundary methods~\cite{Boffi:2015} for instance. It is also worth mentioning the augmented Fourier frame~\cite{Adcock:2011}, where the Fourier basis is completed with a family of polynomial functions, in the purpose of mitigating the Gibbs phenomenon arising at the ends of non-periodic one-dimensional domains. The use of an over-complete frame to perform modal expansions is the most crucial element of the proposed method. Let us precise that the frame $(\phi_n)_{n \geqslant 1}$ could be built from the concatenation of other sets of eigenmodes, as long as those do not \textit{a priori} impose any constraint between pressure and velocity at the boundary $S_{ci}$. However, using the concatenation of the rigid-wall basis $(\xi_m)_{m \geqslant 1}$ and the open atmosphere basis $(\zeta_k)_{k \geqslant 1}$ has two main advantages: (1) they are usually the easiest to obtain analytically or numerically, and (2) the frame $(\phi_n)_{n \geqslant 1}$ formed by their concatenation verifies a generalized Parseval's identity which ensures the well-posedness of modal expansions~\cite{daubechies1986}. Note that for geometrically simple subdomains, the frame can be obtained analytically, whereas for complex subdomains it is generated in a preliminary step thanks to a FEM solver.\par

In the following, a compact vectorial notation is introduced to avoid the use of multiple summation symbols: for any indexed quantity $(f_i)_{i \geqslant 1}$ we denote $\mathbf{f}$ the column-vector such that: ${}^t \mathbf{f} = (f_0 \ \ f_1 \ \ f_2 \ \ ...)$, where ${}^t()$ designates the vector transpose. For a doubly indexed quantity $(f_{ij})_{i,j \geqslant 1}$,  we denote $\mathbf{ f}$ the matrix whose coefficients are the $f_{ij}$. Conversely, $[\mathbf{f}]_n$ is the n\textsuperscript{th} component of the vector $\mathbf{f}$. Similarly to Sec.~\ref{sec:galerkin_expansion}, the pressure modal expansion is sought under the form $\hat{p}(\vec{x},\omega) = \sum_{n} \hat{\gamma}_n (\omega) \phi_n (\vec{x})$, that is, using the vectorial notation $\hat{p}(\vec{x},\omega) = {}^t \hat{\pmb{\gamma}}(\omega)  \pmb{\phi}(\vec{x})$.

As in the previous section, a solution of the pressure Helmholtz equation is obtained thanks to the associated Green's function, which is the solution of the same equation Eq.~\eqref{eq:F_green_ss0}, except that the boundary condition on the connection surface $S_{ci}$ is not specified \textit{a priori} and will only be deduced later from the knowledge of its modal components. This equation is recalled below for convenience:
\begin{align}
\label{eq:F_green_modif}
\left\{ \begin{aligned}
&c_0^2 \nabla^2 \hat{G} (\vec{x},\omega|\vec{x}_0) - j  \omega \alpha  \hat{G} (\vec{x},\omega|\vec{x}_0) + \omega^2 \hat{G} (\vec{x},\omega|\vec{x}_0) = \delta(\vec{x}-\vec{x}_0) \  \mathrm{ for } \ \vec{x} \in \Omega_i \\
&\nabla_s \hat{G} = 0 \ \ \mathrm{ for } \ \ \vec{x} \in S_{wi} \ \ , \ \ \hat{G} = 0 \ \ \mathrm{ for } \ \ \vec{x} \in S_{ai}
\end{aligned}\right.
\end{align}
An expansion of the Green's function onto the modal frame $(\phi_n)_{n \geqslant 0}$ is sought in the form:
\begin{align}
\label{eq:Green_expand_revisisted}
\hat{G}(\vec{x},\omega | \vec{x}_0) = \sum_{n = 0}^{\infty} \hat{\eta}_n(\omega | \vec{x}_0) \phi_n (\vec{x})
\end{align}
Injecting the expansion of Eq.~\eqref{eq:Green_expand_revisisted} into Eq.~\eqref{eq:F_green_ss0} and forming the scalar product with the mode $\phi_m(\vec{x})$ leads to:
\begin{align}
\label{eq:Green_expand_revisisted_3}
\sum_{n=0}^{\infty} \Lambda_{mn} (\omega^2 - j \omega \alpha - \omega_n^2) \hat{\eta}_n (\omega | \vec{x}_0) =  \phi_m(\vec{x}_0)
\end{align}
where $\Lambda_{mn} = \langle \phi_m , \phi_n \rangle$ is the scalar product between modes $m$ and $n$. For an orthogonal modal basis the cross-terms ($m \neq n$) are zero and only the terms $\Lambda_{nn}$ remain. With the compact vectorial notation, Eq.~\eqref{eq:Green_expand_revisisted_3} can be written:
\begin{align}
\label{eq:Green_expand_revisisted_4}
\mathbf{\Lambda} \ \mathbf{D}(\omega) \ \hat{\pmb{\eta}} = \pmb{\phi}(\vec{x}_0)
\end{align}
where $\mathbf{D}(\omega)$ is the diagonal matrix  $diag((\omega^2 - j \omega \alpha - \omega_n^2))_{n \geqslant 0}$, and $\mathbf{\Lambda}$ is the matrix containing the scalar-products $\Lambda_{mn} = \langle \phi_m,\phi_n \rangle$. This matrix is the Gram matrix associated to the frame $(\phi_n)_{n \geqslant 0}$, sometimes also called the acoustic mass matrix. Obtaining a modal expansion for the Green's function then requires to invert Eq.~\eqref{eq:Green_expand_revisisted_4} to express the components vector $\hat{\pmb{\eta}}$. However, as the eigenmodes family $(\phi_n)_{n \geqslant 0}$ is over-complete, its associated Gram matrix $\mathbf{\Lambda}$ is highly ill-conditioned. This inversion therefore requires a specific numerical treatment that  is discussed below and in Sec~\ref{sec:frame_spurious_modes}. Inverting Eq.~\eqref{eq:Green_expand_revisisted_4} then yields:
\begin{align}
\label{eq:Green_expand_revisisted_final}
\hat{G}(\vec{x},\omega | \vec{x}_0) = {}^t \hat{\boldsymbol{\eta}}(\omega | \vec{x}_0) \  \pmb{\phi}(\vec{x}) =  {}^t \pmb{\phi}(\vec{x}_0) \ \mathbf{\Lambda}^{-1}  \mathbf{D}^{-1}(\omega)   \pmb{\phi}(\vec{x})
\end{align}
The modal expansion of Eq.~\eqref{eq:G_decomp_fin_ss0} obtained with the rigid-wall basis appears as a particular case of Eq.~\eqref{eq:Green_expand_revisisted_final} for a diagonal Gram matrix $\mathbf{\Lambda}$. Equation~\eqref{eq:Green_to_press} relating the pressure to the Green's function is general and still valid here. The surface integral on the right-hand side vanishes on $S_{wi}$ and $S_{ai}$, and only a surface integral on $S_{ci}$ remains:
\begin{align}
\label{eq:Green_to_press_revisited}
\begin{aligned}
\hat{p}(\vec{x},\omega) = & \iint_{S_{ci}} c_0^2 \left( \hat{p}(\vec{x}_{0s} ,\omega) \vec{\nabla}_{0,s} \hat{G}(\vec{x}, \omega | \vec{x}_{0s})  -  \hat{G}(\vec{x}, \omega | \vec{x}_{0s}) \vec{\nabla}_{0,s}   \hat{p}(\vec{x}_{0s},\omega ) \right) \ d^2 \vec{x}_{0s} \\
& + \iiint_{\Omega_i} \hat{G}(\vec{x}, \omega | \vec{x}_0) \hat{h} (\vec{x}_0,\omega) d^3 \vec{x}_0 
\end{aligned}
\end{align}

To evaluate the surface integral in Eq.~\eqref{eq:Green_to_press_revisited} through piece-wise approximations, the decomposition of the connection boundary $S_{ci}$ into $M_S$ surface elements $ S_{0j}$ of area $\Delta S_{0j}$ located at $\vec{x}_{0j}$ is once again introduced, and the corresponding adjacent subdomains are noted $\Omega_j$. The inhomogeneous Neumann boundary condition imposed on $S_{ci}$ (Eq.~\eqref{eq:press_ss0}) can be injected into Eq.~\eqref{eq:Green_to_press_revisited} to replace $\nabla_{0,s}   \hat{p}(\vec{x}_{0j} )$ with $ \hat{f}(\vec{x}_{0j})$. Unlike in the previous section, the term $\nabla_{0,s} \hat{G}(\vec{x} , \omega | \vec{x}_{0j}) \hat{p}(\vec{x}_{0j})$ is non-zero and needs to be evaluated. The Green's function gradient can directly be calculated from its modal expansion (Eq.~\eqref{eq:Green_expand_revisisted_final}). The pressure $ \hat{p}(\vec{x}_{0j})$ can be related to $\hat{f}(\vec{x}_{0j})$ through the acoustic impedance $Z(\vec{x}_{0j}) = \hat{p} (\vec{x}_{0j}) / (\rho_0 c_0 \hat{u}_s (\vec{x}_{0j}))$ at the boundary $S_{ci}$ (assuming the mean flow is at rest, \textit{viz.} $\vec{u} = - \vec{\nabla}p/\rho_0$):
\begin{align}
\label{eq:impedance_derivation_interim}
\hat{p}(\vec{x}_{0j}) = -\dfrac{c_0}{j \omega} Z(\vec{x}_{0j}) \nabla_s \hat{p}(\vec{x}_{0j}) = -\dfrac{c_0}{j \omega} Z(\vec{x}_{0j}) \hat{f}(\vec{x}_{0j})
\end{align}
The modal expansion of the Green's function derived in Eq.~\eqref{eq:Green_expand_revisisted_final}, as well as Eq.~\eqref{eq:impedance_derivation_interim} is now injected into Eq.~\eqref{eq:Green_to_press_revisited} to give:
\begin{align}
\label{eq:press_sum_Sj_revisited_0}
\begin{aligned}
\hat{p}(\vec{x},\omega)  =   - \dfrac{c_0}{j \omega} & \sum_{j=1}^{M_S} c_0^2 \Delta S_{0j}\ {}^t \boldsymbol{\nabla_s} \pmb{\phi}(\vec{x}_{0j})  \mathbf{\Lambda}^{-1}   \mathbf{D}^{-1}(\omega)  \pmb{\phi}(\vec{x}) \ Z(\vec{x}_{0j}) \hat{f}(\vec{x}_{0j}) \\
 - & \sum_{j=1}^{M_S} c_0^2 \Delta S_{0j} \  {}^t \pmb{ \phi  }(\vec{x}_{0j})  \mathbf{\Lambda}^{-1}   \mathbf{D}^{-1}(\omega)   \pmb{\phi}(\vec{x}) \  \hat{f}(\vec{x}_{0j}) \\
- & j \omega \left( \gamma - 1 \right) {}^t  \mathcalbf{H}_i  \ \mathbf{\Lambda}^{-1}   \mathbf{D}^{-1}(\omega)   \pmb{\phi}(\vec{x}) \  \hat{Q}(\omega) 
\end{aligned}
\end{align}
The column vector $\mathcalbf{H}_i $ contains all the projections of the flame shape $\mathcal{H}_i (\vec{x}_{0})$ onto the elements of the over-complete frame $(\phi_n (\vec{x}_{0}) )_{n \geqslant 0}$. Similarly to the previous section, the surface forcing $\hat{f}(\vec{x}_{0j})$ can be related to the normal acoustic velocity $\hat{u}_s^{\Omega_j} (\vec{x}_{0j})$ in the adjacent subsystem $\Omega_j$ as $\hat{f}(\vec{x}_{0j}) = - \rho_0 j \omega \hat{u}_s^{\Omega_j} (\vec{x}_{0j})$. At this stage, the impedance $Z(\vec{x}_{0j})$ at the boundary element $\Delta S_{0j}$ is still unknown, but it can be linked to both the normal acoustic velocity $\hat{u}_s^{\Omega_j} (\vec{x}_{0j})$ and the acoustic potential $\hat{\varphi}^{\Omega_j} (\vec{x}_{0j})$ in the adjacent subsystem, thanks to the relation $Z(\vec{x}_{0j}) = -j \omega \hat{\varphi}^{\Omega_j} (\vec{x}_{0j})/(c_0 \hat{u}_s^{\Omega_j} (\vec{x}_{0j}))$. Here the acoustic potential $\hat{\varphi}$ is defined such that it satisfies the relation $\vec{u}' = - \vec{\nabla} \varphi'$. Note that the impedance $Z(\vec{x}_{0j})$ could equivalently be expressed as a function of the pressure in the adjacent subsystem ($Z(\vec{x}_{0j}) = - \hat{p}^{\Omega_j} (\vec{x}_{0j}) /(\rho_0 c_0 \hat{u}_s^{\Omega_j} (\vec{x}_{0j}))$), but the introduction of the acoustic potential will prove convenient in a later step of the derivation, where inverse Fourier transform is applied to revert frequency-domain equations into time-domain equations (see Eq.~\eqref{eq:press_sum_Sj_revisited_3_SP}). Substituting these expressions in Eq.~\eqref{eq:press_sum_Sj_revisited_0} gives:
\begin{align}
\label{eq:press_sum_Sj_revisited_2}
\begin{aligned}
\hat{p}(\vec{x},\omega) =
& - \sum_{j=1}^{M_S} \rho_0 c_0^2 \Delta S_{0j} j \omega \  {}^t \boldsymbol{ \nabla_{s}} \pmb{ \phi  }(\vec{x}_{0j})  \mathbf{\Lambda}^{-1}   \mathbf{D}^{-1}(\omega)   \pmb{\phi}(\vec{x}) \  \hat{\varphi}^{\Omega_j} (\vec{x}_{0j}) \\
& +  \sum_{j=1}^{M_S} \rho_0 c_0^2 \Delta S_{0j} j \omega \ {}^t \pmb{\phi}(\vec{x}_{0j})  \mathbf{\Lambda}^{-1}   \mathbf{D}^{-1}(\omega)  \pmb{\phi}(\vec{x}) \ \hat{u}_s^{\Omega_j}(\vec{x}_{0j}) \\
& -  j \omega \left( \gamma - 1 \right) {}^t  \mathcalbf{H}_i  \ \mathbf{\Lambda}^{-1}   \mathbf{D}^{-1}(\omega)   \pmb{\phi}(\vec{x}) \  \hat{Q}(\omega) 
\end{aligned}
\end{align}
Introducing the decomposition of the pressure field on the modal frame $(\phi_n)_{n \geqslant 0}$ as $\hat{p}(\vec{x}) = \sum_{n} \hat{\gamma}_n (\omega) \phi_n (\vec{x}) = {}^t \hat{\pmb{\gamma}} (\omega) \ \pmb{\phi} (\vec{x})$ shows that the components vector $ \hat{\pmb{\gamma}} $ is such that (since $\mathbf{D}$ is diagonal and $\mathbf{\Lambda}$ and $\mathbf{\Lambda}^{-1}$ are both symmetric):
\begin{align}
\label{eq:press_sum_Sj_revisited_3_SP}
\begin{aligned}
\dfrac{1}{j \omega} \mathbf{D}(\omega) \   \hat{\pmb{\gamma}} (\omega)    =
& - \sum_{j=1}^{M_S} \rho_0 c_0^2 \Delta S_{0j}  \ \mathbf{\Lambda}^{-1}    \boldsymbol{ \nabla_{s}} \pmb{ \phi  }(\vec{x}_{0j})  \  \hat{\varphi}^{\Omega_j} (\vec{x}_{0j}) \\
& +  \sum_{j=1}^{M_S} \rho_0 c_0^2 \Delta S_{0j} \ \mathbf{\Lambda}^{-1} \pmb{\phi}(\vec{x}_{0j})  \ \hat{u}_s^{\Omega_j} (\vec{x}_{0j}) \\
& -   \left( \gamma - 1 \right) \ \mathbf{\Lambda}^{-1}  \mathcalbf{H}_i  \  \hat{Q}(\omega) 
\end{aligned}
\end{align}
Recasting the equation into the time-domain, and introducing $\Gamma_n(t)$ such that $\dot{\Gamma}_n(t) = \gamma_n(t)$, finally leads to the dynamical system governing the evolution of the pressure field in the subdomain $\Omega_i$, under the velocity surface forcing $u_s^{\Omega_j} (\vec{x}_{0j},t)$, the acoustic potential surface forcing $\varphi^{\Omega_j} (\vec{x}_{0j},t)$, and the heat release volume forcing $Q(t)$:
\begin{align}
\label{eq:press_final_equation_revisited}
\left\{ \begin{aligned}
& p \td = \sum_{n=0}^{N} \dot{\Gamma}_n(t) \phi_n (\vec{x}) \\
& \begin{aligned}
& \ddot{\Gamma}_n(t) =  -  \alpha \dot{\Gamma}_n (t) - \omega_n^2 \Gamma_n (t)  -  \sum_{j=1}^{M_S} \rho_0 c_0^2 \ \Delta S_{0j} \left[ \mathbf{ \Lambda }^{-1} \pmb{\phi} (\vec{x}_{0j}) \right]_n  u_s^{\Omega_j} (\vec{x}_{0j},t) \\
& + \sum_{j=1}^{M_S} \rho_0 c_0^2 \Delta S_{0j} \ \left[ \mathbf{ \Lambda }^{-1} \bm{\nabla_s} \pmb{\phi } (\vec{x}_{0j}) \right]_n  \varphi^{\Omega_j} (\vec{x}_{0j},t) \ + \  (\gamma -1) \left[ \mathbf{ \Lambda }^{-1} \mathcalbf{H}_{i}  \right]_n \ Q(t)
\end{aligned}
\end{aligned} \right.
\end{align}	
which is an extension of Eq.~\eqref{eq:press_final_equation_original} when the expansion is performed on the over-complete frame $(\phi_n )_{n \geqslant 1}$ instead of the orthogonal basis $(\psi_n )_{n \geqslant 1}$. This dynamical system governs the temporal evolution of the pressure field in the domain $\Omega_i$, under normal velocity forcing $u_s^{\Omega_j}(\vec{x}_{0s},t)$ and acoustic potential forcing $\varphi^{\Omega_j}(\vec{x}_{0s},t)$ imposed by adjacent subsystems, as well as the volume forcing $Q(t)$ due to the presence of active flames within $\Omega_i$. In the case of the orthogonal modal basis $(\psi_n)_{n \geqslant 1}$ used in the previous section, the terms $\bm{\nabla_s} \pmb{\psi} (\vec{x}_{0j})$ vanish (since the eigenmodes satisfy rigid-wall boundary conditions on $S_{ci}$), and we have the simple relations $ [ \mathbf{\Lambda}^{-1} \pmb{\psi} (\vec{x}_{0j}) ]_n =  \psi_n (\vec{x}_{0j}) / \Lambda_{n}$ and $[\mathbf{\Lambda}^{-1} \mathcalbf{H}_{i} ]_n = \mathcal{H}_{i,n} / \Lambda_{n}$. Equation~\eqref{eq:press_final_equation_revisited} can be used to formulate a state-space realization for a subdomain in an acoustic network. This representation is provided in Appendix~\ref{sec:ss_realization_subdomain}.\par

Because of the use of the over-complete frame $(\phi_n)_{n \geqslant 1}$, the acoustic pressure and velocity are free to evolve independently on the boundary $S_{ci}$, which is the major improvement of the method compared to classical formalism, where the normal acoustic velocity $\vec{u}.\vec{n}_s$ is necessarily zero on the connection boundary $S_{ci}$.\par

Finally, the governing dynamical system of Eq.~\eqref{eq:press_final_equation_revisited} is a projection of the wave equation (Eq.~\eqref{eq:press_ss0}) onto the modal frame $(\phi_n)_{n \geqslant 0}$. However, as this frame is over-complete such projection is ill-conditioned, which constitutes one of the major pitfalls of the proposed method. Concretely, this results in two major consequences:
\begin{itemize}
\item{The first one is a numerical difficulty to compute the inverse of the frame Gram matrix $\mathbf{\Lambda}^{-1}$. In most cases presented in this paper, the condition number $C(\mathbf{\Lambda}) = |\mathbf{\Lambda} | \ | \mathbf{\Lambda}^{-1} |$ (where $|\mathbf{\Lambda} | $ is the Frobenius matrix norm) increases with the size $N$ of the expansion, up to values ranging from 10\textsuperscript{18} to 10\textsuperscript{19}. The inversion of this poorly conditioned matrix can be achieved thanks to the use of an adequate numerical algorithm, based on extra-precision iterative refinement (performed in floating-point quadruple precision). However, a preferable approach consists in computing the Moore-Penrose pseudoinverse through a Singular Value Decomposition~\cite{golub1965} (see Sec.~\ref{sec:spurious_svd_attenuation} for more details). Errors stemming from this inversion are systematically computed \textit{a posteriori} and verified to remain low. Note that since the modal basis size remains in practice limited (typically a few dozens elements), the specific inversion procedure used to calculate $\mathbf{\Lambda}^{-1}$ does not noticeably increase the computational cost in comparison to the classical rigid-wall modal expansion.}
\item{Secondly, even though $\mathbf{\Lambda}$ is accurately inverted, the frame over-completeness may still result in poorly conditioned spurious components in Eq.~\eqref{eq:press_final_equation_revisited}. However, the approach described above produces well-behaved expansions for the pressure, that is expansions where the terms with the highest energy are physically meaningful while low energy terms represent spurious fluctuations. The automatic and robust identification of these spurious modes is therefore an important issue that needs to be addressed. This is the object of Sec.~\ref{sec:frame_spurious_modes}.}
\end{itemize}

%

\section{Convergence assessment in elementary cases} \label{sec:frame_convergence}

Both modal expansions presented in Sec.~\ref{sec:galerkin_expansion} and Sec.~\ref{sec:frame_expansion} are implemented to build a LOM based on the state-space formalism introduced in Sec.~\ref{sec:intro_lom_network_state_space}. The goal is to compare these methods on a series of canonical test cases, in order to show the limits of the rigid-wall modes decomposition and to prove the performances of the over-complete frame approach. Throughout this section, superscripts \textsuperscript{OB} (resp.~\textsuperscript{FR}) refer to results obtained with the use of the orthogonal basis $(\psi_n)_{n \geqslant 1}$ introduced in Sec.~\ref{sec:galerkin_expansion} (resp.~the over-complete frame $(\phi_n)_{n \geqslant 1}$ introduced in Sec.~\ref{sec:frame_expansion}). Superscripts  \textsuperscript{A} designate analytical solutions used for comparison.\par

\subsection{The Gibbs phenomenon at the cross-section change in a duct} \label{sec:example_1Dduct_cross_section_change}

\begin{figure}[h!]
\centering
\includegraphics[width=0.80\textwidth]{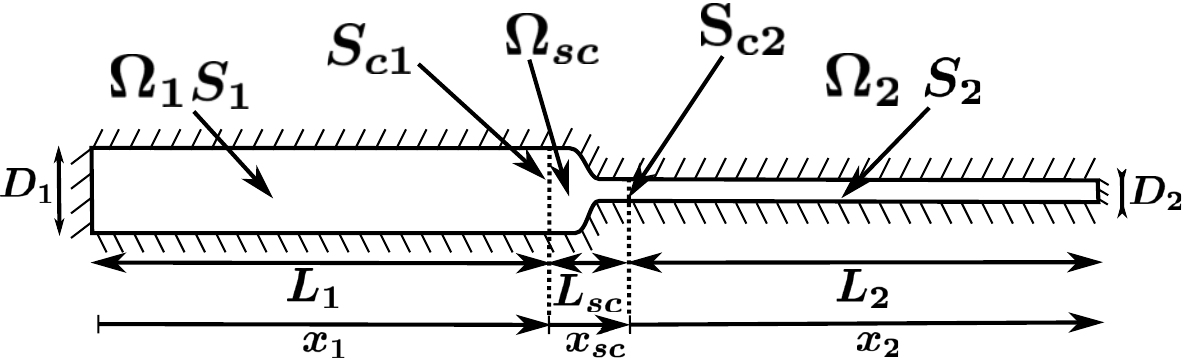}
\caption{A long duct comprising a sharp cross-section change. Hashed lines boundaries are rigid walls. The geometry is decomposed into 3 subdomains. Two of those are long ducts ($\bm{\Omega_1}$ and $\bm{\Omega_2}$), of respective lengths $\bm{L_1}$ and $\bm{L_2}$, cross-sections $\bm{S_1}$ and $\bm{S_2}$, and equivalent diameters $\bm{D_1}$ and $\bm{D_2}$. The third subdomain $\bm{\Omega_{sc}}$ is a small region ($\bm{L_{sc} \ll L_1,L_2}$) located around the cross-section change. $\bm{S_{c1}}$ and $\bm{S_{c2}}$ are connection surfaces between $\bm{\Omega_{sc}}$ and the two tubes.}
\label{fig:1Dduct}
\end{figure}

In this first example, the long duct with a sudden cross-section change represented in Fig.~\ref{fig:1Dduct} is considered. Both ends of the duct are closed by rigid walls. It is decomposed into 3 subsystems, including 2 long ducts ($\Omega_1$ and $\Omega_2$) with constant cross-sections $S_1$ and $S_2$, and a third subsystem $\Omega_{sc}$ of length $L_{sc}$ enclosing the region in the neighborhood of the cross-section variation.

Since both tubes $\Omega_1$ and $\Omega_2$ are long ($D_1 \ll L_1$, $D_2 \ll L_2$), only plane longitudinal acoustic waves are considered here. The rigid-wall orthogonal bases of both ducts are then:
\begin{align}
\label{eq:1Dduct_bases1}
\left\{
\begin{aligned}
&\left(\psi_n^{(1)}(x_1) \right)_{ n \leqslant N_1} = \left( \cos(\dfrac{  n \pi x_1}{L_1}) \right)_{  n \leqslant N_1} \\
&\left(\psi_n^{(2)}(x_2) \right)_{  n \leqslant N_2} = \left( \cos(\dfrac{n \pi x_2}{L_2}) \right)_{  n \leqslant N_2}
\end{aligned}
\right.
\end{align}
where superscript ${}^{(1)}$ (resp.~${}^{(2)}$) refers to the modal basis in $\Omega_1$ (resp.~$\Omega_2$), $x_1$ and $x_2$ are the longitudinal coordinates in the two ducts (Fig.~\ref{fig:1Dduct}), and $N_1$ (resp.~$N_2$) is the number of modes used for the pressure modal expansion in $\Omega_1$ (resp.~$\Omega_2$). Similarly the over-complete frames introduced in Sec.~\ref{sec:frame_expansion} for both ducts are given by:

{\footnotesize  
\setlength{\abovedisplayskip}{6pt}
\setlength{\belowdisplayskip}{\abovedisplayskip}
\setlength{\abovedisplayshortskip}{0pt}
\setlength{\belowdisplayshortskip}{3pt}
 \begin{align}
\label{eq:1Dduct_bases2}
\left\{
\begin{aligned}
&\left(\phi_n^{(1)}(x_1) \right)_{  n \leqslant N_1} = \left( \cos \left(\dfrac{n \pi x_1}{L_1} \right) \right)_{  n \leqslant N_1/2} \bigcup \left( \cos \left(\dfrac{(2n+1) \pi x_1}{2L_1} \right) \right)_{ n \leqslant N_1/2} \\
&\left(\phi_n^{(2)}(x_2) \right)_{ n \leqslant N_2} = \left( \cos \left(\dfrac{ n \pi x_2}{L_2} \right)  \right)_{ n \leqslant N_2/2}  \bigcup \left( \sin \left(\dfrac{(2n+1) \pi x_2}{2L_2} \right) \right)_{  n \leqslant N_2/2}
\end{aligned}
\right.
\end{align}}%

These orthogonal bases (Eq.~\eqref{eq:1Dduct_bases1}) and over-complete frames (Eq.~\eqref{eq:1Dduct_bases2}) are consistent with rigid-wall conditions at both ends of the long duct ($x_1 = 0$ and $x_2=L_2$). However, the orthogonal bases also impose zero velocity near the cross section change (at $x_1 = L_1$ and $x_2=0$), while the over-complete frames retain an additional degree of freedom such that both velocity and pressure are \textit{a priori }undetermined and free to evolve independently near the cross-section change. In the following, the same number $N$ of eigenmodes are used for modal expansions in both ducts ($N_1=N_2=N$). Additionally, all comparisons between orthogonal basis and over-complete frame expansions are carried out with \textit{the same total numbers of modes} $N$: in other words results from any orthogonal basis containing an even number of vectors $N$ are compared to results from a frame composed of two subfamilies of size $N/2$. Note also that, since only plane longitudinal waves are considered here, the connection boundaries $S_{c1}$ and $S_{c2}$ do not need to be discretized into several surface elements $ S_{0j}$: therefore $M_S = 1$ for each subdomain $\Omega_1$ and $\Omega_2$, and $M_S = 2$ for $\Omega_{sc}$.\par

Modal expansion is not performed for the subdomain $\Omega_{sc}$ enclosing the cross-section change as its exact geometry has only very little effect on the global eigenmodes of the long duct: instead, volume-averaged conservation equations are used to derive a state-space representation of this subsystem. More details are given in Appendix~\ref{sec:ss_realization_area_jump}. The Redheffer star-product defined in Eq.~\eqref{eq:redheffer_R_1} and Eq.~\eqref{eq:redheffer_R_2} is then applied recursively to connect state-space representations of subsystems $\Omega_1$, $\Omega_{sc}$, and $\Omega_2$. The state-space realization of the full geometry is then obtained as in Eq.~\eqref{eq:statespace_full}:
\begin{align}
\label{eq:1Dduct_statespace_full}
\dfrac{d}{dt}
\underbrace{
\begin{pmatrix}
\mathbf{X}^{(1)}(t) \\
\mathbf{X}^{(sc)}(t) \\
 \mathbf{X}^{(2)}(t)
\end{pmatrix}}_{\mathbf{X}^{f}(t)} =
\underbrace{
\begin{pmatrix}
\mathbf{A}^{(1)} & \mathbf{B}^{(1)}  \mathbf{C}_{1}^{(sc)} & \mathbf{0} \\
\mathbf{B}_{1}^{(sc)} \mathbf{C}^{(1)} & \mathbf{A}^{(sc)}  & \mathbf{B}_{2}^{(sc)} \mathbf{C}^{(2)} \\
\mathbf{0} & \mathbf{B}^{(2)} \mathbf{C}_{2}^{(sc)} & \mathbf{A}^{(2)}
\end{pmatrix}}_{\mathbf{A}^{f}}
\begin{pmatrix}
\mathbf{X}^{(1)}(t) \\
\mathbf{X}^{(sc)}(t) \\
 \mathbf{X}^{(2)}(t)
\end{pmatrix}
\end{align}
In this equation, matrices $\mathbf{A}^{(1)}$, $\mathbf{B}^{(1)} $, $\mathbf{C}^{(1)} $, $\mathbf{A}^{(2)}$, $\mathbf{B}^{(2)} $, and $\mathbf{C}^{(2)} $ are the state-space representations of subdomains $\Omega_1$ and $\Omega_2$, as defined in Sec.~\ref{sec:intro_lom_network_state_space} and Appendix~\ref{sec:ss_realization_subdomain}. Matrices $\mathbf{A}^{(sc)}$, $\mathbf{B}_1^{(sc)} $, $\mathbf{B}_2^{(sc)} $,  $\mathbf{C}_1^{(sc)} $ and $\mathbf{C}_2^{(sc)} $  are used in the state-space representation of the subdomain $\Omega_{sc}$ and are given in Appendix~\ref{sec:ss_realization_area_jump}. The extra-diagonal blocks of the matrix $\mathbf{A}^f$ represent coupling between the two ducts $\Omega_1$ and $\Omega_2$ and the subdomain $\Omega_{sc}$ enclosing the section change. In the present case, the matrices $\mathbf{A}^{(1)}$ and $\mathbf{A}^{(2)}$ share the same size $2N$ (see Eq.~\eqref{eq:statespace_subdomain_dyn_1}), and from Eq.~\eqref{eq:1Dduct_state_space_1}, $\mathbf{A}^{(sc)}$ is a $3 \times 3$ matrix. Thus, the dynamical system governing the pressure evolution in the whole geometry is of size $4N+3$. The corresponding acoustic eigenfrequencies and eigenmodes are obtained by solving for the eigenvalues and eigenvectors of $\mathbf{A}^{f}$. Resulting eigenfrequencies are noted $f_n^{OB}(N)$ if the rigid-wall orthogonal basis is used, and $f_n^{FR}(N)$ if the pressure is expanded onto the over-complete frame of Sec.~\ref{sec:frame_expansion}. The corresponding pressure mode shapes are $\Upsilon_{n,p}^{OB}(x;N)$ and  $\Upsilon_{n,p}^{FR}(x;N)$, and finally the velocity mode shapes are $\Upsilon_{n,u}^{OB}(x;N)$ and $\Upsilon_{n,u}^{FR}(x;N)$.\par

In the following example the cross-section change is located at 1/3 of the duct length ($L_1 = L_2/2 = L$), and the cross-section ratio is such that $S_1 = 2S_2$. Under the compactness assumption ($L_{sc} \ll L_1,L_2$) the analytical frequency reads:

{\footnotesize
\setlength{\abovedisplayskip}{6pt}
\setlength{\belowdisplayskip}{\abovedisplayskip}
\setlength{\abovedisplayshortskip}{0pt}
\setlength{\belowdisplayshortskip}{3pt}
\begin{align}
\label{eq:1Dduct:analytical_frequency}
f_n^A = \dfrac{n c_0}{2L}\ \mathrm{if} \ n=0 \ modulo \ 3, \ \mathrm{or} \ f_n^A = \dfrac{n c_0}{L} \pm \dfrac{c_0}{\pi L} \arctan((2 \pm \sqrt{3})^{1/2}) \ \mathrm{otherwise}
\end{align}}%

For the first mode ($n=1$), the analytical pressure and velocity modal shapes write:
\begin{align}
\label{eq:1Dduct:analytical_pressure_1}
\Upsilon_{1,p}^A(x) = 
\left\{
\begin{aligned}
-&\dfrac{1}{\sqrt{3}}\cos \left( \dfrac{\lambda x_1}{L} \right) \ , \ \mathrm{in } \ \Omega_1\\[5pt]
&\cos \left( \dfrac{2 \lambda (x_2- 2 L)}{2L} \right) \ , \ \mathrm{in } \ \Omega_2
\end{aligned}
\right.
\end{align}
\begin{align}
\label{eq:1Dduct:analytical_velocity_1}
\Upsilon_{1,u}^A(x) = 
\left\{
\begin{aligned}
&\dfrac{1}{\sqrt{3}\rho_0 c_0}\sin \left( \dfrac{\lambda x_1}{L} \right) \ , \ \mathrm{in } \ \Omega_1\\[5pt]
&-\dfrac{1}{\rho_0 c_0} \sin \left( \dfrac{2 \lambda (x_2-2L)}{2L} \right) \ , \ \mathrm{in } \ \Omega_2
\end{aligned}
\right.
\end{align}
where $\lambda = 2 \arctan ((2-\sqrt{3})^{1/2})$. Note that for the first mode both pressure and velocity are non-zero at the cross-section change ($x_1 = L$, $x_2 = 0$).\par

\begin{figure}[h!]
\centering
\includegraphics[width=0.99\textwidth]{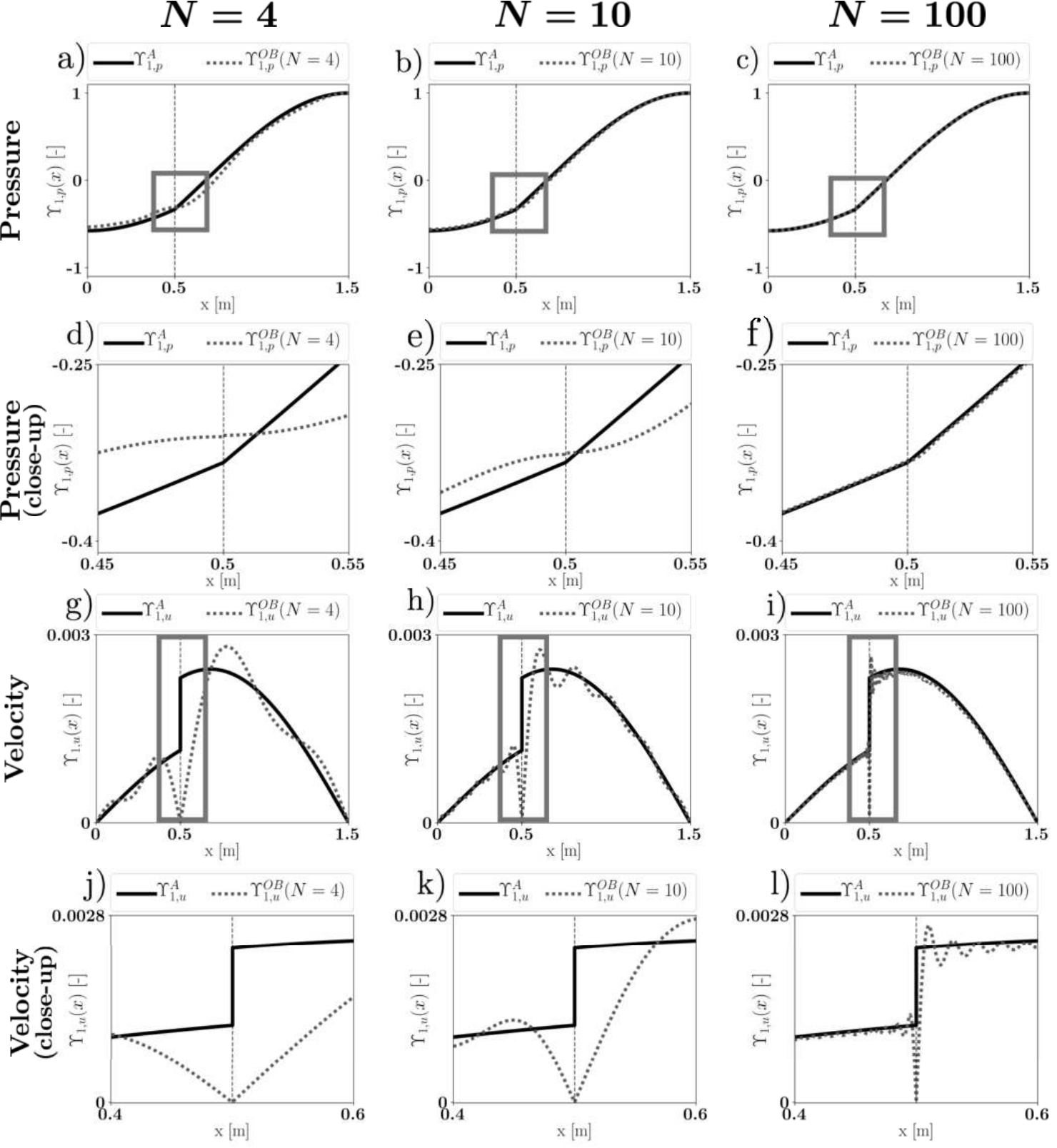}
\caption{Pressure mode shape and velocity mode shape of the first mode ($\bm{n=1}$), for $\bm{N=4,10,100}$ vectors in the OB expansion. Computed solutions (grey dashed lines) are compared to the analytical solutions of Eq.~\eqref{eq:1Dduct:analytical_pressure_1} and Eq.~\eqref{eq:1Dduct:analytical_velocity_1} (thick dark lines). Closeup views (corresponding to the rectangles) of the pressure and the velocity mode shapes are given on the second row ((d)-(f)) and the fourth row ((j)-(l)) respectively.}
\label{fig:1Dduct_mode_shapes_OB}
\end{figure}

In Fig.~\ref{fig:1Dduct_mode_shapes_OB}, the pressure and velocity mode shapes for the first mode ($n=1$) are displayed and compared to the analytical solutions of Eq.~\eqref{eq:1Dduct:analytical_pressure_1} and Eq.~\eqref{eq:1Dduct:analytical_velocity_1}.
The use of the rigid-wall eigenmodes expansions defined in Eq.~\eqref{eq:1Dduct_bases1} provides numerical solutions for the pressure mode shape that converge towards the analytical solution when the number $N$ of modes in the orthogonal basis increases (Fig.~\ref{fig:1Dduct_mode_shapes_OB}-(a) to (c)). However, the close-up view on the region near the cross-section change (Fig.~\ref{fig:1Dduct_mode_shapes_OB}-(d) to (f)) shows a slight discrepancy in the pressure modal shape. Figures~\ref{fig:1Dduct_mode_shapes_OB}-(g) to (l) support this observation by evidencing a strong singularity in the velocity mode shape, which results in an erroneous representation of the velocity mode over a large region of the domain. This singularity is the direct consequence of the use of the rigid-wall eigenmodes expansion: the numerical velocity $\Upsilon_{1,u}^{OB}(x)$ is indeed zero at the cross-section change (because the derivative of $\psi_n$ in Eq.~\eqref{eq:1Dduct_bases1} is zero for $x_1=L$ and $x_2=0$), in contradiction with the analytical solution $\Upsilon_{1,u}^{A}(x)$ at this point (see Eq.~\eqref{eq:1Dduct:analytical_velocity_1} for $x_1=L$ or $x_2 = 0$). Most importantly, increasing the number $N$ of modes in the modal basis fails to suppress this singularity, but rather results in higher frequency oscillations around the cross section area change. These spatial fluctuations suggest a Gibbs-like phenomenon affecting the convergence of the velocity representation, typical of Fourier series expansions of irregular functions. These oscillations on the velocity in the vicinity of the duct contraction may be critical in thermoacoustics, since the velocity is the input of the classical Flame Transfer Function used to model flame-acoustic coupling.

\begin{figure}[h!]
\centering
\includegraphics[width=0.99\textwidth]{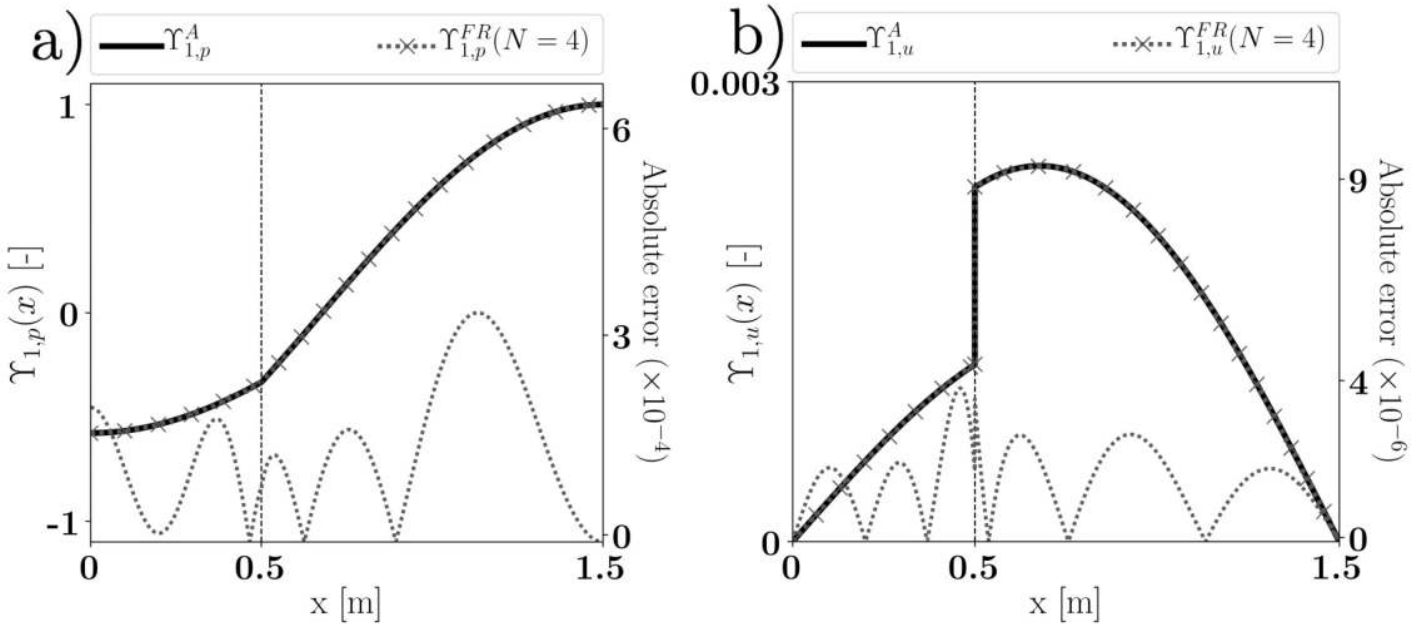}
\caption{Pressure mode shape (a) and velocity mode shape (b) of the first mode ($\bm{n=1}$) for $\bm{N=4}$ vectors in the frame. Numerical solutions (grey lines with $\bm{\times}$) are compared to the analytical solutions of Eq.~\eqref{eq:1Dduct:analytical_pressure_1} and Eq.~\eqref{eq:1Dduct:analytical_velocity_1} (thick dark lines). The local absolute errors for the mode shapes with $\bm{N=4}$ are also plotted (dashed line), with values indicated on the right axis.}
\label{fig:1Dduct_mode_shapes_OC}
\end{figure}

On the contrary, as shown in Fig.~\ref{fig:1Dduct_mode_shapes_OC}, modal expansions onto the over-complete frames of Eq.~\eqref{eq:1Dduct_bases2} accurately represent the analytical solutions $\Upsilon_{1,p}^{A}(x)$ and $\Upsilon_{1,u}^{A}(x)$, even for a number of modes as low as $N=4$, as the local absolute error does not exceed $3 \times 10^{-4}$ for the pressure, and $5 \times 10^{-6}$ for the velocity.\par

The relative errors for the  n\textsuperscript{th} global eigenfrequency and pressure mode shapes are defined as (with similar definition for FR):
\begin{align}
\label{eq:error_frequency_def}
E_{f_n}^{OB} (N) = \dfrac{| f_n^A - f_n^{OB}(N)|}{f_n^A} \ , \ E_{n,p}^{OB} (N)
 = \dfrac{\lVert \Upsilon_{n,p}^{A} - \Upsilon_{n,p}^{OB} (N)  \rVert}{\lVert \Upsilon_{n,p}^{A} \rVert}
 \end{align}
The relative error $E_{n,u}^{OB} (N)$ for the velocity mode shape is defined in the same fashion.

\begin{figure}[h!]
\centering
\includegraphics[width=0.99\textwidth]{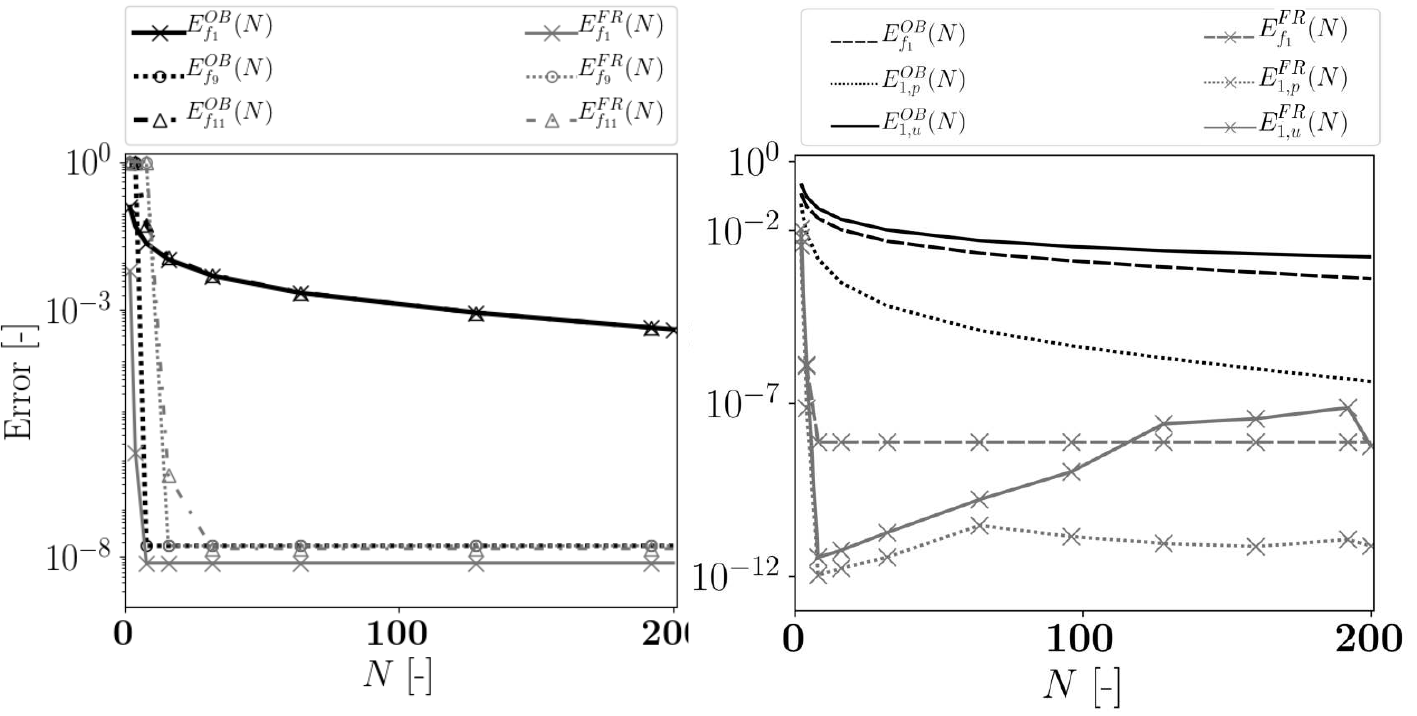}
\caption{Left: Comparison between the OB and the FR results in terms of frequency relative error ($\bm{E_{f_n}^{OB/FR}}$), for global modes 1, 9 and 11, in function of the number $\bm{N}$ of vectors used in the modal expansions. Dark lines are results obtained with rigid-wall eigenmodes expansions (Eq.~\eqref{eq:1Dduct_bases1}), while gray lines are results obtained with frames modal expansions (Eq.~\eqref{eq:1Dduct_bases2}). Right: Comparison between the OB and the FR results in terms of frequency relative error ($\bm{E_{f_1}^{OB/FR}}$), pressure mode shape relative error ($\bm{E_{1,p}^{OB/FR}}$), and velocity mode shape relative error ($\bm{E_{1,u}^{OB/FR}}$). Curves corresponding to the results obtained with rigid-wall eigenmodes expansions (Eq.~\eqref{eq:1Dduct_bases1}) are in dark, while those obtained with expansions onto over-complete frames (Eq.~\eqref{eq:1Dduct_bases2}) are in gray with $\bm{\times}$ marks.}
\label{fig:1Dduct_errors}
\end{figure}

Figure~\ref{fig:1Dduct_errors}(left) provides the convergence behavior for the global eigenfrequencies of modes 1, 9, and 11. It appears that the method based on orthogonal bases expansions present a relatively slow convergence speed for eigenfrequencies $f_1$ and $f_{11}$. This poor convergence is due to the Gibbs phenomenon affecting the numerical solution for the velocity mode $\Upsilon_{1,u}^{OB}(x)$ in the vicinity of the cross-section change. Mode 9 however, is accurately resolved even for small values of $N$; this is because it naturally has a velocity node at the cross-section change (Eq.~\eqref{eq:1Dduct:analytical_velocity_1}) and is therefore, by chance, not subjected to the Gibbs phenomenon.  In contrast, the method relying on modal expansions onto over-complete frames results in low relative errors, ranging between $10^{-7}$ and $10^{-8}$ for all modes 1, 9 and 11, even with small values of $N$. For instance, only about 10 modes in each frames are necessary to accurately capture the eigenfrequency $f_{11}$. The condition number of the frame Gram matrix, which is an indicator of the expansion over-completeness, is $C(\mathbf{\Lambda}) = 10^{8}$ for $N=6$. It then progressively deteriorates and reaches $10^{19}$ at $N=50$, and saturates to this value for large $N$. This deterioration of the frame conditioning does not result into a degradation of the numerical results. However, it can be related to a saturation of the error, since increasing the size of the frame beyond $N=20$ does not result in smaller errors.\par  

The Gibbs phenomenon effect on the convergence behavior of the 1\textsuperscript{st}-mode pressure ($\Upsilon_{1,p}^{OB}(x)$) and velocity ($\Upsilon_{1,u}^{OB}(x)$) is quantitatively evaluated in Fig.~\ref{fig:1Dduct_errors} (right). The method based on rigid-wall eigenmodes expansions yields large errors on the pressure and velocity modes when a small number $N$ of basis eigenmodes is used. When $N$ is increased, these two errors decrease at different speeds. The pressure mode error $E_{1,p}^{OB}$ declines relatively quickly: a number of basis eigenmodes as low as $N=10$ is sufficient to achieve a $0.1\%$ relative error on the pressure mode. On the contrary, the velocity convergence is significantly slower: more than 180 eigenmodes in the rigid-wall basis are indeed necessary to reach a $0.1\%$ relative error on the velocity mode shape. In contrast, the method relying on modal expansions onto over-complete frames results in notably better convergence properties. Even for a low order frame ($N=4$), relative errors on the pressure and velocity mode are unimportant ($10^{-12}$, and $10^{-11}$ respectively). As previously observed on the frequencies convergence, further increasing the size $N$ of the expansion frames does not improve the accuracy. This effect is clearly visible on the pressure and velocity errors, that even increase when $N$ increases. However, they both stay below $10^{-7}$ and are several orders of magnitude better than that obtained with the orthogonal basis expansion. This is explained by the fact that the FR method precision is not limited by the modal expansion itself, but rather by round-off errors resulting from the numerical inversion of the Gram matrix $\mathbf{\Lambda}$ and from the numerical computation of the surface gradients in Eq.~\eqref{eq:press_final_equation_revisited}.  Finally, even large orthogonal basis expansions ($N=200$) are still outperformed by low-order over-complete frames expansions with $N$ as small as 4. Similar trends are observed for all the eigenmodes (not shown here).

\subsection{Flame induced Gibbs phenomenon}

The previous non-reactive example showed that the classical Galerkin expansion results in a Gibbs phenomenon affecting the velocity field near boundaries that are not rigid-walls. Gibbs oscillations were also reported in a number of previous works (for example Sayadi \textit{et al.}~\cite{sayadi2014}) using the orthogonal basis modal expansion to study systems comprising active flames. The object of this second example is therefore to discuss the occurrence of the Gibbs phenomenon in a one-dimensional reactive case.
\begin{figure}[h!]
\centering
\includegraphics[width=0.6\textwidth]{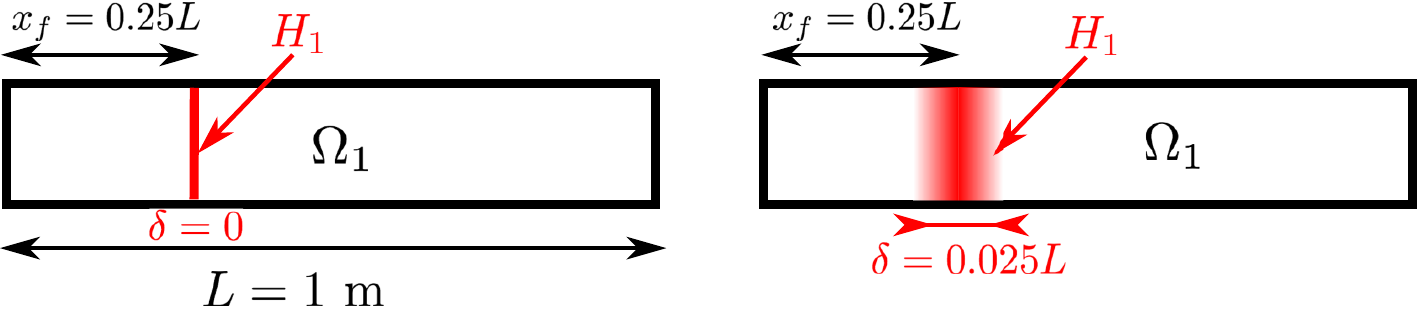}
\caption{Left: Schematic of the system of interest, which consists of a long duct of length $\bm{L}$ and section $\bm{S_0 = 10^{-4}}$~m\textsuperscript{2} comprising an infinitely thin flat flame located at $\bm{x_f = 0.25 L}$. All the boundaries are rigid-walls. The mean sound speed and density fields are uniform with $\bm{c_0 = 347}$~m/s and $\bm{\rho_0 = 1.17}$ kg/m\textsuperscript{3}. As the geometry is not split, the acoustic network comprises only two subsystems: the duct $\bm{\Omega_1}$ and the heat release source $\bm{H_1}$. Right: same case, with a flame of finite thickness $\bm{\delta = 0.025L}$ distributed over a Gaussian profile.}
\label{fig:schematic_1Dflame_Gibbs}
\end{figure}\par

In a system similar to the one represented in Fig.~\ref{fig:schematic_1Dflame_Gibbs} (left) comprising an infinitely thin flat flame, Sayadi \textit{et al.}~\cite{sayadi2014} expended the pressure onto the rigid-wall orthogonal basis as: $p(x,t) = \sum_{n=1}^{N}a_n(t) \cos(n \pi x  /L)$.  This expansion was compared to a direct discretization method: Gibbs fringes could still be observed with $N=128$ Galerkin modes, whereas those could be completely avoided in the direct discretization approach thanks to the use of a specific numerical scheme in the vicinity of the velocity discontinuity. Generally speaking, Gibbs fringes may appear at any location where the modal expansion lacks the ability to represent the \textit{exact} solution of the problem under consideration. In thermoacoustics, such situation may occur due to two factors:
\begin{enumerate}
\item{\textbf{At the boundaries} of the (sub-)domains, that are neither rigid-walls ($u'=0$) nor pressure release ($p'=0$).}
\item{\textbf{In the interior} of the (sub-)domains, where regions of fluctuating heat release cause a sharp variation of the acoustic velocity field.}
\end{enumerate}
Sayadi \textit{et al.}, and many other similar studies, reported a Gibbs phenomenon stemming from the second factor. Indeed, they employ for convenience an infinitely thin region of fluctuating heat release, modeled as a Dirac function ($\mathcal{H}_1 (x) =  \delta(x-x_f)$, where $x_f$ is the flame location). This assumption yields an \textit{exact} solution for the velocity field with a discontinuity at $x_f$, which is the cause of the Gibbs phenomenon (since cos/sin series cannot accurately represent a discontinuous field). On the contrary, the frame expansion proposed earlier is intended to tackle the first point, by accurately representing any boundary condition at domains frontiers or subdomains interfaces. Consequently, it \textit{does not have the ability to suppress Gibbs fringes due to active flames located within the subdomains}, that have a fundamentally different origin.\par

There is however a simple workaround to mitigate the Gibbs phenomenon caused by an interior discontinuity. Replacing the infinitely thin flame with a slightly thicker and smoother profile yields an exact solution for the velocity field that has a sharp yet continuous spatial variation in the vicinity of the flame. Let us illustrate this point with the second case represented in Fig.~\ref{fig:schematic_1Dflame_Gibbs} (right), where the heat source $H_1$ has a Gaussian profile of thickness $\delta$:
\begin{align}
\label{eq:thicker_flame_shape}
\mathcal{H}_1 (x) = h_0 \exp \left( -5.545 \dfrac{(x - x_f)^2}{\delta^2} \right)
\end{align}
where $h_0$ is a normalization parameter enforcing a unity value of the integral of $\mathcal{H}_1 (x)$ over $\Omega_1$. The flame response is modeled by a simple constant $n - \tau $ FTF:
\begin{align}
\label{eq:FTF_constant_n_tau}
\hat{Q}(\omega) = \overline{Q} e^{-j \omega \tau}  \dfrac{\hat{u}( x_{ref},\omega)}{\overline{u}} 
\end{align}
where $\overline{Q} = 40$~W is the mean flame power, $x_{ref} = 0.22L$ is the reference point, $\tau = 1$~ms is the time-delay, and $\overline{u} = 1$~m/s the mean bulk velocity at this reference location. The state-space realization for this FTF is detailed in Sec.~\ref{sec:frame_application_annular} and in Appendix~\ref{sec:ss_realization_flame_PBF}.\par

\begin{figure}[h!]
\centering
\includegraphics[width=\textwidth]{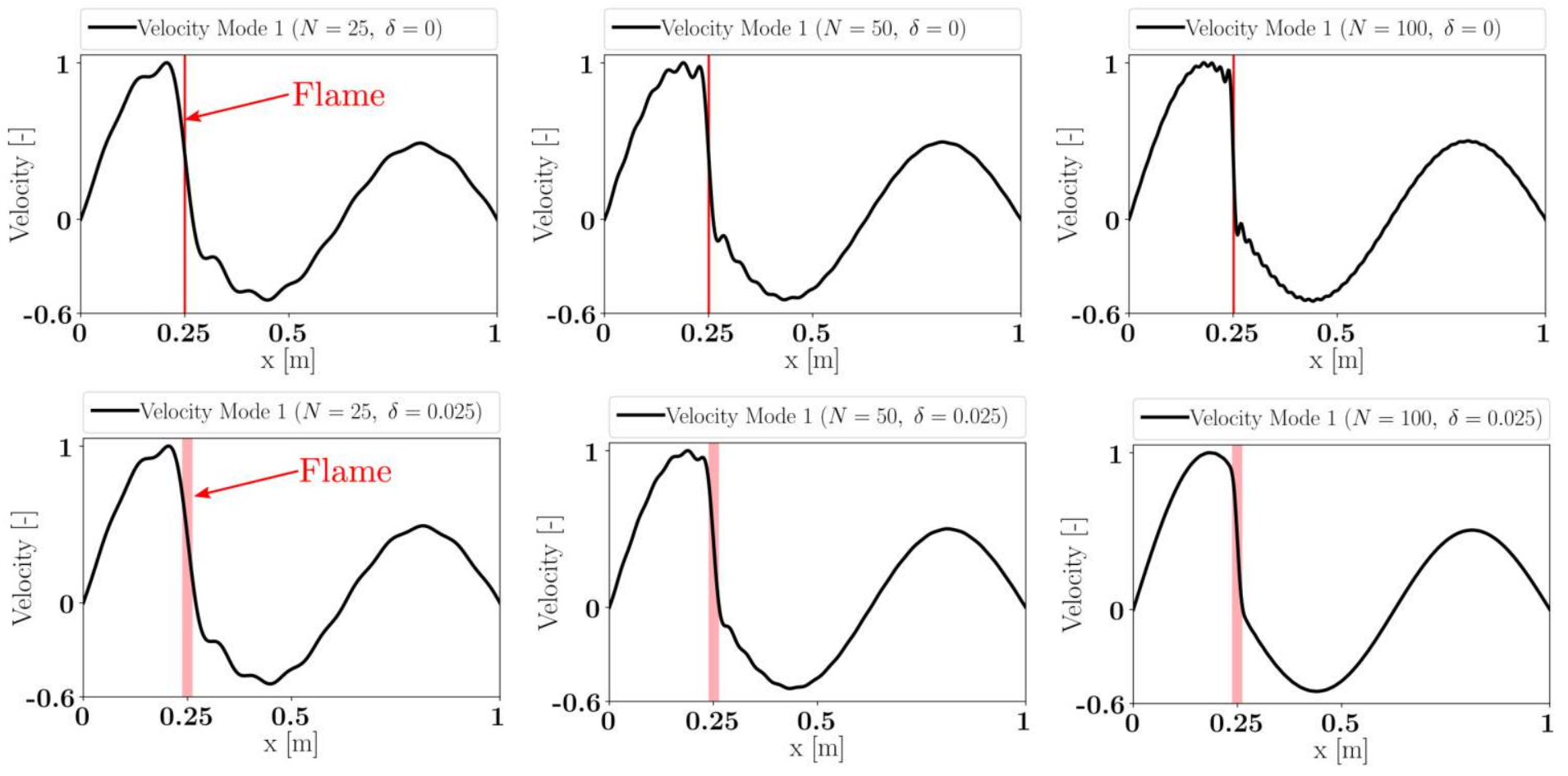}
\caption{Velocity field for the first unstable mode of a long duct of length $\bm{L=1}$~m with a flame located at $\bm{x_f = 0.25}$~m. First row: effect of increasing the number of Galerkin modes ($\bm{N = 25, \ 50, \ 100}$) in the case of an infinitely thin flame region ($\bm{\delta =  0}$). Second row: same results for a flame region of finite thickness $\bm{\delta = 0.025L}$.}
\label{fig:gibbs}
\end{figure}

The frame expansion is not used here: instead the pressure is expanded onto the rigid-wall Galerkin basis $ \left( \cos(n \pi x  /L) \right)_{1 \leq n \leq N}$. Results obtained with the infinitely thin flame represented by a Dirac distribution are compared to that obtained with the thicker heat release region. The first unstable mode of the system is displayed in Fig.~\ref{fig:gibbs}.
In the case of an infinitely thin flame region, the Gibbs fringes do not vanish with increasing number of Galerkin modes. Significant oscillations are still visible with $N=100$, consistently with previous observations from other groups. On the contrary, in the case of a smooth and finite thickness flame region these oscillations are significantly reduced for $N=50$, and are even completely suppressed for $N=100$. In this latter case, the velocity has a sharp yet continuous variation in the vicinity of the heat release region. Thus considering a smooth heat release region of small but finite spatial extent is sufficient to regularize the problem and yield a numerical solution without Gibbs fringes. Finally, note that this difference in flame thickness has only little effect on the frequencies and growth rates: $f=465.6$~Hz, $\sigma = 27.9$~s\textsuperscript{-1} for $\delta = 0$, while $f=466.0$~Hz, $\sigma = 26.9$~s\textsuperscript{-1} for $\delta = 0.025L$.\par

Thus, infinitely thin flames are interesting conceptually and from an analytical perspective, since they translate into a "jump" relation on the Riemann invariants $A^{+}/A^{-}$. However, they should be avoided in studies based on modal expansions, since those cannot, by construction, properly represent discontinuous functions. For the latter, flame of finite thickness distributed over a sufficiently regular profile (at least differentiable) should be preferred. Note also that physical flames are always finite and sometimes thick enough to produce non-compact effects on thermoacoustic instabilities~\cite{subramanian2015}. It is worth mentioning that the over-complete frame expansion is nonetheless able to represent infinitely thin flames by \textit{artificially} splitting the duct of Fig.~\ref{fig:schematic_1Dflame_Gibbs} into two shorter tubes located on the left side and the right side of the reaction region, respectively. The flame can then be considered as a boundary for these subdomains, and the velocity discontinuity it induces can be dealt with thanks to the frames of Eq.~\eqref{eq:1Dduct_bases2}.

\subsection{Intrinsic thermoacoustic modes in a duct} \label{sec:example_intrinsic}

Increasing the acoustic losses in a combustor is usually thought as a simple and efficient strategy to prevent the apparition of thermoacoustic instabilities. However, a number of recent studies proved that under some specific conditions, increasing the damping leads to the apparition of a family of unstable thermoacoustic eigenmodes~\cite{Silva:2015,Courtine:2015,Emmert:2015b,Hoeijmakers:2014}. These so-called Intrinsic Themoacoustic instabilities (ITA) were observed in longitudinal configurations~\cite{Sogaro:2019,Ghani:2019,Orchini:2020}, and more recently in annular systems~\cite{Buschmann:2020}, by inducing significant acoustic losses at non-reflective boundary conditions. Since the frame expansion proposed in Sec.~\ref{sec:frame_expansion} is precisely intended to enable the modeling of boundaries that are neither rigid-wall nor pressure release in modal expansion-based LOMs, it is natural to assess its ability to capture ITA in a one-dimensional system with anechoic ends.\par

\begin{figure}[h!]
\centering
\includegraphics[width=0.55\textwidth]{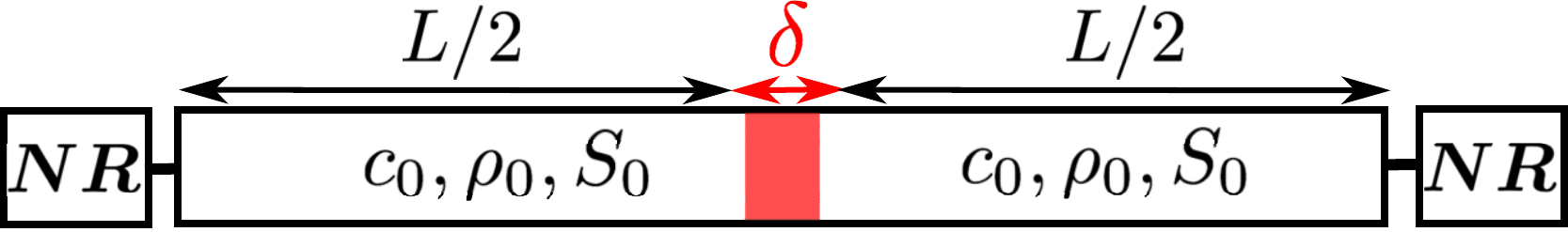}
\caption{A ducted burner, with an active flame of thickness $\delta$ located in its middle. Both ends of the duct are acoustically non-reflecting (NR). The acoustic network comprises 4 subsystems: the duct, the heat release source, and the two non-reflecting boundaries.}
\label{fig:tube}
\end{figure}

Let us consider the one-dimensional duct shown in Fig.~\ref{fig:tube}. Its physical properties are the same as in the previous example. A region of heat release of finite thickness $\delta = 2$ cm distributed over a Gaussian profile (see Eq.~\eqref{eq:thicker_flame_shape}) is located in the middle of the duct. Its response is modeled with a constant $n-\tau$ FTF similar to the one of Eq.~\eqref{eq:FTF_constant_n_tau}. The values for the flame interaction index and the time-delay are: $n = (\gamma-1)\overline{Q}/(S_0 \rho_0 c_0^2 \overline{u}) = 2.064$ and $\tau = 5$ms.
In order to promote the apparition of intrinsic thermoacoustic modes, both ends of the domain are considered as acoustically non-reflecting. This type of boundary condition is obviously not rigid-wall ($u'=0$) nor pressure release ($p'=0$), and it cannot be treated with the classical orthogonal basis expansion. A modal expansion of the pressure onto an over-complete frame is therefore performed. This frame is:
\begin{align}
\label{eq:1Dduct_NR_frame}
\left(\phi_n (x) \right)_{  n \leqslant N} = \left( \cos \left(\dfrac{n \pi x}{L} \right) \right)_{  n \leqslant N/2} \bigcup \left( \sin \left(\dfrac{ n \pi x}{L} \right) \right)_{ n \leqslant N/2}
\end{align}
It has the ability to verify any type of prescribed boundary conditions at both ends of the duct. The non-reflecting relation, which writes $p' = \rho_0 c_0 u'_s$, is enforced at the tube extremities thanks to a state-space representation that is given in Appendix~\ref{sec:ss_realization_NR}.

\begin{figure}[h!]
\centering
\includegraphics[width=\textwidth]{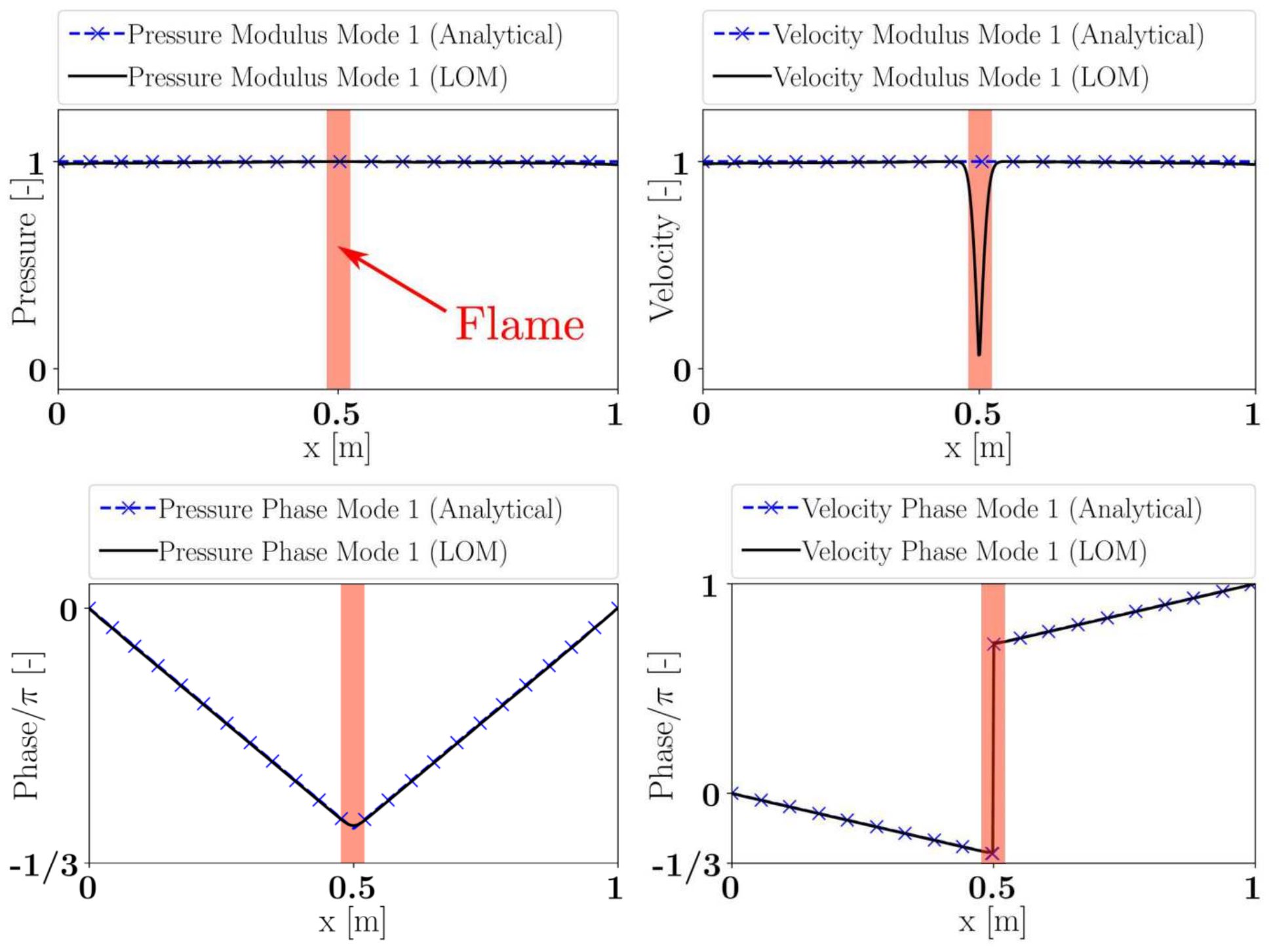}
\caption{Pressure and velocity spatial shapes of the first unstable intrinsic thermoacoustic mode of the ducted burner shown in Fig.~\ref{fig:tube}. Low Order Model (LOM) results based on frame expansion (plain dark lines) are compared to theoretical results based on Riemann invariants (blue lines with symbols).}
\label{fig:intrinsic}
\end{figure}

The first unstable intrinsic thermoacoustic mode resolved with the frame expansion method is displayed in Fig.~\ref{fig:intrinsic}. It is compared to theoretical results based on Riemann invariants. Note that in this analytical resolution the flame is assumed infinitely thin, whereas it has finite thickness $\delta$ in the case treated with the frame expansion.
An excellent agreement is found for both the pressure and velocity mode shapes. The only significant difference is found in the flame region, due to the thickness difference between the two cases. The theoretical frequency and growth rate are respectively $f = 100$~Hz, and $\sigma = 1$~s\textsuperscript{-1}. The frame LOM yields a frequency of $100.00$ Hz and a growth rate of $1.07$, which constitutes a good agreement with the analytical values. Note that the results presented here were obtained with $N=150$ frame modes. Increasing the size of the expansion did not significantly affect the results. However, smaller expansion frames yield erroneous results, particularly on the velocity mode shape and the growth rate. Indeed, ITA modes typically present an important velocity jump across the flame (due to a large interaction index $n$). As explained in the previous example, a large number of modes are necessary to avoid Gibbs fringes due to large velocity variations across a flame located \textit{within} the domain, and the frame method is not intended to circumvent this issue. The simplest way to avoid using a large frame is therefore to consider a slightly thicker region of heat release (but results will inevitably deviate from those obtained through Riemann invariants with an infinitely thin flame).

\subsection{A multi-perforated liner in a duct}

The example of Sec.~\ref{sec:example_1Dduct_cross_section_change} showed that the rigid-wall modal basis yields  Gibbs oscillations near a velocity discontinuity induced by a cross-section change located between two long ducts. On the contrary the frame expansion does not produce Gibbs fringes and accurately captures the velocity jump. The present example aims at proving that the frame expansion can equally resolve a \textit{pressure} discontinuity. Such discontinuities are typically caused by the acoustic losses occurring through multi-perforated liners.\par

Let us then consider an acoustic network comprising three components: two long ducts $\Omega_1$ and $\Omega_2$ of equal cross-section area $S_0$ and respective lengths $L_1 = 1$~m and $L_2 = 2$~m, that are separated by a multi-perforated plate. This latter component imposes a pressure jump between the two ducts, following the relation:
\begin{align}
\label{eq:pressure_jump_1D}
K_R (j \omega) \left( p^{\Omega_2} (x_2 = 0) - p^{\Omega_1} (x_1 = L_1) \right) = - j \omega \rho_0 u^{\Omega_1} (x_1 = L_1)
\end{align}
where $K_R (j \omega)$ is the Rayleigh conductivity. In the present case, it is based on a 2\textsuperscript{nd}-order approximation of the classical Howe's model~\cite{howe1979}, valid for low frequencies, which writes:
\begin{align}
\label{eq:howe_approximation_1d_comparison}
K_R (j \omega) = -K_R^A j \omega + K_R^B \omega^2 , \ \textrm{with} \ K_R^A = - \dfrac{\pi a^2}{2 U} \ , \ K_R^B = \dfrac{2 a^3}{3 U^2} + \dfrac{\pi a^2 h}{4 U^2}
\end{align}
where $a$ is the aperture radius and $U$ the bias flow speed. The state-space realization of this Rayleigh conductivity is detailed in Appendix~\ref{sec:ss_realization_conductivity}. In order to satisfy the pressure jump imposed by Eq.~\eqref{eq:pressure_jump_1D} at the ducts junction, the same frames as in Eq.~\eqref{eq:1Dduct_bases2} are used. The LOM results are compared to reference solutions based on Riemann invariants, that involve the resolution of the following dispersion relation in the complex plane:
\begin{align}
\label{eq:dispesrion_mlpf_1D}
\sin \left( \dfrac{\omega (L_1+L_2)}{c_0} \right) - \dfrac{ \omega d^2}{c_0 K_R (j \omega)} \sin \left( \dfrac{\omega L_1}{c_0} \right)   \sin \left( \dfrac{\omega L_2}{c_0} \right) = 0  
\end{align}\par

\begin{figure}[h!]
\centering
\includegraphics[width=0.99\textwidth]{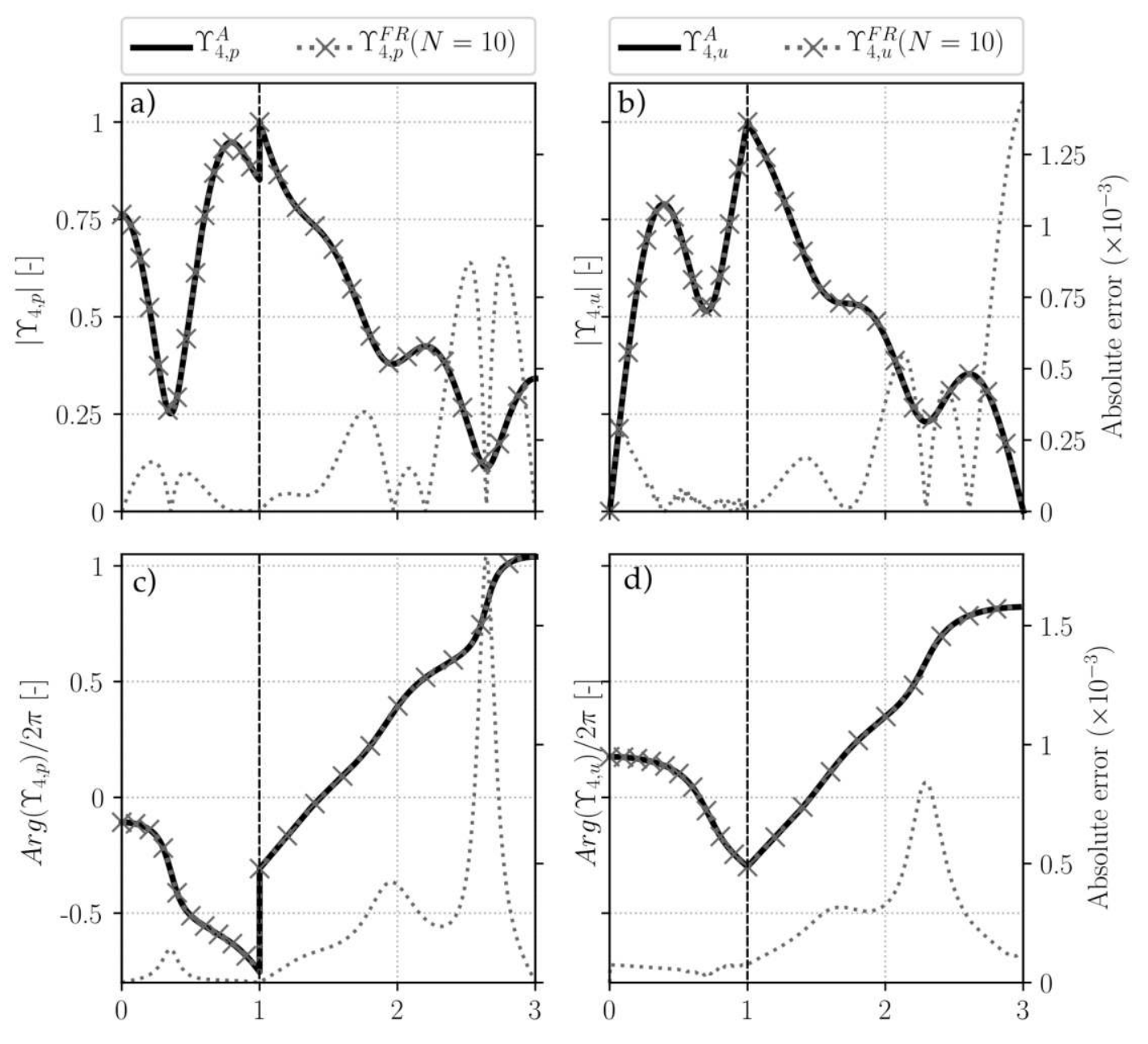}
\caption{Pressure mode shape and velocity mode shape of the 4\textsuperscript{th} mode for $\bm{N=10}$ vectors in the frame. (a) Pressure modulus, (b) velocity modulus, (c) pressure phase, (q) velocity phase. Numerical solutions (grey lines with $\bm{\times}$) are compared to the analytical solutions. The local absolute errors for the mode shapes with $\bm{N=10}$ are also plotted (dashed line), with values indicated on the right axes.}
\label{fig:1Dduct_mode_shapes_mlpf}
\end{figure}

Figure~\ref{fig:1Dduct_mode_shapes_mlpf} compares the shape of  the 4\textsuperscript{th} mode obtained with the frame expansion to the reference solution.
An overall excellent agreement between the LOM results and the analytical solution is observed. The local absolute error does not exceed $10^{-3}$, on both the velocity and the pressure mode shapes. In addition, the numerical frequency and growth rate for the 4\textsuperscript{th} mode are within a $10^{-6}$ relative error margin for $N \geq 10$. Other modes display a convergence behavior highly similar to that observed in Sec.~\ref{sec:example_1Dduct_cross_section_change}. These observations prove the ability of the frame modal expansion to not only accurately resolve a velocity discontinuity, but also a pressure discontinuity.

\section{Application to thermoacoustic instabilities in a model annular combustor} \label{sec:frame_application_annular}

\begin{figure}[h!]
\centering
\includegraphics[width=144mm]{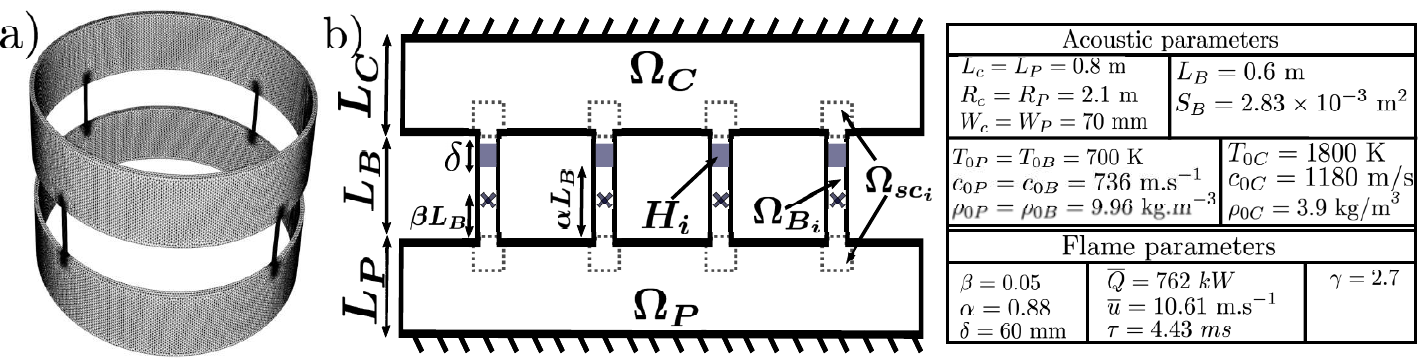}
\caption{(a) The unstructured mesh used in the AVSP FE solver. (b) The low-order thermoacoustic network representing this system, which consists of an annular plenum $\bm{\Omega_P}$, 4 burners $\bm{\Omega_{B_i}}$, an annular chamber $\bm{\Omega_C}$, and a set of 8 subdomains $\bm{\Omega_{sc_i}}$ containing the cross-section changes. Thick dark lines represent rigid-wall boundary conditions. Grey area ($\bm{H_i}$) are the active flames, and the crosses represent the reference points used in the definition of the flame response. The width of the plenum (resp.~chamber) in the radial direction is $\bm{W_P}$ (resp.~$\bm{W_C}$). All required numerical values for acoustic and flame parameters are indicated in the table.}
\label{fig:atamac_schema}
\end{figure}

In order to demonstrate its ability to predict thermoacoustic instabilities in complex geometries, the state-space LOM based on generalized frame modal expansions (FR) is now applied to a more advanced academic configuration comprising active flames in an annular chamber. Results obtained with the classical orthogonal rigid-wall basis (OB) are also provided and compared to those computed the 3D Finite Element (FE) solver AVSP~\cite{nicoud2007}. The geometry studied is displayed in Fig.~\ref{fig:atamac_schema}-(a), and the corresponding low-order acoustic network is represented in Fig.~\ref{fig:atamac_schema}-(b). It comprises an annular plenum (denoted with the subscript \textsubscript{P}), an annular chamber (subscript \textsubscript{C}), and four identical ducted burners (subscript \textsubscript{B}) of length $L_B$ where the active flames are located. Rigid-wall boundary conditions are assumed at the plenum backplane and the chamber outlet plane.
The AVSP unstructured mesh consists of $3 \times 10^6$ tetrahedral cells, while the acoustic network contains 14 subdomains (1 plenum $\Omega_P$, 1 chamber $\Omega_C$, 4 burners $\Omega_{Bi}$, and 8 cross-section changes $\Omega_{sc_i}$), with the addition of 4 active flames. The flames $H_i$ are located at the coordinate $\alpha L_B$ within each burner, and are considered as planar volume source of thickness $\delta$. Note that the cross-section area $S_B$ of the ducted burners is much smaller than the area of the plenum exit plane and of the chamber backplane. This allows for the simplification of the plenum-burner and  burner-chamber junctions, by only considering discrete point-like connections. In other words, the chamber (or the plenum)  is connected to the burner $\Omega_{B_i}$ through the subdomain $\Omega_{sc_i}$ at a single point, implying that $M_S = 4$ for the chamber and the plenum (a single discrete connection surface $\Delta S_{0j} = S_B$ is used for each one of the 4 burners). The subdomain $\Omega_{sc_i}$ is derived from the cross-section change described in Sec.~\ref{sec:frame_convergence}. Its state-space realization, given in Appendix~\ref{sec:ss_realization_area_jump_1D3D}, essentially enforces continuity of pressure and acoustic flux between the burner end and the backplane of the chamber (or the exit plane of the plenum). This simplification also allows us to consider rigid-wall boundary conditions at the chamber backplane and at the plenum exit plane when defining the eigenmodes for these subdomains (since the velocity should actually be non-zero only at 4 point-like locations of infinitely small spatial extent). The Gibbs phenomenon evidenced in Section~\ref{sec:example_1Dduct_cross_section_change} is then not expected to appear in the chamber and in the plenum, and it is therefore valid to employ orthogonal rigid-wall bases in these two subdomains. On the contrary, boundary conditions at both ends of the burners are expected to differ from rigid-wall or open atmosphere, and it is therefore necessary to employ over-complete frame expansions in the burners in order to mitigate the Gibbs phenomenon that may appear. Thus, the plenum is modeled as a 2D annular subdomain of coordinates $(x_P,\theta_P)$, whose orthogonal basis is:
\begin{align}
\label{eq:plenum_basis}
\psi_{n,m}^{(P)}(x_P,\theta_P)  = \left( \cos(\dfrac{  n \pi x_P}{L_P})  \cos(  m \theta_P),  \cos(\dfrac{  n \pi x_P}{L_P})  \sin(  m \theta_P)  \right)
\end{align}
Pressure in the chamber could be expanded onto an analogous analytical modal basis. However, this one is deliberately assumed analytically unknown, and a modal basis computed thanks to a preliminary AVSP simulation of the isolated chamber (without the burners and the plenum) is used instead. Note that it is not necessary to perform an AVSP simulation for each LOM simulation: it is indeed preferable to generate the chamber modal basis and all related quantities (including the matrix $\mathbf{\Lambda}^{-1}$) in a single preliminary AVSP simulation, and to assemble the state-space realization of this subdomain, which can then be employed in as many LOM simulations as desired. In the present example, the use of a numerically computed modal basis demonstrates the ability of the present framework to combine in an acoustic network subdomains of different types and thus to handle efficiently arbitrarily complex systems.\par
In the following, the size of the modal bases in the chamber and in the plenum is fixed to $N=12$ modes. For cases where the rigid-wall basis is used in the ducted burners, this one is the same as in the previous section (Eq.~\eqref{eq:1Dduct_bases1}). If an over-complete frame is used, it is given by:
\begin{align}
\label{eq:1Dburner_frame}
\left(\phi_n^{(B)}(x_B) \right)_{ n \leqslant N} = \left( \cos \left(\dfrac{n \pi x_B}{L_B} \right) \right)_{ n \leqslant N/2} \bigcup \left( \sin \left(\dfrac{n \pi x_B}{L_B} \right) \right)_{ n \leqslant N/2}
\end{align}
Note the difference with the over-complete frame of Eq.~\eqref{eq:1Dduct_bases2}, as the present one allows pressure and velocity to evolve independently from one another at \textit{both ends} of the duct.\par
Active flames are located within each burner, and the flame shape $\mathcal{H}_i(x_B)$ is the Gaussian function of thickness $\delta$ centered around $x_B = \alpha L_B$, as defined in Eq.~\eqref{eq:thicker_flame_shape}. The flame response is modeled thanks to a classical Flame Transfer Function (FTF), relating the fluctuations of heat release to the fluctuations of acoustic velocity at a reference point located at $x_B^{(ref)} = \beta \L_B$. The flame reference point is located in the burners, near the plenum exit-plane ($\beta = 0.05$). A simple constant-delay FTF is assumed, such that in the frequency domain the fluctuating heat release rate reads:
\begin{align}
\label{eq:FTF_model}
\hat{Q}(\omega) = \overline{Q} e^{-j \omega \tau} \left( \dfrac{\hat{u}( x_B = \beta \L_B ,\omega)}{\overline{u}} \right)
\end{align}
where $\overline{Q}$ is the flame power, $\tau$ is the flame delay, and $\overline{u}$ is the mean flow speed through the injector. A state-space realization of the time-delay $e^{-j \omega \tau}$ is generated thanks to a Multi-Pole expansion:
\begin{align}
\label{eq:FTF_multi_pole}
e^{-j \omega \tau} \approx \sum_{k = 1}^{M_{PBF}} \dfrac{-2 a_k j \omega}{\omega^2 + 2 c_k j \omega - \omega_{0k}^2}
\end{align}
where each term in the sum is called a Pole Base Function (PBF). The coefficients $a_k,c_k,w_{0k}$ are determined thanks to a recursive fitting algorithm proposed by Douasbin \textit{et al.}~\cite{douasbin2018}. By making use of the inverse Fourier transform, it is then straightforward to convert this frequency domain transfer function into a time-domain state-space realization of size $2M_{PBF} \times 2M_{PBF}$, whose expression is provided in Appendix~\ref{sec:ss_realization_flame_PBF}. This procedure to generate a state-space realization of a FTF was already used by Ghirardo \textit{et al.}~\cite{ghirardo2015}. For the flame-delay considered here, 12 PBFs were observed to be sufficient to accurately fit the term $e^{-j \omega \tau}$, yielding 24 DoF for each flame in the state-space representation of the whole system.  Finally, the 18 state-space representations of the 14 acoustic subdomains and 4 active flames are connected together.\par

\begin{figure}[h!]
\centering
\includegraphics[width=1.0\textwidth]{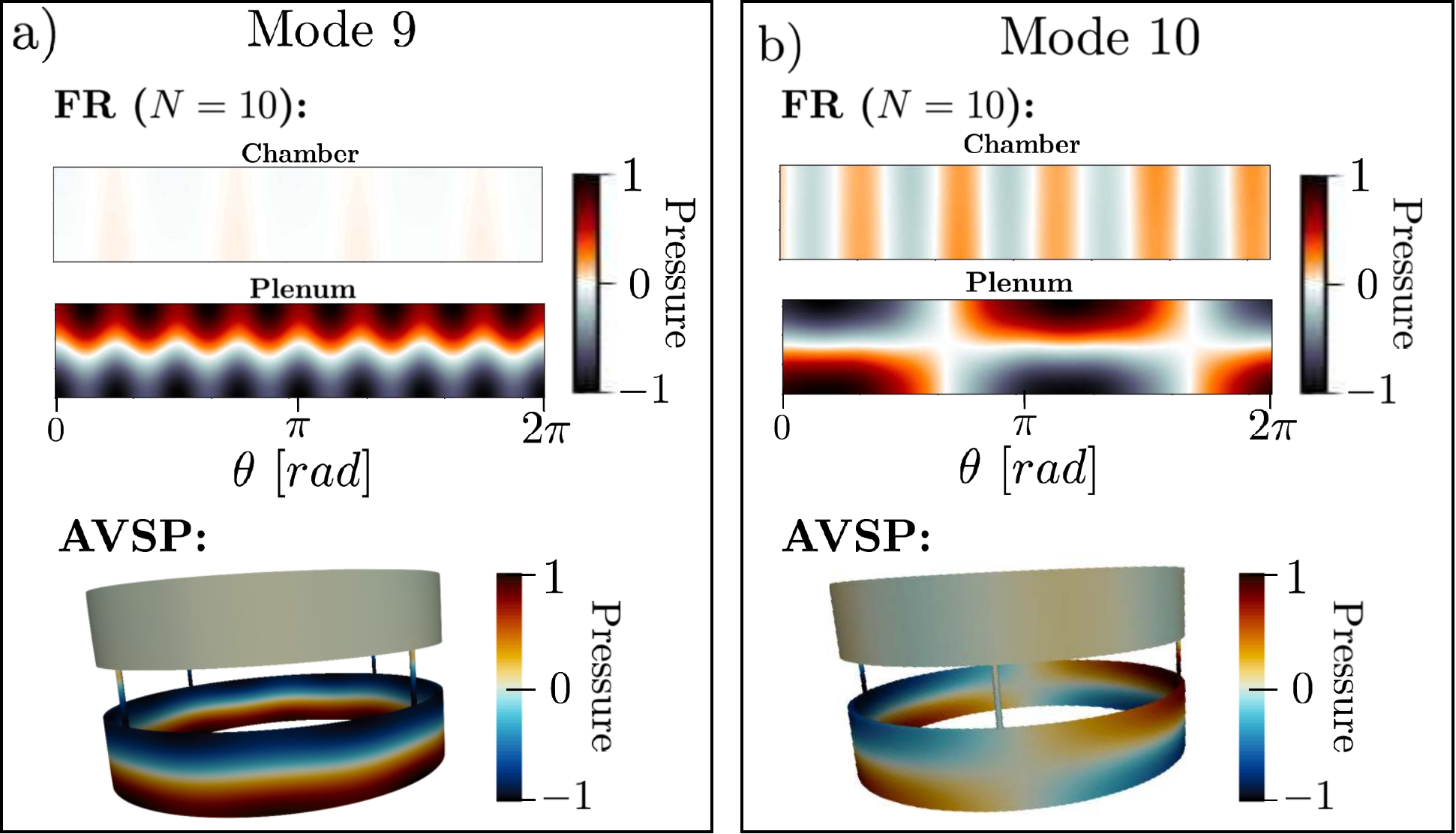}
\caption{The mixed modes 9 and 10. Pressure mode shapes obtained with the frame expansion of size $\bm{N=10}$ are compared to spatial structures solved with the FE solver AVSP.}
\label{fig:mixed_modes}
\end{figure}

As mentioned earlier, the Gibbs phenomenon is expected to occur at both ends of the ducted burners. A particular attention is therefore paid to the type of modal expansion carried out in these subdomains. A total of 4 LOM simulations are performed, and results are compared to the ones computed with the FE solver AVSP, which are used as reference. Results for 10 of the first modes of the combustor are summarized in Tab.~\ref{tab:atacamac_results}. First of all, FR and OB expansions with $N=10$ modes are both compared to AVSP. Then these computations are repeated with a number of modes increased to $N=30$. Mode 1 is the combustor Helmholtz mode, and its frequency and growth rate were observed to be very sensitive to the addition of a correction length to the ducted burners. As the determination of the optimal correction length is out of the scope of this paper, differences regarding the Helmholtz mode are not further discussed. The FR expansion with $N=10$ appears to successfully resolve the frequencies and growth rates of the modes considered, with relative errors compared to AVSP below 10\%. On the contrary, the OB expansion onto $N=10$ modes largely fails at resolving the growth rates of all but one of the considered modes, with relative errors up to 348\% (Mode 5). Mode 3 is the first unstable azimuthal mode of the combustor, and is therefore of particular interest: FR expansion yields an error of only 6\% for the growth rate of this mode, whereas OB expansion produces an error of 45\%. When the size of the expansion basis/frame is increased, results are globally improved for both FR and OB cases. Yet, even with $N=30 $ the OB approach still fails at achieving an acceptable accuracy for Modes 3, 5, and 7. The error on the growth rate of the first unstable azimuthal mode is of 0.3\% in the FR case, while it is still overestimated by 18\% in the OB case. Modes 9 and 10 are mixed modes, similar to those described by Evesque \textit{et al.}~\cite{evesque2002}. Their spatial structures computed with the FR method are compared to FE results in Fig~\ref{fig:mixed_modes}.
In mode 9 the plenum first longitudinal mode prevails, while mode 10 is the mixed 1\textsuperscript{st}-azimuthal-1\textsuperscript{st}-longitudinal plenum mode, coupled with the 5\textsuperscript{th} azimuthal chamber mode. Previous comments also apply to these mixed modes: the over-complete frame expansion of size $N=10$ yields an excellent agreement with the FE solver, and outperforms even larger orthogonal basis expansions. This point proves that this LOM is not limited to azimuthal modes, but can also capture longitudinal and mixed modes. Note that in the FR case, the condition number of the frame Gram matrix used in the 4 ducted burners is of order $10^{8}$ for $N = 10$, and increases to $10^{18}$ for $N=30$. It was verified that increasing the frame size does not further deteriorate its conditioning: it instead saturates at $10^{18}$ even for large $N$.

\begin{table}[h]
\setlength\arrayrulewidth{1pt} 
\centering
\resizebox{\linewidth}{!}{%
\begin{tabular}{c c c c c c c c c c c}

\cline{1-11}
\multicolumn{1}{|c}{ Mode} &  \multicolumn{2}{|c}{ FE (AVSP) } & \multicolumn{2}{|c}{ FR ($N=10$) } & \multicolumn{2}{|c}{ OB ($N=10$) } & \multicolumn{2}{|c}{ FR ($N=30$) } & \multicolumn{2}{|c|}{ OB ($N=30$) } \\

\cline{2-11}
\multicolumn{1}{|c}{ } & \multicolumn{1}{|c}{ f (Hz)} &  \multicolumn{1}{c}{ $\sigma$ (s\textsuperscript{-1})} & \multicolumn{1}{|c}{f (Hz)} & \multicolumn{1}{c}{$\sigma$ (s\textsuperscript{-1})} & \multicolumn{1}{|c}{ f (Hz)} &  \multicolumn{1}{c}{ $\sigma$ (s\textsuperscript{-1})} & \multicolumn{1}{|c}{ f (Hz)} &  \multicolumn{1}{c}{ $\sigma$ (s\textsuperscript{-1})} & \multicolumn{1}{|c}{ f (Hz)} &  \multicolumn{1}{c|}{ $\sigma$ (s\textsuperscript{-1})} \\

\hline
\multicolumn{1}{|c}{1} & \multicolumn{1}{|c}{ $37.5$} &  \multicolumn{1}{c}{ $6.05$} & \multicolumn{1}{|c}{ $38.0$ } & \multicolumn{1}{c}{ $6.37$ } & \multicolumn{1}{|c}{$37.0$} &  \multicolumn{1}{c}{\cellcolor{black!25} $10.4$ ($72$\%)} & \multicolumn{1}{|c}{ $38.0$ } &  \multicolumn{1}{c}{\cellcolor{black!25} $7.0$ ($16$\%)} & \multicolumn{1}{|c}{$40.3$ } &  \multicolumn{1}{c|}{ \cellcolor{black!25} $7.7$ ($27$\%)} \\

\hline
\multicolumn{1}{|c}{2} & \multicolumn{1}{|c}{ $56.5$} &  \multicolumn{1}{c}{ $-1.08$} & \multicolumn{1}{|c}{ $56.4$} & \multicolumn{1}{c}{$-1.11$ } & \multicolumn{1}{|c}{$57.3$} &  \multicolumn{1}{c}{\cellcolor{black!25} $-1.9$ ($76$\%)} & \multicolumn{1}{|c}{ $56.5$} &  \multicolumn{1}{c}{ $-1.1$} & \multicolumn{1}{|c}{$56.4$} &  \multicolumn{1}{c|}{ $-1.07$} \\

\hline
\multicolumn{1}{|c}{3} & \multicolumn{1}{|c}{ $91.9$} &  \multicolumn{1}{c}{ $14.25$} & \multicolumn{1}{|c}{ $91.7$} & \multicolumn{1}{c}{$15.1$ } & \multicolumn{1}{|c}{$86.1$} &  \multicolumn{1}{c}{\cellcolor{black!25} $7.9$ ($46$\%)} & \multicolumn{1}{|c}{ $90.6$} &  \multicolumn{1}{c}{ $14.3$} & \multicolumn{1}{|c}{$91.5$} &  \multicolumn{1}{c|}{\cellcolor{black!25} $16.8$ ($18$\%)} \\

\hline
\multicolumn{1}{|c}{4} & \multicolumn{1}{|c}{ $111.6$} &  \multicolumn{1}{c}{ $0.0$} & \multicolumn{1}{|c}{ $111.6$} & \multicolumn{1}{c}{$0.0$ } & \multicolumn{1}{|c}{$111.6$} &  \multicolumn{1}{c}{ $0.0$} & \multicolumn{1}{|c}{ $111.6$} &  \multicolumn{1}{c}{ $0.0$} & \multicolumn{1}{|c}{$111.6$} &  \multicolumn{1}{c|}{ $0.0$} \\

\hline
\multicolumn{1}{|c}{5} & \multicolumn{1}{|c}{ $116.9$} &  \multicolumn{1}{c}{ $13.7$} & \multicolumn{1}{|c}{ $117.0$} & \multicolumn{1}{c}{$13.9$ } & \multicolumn{1}{|c}{$115.6$} &  \multicolumn{1}{c}{\cellcolor{black!25} $-34.0$ ($348$\%)} & \multicolumn{1}{|c}{ $117.5$} &  \multicolumn{1}{c}{ $12.7$} & \multicolumn{1}{|c}{$116.3$} &  \multicolumn{1}{c|}{\cellcolor{black!25} $15.1$ ($10$\%)} \\

\hline
\multicolumn{1}{|c}{6} & \multicolumn{1}{|c}{ $167.6$} &  \multicolumn{1}{c}{ $0.37$} & \multicolumn{1}{|c}{ $167.6$} & \multicolumn{1}{c}{$0.37$ } & \multicolumn{1}{|c}{$167.3$} &  \multicolumn{1}{c}{\cellcolor{black!25} $0.47$ ($27$\%)} & \multicolumn{1}{|c}{ $167.6$} &  \multicolumn{1}{c}{ $0.37$} & \multicolumn{1}{|c}{$167.6$} &  \multicolumn{1}{c|}{ $0.35$} \\

\hline
\multicolumn{1}{|c}{7} & \multicolumn{1}{|c}{ $271.2$} &  \multicolumn{1}{c}{ $1.44$} & \multicolumn{1}{|c}{ $271.4$} & \multicolumn{1}{c}{$1.5$ } & \multicolumn{1}{|c}{$270.7$} &  \multicolumn{1}{c}{\cellcolor{black!25} $2.95$ ($105$\%)} & \multicolumn{1}{|c}{ $271.4$} &  \multicolumn{1}{c}{ $1.5$} & \multicolumn{1}{|c}{$271.8$} &  \multicolumn{1}{c|}{\cellcolor{black!25} $1.8$ ($25$\%)} \\

\hline
\multicolumn{1}{|c}{8} & \multicolumn{1}{|c}{ $328.4$} &  \multicolumn{1}{c}{ $27.9$} & \multicolumn{1}{|c}{ $330.2$} & \multicolumn{1}{c}{$27.4$ } & \multicolumn{1}{|c}{$333.0$} &  \multicolumn{1}{c}{\cellcolor{black!25} $0.09$ ($100$\%)} & \multicolumn{1}{|c}{ $330.0$} &  \multicolumn{1}{c}{ $26.8$} & \multicolumn{1}{|c}{$329.5$} &  \multicolumn{1}{c|}{ $29.8$} \\

\hline
\multicolumn{1}{|c}{9} & \multicolumn{1}{|c}{ $454.9$} &  \multicolumn{1}{c}{ $0.02$} & \multicolumn{1}{|c}{ $454.8$} & \multicolumn{1}{c}{$0.02$ } & \multicolumn{1}{|c}{$454.7$} &  \multicolumn{1}{c}{\cellcolor{black!25} $0.04$ ($100$\%)} & \multicolumn{1}{|c}{ $454.9$} &  \multicolumn{1}{c}{ $0.02$} & \multicolumn{1}{|c}{$454.9$} &  \multicolumn{1}{c|}{\cellcolor{black!25} $0.03$ ($50$\%)} \\

\hline
\multicolumn{1}{|c}{10} & \multicolumn{1}{|c}{ $458.3$} &  \multicolumn{1}{c}{ $0.06$} & \multicolumn{1}{|c}{ $458.2$} & \multicolumn{1}{c}{$0.065$ } & \multicolumn{1}{|c}{$458.0$} &  \multicolumn{1}{c}{\cellcolor{black!25} $0.09$ ($50$\%)} & \multicolumn{1}{|c}{ $458.2$} &  \multicolumn{1}{c}{ $0.065$} & \multicolumn{1}{|c}{$458.2$} &  \multicolumn{1}{c|}{ $0.064$ } \\

\hline

\end{tabular}}
\caption{Frequencies ($\bm{f}$) and growth rates ($\bm{\sigma}$) for 10 of the first thermoacoustic eigenmodes of the annular combustor considered. FE (AVSP) results serve as reference. Low-order simulations are performed with FR/OB expansions, first onto $\bm{N=10}$ modes, and then onto $\bm{N=30}$ modes. Grey cells indicate frequencies/growth rates for which the relative error in comparison to AVSP is greater than 10\%. The corresponding values of the relative errors are written between parentheses.}\label{tab:atacamac_results}
\end{table}

\begin{figure}[h!]
\centering
\includegraphics[width=144mm]{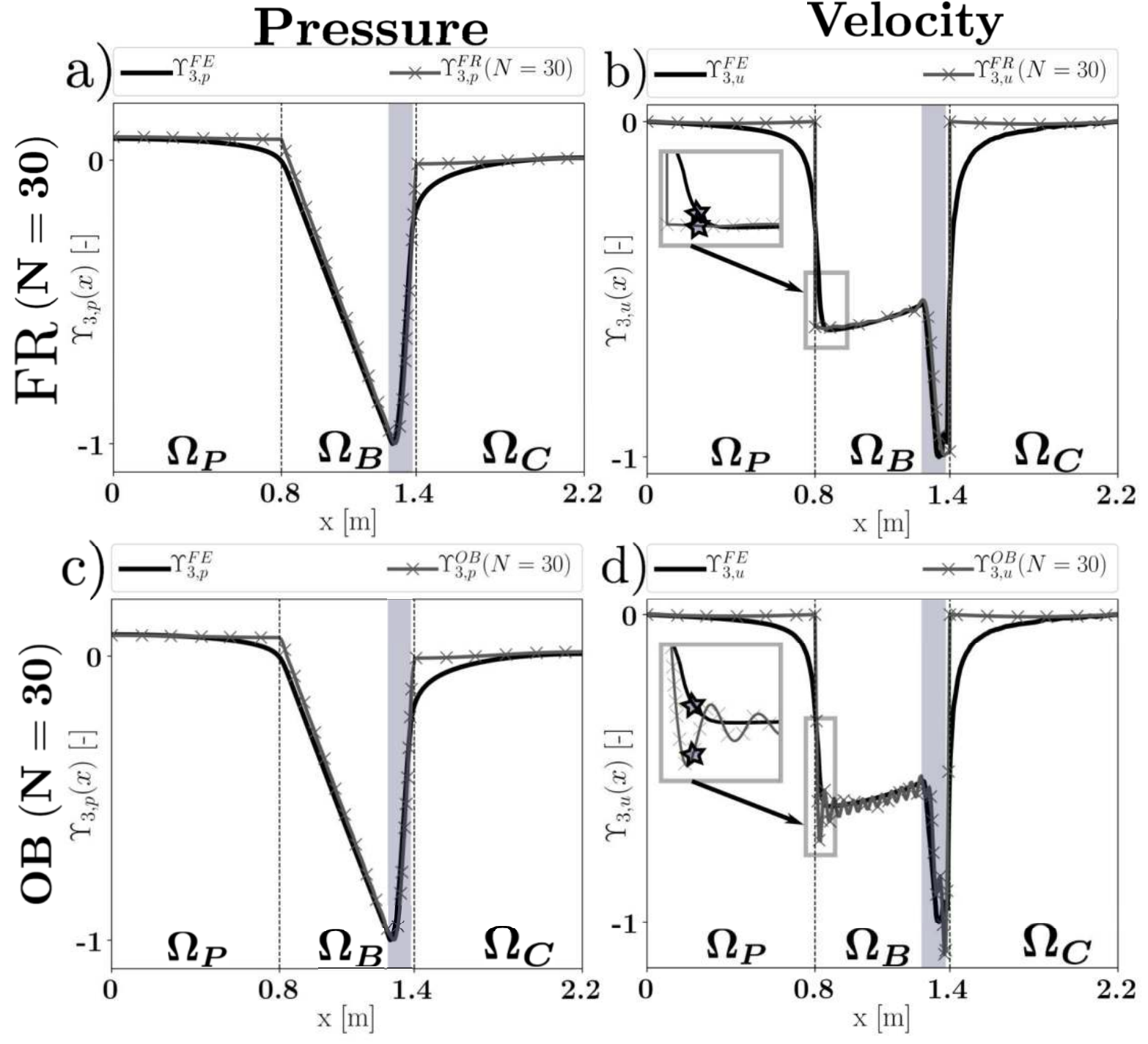}
\caption{Pressure and velocity mode shapes for Mode 3, plotted over a line extending from the bottom of the plenum to the end of the chamber and passing through a burner. Thick dark lines are results of AVSP computations. Grey lines with crosses are results of the LOM computations with $\bm{N=30}$ modes. Vertical dashed lines are the separation between the subdomains in the LOM network. Grey area represent the location of the active flames. First column: (a) pressure mode shape in the FR case; (c) pressure mode shape in the OB case. Second column: (b) velocity mode shape in the FR case, with a closeup view around the location of the flame reference point, which is represented by a star; (d) velocity mode shape in the OB case, with the same closeup view.}
\label{fig:atamac_modes}
\end{figure}

The relatively poor accuracy of the OB expansion LOM is explained by a closer examination of the modes shapes in the burners. Figure~\ref{fig:atamac_modes} shows the shape of the first unstable azimuthal mode of the combustor (Mode 3), plotted over a line starting from the bottom of the plenum, passing through a burner, and ending at the chamber outlet plane. With $N=30$, both FR and OB pressure mode shapes (Fig.~\ref{fig:atamac_modes}-(a),(c)) are relatively close to the AVSP computation, except for 3D effects in the neighborhood of the subdomains connections that cannot be captured. On the contrary, Fig.~\ref{fig:atamac_modes}-(d) shows that the OB expansion produces significant Gibbs oscillations of the velocity at both ends of the burner. Note that the Gibbs phenomenon is only present at the ends of the burners, and does not affect the velocity field within the chamber and the plenum. Conversely, Fig.~\ref{fig:atamac_modes}-(b) shows that the frame expansion successfully mitigates this Gibbs phenomenon. 
The conjugation of these spurious oscillations with the presence of active flames responding to velocity fluctuations explains the large discrepancies observed in the growth rates. Indeed, the closeup view displayed in Fig.~\ref{fig:atamac_modes}-(d) reveals that the flame reference point (represented by a star) lies in a region where the velocity is strongly affected by Gibbs oscillations. In contrast, the FR expansion (Fig.~\ref{fig:atamac_modes}-(b)) yields a reference velocity close to the AVSP reference velocity. As heat release fluctuations are directly proportional to the reference velocity (Eq.~\eqref{eq:FTF_model}), any misprediction of the velocity in the burner results in a potentially erroneous growth rate. Thus, should the point of reference lie in a region where numerical oscillations are present, the  computed thermoacoustic modes may strongly depend on unphysical and uncontrolled details such as the relative position of the point of reference and the Gibbs oscillations. An indicator for such dependence is the sensitivity of the LOM growth rates with respect to the flame reference point, defined as:
\begin{align}
\label{eq:ref_velocity_sensitivity}
\mathcal{S}_{\sigma}(\beta_0) = \dfrac{\beta_0}{\sigma_0} \left( \dfrac{\partial \sigma}{\partial \beta} \right)_{\beta = \beta_0}
\end{align}
This sensitivity coefficient is computed by performing another set of simulations with a slightly shifted flame reference point. For Mode 3, the FE solver and FR expansion LOM are weakly sensitive to the location of the reference point: $\mathcal{S}_{\sigma}^{(FE)} = 0.012$ and $\mathcal{S}_{\sigma}^{(FR)} = 0.017$. On the contrary, the OB expansion is highly sensitive to the location of the reference point: $\mathcal{S}_{\sigma}^{(OB)} = 0.27$. This sensitivity is not physical, but is rather a numerical artifact consequent of the Gibbs oscillations affecting the velocity field. Consequently, the orthogonal rigid-wall basis expansion results in a LOM that is strongly sensitive to the location of the flame reference point, which is a highly undesirable feature of a numerical model.\par

To conclude, the example considered in this section demonstrates the modularity of the LOM proposed, which can combine in a same thermoacoustic network active flames, one-dimensional subdomains (the burners), 2D subdomains (the plenum), and complex 3D subdomains of arbitrary shape (the chamber) for which the modal basis is numerically computed. Obviously, the approach is not limited to azimuthal eigenmodes, but is also able to capture any other form of thermoacoustic eigenmodes. It is also worth comparing the cost associated with the over-complete frame expansion LOM to existing low-order models. As shown above, $N=10$ modes were sufficient to achieve a satisfactory resolution (with error below 10\%) of the first 20 modes of the combustors (not all shown in Tab.~\ref{tab:atacamac_results}). The state-space of the whole system comprises 248 DoF: $2 \times 24$ for the plenum and the chamber, $4 \times 20$ for the straight ducts, $8 \times 3$ for the cross-section changes, and $4 \times 24$ for the active flames. After the preliminary computation of the chamber modal basis with AVSP (160 CPU seconds for 12 modes), the LOM computation of all the eigenmodes was performed in a few CPU seconds only. This is comparable to the 300 DoF necessary to treat a similar annular configuration in the work of Schuermans \textit{et al.}~\cite{bellucci2004,schuermans2003}. However, unlike this latter method, the present example did not assume acoustically compact injectors represented as lumped elements, and the acoustic field is fully resolved within the burners. LOMs relying on direct discretization of the flow domain, although more straightforward to put into application, result in more DoF and higher cost: Emmert \textit{et al.}~\cite{emmert2016} required $10^5$ DoF and 38 seconds on 8 CPUs to compute 5 eigenmodes of a 12-injectors annular geometry without any active flames. Finally, in contrast to mixed-method LOMs based on modal expansions alongside Riemann invariants $A^{+}/A^{-}$~\cite{li2015}, the proposed low-order model (Eq.~\eqref{eq:statespace_full}) can be directly integrated in time for temporal simulations of thermoacoustic acoustic instabilities, whereas the time-domain translation of Riemann invariants appears to be somehow constraining~\cite{meindl2016}.

\section{The spurious eigenmodes} \label{sec:frame_spurious_modes}

As mentioned in Sec.~\ref{sec:frame_expansion}, using the frame expansion comes at a price: its over-completeness may indeed result in poorly conditioned spurious components of the governing dynamical system (Eq.~\eqref{eq:press_final_equation_revisited}). This section discusses strategies that were implemented to automatically identify these spurious eigenmodes, and to ensure that they do not interact with other elements in the network.

\subsection{Energy-based identification criterion}
 
It is primordial to distinguish spurious modes from the physically meaningful ones, and this process needs to be automatic and robust. The interested reader is reported to~\cite{boyd2001} for further discussion regarding spurious components identification. The simplest method to achieve it is the \textit{brute force} approach: after obtaining the eigenmodes of the full-system with an expansion of size $N$, the computation is repeated with a different value of $N$. Eigenmodes that are very sensitive to the number of DoF are then considered as spurious.
However, the over-complete frame expansion presented in Sec.~\ref{sec:frame_expansion} was observed to produce only low energy spurious components. Therefore, a more efficient identification procedure based on an energetic criterion was implemented. This one is described below.\par

Let us consider an eigenvector $\mathbf{v} = {}^t (... \mathbf{v}_i  ...)$ of the  whole system dynamics matrix $\mathbf{A}^f$ (see Eq.~\eqref{eq:statespace_full}). The vector $\mathbf{v}_i = {}^t ( a_1 \ b_1 \ ...  \ a_N \ b_N )$ is the portion of the eigenvector $\mathbf{v}$ corresponding to the subdomain $\Omega_i$ state-space realization. More precisely, the coefficients $a_1,...,a_N$ are the state variables involved in the velocity mode shape in the subdomain $\Omega_i$ (they correspond to the time-dependent modal amplitudes $\Gamma_n(t)$), and the coefficients $b_1,...,b_N$ are the state variables associated to the pressure mode shape in the subdomain $\Omega_i$ (they correspond to the time-dependent modal amplitudes $\dot{\Gamma}_n(t)$). Then, the eigenmode pressure in the subdomain $\Omega_i$ is reconstructed as $\Upsilon_p(\vec{x}) = \sum_n b_n \phi_n (\vec{x}) = {}^t \mathbf{b} \ \pmb{\phi}(\vec{x}) $. The L-2 norm of the restriction of the full system's eigenmode $\Upsilon_p(\vec{x})$ to the subdomain $\Omega_i$ (\textit{i.e.} its energy) is given by:
\begin{align}
\label{eq:spurious_id_1}
||\Upsilon_p||_2^2 = \langle \Upsilon_p(\vec{x}) , \Upsilon_p(\vec{x}) \rangle = {}^t \mathbf{b} \ \langle \pmb{\phi}(\vec{x}) ,  {}^t \pmb{\phi}(\vec{x}) \rangle \ \mathbf{b} = {}^t \mathbf{b} \ \mathbf{\Lambda} \ \mathbf{b}
\end{align}
where $\mathbf{\Lambda}$ is the Gram matrix of the over-complete frame $(\phi_n(\vec{x}))$ for the subdomain $\Omega_i$. Therefore, an eigenmode $\Upsilon_p$ of the  whole system has a small energy, say lower than a threshold $\varepsilon_e$, if and only if:
\begin{align}
\label{eq:spurious_id_2}
\dfrac{1}{ | \mathbf{\Lambda} | \ | \mathbf{b} |^2} \  {}^t \mathbf{b} \ \mathbf{\Lambda} \ \mathbf{b} < \varepsilon_e
\end{align}
where the left-hand side has been non-dimensionalized by $| \mathbf{\Lambda}|$ and $| \mathbf{b} |^2$, the Frobenius norms of the Gram matrix and of the eigenvector $\mathbf{b}$, respectively.  Equation~\eqref{eq:spurious_id_2} shows that low energy eigenmodes of the full system are directly related to the poor conditioning of the Gram matrix $\mathbf{\Lambda}$. Indeed, if an orthogonal basis is used for the modal expansion, $\mathbf{\Lambda}$ is well-conditioned (it is in fact diagonal), and no eigenmodes can possibly satisfy Eq.~\eqref{eq:spurious_id_2}: there is therefore no spurious eigenmodes in this case. On the contrary, if an overcomplete frame is used, $\mathbf{\Lambda}$ becomes ill-conditioned, and vectors $\mathbf{b}$ lying in singular regions of the spectrum of the quadratic form associated to $\mathbf{\Lambda}$ can exist. Those satisfy the relation of Eq.~\eqref{eq:spurious_id_2} and are then considered as spurious eigenmodes. The threshold was empirically fixed to $\varepsilon_e = 10^{-4}$. However, the procedure showed little sensitivity to this parameter: changing the value of $\varepsilon_e$ to $10^{-5}$ or $10^{-3}$ did not affect the modes identified as spurious. This methodology was validated on a number of cases with available reference solutions. It is however not formally proved that it is able to differentiate spurious modes for any given system; if it happens to misidentify those for a particular case, the brute force method should be preferred.

 \subsection{Decoupling spurious eigenmodes from active flames} \label{sec:spurious_decoupling}
 
The energy-based criterion detailed above successfully identifies spurious eigenmodes in the absence of an active flame in the acoustic network. However, when the system comprises sources of heat-release, the spurious modes cannot be systematically distinguished from the physically-meaningful ones. In the worst cases, the frame expansion may even lead to significant errors on the physical thermoacoustic eigenmodes. This deterioration has a simple explanation: the heat-release source term essentially couples the \textit{pure} acoustic eigenmodes together (see right-hand side of Eq.~\ref{eq:press_final_equation_revisited}), therefore, if spurious features are included in the pure acoustic eigenmodes, the presence of a flame erroneously couple them to physical ones, which in turn corrupts the resolution of the thermoacoustic eigenmodes. A heat-release source can also feed energy to a spurious mode, which can render it strongly unstable. Thus, a \textit{decoupling} operation was implemented to alleviate the interference between the spurious modes and the flames.\par

Henceforth, the network components are divided into two distinct categories: the acoustic elements, denoted with a subscript ${}_{a}$, (\textit{e.g.} subdomains, complex impedance boundaries, multi-perforated liners, cross-section changes, etc.), and the flames, denoted with a subscript ${}_f$. In addition, the network components are ordered with the acoustic elements first, followed by the flames. Under this condition, the dynamics matrix of the whole system has the following block structure:
\begin{align}
\label{eq:dynamics_matrix_acous_flame}
\mathbf{A}^f = 
\left(
\begin{array}{cc}
\mathbf{A}_a & \mathcalbf{C}_{fl \rightarrow a} \\
\mathcalbf{C}_{a \rightarrow fl} & \mathbf{A}_{fl}
\end{array}
\right)
\end{align}
where $\mathbf{A}_a$ and $\mathbf{A}_{fl}$ are the dynamics matrices of the acoustic and flame elements, respectively, $\mathcalbf{C}_{fl \rightarrow a}$ contains the coupling terms from flames to acoustic elements (\textit{i.e.} the heat-release source terms), while $\mathcalbf{C}_{a \rightarrow fl}$ represents the coupling from the acoustic elements to the flames (\textit{i.e.} velocity or pressure fluctuations at the reference points). As previously explained, the spurious modes present in $\mathbf{A}_a$ corrupt the eigenvalues and eigenvectors of the dynamics matrix $\mathbf{A}^f$ through the coupling matrices $\mathcalbf{C}_{fl \rightarrow a}$ and $\mathcalbf{C}_{a \rightarrow fl}$. The decoupling operation therefore consists in applying correction terms to these latter matrices in order to dissociate the flames from the spurious acoustic eigenmodes. This procedure is articulated around 4 successive steps:
\begin{enumerate}
\item{The eigenvectors $\mathbf{v}_{a,n}$ and eigenvalues of $\mathbf{A}_a$ are resolved. The eigenvectors are gathered in the columns of a matrix $\mathbf{V}_{a}$. Note that $\mathbf{V}_{a}$ is a change-of-basis matrix that maps the natural state-space coordinates to coordinates in the spectral domain of $\mathbf{A}_{a}$. More precisely, for any vector $\mathbf{X}_{a}$ expressed in the natural state-space coordinates, $\left[ \mathbf{X}'_{a} \right]_n = \left[ \mathbf{V}_{a}^{-1} \mathbf{X}_{a}  \right]_n $ is the component of $\mathbf{X}_{a}$ on the n\textsuperscript{th} eigenvector $\mathbf{v}_{a,n}$ of $\mathbf{A}_a$.}
\item{The spurious acoustic eigenmodes are identified by applying the energy-based criterion to the columns of $\mathbf{V}_a$. We define the projection matrix $\mathcalbf{P}_{sp}$ that is diagonal, and whose elements are $0$ for physical modes and $1$ for spurious modes. A multiplication by $\mathcalbf{P}_{sp}$ corresponds to a projection onto the subspace spanned by the spurious eigenmodes.}
\item{The heat-release onto spurious modes is isolated; its expression in the $\mathbf{A}_a$-spectral coordinates system is $\mathcalbf{C}_{fl \rightarrow a}' = \mathcalbf{P}_{sp} \mathbf{V}_{a}^{-1} \mathcalbf{C}_{fl \rightarrow a}$. Similarly, the spurious velocity or pressure passed to the flames are expressed as $\mathcalbf{C}_{a \rightarrow fl}' = \mathcalbf{C}_{a \rightarrow fl} \mathbf{V}_{a} \mathcalbf{P}_{sp}$ in spectral coordinates.}
\item{The corrections on the coupling matrices in the natural state-space coordinates are defined by mapping $\mathcalbf{C}_{fl \rightarrow a}'$ and $\mathcalbf{C}_{a \rightarrow fl}'$ to the natural state-space coordinates, such that in the corrected dynamics matrix $\widetilde{\mathbf{A}}^f$ the off-diagonal blocks must be replaced with:
\begin{align}
& \widetilde{\mathcalbf{C}}_{fl \rightarrow a} = \mathcalbf{C}_{fl \rightarrow a} - \Re \left( \mathbf{V}_{a} \mathcalbf{P}_{sp} \mathbf{V}_{a}^{-1} \mathcalbf{C}_{fl \rightarrow a} \right)  \label{eq:decoupling_spurious_flames_1}\\
& \widetilde{\mathcalbf{C}}_{a \rightarrow fl} = \mathcalbf{C}_{a \rightarrow fl} - \Re \left( \mathcalbf{C}_{a \rightarrow fl} \mathbf{V}_{a} \mathcalbf{P}_{sp} \mathbf{V}_{a}^{-1} \right) \label{eq:decoupling_spurious_flames_2}
\end{align}}
\end{enumerate}
The first line essentially eliminates the receptivity of the spurious eigenmodes to the heat-release fluctuations. Conversely, the second line eliminates the flame receptivity to acoustic fluctuations stemming from spurious eigenmodes. Note the real parts in the right-hand side of Eq.~\eqref{eq:decoupling_spurious_flames_1} and Eq.~\eqref{eq:decoupling_spurious_flames_2}. Indeed, even though the original dynamics matrix $\mathbf{A}^f$ is real, its eigenvectors are complex-valued, such that the correction terms may also be complex. Therefore, with the purpose of conserving a real dynamics matrix, only the real parts of the correction terms are used.\par

This decoupling operation is essential for the accurate resolution of thermoacoustic eigenmodes in combustion systems comprising active flames. It is therefore applied in every reactive case presented in this work.

 \subsection{Artificial damping of spurious eigenmodes}
 
Uncoupling the spurious acoustic modes from the flames is usually sufficient to accurately resolve the thermoacoustic problem. However, for reactive cases where the flames have a large interaction index $n$ (e.g. in the example presented in Sec.~\ref{sec:example_intrinsic}), the decoupling procedure is unable to completely alleviate spurious features. A supplemental strategy was therefore implemented to artificially and selectively damp the spurious modes.\par

This artificial damping operation also relies on the block structure of the dynamics matrix $\mathbf{A}^{f}$ (see Eq.~\eqref{eq:dynamics_matrix_acous_flame}). It proceeds in 3 steps:
\begin{enumerate}
\item{As in Sec.~\ref{sec:spurious_decoupling} the eigenvectors $\mathbf{V}_a$ are computed, and the spurious modes are identified thanks to the energy-based criterion, which leads to the definition of the projection matrix $\mathcalbf{P}_{sp}$.}
\item{An artificial dissipation matrix $\pmb{\alpha}'_{ad}$ is defined in the $\mathbf{A}_a$-spectral coordinate system as $\pmb{\alpha}_{ad}' = \mathcalbf{P}_{sp} \pmb{\sigma}'_{ad}  \mathcalbf{P}_{sp}$, where $\pmb{\sigma}'_{ad}$ is diagonal with coefficients that can be set to arbitrarily large values.}
\item{The artificial damping is applied to the dynamics matrix $\mathbf{A}^{f}$ by mapping $\pmb{\alpha}_{ad}'$ to the natural state-space coordinates; $\mathbf{A}_{a}$ is then replaced with:
\begin{align}
\label{eq:artificial_damping}
\widetilde{\mathbf{A}}_a = \mathbf{A}_{a} - \Re \left(  \mathbf{V}_a \pmb{\alpha}_{ad}' \mathbf{V}_a^{-1}   \right) = \mathbf{A}_{a} - \Re \left(  \mathbf{V}_a \mathcalbf{P}_{sp} \pmb{\sigma}'_{ad}  \mathcalbf{P}_{sp} \mathbf{V}_a^{-1}   \right)
\end{align}}
\end{enumerate}
Ultimately, this artificial damping operation affects the growth-rates of the spurious eigenmodes by setting them to negative values of arbitrarily large magnitude, without affecting the physical modes. As a result, even though spurious modes may still be weakly coupled to active flames after applying the dissociation of Sec.~\ref{sec:spurious_decoupling}, the energy that they exchange with those is not sufficient to significantly affect their growth-rates, and they therefore remain strongly damped. Note that the real-parts in the right-hand side of Eq.~\eqref{eq:artificial_damping} are not mandatory, but are used here for convenience in order to retain a real-valued dynamics matrix.\par

Finally, after applying consecutively the decoupling operation of Sec.~\ref{sec:spurious_decoupling} and the artificial damping, the thermoacoustic problem comprising heat-release sources consists in solving for the eigenvalues and eigenvectors of the corrected dynamics matrix $\widetilde{\mathbf{A}}^f$, defined as :
\begin{align}
\label{eq:corrected_dynamics_matrix_full}
\widetilde{\mathbf{A}}^f  = 
\left(
\begin{array}{cc}
\widetilde{\mathbf{A}}_a & \widetilde{\mathcalbf{C}}_{fl \rightarrow a} \\
\widetilde{\mathcalbf{C}}_{a \rightarrow fl} & \mathbf{A}_{fl}
\end{array}
\right)
\end{align}
where $\widetilde{\mathbf{A}}_a$ is defined in Eq.~\eqref{eq:artificial_damping}, $\widetilde{\mathcalbf{C}}_{fl \rightarrow a}$ in Eq.~\eqref{eq:decoupling_spurious_flames_1}, and $\widetilde{\mathcalbf{C}}_{a \rightarrow fl}$ in Eq.~\eqref{eq:decoupling_spurious_flames_2}. Experience proved the addition of an artificial damping to be crucial in many systems including large interaction index heat-sources, and it is therefore systematically applied to reactive cases.

 \subsection{Singular Value Decomposition-based mitigation} \label{sec:spurious_svd_attenuation}

A fourth procedure is implemented to further allievate the spurious modes undesirable effects. However, since this one leads to notably improved results only in a few cases, it was not used in the results presented in this chapter, and was only used in some of the computations performed in Chapter~\ref{chap:complex_boundaries}. It is nonetheless described here to illustrate alternative mitigation strategies. Its principle is based on the Singular Value Decomposition (SVD) that is used to compute the Moore-Penrose pseudo-inverse~\cite{golub1965} of the frame Gram matrix $\mathbf{\Lambda}$.\par

Let us start by recalling the Gram matrix pseudo-inverse computation, for an over-complete frame of size $N$. As $\mathbf{\Lambda}$ is a symmetric positive matrix, its SVD writes:
\begin{align}
\label{eq:spurious_SVD_def}
\mathbf{\Lambda} = \mathcalbf{U} \mathbf{\Sigma} {}^t \mathcalbf{U} = 
\left( \mathcalbf{U}_1 \ \ \mathcalbf{U}_2  \right)
\left(
\begin{array}{cc}
\mathbf{\Sigma}_1 & \mathbf{0} \\
\mathbf{0} & \mathbf{\Sigma}_2
\end{array}
\right)
\begin{pmatrix}
{}^t \mathcalbf{U}_1 \\
{}^t \mathcalbf{U}_2
\end{pmatrix}
\end{align}
where $\mathcalbf{U}$ is a $N \times N$ unitary matrix whose columns are the singular vectors, and $\mathbf{\Sigma}$ is a $N \times N$ diagonal matrix with positive coefficients $\sigma_n$ called singular values, that are in addition sorted in decending order ($\sigma_1 \geq  \sigma_2 \geq \hdots \ \geq \sigma_N \geq 0$). In the right-hand side of Eq.~\eqref{eq:spurious_SVD_def} the matrices have been split into blocks such that $\mathbf{\Sigma}_1$ contains all the singular values greater than a given threshold $\varepsilon_{sv}$, and $\mathbf{\Sigma}_2$ contains the singular values lower than $\varepsilon_{sv}$. Accordingly, the columns of $\mathcalbf{U}_1$ are the singular vectors associated to the singular values in  $\mathbf{\Sigma}_1$. With these notations, $\mathbf{\Sigma}_1$ is of size $r \times r$ and $\mathcalbf{U}_1$ of size $N \times r$, where $r$ is the $\varepsilon_{sv}$-rank of $\mathbf{\Lambda}$. Usually the threshold $\varepsilon_{sv}$ is set to $\varepsilon_{sv} = \varepsilon_{w} N \sigma_1$, with $\varepsilon_{w}$ the working machine precision; this choice is followed here. A standard interpretation of the decomposition of Eq.~\eqref{eq:spurious_SVD_def} consists in considering that $\mathbf{\Sigma}_1$ and $\mathcalbf{U}_1$ are the dominant, well-conditioned singular values and vectors, whereas $\mathbf{\Sigma}_2$ and $\mathcalbf{U}_2$ are the noisy, ill-conditioned components~\cite{Brunton:2019}. More specifically, is be can shown that the Gram matrix condition number based on the Frobenius matrix norm is equal to $C(\mathbf{\Lambda}) = \sigma_1 / \sigma_N$, which indicates that the smaller singular values of $\mathbf{\Lambda}$ are responsible for the frame poor conditioning. The Moore-Penrose pseudo-inverse is then computed by considering the truncated SVD:
\begin{align}
\label{eq:Moore_penrose_pseudo_inv}
\mathbf{\Lambda}^+ = \mathcalbf{U}_1 \ \mathbf{\Sigma}_1^{-1} \  {}^t \! \mathcalbf{U}_1
\end{align}
Note that utilizing a truncated SVD in the pseudo-inverse computation is equivalent to considering a low-rank approximation of the frame $(\phi_n (\vec{x}))$. In the examples presented in this work, the Gram matrix inverse $\mathbf{\Lambda}^{-1}$ was simply replaced with the pseudo-inverse $\mathbf{\Lambda}^+$. Note the similarity of this procedure with the \textit{frame regularization} proposed in~\cite{adcock2019}, which essentially consists in discarding the frame ill-conditioned features.\par
 
Since $\mathcalbf{U}$ is a unitary matrix, the vector space $\mathbb{R}^n$ can be decomposed into two distinct subspaces: the subspace $Span(\mathcalbf{U}_1)$ spanned by the dominant singular vectors $\mathcalbf{U}_1$, and the space $Span(\mathcalbf{U}_2)$ spanned by the noisy components $\mathcalbf{U}_2$. The dynamical system of Eq.~\eqref{eq:press_final_equation_revisited} shows that the pressure evolution in the subdomain $\Omega_i$ is governed by surface source terms that includes the matrices $\pmb{\phi} (\vec{x}_s)$ and $\pmb{\nabla_s \phi} (\vec{x}_s)$. Conversely, the state-space realization of $\Omega_i$ has to output the velocity and pressure (or acoustic potential) on its boundaries, whose computations also involve these two matrices (see Appendix~\ref{sec:ss_realization_subdomain}). As $\pmb{\phi} (\vec{x}_s)$ and $\pmb{\nabla_s \phi} (\vec{x}_s)$ involve all the frame modes $\phi_n$, they include components on both  $Span(\mathcalbf{U}_1)$ and $Span(\mathcalbf{U}_2)$. Components on the second subspace are prone to interact with ill-conditioned features of the frame (\textit{e.g.} these source terms feed energy to modes or combination of modes lying in $Span(\mathcalbf{U}_2)$), and they therefore promote the apparition of spurious eigenmodes. Thus, the proposed mitigation strategy naturally consists in discarding the ill-conditioned components of the surface matrices $\pmb{\phi} (\vec{x}_s)$ and $\pmb{\nabla_s \phi} (\vec{x}_s)$, thanks to the projection operator on the subspace $Span(\mathcalbf{U}_1)$ given by $\mathcalbf{P}_{sv1} = \mathcalbf{U}_1 {}^t \mathcalbf{U}_1$. The surface matrices are then replaced with their respective projections on $Span(\mathcalbf{U}_1)$:
\begin{align}
\label{eq:spurious_svd_projections}
\widetilde{\pmb{\phi}} (\vec{x}_s) = \mathcalbf{U}_1 {}^t \mathcalbf{U}_1  \pmb{\phi} (\vec{x}_s) \ , \ \widetilde{\pmb{\nabla_s \phi}} (\vec{x}_s)  = \mathcalbf{U}_1 {}^t \mathcalbf{U}_1 \pmb{\nabla_s \phi} (\vec{x}_s)
\end{align}
This operation nullify the components of $\pmb{\phi} (\vec{x}_s)$ and $\pmb{\nabla_s \phi} (\vec{x}_s)$ on $Span(\mathcalbf{U}_2)$, without affecting those on $Span(\mathcalbf{U}_1)$. The projected surface matrices $\widetilde{\pmb{\phi}} (\vec{x}_s)$ and $\widetilde{\pmb{\nabla_s \phi}} (\vec{x}_s)$ can be used in the computation of both the inputs and the outputs of the subdomain $\Omega_i$. It can be seen as a decoupling procedure, comparable to the one presented in Sec.~\ref{sec:spurious_decoupling}, with the difference that it operates on the surface sources terms rather than the volume source terms, and that it is applied before any eigenmode resolution.\par

\paragraph{Remark 1} \mbox{}\\[-5mm]

\noindent The SVD can also be used to give an alternative insight into the energy-based spurious modes identification criterion of Eq.~\eqref{eq:spurious_id_2}. Using the fact that $\mathcalbf{U}$ is unitary, a change of basis is applied to the vector $\mathbf{b}$ which writes $\mathbf{b} = \mathcalbf{U} \mathbf{b}' =  \mathcalbf{U}_1 \mathbf{b}_1' + \mathcalbf{U}_2 \mathbf{b}_2'$, where $ \mathbf{b}_1'$ (resp. $\mathbf{b}_2'$) are the components of $\mathbf{b}'$ located in the subspace $Span(\mathcalbf{U}_1)$ (resp. $Span(\mathcalbf{U}_2)$). We then have the relation:
\begin{align}
\label{eq:spurious_svd_subspaces_relation}
 {}^t \mathbf{b} \ \mathbf{\Lambda} \ \mathbf{b} = {}^t \mathbf{b}'_1 \ \mathbf{\Sigma}_1 \ \mathbf{b}'_1 + {}^t \mathbf{b}'_2 \ \mathbf{\Sigma}_2 \ \mathbf{b}'_2 = \sum_{n = 1}^{r} (b'_n)^2 \sigma_n + \sum_{n = r+1}^{N} (b'_n)^2 \sigma_n
\end{align} 
If $\mathbf{b}$ verifies the criterion of Eq.~\eqref{eq:spurious_id_2}, using the facts that the singular values are sorted in descending order and that the columns of $\mathcalbf{U}$ are orthonormal vectors leads to:
\begin{align}
\label{eq:spurious_svd_criterion}
\vert \mathbf{b}'_2 \vert^2 > \dfrac{\sigma_r - \varepsilon_e \vert \mathbf{\Lambda} \vert }{\varepsilon_e \vert \mathbf{\Lambda} \vert - \sigma_N} \vert \mathbf{b}'_1 \vert^2
\end{align}
under the necessary condition $\varepsilon_e > \sigma_N / \vert \mathbf{\Lambda} \vert$. This inequality evidences that the spurious modes identified by the energy-based criterion essentially corresponds to eigenvectors with a significant portion of their weight lying in the noisy components subspace $Span(\mathcalbf{U}_2)$. This observation strengthens the coherence of the connection relating the apparition of low-energy spurious-modes to ill-conditioned features of the SVD.

\paragraph{Remark 2} \mbox{}\\[-5mm]

\noindent Let us consider the set of $r$ functions $(\Xi_n (\vec{x}) )_{1 \leq n \leq r }$ defined from the truncated SVD:
\begin{align}
\label{eq:spurious_symmetric orthogonalization}
\mathbf{\Xi} (\vec{x}) = \mathbf{\Sigma}_1^{-1/2} \ {}^t \mathcalbf{U}_1 \pmb{\phi} (\vec{x})
\end{align}
It is possible to prove that the functions $(\Xi_n (\vec{x}) )_{1 \leq n \leq r }$ are orthogonal with respect to the scalar product $< ., . >$. The linear transformation $\mathbf{\Sigma}_1^{-1/2} \ {}^t \mathcalbf{U}_1$ is therefore an orthogonalization operator, sometimes referred as symmetric or democratic L\"owdin orthogonalization in the field of quantum chemistry~\cite{Piela:2006}. Equation~\eqref{eq:spurious_symmetric orthogonalization} therefore shows that the over-complete frame $(\phi_n (\vec{x}))$ of size $N$ can be mapped to an orthogonal basis $(\Xi_n (\vec{x}) )_{1 \leq n \leq r }$ of smaller size $r$. Thus, the frame expansion method could be entirely reformulated by decomposing the acoustic pressure and velocity onto this orthogonal basis. Note however that unlike the functions $\phi_n (\vec{x})$, the elements $\Xi_n (\vec{x})$ are not acoustic modes of $\Omega_i$ (they are instead linear combinations of such eigenmdoes), and the reasoning of Sec.~\ref{sec:galerkin_expansion} would therefore require significant adaptations to be applied to the orthogonal basis $(\Xi_n (\vec{x}) )_{1 \leq n \leq r }$.

\section{Conclusion} \label{sec:frame_method_conclusion}

The misrepresentation of the acoustic field arising from modal expansions onto rigid-wall eigenmodes bases is a long-known issue in LOMs for thermoacoustics~\cite{culick2006}. This work proposes for the first time a reformulation of the classical Galerkin modal expansion making use of over-complete frames of eigenmodes. The analysis of canonical one-dimensional cases evidences a limitation of the rigid-wall modal expansion due to a Gibbs phenomenon affecting the velocity field near non-rigid-wall boundaries such as interfaces between subdomains composing the acoustic network. The Gibbs phenomenon significantly deteriorates the convergence and accuracy of the classical Galerkin expansion. This point is therefore a serious limitation that prevents the use of the rigid-wall Galerkin basis to construct acoustic networks composed of multiple subdomains representing complex geometries. On the contrary, the frame expansion does not produce Gibbs oscillations and displays superior convergence properties. It also has the potential to accurately represent the acoustic fields near any type of complex impedance boundaries, a characteristic that can be for instance used to impose anechoic conditions in order to study intrinsic thermoacoustic instabilities. Combined with the state-space formalism, the frame modal expansion can also be used to build elaborate and modular acoustic networks, thus enabling the low-order modeling of thermoacoustic instabilities in complex combustors. More precisely, it has the ability to combine in a same acoustic network highly heterogeneous classes of elements such as active flames, one-dimensional ducts, or complex 3D cavities. From a practical point of view, the implementation of the proposed method only requires minimal changes to existing algorithms based on rigid-wall expansions.\par

The main pitfall stemming from the frame expansion, lies in its over-completeness that may entail ill-conditioned features, which may produce spurious, non-physical dynamics of the pressure evolution. Thus, a specific inversion procedure is used to compute $\mathbf{\Lambda}^{-1}$, the inverse of the frame Gram matrix. This approach ensures that these spurious components stay negligible in comparison to physically meaningful components. More rigorously, it is demonstrated that spurious eigenmodes arising from the frame over-completeness have low energy, and a criterion is derived to systematically identify them. As these spurious eigenmodes may plague the entire outcome of the low-order modeling approach, a range of strategies intended to mitigate their undesirable effects is implemented.\par

\begin{figure}[h!]
\centering
\includegraphics[width=0.90\textwidth]{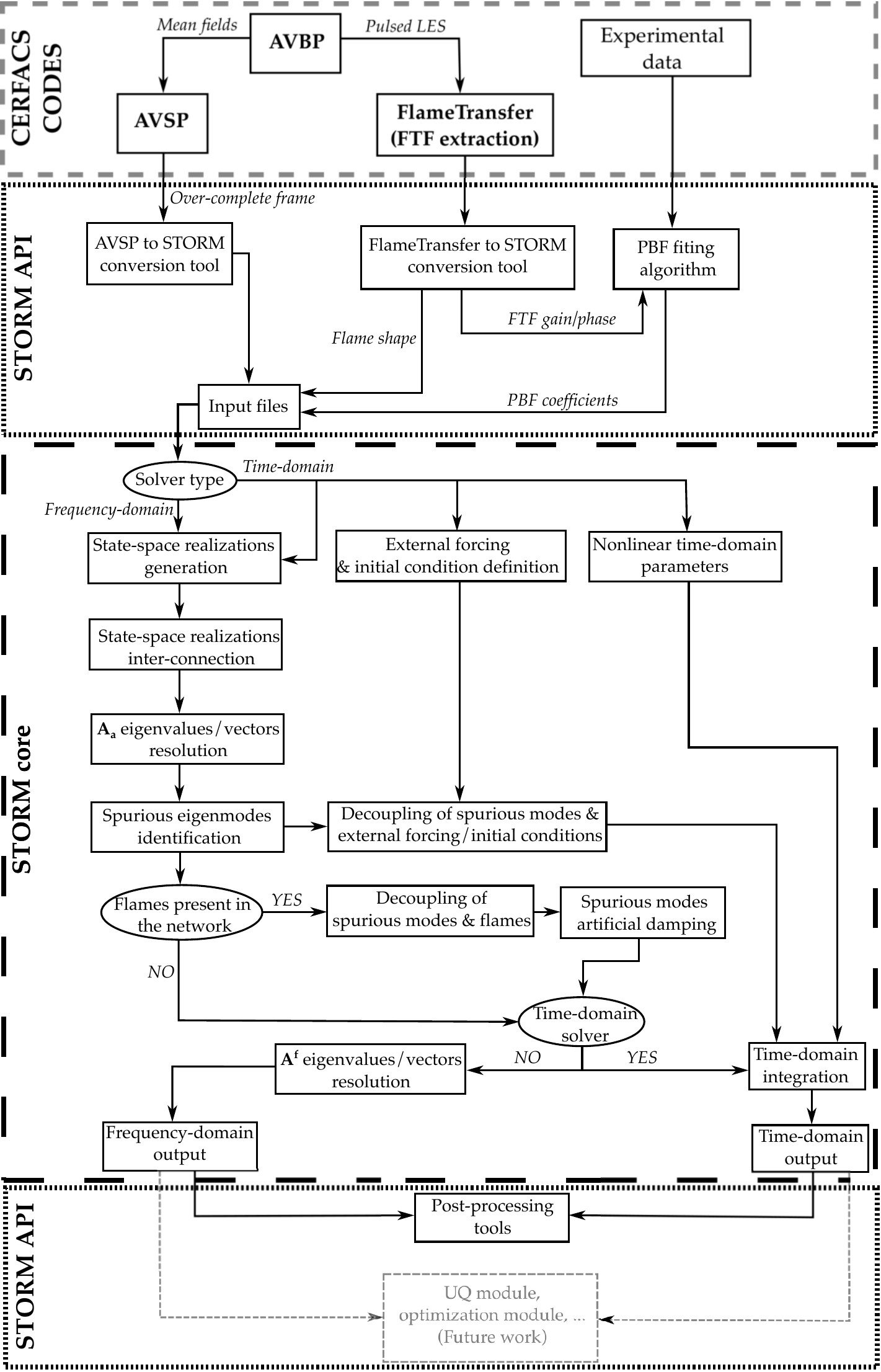}
\caption{Flowchart of the LOM code STORM, including the core numerical solvers, as well as its API through which pre- and post-processing tools, as well as interaction with other CERFACS codes, are managed.}
\label{fig:storm_flowchart}
\end{figure}

The frame modal expansion, as well as a library of state-space realizations for diverse acoustic elements (some of which are in Appendix~\ref{Appendix:A}), are implemented in a LOM code dubbed STORM. This code can be used as a frequency-domain or a time-domain solver (although not shown here) to resolve thermoacoustic instabilities in complex combustors. It is accompanied by a package of pre- and post-processing tools, intended to both ensure compatibility with existing software developed at CERFACS, and enhance its handiness, flexibility and modularity.  The use of these tools and their intercation with the numerical solvers is managed through an Application Programming Interface (API) currently under development. The flowchart displayed in Fig.~\ref{fig:storm_flowchart} summarizes this structure as well as the main steps in the core algorithm resolving the thermoacoustic problem.

				
\chapter{Including geometrically complex boundaries in thermoacoustic low-order models} \label{chap:complex_boundaries}
\minitoc				

\begin{chapabstract}
The previous chapter introduced the frame modal expansion, devised to satisfy any prescribed boundary condition that linearly relates the pressure and the velocity. However, the geometrical complexity inherent to boundaries usually encountered in industrial combustors, that can consists of wide and curved panels, is a major difficulty that was overlooked. This chapter therefore presents a novel method to include topologically complex boundary conditions in frame expansion-based LOMs. It starts by briefly reviewing previous studies that attempted to include complex boundaries in thermoacoustic LOMs. Then, the main issue in the low-order modeling of complex boundaries is evidenced through a two-dimensional example. More specifically, it shows that evaluating surface integrals that may appear in the model thanks to piece-wise approximations is the source of imprecisions that can be considerably amplified through the frame ill-conditioning. The core of the proposed method is then introduced: it lies in the modeling of complex shape boundaries as two-dimensional manifolds, where surface modal projections are used to expand acoustic variables onto an orthogonal basis of eigenmodes solutions of a curvilinear Helmholtz equation. The inclusion of acoustic impedance or Rayleigh conductivity into these surface projections enforces the proper conservation equations at the frontiers of the adjacent volumes. Resulting equations are formulated in the state-space formalism to be embedded into acoustic networks. A first non-reacting canonical test case, consisting of a multi-perforated liner in a cylindrical geometry is studied to assess the convergence and precision of the method. In a second example, the surface modal expansion method is used to model the effect on thermoacoustic instabilities of a partially reflecting outlet in an annular combustor, which is characteristic of an industrial gas turbine. This technique is expected to be one of the first to enable the inclusion of boundaries and liners of arbitrary geometrical complexity in modal projection-based thermoacoustic LOMs.
\end{chapabstract}

\section{Overview of existing methods} \label{sec:complex_BCs_overview}

As FEM Helmholtz solvers perform a relatively fine spatial discretization of the combustor geometry, the representation of complex shaped boundaries is not a particular difficulty. Many FEM Helmholtz solvers have the ability to account for complex impedance boundaries, the most often at the combustor inlet or outlet~\cite{nicoud2007,camporeale2011,laera2017,mensah2016,krebs2001}. State-of-the-art FEM Helmholtz solvers can also include multi-perforated liners located \textit{within} the chamber~\cite{gullaud2012,andreini2011,giusti2013}.\par

The representation of geometrically complex boundaries has proved to be more challenging in lower-order thermoacoustic models, and the related literature is rather sparse. The simplest approaches include pointwise impedances representing Helmholtz resonators fixed to a combustion chamber, a method that was used in Galerkin expansion-based LOMs~\cite{noiray2012,schuermans2003_th,bellucci2004,yang2019}. Most LOMs using the Riemann invariants, such as LOTAN~\cite{stow2001,Dowling2003_lotan} or OSCILOS~\cite{li2014,xia2018,yang2018}, also have the ability to incorporate generic impedance boundary conditions that can represent a chocked inlet or outlet for instance. However, as previously emphasized, they  suffer severe restrictions since they are limited to idealized perfectly annular combustors. Note that the modeling of multi-perforated liners is not reported in these works. Galerkin-expansion based LOMs, on the other hand, are restricted to rigid-wall conditions, and cannot accurately represent a non-trivial impedance prescribed on a complex surface. This limitation was relaxed by Ghirardo~\textit{et al.}~\cite{ghirardo2018}, who formulated the Galerkin projection in an annular combustor with a complex inlet impedance as a perturbation problem of that with a rigid-wall boundary. This technique is however only applicable to cases where the impedance value slightly deviates from a homogeneous Neumann boundary condition.\par

The frame expansion proposed in Chapter~\ref{chap:FRAME} is specifically designed to overcome the restrictions of the classical Galerkin expansion by accurately resolving the acoustic fields near any boundary where a non-trivial impedance is imposed. Although providing significant improvement in comparison with the classical rigid-wall modal expansion, it does not directly address the geometrical complexity of boundaries encountered in realistic combustion systems, which are usually curved surfaces of large extent. Interestingly, in the related fields of Vibro-acoustics and Structural acoustics, the use of the classical Galerkin expansion is the basis of the popular Acoustoelastic method~\cite{dowell1977,laulagnet1989,missaoui1997}, where cavities are coupled through thin vibrating membranes. On these surfaces, which are evidently not rigid-walls, the variables describing the solid vibrations are expanded onto an orthogonal basis of surface modes. However, most of these authors employ rigid-wall orthogonal bases to expend the acoustic variables in the fluid domain, and little attention is paid to the singularities that it may produce in the vicinity of the vibrating membranes. The present modeling strategy is strongly inspired by the Acoustoelastic method, as it combines the benefits of the surface modal expansion on the boundaries, to that of the over-complete frame expansion in the fluid domain interior. To the knowledge of the author, this is one of the first attempt to include advanced boundary conditions in LOMs utilizing modal expansions for the prediction of thermoacoustic instabilities in complex geometries.\par

\section{The surface modal expansion method} \label{sec:surface_modal_expansion}

\begin{figure*}[h!]
\centering
\includegraphics[width=80mm]{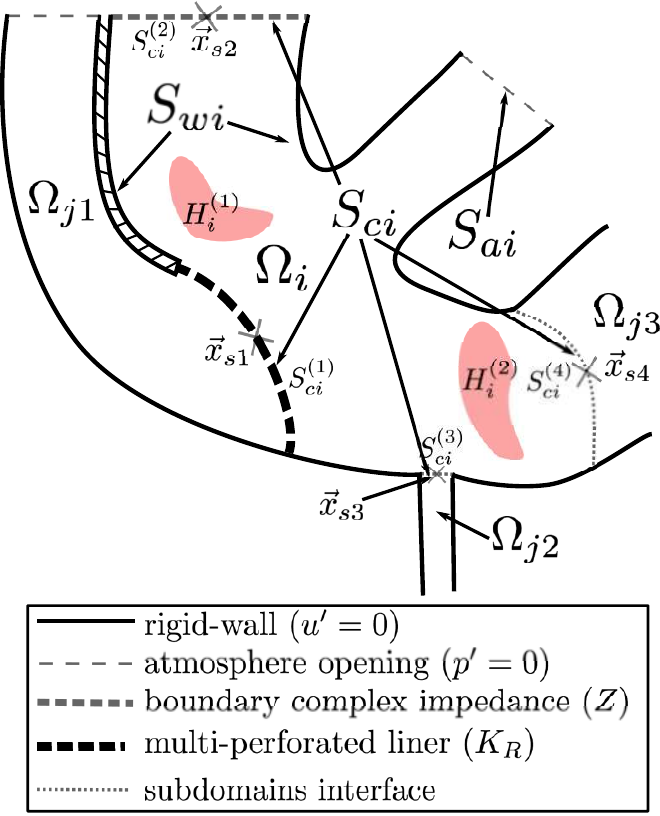}
\caption{Schematic of an acoustic network. The system is divided into a set of simpler subsystems comprising volume subdomains ($\bm{\Omega_i}$, $\bm{\Omega_{j1}}$, $\bm{\Omega_{j2}}$), heat sources ($\bm{H_i^{(1)}}$, $\bm{H_i^{(2)}}$), and subdomains complex boundaries ($\bm{S_{ci}^{(1)}}$,
$\bm{S_{ci}^{(2)}}$, $\bm{S_{ci}^{(3)}}$, $\bm{S_{ci}^{(4)}}$). The subdomain $\bm{\Omega_i}$ frontier is split into its rigid-wall boundary $\bm{S_{wi}}$, its boundary opened to the atmosphere $\bm{S_{ai}}$, and its connection boundary $\bm{S_{ci}}$. This latter consists of $\bm{M_S=4}$ subsurfaces including simple interfaces ($\bm{S_{ci}^{(3)}}$, $\bm{S_{ci}^{(4)}}$), multi-perforated liners ($\bm{S_{ci}^{(3)}}$), complex outlet impedance ($\bm{S_{ci}^{(3)}}$), and any other type of boundary that is neither a rigid-wall nor an opening to the atmosphere. The center of each subsurface $\bm{S_{ci}^{(m)}}$ is located at $\bm{\vec{x}_{sm}}$.}
\label{fig:network}
\end{figure*}

For convenience, the decomposition of a complex thermoacoustic system into a network of smaller subsystems is recalled in Fig.~\ref{fig:network}).
These subsystems can be sorted in distinct classes: volume subdomains (\textit{e.g.} $\Omega_i$), complex subdomains boundaries (\textit{e.g.} $S_{ci}^{(1)}$ to $S_{ci}^{(4)}$, located on the frontier of $\Omega_i$), and heat sources (\textit{e.g.} $H_i^{(1)}$ and $H_i^{(2)}$, contained within $\Omega_i$).\par

Let us consider a volume subdomain $\Omega_i$ similar to the one in Fig.~\ref{fig:network}. In the most general case, its boundary $\partial \Omega_i$ can be decomposed as: $\partial \Omega_i = S_{wi} \cup S_{ai} \cup S_{ci}$, where $S_{wi}$ is a rigid-wall ($u' = 0$), $S_{ai}$ is opened to the atmosphere ($p'=0$), and $S_{ci}$ is a connection boundary containing any complex frontier that is neither rigid-wall nor pressure release. More precisely, $S_{ci}$ contains boundaries with the exterior characterized by a finite impedance $Z$ (\textit{e.g.} chocked inlet or outlet), boundaries between two subdomains with a Rayleigh conductivity $K_R$ (\textit{e.g.} multi-perforated liner), and simple subdomains interfaces. The boundary $S_{ci}$ can be further split into $M_S$ subsurfaces $S_{ci}^{(m)}$, as depicted in Fig.~\ref{fig:network}. We recall that the expansion of the acoustic pressure and velocity in $\Omega_i$ onto an over-complete frame $(\phi_n (\vec{x}))$ writes:
\begin{align}
\label{eq:pressure_expansion_reminder}
\begin{aligned}
&p \td = \sum_{n=1}^{\infty} \dot{\Gamma}_n(t) \phi_n (\vec{x}) = {}^t \dot{\mathbf{\Gamma}}(t) \pmb{\phi} (\vec{x}) \\
& \vec{u} \td = - \dfrac{1}{\rho_0} \sum_{n=1}^{\infty} \Gamma_n(t) \vec{\nabla} \phi_n (\vec{x}) = - \dfrac{1}{\rho_0} {}^t \mathbf{\Gamma} (t) \boldsymbol{\vec{\nabla}} \pmb{\phi} (\vec{x})
\end{aligned}
\end{align}
We also remind the outcome of the frame expansion applied to the acoustic pressure in $\Omega_i$. It yields the dynamical system governing the temporal evolution of the pressure modal amplitudes $\Gamma_n(t)$, under the surface forcing from subsystems adjacent to $\Omega_i$, and the volume forcing from the $M_H$ heat sources $H_i^{(l)}$ contained within $\Omega_i$:
\begin{align}
\label{eq:subdomain_dynamical_system}
\begin{aligned}
& \ddot{\Gamma}_n(t) =  -  \alpha \dot{\Gamma}_n (t) - \omega_n^2 \Gamma_n (t)\\
& +  \sum_{m=1}^{M_S}  \iint_{S_{ci}^{(m)}} \rho_0 c_0^2 \left(  \varphi^{S_{ci}^{(m)}} (\vec{x}_s,t)  \left[ \mathbf{\Lambda}^{-1}  \boldsymbol{\nabla_s} \pmb{\phi} (\vec{x}_s) \right]_n - u_s^{S_{ci}^{(m)}} (\vec{x}_s,t) \left[ \mathbf{\Lambda}^{-1} \pmb{\phi} (\vec{x}_s) \right]_n  \right) d^2 \vec{x}_s   \\
& + \sum_{l=1}^{M_H} (\gamma -1) Q_l(t) \left[  \mathbf{\Lambda}^{-1} \langle \pmb{\phi} , \mathcal{H}_{i}^{(l)} \rangle   \right]_n
\end{aligned}
\end{align}
In this equation, $\varphi^{S_{ci}^{(m)}} (\vec{x}_s,t)$ and $u_s^{S_{ci}^{(m)}} (\vec{x}_s,t)$ are respectively the acoustic potential and normal velocity in the adjacent complex boundary subsystems $S_{ci}^{(m)}$. Note a major difference with Eq.~\eqref{eq:press_final_equation_revisited}: the surface integral in the right-hand side of Eq.~\eqref{eq:subdomain_dynamical_system} is indeed not evaluated through a piece-wise approximation. The accurate representation of these boundary terms is the object of the method proposed in this chapter.

\subsection{The inadequacy of the surface integral piece-wise approximation} \label{sec:inadequacy_surface_discretization}

Let us now consider a complex boundary subsystem $S_{ci}^{(m)}$ that is located at a frontier of the subdomain $\Omega_i$ either with another subdomain (\textit{e.g.} $S_{ci}^{(1)}$, $S_{ci}^{(3)}$, $S_{ci}^{(4)}$ in Fig.~\ref{fig:network}) or with the exterior (\textit{e.g.} $S_{ci}^{(2)}$ in Fig.~\ref{fig:network}). As shown through Eq.~\eqref{eq:subdomain_dynamical_system}, the potential and velocity in $S_{ci}^{(m)}$ partly govern the acoustics in $\Omega_i$; it is therefore necessary to accurately evaluate the surface integrals in the right-hand side to enforce the proper subsystem coupling. In the derivation presented in Chapter~\ref{chap:FRAME}, the surface area $\Delta S_{ci}^{(m)}$ of the connection boundaries $S_{ci}^{(m)}$ were assumed small (which is true in the case of a narrow duct giving on a large cavity such as $S_{ci}^{(3)}$ in Fig.~\ref{fig:network}). The acoustic variables can then be considered uniform over the boundary, and the surface integrals in Eq.~\eqref{eq:subdomain_dynamical_system} can be approximated with their value at the center-point $\vec{x}_{sm}$:

{\footnotesize  
\setlength{\abovedisplayskip}{6pt}
\setlength{\belowdisplayskip}{\abovedisplayskip}
\setlength{\abovedisplayshortskip}{0pt}
\setlength{\belowdisplayshortskip}{3pt}
\begin{align}
\label{eq:surf_int_approximation}
\iint_{S_{ci}^{(m)}} \rho_0 c_0^2 u_s^{S_{ci}^{(m)}} (\vec{x}_s,t) \left[ \mathbf{\Lambda}^{-1} \boldsymbol{\phi} (\vec{x}_s) \right]_n  d^2 \vec{x}_s  \approx  \rho_0 c_0^2 \Delta S_{ci}^{(m)} u_{s}^{S_{ci}^{(m)}} (\vec{x}_{sm},t) \left[ \mathbf{\Lambda}^{-1} \boldsymbol{\phi} (\vec{x}_{sm}) \right]_n
\end{align}}%
and similarly for the second surface integral. If $S_{ci}^{(m)}$ is large, it was proposed to further subdivide it into smaller surface elements, which leads to a piece-wise approximation of the surface integral of Eq.~\eqref{eq:surf_int_approximation}.\par

\begin{figure}[h!]
\centering
\includegraphics[width=0.8\textwidth]{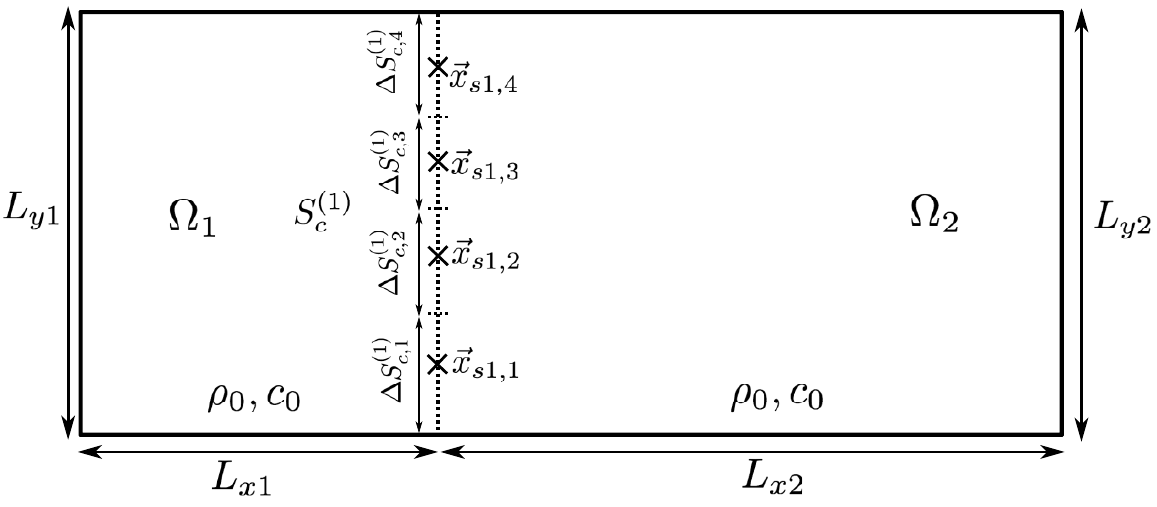}
\caption{An acoustic network comprising 2 rectangular subdomains $\bm{\Omega_1}$ and $\bm{\Omega_2}$, separated by an interface $\bm{S_c^{(1)}}$, which is subdivided into 4 surface elements $\bm{\Delta S_{c,j}^{(1)}}$ located at the points $\bm{\vec{x}_{s1,j}}$. The density and sound speed fields are uniform over the entire domain, and the outer boundaries are rigid-walls.} 
\label{fig:schematic_bad_results}
\end{figure}

This approach is illustrated on the simple two-dimensional rectangular system represented in Fig.~\ref{fig:schematic_bad_results}.\par
For both subdomains $\Omega_1$ and $\Omega_2$, the surface source terms on the interface $S_{c}^{(1)}$ are expressed thanks to a piece-wise integral approximation similar to that of Eq.~\eqref{eq:surf_int_approximation}. In the present case, $S_{c}^{(1)}$ is subdivided into 4 equal surface elements $\Delta S_{c,j}^{(1)}$ regularly distributed along the vertical direction. The network also integrates 4 components similar to the cross-section change of Sec.~\ref{sec:example_1Dduct_cross_section_change} (with $S_1 = S_2$), to enforce the continuity of pressure and flux at the locations $\vec{x}_{s1,j}$ of the surface elements $\Delta S_{c,j}^{(1)}$. Note that the surface integral piece-wise approximation raises a number of subsequent questions regarding the number, the size, and the locations of the elements necessary to achieve an accurate representation of the surface integrals. Most importantly, even though this spatial discretization may work for simple subdomains $\Omega_i$ where the frame is obtained analytically, it usually fails to produce acceptable results if the frame is generated through an FE solver. 
\begin{figure}[h!]
\centering
\includegraphics[width=0.8\textwidth]{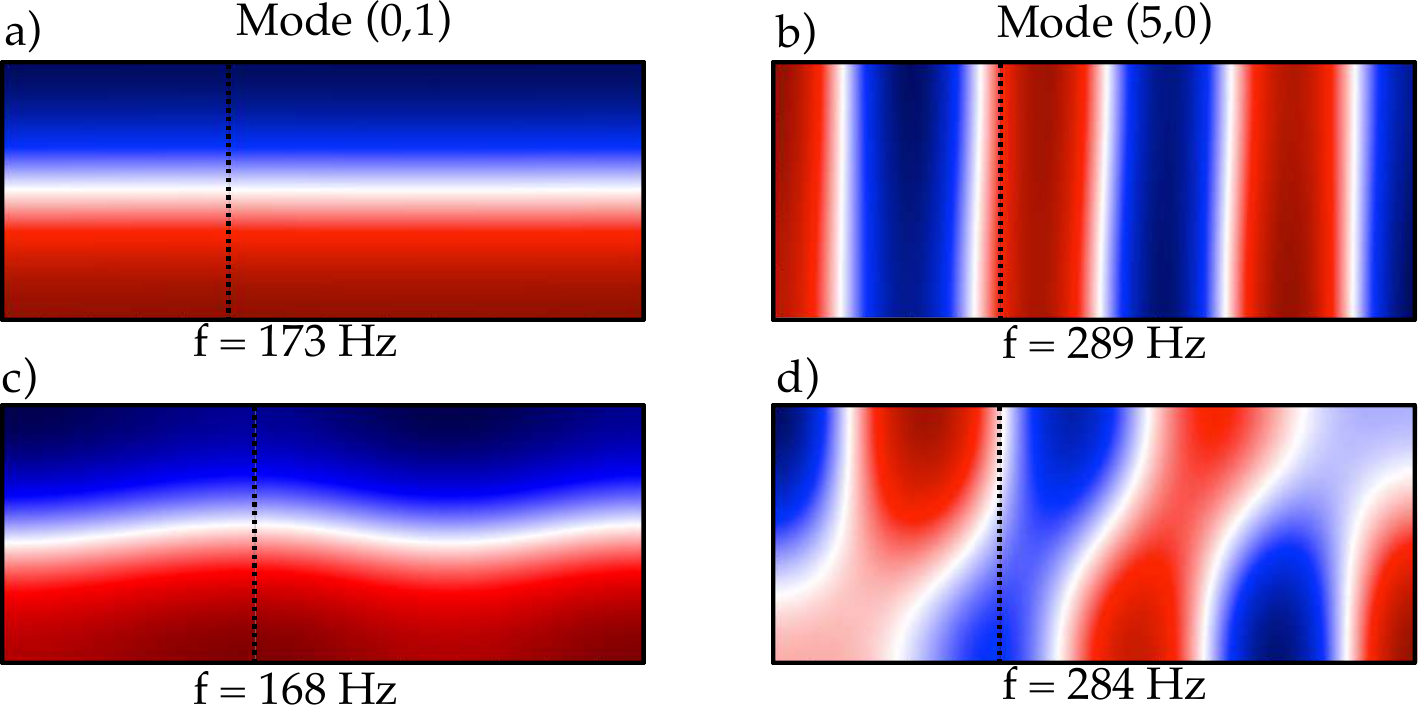}
\caption{Comparison between the pressure mode shapes of two of the first modes in the two-dimensional rectangular cavity of Fig.~\ref{fig:schematic_bad_results}. (a) and (b): the frame expansions in $\bm{\Omega_1}$ and $\bm{\Omega_2}$ are analytical. (c) and (d): the frame expansions are numerically generated with a FE Helmholtz solver. In both cases, the frame size is $\bm{N = 40}$ in $\bm{\Omega_1}$ and $\bm{\Omega_2}$. The theoretical frequencies are $\bm{f_{0,1} = 173}$~Hz and $\bm{f_{5,0} = 289}$~Hz.}
\label{fig:bad_results}
\end{figure}
This point was verified by performing two computations of the acoustic network of Fig.~\ref{fig:schematic_bad_results}: a first one with analytical frames in both subdomains $\Omega_1$ and $\Omega_2$, and a second one with frames numerically generated thanks to a FE Helmholtz solver. In the former case, the frame expansion in $\Omega_1$ writes:
\begin{align}
\label{eq:2Drect_frame}
\begin{aligned}
\left(\phi_n^{(1)} (x) \right)_{  n \leqslant N} & = \left( \cos \left(\dfrac{n \pi x_1}{L_{x_1}} \right) \cos \left(\dfrac{m \pi y_1}{L_{y_1}} \right)  \right)_{  n \leqslant N/2} \\
& \bigcup \left( \sin \left(\dfrac{ n \pi x_1}{L_{x_1}} \right)  \cos \left(\dfrac{m \pi y_1}{L_{y_1}} \right) \right)_{ n \leqslant N/2}
 \end{aligned}
\end{align}
It is devised to verify any type of boundary condition on the right side of the rectangle. An analogous frame is used in $\Omega_2$. The analytical frame approach yields results in excellent agreement with the theory (Fig.~\ref{fig:bad_results}-(a,b)). In contrast, frames constructed with the FE solver lead to a spectacular deterioration of the results (Fig.~\ref{fig:bad_results}-(c,d)). Increasing either the frame size $N$, or the number of surface subdivisions degrades even further the LOM accuracy. This limitation has a simple interpretation: values at the center-points $\vec{x}_{sm}$ used in the surface integral piece-wise evaluation are inevitably affected by noise due to numerical approximations. In addition, the frame modes are themselves subjected to errors stemming from the FEM Helmholtz solver numerical methods. These minute perturbations can then be uncontrollably amplified due to the frame ill-conditioning (we remind that $\mathbf{\Lambda}^{-1}$ contains very large terms), such that Eq.~\eqref{eq:surf_int_approximation} is highly sensitive to the mesh, as well as other numerical parameters used in the computation of the frame $(\phi_n (\vec{x}))_{n \geq 1}$. It is also strongly sensitive to the locations and the sizes of the surface elements used in the approximation of the integrals. Thus, a spatial discretization of $S_{ci}^{(m)}$ into smaller subsurfaces, combined with the poor conditioning of $\mathbf{\Lambda}$, leads to large and erroneous surface source terms in Eq.~\eqref{eq:subdomain_dynamical_system}, which ultimately compromises the entire method. The piece-wise approximation of the surface integrals employed in Chapter~\ref{chap:FRAME} is therefore inadequate to model topologically complex boundaries that are usually encountered in industrial combustors.\par

\subsection{Preliminary result: the curvilinear Helmholtz-equation} \label{sec:curvilinear_helmholtz}

To circumvent this pitfall, we propose to express the surface quantities $u_s^{S_{ci}^{(m)}} (\vec{x}_s,t)$, $\varphi^{S_{ci}^{(m)}} (\vec{x}_s,t) $, $\boldsymbol{\nabla_s} \pmb{\phi} (\vec{x}_{s})$, and $\pmb{\phi} (\vec{x}_{s})$ thanks to a \textit{spectral} discretization rather than a spatial one. This spectral discretization is inspired by the Acoustoelastic method~\cite{laulagnet1989,missaoui1997,dowell1977}, where acoustic variables in cavities are coupled with thin vibrating membranes: dynamics variables in these membranes are projected on a basis of eigenmodes solutions of the Kirchoff-Love equation of shells~\cite{timoshenko1959}. The present approach does not rely on thin vibrating membranes, but instead on a curvilinear Helmholtz equation. The goal of this section is therefore to introduce this curvilinear Partial Differential Equation governing the acoustics in a thin shell-like volume, which constitutes a prerequisite used in the subsequent steps of the method.\par

Let us first consider a shell-like control volume of thickness $2 L$, parameterized by the curvilinear coordinates $(\alpha,\beta)$, and by the normal coordinate $\xi$ along the direction normal to its surface. It is assumed that the curvilinear coordinates orientation corresponds to the principal direction of curvatures. Those are noted $\kappa_1 (\alpha,\beta)$ and $\kappa_2 (\alpha,\beta)$, and the corresponding radii of curvature are $R_1 (\alpha,\beta)$ and $R_1 (\alpha,\beta)$. The mean curvature is $\kappa_m ( \alpha, \beta) = (\kappa_1 + \kappa_2)/2$. The Lam\'{e} parameters $A$ and $B$ are also introduced: they are metric coefficients relating a small variation $(d \alpha, d \beta)$ in curvilinear coordinates to a variation $(ds_1, ds_2)$ in arc length:
\begin{align}
\label{eq:definition_Lame}
\left\{ \begin{aligned}
& ds_1 = A (\alpha, \beta) \left(1 - \dfrac{\xi}{R_1 (\alpha, \beta)} \right) d \alpha  = \left(1 - \dfrac{\xi}{R_1} \right) ds_1^0   \\
& ds_2 = B (\alpha, \beta) \left(1 - \dfrac{\xi}{R_2 (\alpha, \beta)} \right) d \beta = \left(1 - \dfrac{\xi}{R_2} \right) ds_2^0
\end{aligned}\right.
\end{align}
where $ds_1^0 $ and $ds_1^0$ are the arc-lengths on the middle surface located at $\xi = 0$. In addition, it is supposed that the shell-like domain is sufficiently thin, or equivalently weakly curved, such that $L \kappa_1, L \kappa_2 \ll 1$. Terms of order $O (L \kappa_1)$ and $O (L \kappa_1)$ are therefore discarded in the following. In the absence of any dissipation phenomena, the acoustic flow in the control volume is solution of the zero-Mach number linearized Euler equations, and the mean physical parameters $\rho_0 (\alpha ,\beta)$ and $c_0 (\alpha , \beta)$ are assumed independent of the normal coordinated $\xi$. The conservation equations write:
\begin{align}
\label{eq:euler_linearized_forCurv}
\left\{
\begin{aligned}
& \rho_0 \pdv{\vec{u}}{t} = - \vec{\nabla} p \\
& \pdv{\rho}{t} = - \rho_0 \vec{\nabla} . \vec{u}
\end{aligned}
\right.
\end{align}\par

\begin{figure}[h!]
\centering
\includegraphics[width=0.9\textwidth]{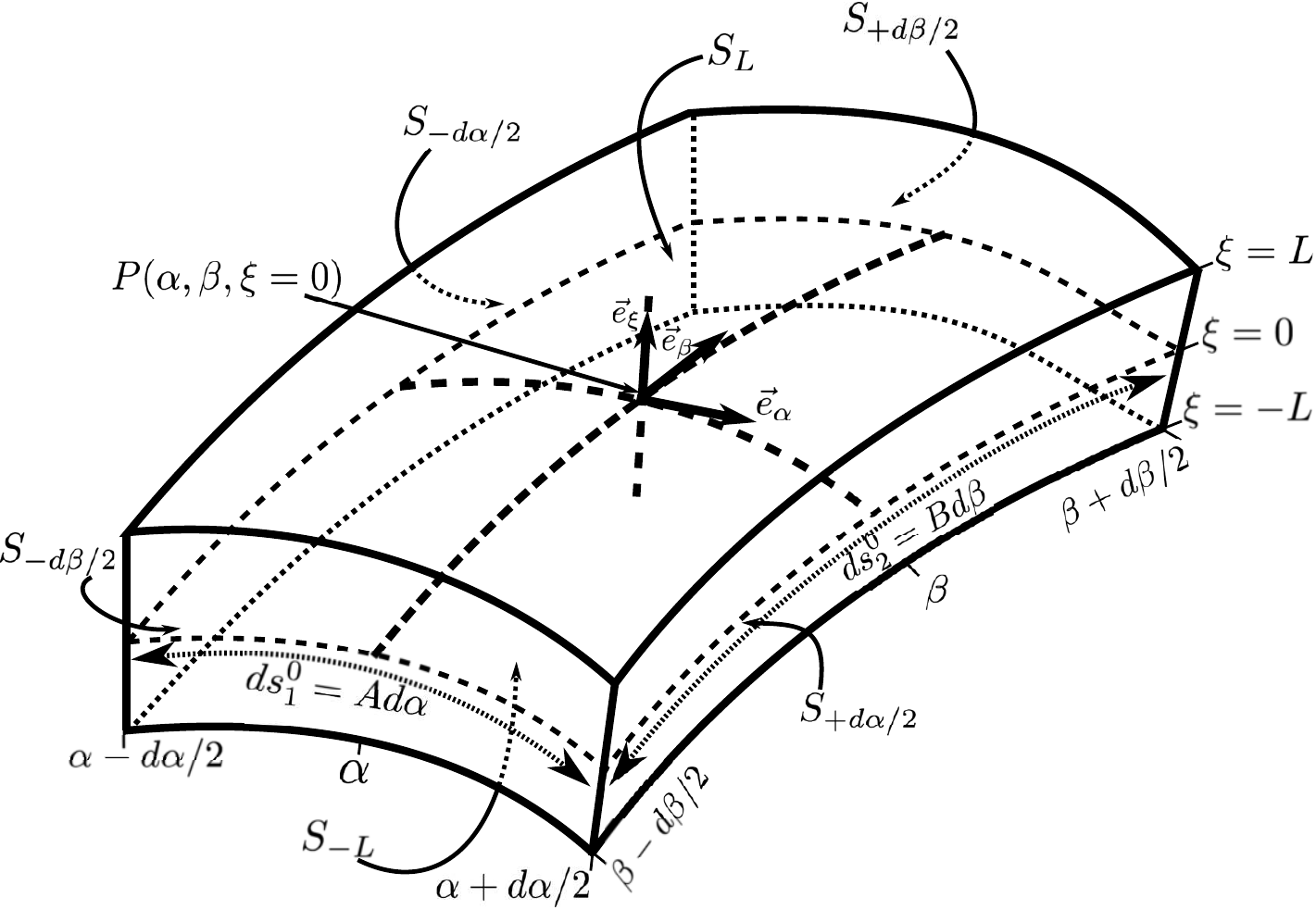}
\caption{Schematic of a volume element $\bm{d \mathscr{V}}$, parameterized by the curvilinear coordinates and by the normal coordinate $\bm{\xi}$. Its center is the point $\bm{P(\alpha,\beta, \xi = 0)}$ located on its middle-surface, and it is comprised in the ranges $\bm{\left[\alpha - d \alpha / 2, \alpha + d \alpha / 2 \right]}$, $\bm{\left[\beta - d  \beta /2 , \beta +d  \beta /2 \right]}$, and $\bm{\left[-L, L  \right]}$. Its lateral boundaries are noted according to their relative positions with respect to the center point $\bm{P}$: $\bm{S_{-d \alpha}}$, $\bm{S_{+d \alpha}}$, $\bm{S_{- d \beta}}$, $\bm{S_{+ d \beta}}$, $\bm{S_{-L}}$, and $\bm{S_{L}}$. The Lam\'{e} parameters relating $\bm{d \alpha}$ and $\bm{d \beta}$ to the middle surface arc lengths $\bm{ds_1^0}$ and $\bm{ds_2^0}$, respectively, are also displayed.}
\label{fig:element_shell}
\end{figure}

As shown in Fig.~\ref{fig:element_shell}, a small portion $d \mathscr{V}$ is now isolated from the shell-like control volume.
The volume of $d \mathscr{V}$ can be approximated at first order by $2ABL d \alpha d \beta$. Applying the divergence definition, the following relations are obtained:
\begin{align}
\label{eq:divergence_unit_vectors}
\nabla . \vec{e}_{\alpha} = \dfrac{1}{AB} \dfrac{\partial B}{\partial \alpha} \ , \ \nabla . \vec{e}_{\beta} = \dfrac{1}{AB} \dfrac{\partial A}{\partial \beta} \ , \ \nabla . \vec{e}_{\xi} = - 2 \kappa_m
\end{align}
The conservation of mass and momentum in the directions $\alpha, \beta, \xi$ are now integrated over $d \mathscr{V}$. \par

\paragraph{$\bm{\xi}$-momentum conservation} \mbox{}\\[-5mm]

The scalar product of the first row of Eq.~\eqref{eq:euler_linearized_forCurv} with $\vec{e}_{\xi}$ is formed and integrated over $d \mathscr{V}$, which leads to:
\begin{align}
\label{eq:xi_momentum_1}
\pdv{}{t} \left(  \iiint_{d \mathscr{V}} \rho_0 u_{\xi} d^3 \vec{x} \right) = - \varoiint_{\partial d \mathscr{V}} p \vec{e}_{\xi} . \vec{n}_s d^2 \vec{x}_s + \iiint_{d \mathscr{V}} p \nabla . \vec{e}_{\xi} d^3 \vec{x}
\end{align}
where the relation $\vec{\nabla}p. \vec{e}_{\xi} = \nabla . (p \vec{e}_{\xi}) - p \nabla . \vec{e}_{\xi}$ and the divergence theorem have been used. Replacing $\nabla. \vec{e}_{\xi}$ in the right-hand side (RHS) of Eq.~\eqref{eq:xi_momentum_1} with the expression of Eq.~\eqref{eq:divergence_unit_vectors} leads to a term of order $O(L \kappa_m)$, and it is therefore dropped. The integral on the left-hand side (LHS) can be approximated by its value at the center point P:
\begin{align}
\label{eq:xi_momentum_int_vol}
\pdv{}{t} \left(  \iiint_{d \mathscr{V}} \rho_0 u_{\xi} d^3 \vec{x} \right) = 2 A B L \rho_0 \pdv{\overline{u_{\xi}}}{t} d \alpha  d \beta
\end{align}
where $\overline{(.)}$ is the spatial averaging along the $\xi$ direction: $\overline{(.)} = 1/2L \int_{-L}^L{} (.) d \xi$. The surface integral on the RHS of Eq.~\eqref{eq:xi_momentum_1} can be split into 6 contributions from each one of the 6 boundaries delimiting $d \mathscr{V}$; only the contributions from $S_{-L}$ and $S_L$ remain, as $\vec{e}_{\xi} . \vec{n}_s = 0$ on the others. The RHS of Eq.~\eqref{eq:xi_momentum_1} then reads:
\begin{align}
\label{eq:xi_momentum_int_surf}
- \varoiint_{\partial d \mathscr{V}} p \vec{e}_{\xi} . \vec{n}_s \ d^2 \! \vec{x}_s = A B  p(-L) d \alpha d \beta - A B  p(L) d \alpha d \beta
\end{align}
Combining Eq~\eqref{eq:xi_momentum_int_vol} and Eq.~\eqref{eq:xi_momentum_int_surf} finally leads to:
\begin{align}
\label{eq:xi_momentum_final}
\pdv{\overline{u_{\xi}}}{t} (\alpha, \beta, t) = \dfrac{1}{2 L \rho_0 (\alpha, \beta )} \left[ p^{S_{-L}} (\alpha, \beta, t) - p^{S_{L}} (\alpha, \beta, t) \right]
\end{align}\par

\paragraph{$\bm{\alpha}$-momentum conservation} \mbox{}\\[-5mm]

Forming the scalar product of $\vec{e}_{\alpha}$ with the first row of Eq.~\eqref{eq:euler_linearized_forCurv} and integrating over  $d \mathscr{V}$ gives an equation analogous to Eq.~\eqref{eq:xi_momentum_1} for the $\alpha$-momentum. Similarly, the volume integral on the LHS writes:
\begin{align}
\label{eq:alpha_momentum_vol}
\pdv{}{t} \left(  \iiint_{d \mathscr{V}} \rho_0 u_{\alpha} d^3 \vec{x} \right) = 2 A B L \rho_0 \pdv{\overline{u_{\alpha}}}{t} d \alpha  d \beta
\end{align}
Injecting the expression of $\nabla . \vec{e}_{\alpha}$ from Eq.~\eqref{eq:divergence_unit_vectors} into the RHS volume integral gives:
\begin{align}
\label{eq:alpha_momentum_rhs_vol}
\iiint_{d \mathscr{V}} p \nabla . \vec{e}_{\alpha} d^3 \vec{x} = 2 L \overline{p} \pdv{B}{\alpha} d \alpha d \beta
\end{align}
The RHS surface integral is split into 6 distinct contributions from the boundaries of $d \mathscr{V}$. As $\vec{e}_{\alpha} . \vec{n}_s = 0$ on every one of those, expect $S_{- d \alpha/2}$ and $S_{+ d \alpha/2}$, this surface integral yields:
\begin{align}
\label{eq:alpha_momentum_int_surf_1}
\begin{aligned}
- \varoiint_{\partial d \mathscr{V}} p \vec{e}_{\alpha} . \vec{n}_s \ d^2 \! \vec{x}_s = & -2 L B \left( \alpha + \dfrac{\alpha}{2}, \beta \right) \overline{p} \left( \alpha + \dfrac{\alpha}{2}, \beta \right) d \beta\\
& + 2 L B \left( \alpha - \dfrac{\alpha}{2}, \beta \right) \overline{p} \left( \alpha - \dfrac{\alpha}{2}, \beta \right) d \beta
\end{aligned}
\end{align}
Combining Eq.~\eqref{eq:alpha_momentum_vol}, Eq.~\eqref{eq:alpha_momentum_rhs_vol}, and Eq.~\eqref{eq:alpha_momentum_int_surf_1} finally leads to:
\begin{align}
\label{eq:alpha_momentum_final}
\begin{aligned}
\pdv{\overline{u_{\alpha}}}{t} (\alpha, \beta, t) = & \dfrac{1}{A B \rho_0} \overline{p} \pdv{B}{\alpha}  - \dfrac{1}{A B \rho_0} \pdv{(B \overline{p})}{\alpha} \\
= &  -\dfrac{1}{A (\alpha, \beta) \rho_0 (\alpha, \beta)} \pdv{\overline{p}}{\alpha} (\alpha, \beta, t)
\end{aligned}
\end{align}
The same reasoning can be applied to the $\beta$-momentum, which results in:
\begin{align}
\label{eq:beta_momentum_final}
\pdv{\overline{u_{\beta}}}{t} (\alpha, \beta, t) = -\dfrac{1}{B (\alpha, \beta) \rho_0 (\alpha, \beta)} \pdv{\overline{p}}{\beta} (\alpha, \beta, t)
\end{align}
\par

\paragraph{Mass conservation} \mbox{}\\[-5mm]

Integrating the second row of Eq.~\eqref{eq:euler_linearized_forCurv}  over the volume element $d \mathscr{V}$ writes:
\begin{align}
\label{eq:mass_first}
\pdv{}{t} \left( \iiint_{d \mathscr{V}} \rho \  d^3 \vec{x}  \right) = - \varoiint_{\partial d \mathscr{V}} \rho_0 \vec{u} . \vec{n}_s \  d^2 \vec{x}_s
\end{align}
As before the LHS volume integral can be evaluated at the center-point $P$:
\begin{align}
\label{eq:mass_volume_int_LHS}
\pdv{}{t} \left( \iiint_{d \mathscr{V}} \rho \  d^3 \vec{x}  \right) = 2  A B L \dfrac{1}{c_0^2} \pdv{\overline{p}}{t} d \alpha d \beta
\end{align}
where $\rho$ has been replaced with $p/c_0^2$ thanks to the isentropic relation. The RHS surface integrals in Eq.~\eqref{eq:mass_first} is once again split into 6 contributions from each one of the boundaries of $d \mathscr{V}$. The contribution from $S_{-L}$ and $S_L$ writes:
\begin{align}
\label{eq:mass_surface_L}
I_{\xi} = A B  \rho_0 u_{\xi} (-L) d \alpha d \beta - A B  \rho_0 u_{\xi} (L) d \alpha d \beta
\end{align}
The contribution from the boundary pairs $S_{- d \alpha}$, $S_{+ d \alpha}$ and $S_{- d \beta}$, $S_{+ d \beta}$ respectively read:
\begin{align}
\label{eq:mass_surface_alpha}
\begin{aligned}
I_{\alpha} & = - 2 L B \ \rho_0 \left(\alpha + \dfrac{d \alpha}{2} \right) \overline{u_{\alpha}} \left(\alpha + \dfrac{d \alpha}{2} \right) d \beta + 2 L B \ \rho \left(\alpha - \dfrac{d \alpha}{2} \right) \overline{u_{\alpha}} \left(\alpha - \dfrac{d \alpha}{2} \right) d \beta \\
I_{\beta} & = - 2 L A \ \rho_0 \left(\beta + \dfrac{d \beta}{2} \right) \overline{u_{\beta}} \left(\beta + \dfrac{d \beta}{2} \right) d \alpha + 2 L A \ \rho \left(\beta - \dfrac{d \beta}{2} \right) \overline{u_{\beta}} \left(\beta - \dfrac{d \beta}{2} \right) d \alpha
\end{aligned}
\end{align}
Finally, combining Eq.~\eqref{eq:mass_first} to Eq.~\eqref{eq:mass_surface_alpha} results in:
\begin{align}
\label{eq:mass_final}
\begin{aligned}
\pdv{\overline{p}}{t}  (\alpha, \beta,t) = & \dfrac{\rho_0 (\alpha, \beta) c_0^2 (\alpha, \beta) }{2 L} \left[ u_{\xi}^{S_{-L}} (\alpha, \beta,t) - u_{\xi}^{S_{L}} (\alpha, \beta,t)  \right] \\
  - & \dfrac{ c_0^2 (\alpha, \beta) }{A (\alpha, \beta) B (\alpha, \beta) } \ \pdv{}{\alpha} \Bigl( \rho_0 (\alpha, \beta) B (\alpha, \beta) \overline{u_{\alpha}} (\alpha, \beta,t)  \Bigr) \\
  - & \dfrac{ c_0^2 (\alpha, \beta) }{A (\alpha, \beta) B (\alpha, \beta) } \ \pdv{}{\beta} \Bigl( \rho_0 (\alpha, \beta) A (\alpha, \beta) \overline{u_{\beta}} (\alpha, \beta,t)  \Bigr) 
\end{aligned}
\end{align}
\par

The next step of the derivation consists in introducing Eq.~\eqref{eq:alpha_momentum_final} and Eq.~\eqref{eq:beta_momentum_final} into the time derivative of Eq.~\eqref{eq:mass_final} to elimininate $\overline{u_{\alpha}}$ and $\overline{u_{\beta}}$ from the problem. Finally, in the resulting equation, the pressure $\overline{p}$ is replaced with the acoustic potential $\overline{\varphi}$. Gathered with Eq.~\eqref{eq:xi_momentum_final} the following system is formed:
\begin{align}
\label{eq:curvilinear_Helmholtz_final}
\left\{
\begin{aligned}
& \pdv{\overline{u_{\xi}}}{t} (\alpha, \beta, t) = \dfrac{1}{2 L \rho_0 } \left[ p^{S_{-L}} (\alpha, \beta, t) - p^{S_{L}} (\alpha, \beta, t) \right]
\\
& c_0^2 \nabla_c^2 \overline{\varphi} (\alpha,\beta,t) - \dfrac{\partial^2  \overline{\varphi}}{\partial t^2}  (\alpha,\beta,t) = \dfrac{c_0^2}{2L} \left[  u_{\xi}^{S_{L}} (\alpha,\beta,t)  - u_{\xi}^{S_{-L}} (\alpha,\beta,t) \right]
\end{aligned}
\right.
\end{align}
where $\nabla_c^2$ is a $\rho_0$-weighted curvilinear Laplacian operator, defined as:
\begin{align}
\label{eq:laplacien_curvilinear_definition}
\nabla_c^2 f  = \dfrac{1}{AB \rho_0} \left[ \pdv{}{\alpha} \left( \dfrac{B}{A} \pdv{}{\alpha} (\rho_0 f)   \right) + \pdv{}{\beta} \left( \dfrac{A}{B} \pdv{}{\beta} (\rho_0 f)   \right)  \right]
\end{align}
Equation~\eqref{eq:curvilinear_Helmholtz_final}, the second row of which is a curvilinear Helmholtz equation, shows that $\xi$-averaged acoustics in the shell-like control volume are entirely determined by the pressure and normal velocity values at the boundaries $S_{-L}$ and $S_{L}$. Later, the curvilinear gradient $\vec{\nabla}_c$ is also employed; it is defined as:
\begin{align}
\label{eq:gradient_curvilinear_definition}
\vec{\nabla}_c f = \dfrac{1}{A} \pdv{f}{\alpha} \vec{e}_{\alpha} + \dfrac{1}{B} \pdv{f}{\beta} \vec{e}_{\beta}
\end{align}

\subsection{Surface modal expansion derivation} \label{sec:surface_modal_expansion_derivation}

\begin{figure}[h!]
\centering
\includegraphics[width=0.99\textwidth]{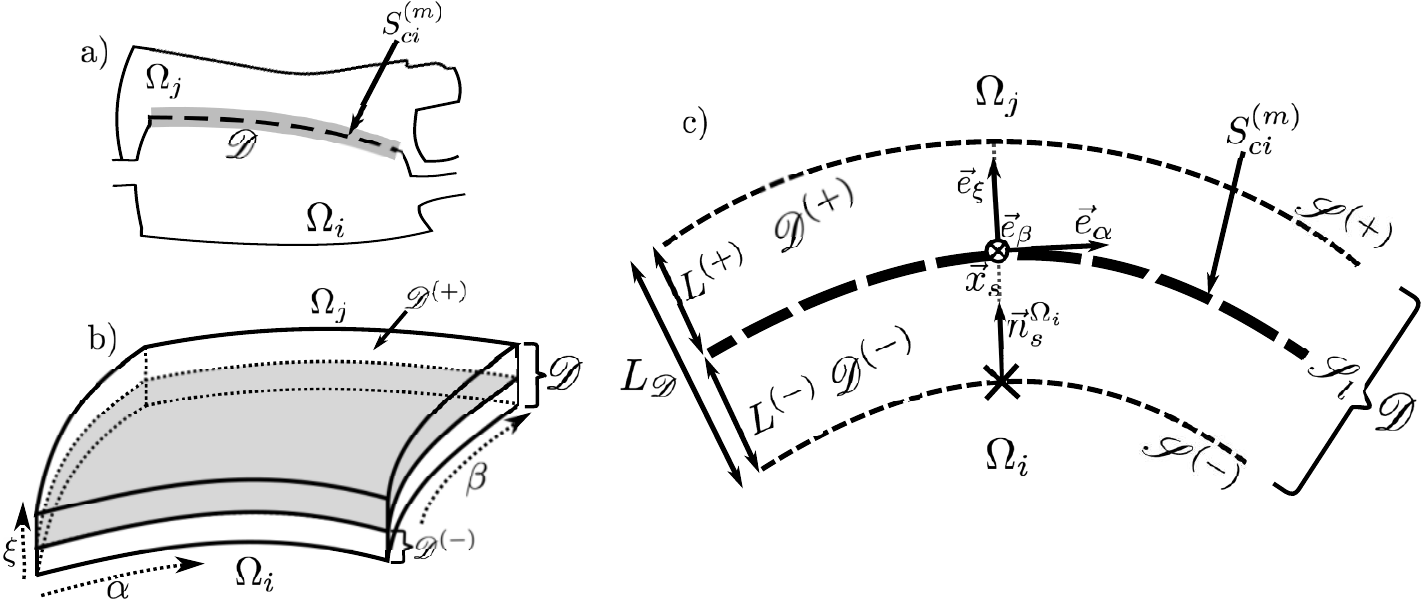}
\caption{(a) Schematic of an acoustic network with a complex boundary $\bm{S_{ci}^{(m)}}$ and a thin control volume $\pmb{\mathscr{D}}$ that encloses it. (b) Three-dimensional representation of the shell-like control volume $\pmb{\mathscr{D}}$ enclosing $\bm{S_{ci}^{(m)}}$, and the 2 sub-layers $\pmb{\mathscr{D}^{(+)}}$ and $\pmb{\mathscr{D}^{(-)}}$. (c) Two-dimensional view of $\pmb{\mathscr{D}}$. The sub-layer $\pmb{\mathscr{D}^{(+)}}$ (resp. $\pmb{\mathscr{D}^{(-)}}$) is delimited by the surfaces $\pmb{\mathscr{S}^{(+)}}$ and $\pmb{\mathscr{S}_l}$ (resp. $\pmb{\mathscr{S}^{(-)}}$ and $\pmb{\mathscr{S}_l}$). $\pmb{\mathscr{D}}$ and the 2 sub-layers are parameterized by a common system of coordinates $\bm{(\alpha,\beta,\xi)}$, where $\bm{\alpha}$ and $\bm{\beta}$ are curvilinear coordinates tangential to the surface, and $\bm{\xi}$ is the coordinate normal to the surface. Location of a surface point $\bm{\vec{x}_s (\alpha,\beta)}$ with the associated normal vector $\bm{\vec{n}_s^{\Omega_i}}$ are indicated.}
\label{fig:shell3d}
\end{figure}

Let us now go back to the case of a complex boundary $S_{ci}^{(m)}$ representing a multi-perforated liner between two subdomains $\Omega_i$ and $\Omega_j$ and characterized by a conductivity $K_R$ (but the reasoning also holds for any other type of complex boundaries, including those connecting two subdomains without jump conditions, \textit{i.e.} $K_R = \infty$). The aim of this section is to make use of the curvilinear Helmholtz equation derived in Sec.~\ref{sec:curvilinear_helmholtz} to obtain a dynamical system governing the acoustics in the vicinity of  $S_{ci}^{(m)}$. In this purpose, Fig.~\ref{fig:shell3d} shows the definition of a shell-like control volume $\mathscr{D}$ enclosing $S_{ci}^{(m)}$.
It is split into two distinct sub-layers: $\mathscr{D}^{(-)}$ (resp. $\mathscr{D}^{(+)}$) of thickness $L^{(-)}$ (resp. $L^{(+)}$) located on the side of the subdomain $\Omega_i$ (resp. $\Omega_j$). Note that the control volume $\mathscr{D}$ is only an intermediate for the current derivation, and its thickness $L_{\mathscr{D}}$ can be chosen arbitrarily small and does not have any physical meaning. Once integrated in the LOM network, the entire control volume $\mathscr{D}$ collapses to a two-dimensional manifold corresponding to $S_{ci}^{(m)}$, or in other words the limit $L_{\mathscr{D}} \rightarrow 0$ is retained. This also allows the mean fields $c_0$ and $\rho_0$ to be considered uniform within $\mathscr{D}$ along the $\xi$ direction; they can then be defined as the averages of their respective values on $\mathscr{S}^{(-)}$ and  $\mathscr{S}^{(+)}$. The jump relations at the middle surface $S_l$ relate pressure and flux through:
\begin{align}
\label{eq:conductivity_relation}
\left\{
\begin{aligned}
& \hat{u}_{\xi}^{ \mathscr{S}_l^{(-)}} \left( \alpha, \beta, \omega \right) =\hat{ u}_{\xi}^{ \mathscr{S}_l^{(+)}} \left( \alpha, \beta,  \omega \right) \\
&  \left[ \hat{p}^{\mathscr{S}_l^{(+)}} \left( \alpha, \beta ,  \omega \right) - \hat{p}^{\mathscr{S}_l^{(-)}} \left( \alpha, \beta ,  \omega \right)  \right] = \dfrac{- j \omega d^2 \rho_u}{K_R (j \omega)} \hat{u}_{\xi}^{ \mathscr{S}_l^{(-)}}  \left( \alpha, \beta ,  \omega \right)
\end{aligned}
\right.
\end{align}
where $\mathscr{S}_l^{(+)}$ (resp. $\mathscr{S}_l^{(-)}$) refers to the values of the variables in $\mathscr{D}^{(+)}$ (resp. $\mathscr{D}^{(-)}$) on the boundary $\mathscr{S}_l$. In Eq.~\eqref{eq:conductivity_relation} the aperture spacing on the multi-perforated liner is noted $d$ and $\rho_u$ is the upstream mean density. For simplicity, the liner physical parameters are here uniform, but the derivation can be extended to non-uniform cases. It is assumed that the inverse Fourier transform of $- j \omega d^2 \rho_u/K_R (j \omega)$ can be represented by a Single-Input-Single-Output (SISO) linear state-space realization \{$\mathbf{A}_{\mathbf{K_R}}$,$\mathbf{B}_{\mathbf{K_R}}$,$\mathbf{C}_{\mathbf{K_R}}$\}, that relates any input signal $U_{K_R}(t)$ (here the velocity) to an output $Y_{K_R} \{ U_{K_R}(t) \}$ (\textit{e.g.} the pressure jump across the liner). Such state-space representation exists for any conductivity of the form $K_R (j \omega) = \mathscr{Q}( j \omega) / \mathscr{R}( j \omega)$, where  $\mathscr{Q}$ and $\mathscr{R}$ are polynomials. This allows for a great flexibility regarding the conductivity, that can be defined from an analytical model, or fitted to match either experimental or numerical data.\par

Equation~\eqref{eq:curvilinear_Helmholtz_final} applied to the control volume $\mathscr{D}^{(-)}$ writes:
\begin{align}
\label{eq:curvilinear_Helmholtz_sublayer}
\left\{
\begin{aligned}
& \dfrac{\partial \overline{u}_{\xi}^{(-)}}{\partial t} (\alpha,\beta,t) = \dfrac{1}{L^{(-)} \overline{\rho_0}} \left[ p^{\Omega_i} (\alpha,\beta,t) - p^{\mathscr{S}_l^{(-)}} (\alpha,\beta,t) \right] \\
& \overline{c_0}^2 \nabla_c^2 \overline{\varphi}^{(-)} (\alpha,\beta,t) - \dfrac{\partial^2  \overline{\varphi}^{(-)}}{\partial t^2}  (\alpha,\beta,t) = \dfrac{\overline{c_0}^2}{L^{(-)}} \left[  u_{\xi}^{\Omega_i} (\alpha,\beta,t)  - u_{\xi}^{\mathscr{S}_l^{(-)}} (\alpha,\beta,t) \right]
\end{aligned}
\right.
\end{align}
where $\nabla_c^2$ is the curvilinear Laplacian operator defined in Eq.~\eqref{eq:laplacien_curvilinear_definition}. Analogous relations are verified in $\mathscr{D}^{(+)}$. The $\xi$-averaging across the entire control volume $\mathscr{D}$ (excluding the middle surface $\mathscr{S}_l$) is expressed as $\overline{(.)} =( L^{(+)} \overline{(.)}^{(+)} + L^{(-)} \overline{(.)}^{(-)} ) / L_{\mathscr{D}}$. In addition, taking the limit $L^{(+)},L^{(-)} \rightarrow 0$ is used to replace $\hat{u}_{\xi}^{\mathscr{S}_l^{(-)} }$ in Eq.~\eqref{eq:conductivity_relation} thanks to the 1\textsuperscript{st}-order approximation $ \hat{u}_{\xi}^{\mathscr{S}_l^{(-)} } \approx  \hat{u}_{\xi}^{ \Omega_i}$ Thus, combining Eq.~\eqref{eq:curvilinear_Helmholtz_sublayer}, its counterpart for $\mathscr{D}^{(+)}$, and the inverse Fourier transform of Eq.~\eqref{eq:conductivity_relation} gives:
\begin{align}
\label{eq:curvilinear_Helmholtz_full}
\left\{
\begin{aligned}
& \overline{c_0}^2 \nabla_c^2 \overline{\varphi} (\alpha,\beta,t) - \dfrac{\partial^2  \overline{\varphi}}{\partial t^2}  (\alpha,\beta,t) = \dfrac{\overline{c_0}^2}{L_{\mathscr{D}}} \left[  u_{\xi}^{\Omega_i} (\alpha,\beta,t)  - u_{\xi}^{\Omega_j} (\alpha,\beta,t) \right] \\
&\begin{aligned} \dfrac{\partial \overline{u}_{\xi}}{\partial t} (\alpha,\beta,t) = \dfrac{1}{L_{\mathscr{D}} \overline{\rho_0}}
\biggl[ p^{\Omega_i} (\alpha,\beta,t) - p^{\Omega_j} (\alpha,\beta,t) + Y_{K_R} \left\{  u_{\xi}^{ \Omega_i}  \left( \alpha, \beta ,  t \right)  \right\} \biggr] \end{aligned}
\end{aligned} \right.
\end{align}\par

In order to convert Eq.~\eqref{eq:curvilinear_Helmholtz_full} into a state-space representation for the complex boundary $S_{ci}^{(m)}$, a set of surface modes $(\mathscr{K}_k (\vec{x}_s) )_{k \geq 1}$ is introduced. Those are solutions of the following curvilinear Helmholtz eigen-problem in the two-dimensional manifold $S_{ci}^{(m)}$:
\begin{align}
\label{eq:surface_eigenproblem}
\left\{
\begin{aligned}
& \overline{c_0}^2 \nabla_c^2 \mathscr{K}_k (\alpha, \beta) + \omega_k^2 \mathscr{K}_k (\alpha, \beta) = 0 \ \text{in} \ S_{ci}^{(m)}\\
& \mathscr{K}_k = 0 \ \  \text{or}  \ \ \vec{\nabla}_c \mathscr{K}_k .\vec{n}_s = 0 \  \text{on}  \  \partial S_{ci}^{(m)}
\end{aligned}
\right.
\end{align}
where the curvilinear Laplacian operator $\nabla_c^2$ is defined in Eq.~\eqref{eq:laplacien_curvilinear_definition}, and the curvilinear gradient $\vec{\nabla}_c$ is given in Eq.~\eqref{eq:gradient_curvilinear_definition}. In Eq.~\eqref{eq:surface_eigenproblem}, the homogeneous Neumann and Dirichlet conditions on the one-dimensional contour $\partial S_{ci}^{(m)}$ are chosen to match those of the subdomains $\Omega_i$ and $\Omega_j$. It can be demonstrated that $(\mathscr{K}_k (\vec{x}_s) )_{k \geq 1}$ is an orthogonal basis of $S_{ci}^{(m)}$ for the surface scalar product:
\begin{align}
\label{eq:surface_scalar_prod}
\left( f | g \right) = \iint_{S_{ci}^{(m)}} f(\vec{x}_s)  g(\vec{x}_s) \  d^2 \vec{x}_s
\end{align}
The squared L-2 norm of the surface modal basis vector $\mathscr{K}_k (\vec{x}_s)$ is noted $\lambda_k = \left( \mathscr{K}_k | \mathscr{K}_k \right)$. Similarly to the subdomain frame $(\phi_n)_{n \geq 1}$, the surface modal basis  is obtained analytically for a topologically simple surface $S_{ci}^{(m)}$. For a more complex geometry where an analytical treatment is impossible, $( \mathscr{K}_k)_{k \geq 1}$ can be generated in a preliminary step by resolving Eq.~\eqref{eq:surface_eigenproblem} thanks to a curvilinear FE solver. Another possibility, which is the one adopted in this work, consists in building a thin three-dimensional shell domain enclosing $S_{ci}^{(m)}$, and to resolve the classical homogeneous Helmholtz equation thanks to a 3D FE solver. Extracting the values on the middle surface then provides good approximations to the solutions of Eq.~\eqref{eq:surface_eigenproblem}. Note that this shell domain is not related to $\mathscr{D}$, which is only a control volume serving as an intermediate in the derivation of Eq.~\eqref{eq:curvilinear_Helmholtz_full}. Surface modal expansions of $\overline{\varphi}$ and $\overline{u}_{\xi}$ are sought under the form:
\begin{align}
\label{eq:surface_modal_expansion}
\begin{aligned}
& \overline{\varphi} (\alpha,\beta,t) = \sum_{k=1}^{\infty} \nu_k (t) \mathscr{K}_k (\alpha, \beta) = {}^t \boldsymbol{\nu} (t) \mathscrbf{K} (\vec{x}_s) \\
& \overline{u}_{\xi} (\alpha,\beta,t) = \sum_{k=1}^{\infty} \mu_k (t) \mathscr{K}_k (\alpha, \beta) = {}^t \boldsymbol{\mu} (t) \mathscrbf{K} (\vec{x}_s)
\end{aligned}
\end{align}
Mimicking the process of the classical Galerkin expansion, Eq.~\eqref{eq:surface_modal_expansion} is then injected into Eq.~\eqref{eq:curvilinear_Helmholtz_full}, and both the surface scalar product of Eq.~\eqref{eq:surface_scalar_prod} and the fact that the orthogonal basis is a solution of the curvilinear Helmholtz eigen-problem of Eq.~\eqref{eq:surface_eigenproblem} are used. At this point, it is also useful to express the source terms in Eq.~\eqref{eq:curvilinear_Helmholtz_full} thanks to frame expansions (Eq.~\eqref{eq:pressure_expansion_reminder}) in the subdomains $\Omega_i$ and $\Omega_j$ to obtain:
\begin{align}
\label{eq:surface_dynamical_system}
\left\{
\begin{aligned}
& \begin{aligned} \ddot{\nu}_k(t) + \omega_k^2 \nu_k (t) =  -\dfrac{\overline{c_0}^2}{\lambda_k L_{\mathscr{D}}} & \biggl[ \mathpzc{s}_i \sum_{n=1}^{N^{(i)}} -\dfrac{1}{\rho_0^{(i)}} \left( \nabla_s \phi_n^{(i)} | \mathscr{K}_k \right) \Gamma_{n}^{(i)} (t) \\
-  & \mathpzc{s}_j \sum_{n=1}^{N^{(j)}} -\dfrac{1}{\rho_0^{(j)}} \left( \nabla_s \phi_n^{(j)} | \mathscr{K}_k \right) \Gamma_{n}^{(j)} (t) \biggr] \end{aligned} \\
&\begin{aligned} \dot{\mu}_k (t) =  \dfrac{1}{\lambda_k L_{\mathscr{D}} \overline{\rho_0}} &
\biggl[  \mathpzc{s}_i \sum_{n=1}^{N^{(i)}} \left( \phi_n^{(i)} | \mathscr{K}_k  \right) \dot{\Gamma}_n^{(i)} (t)   + \mathpzc{s}_j \sum_{n=1}^{N^{(j)}} \left( \phi_n^{(j)} | \mathscr{K}_k  \right) \dot{\Gamma}_n^{(j)} (t)  \\
  + & \mathpzc{s}_i  \sum_{n=1}^{N^{(i)}} -\dfrac{1}{\rho_0^{(i)}} \left( \nabla_s \phi_n^{(i)} | \mathscr{K}_k \right) Y_{K_R} \left\{ \Gamma_n ^{(i)} (t)  \right\}  \biggr] \end{aligned}
\end{aligned} \right.
\end{align}
where the linearity of the operator $Y_{K_R}$ has been used, and where $\mathpzc{s}_i = \vec{n}_s^{\Omega_i} . \vec{e}_{\xi} = \pm 1$ and  $\mathpzc{s}_j = \vec{n}_s^{\Omega_j} . \vec{e}_{\xi} = \mp 1$ define the orientation of the subdomains surface normal vector with restpect to the orientation chosen for $\mathscr{D}$. The superscripts ${}^{(i)}$ and ${}^{(j)}$ refer to quantities evaluated in $\Omega_i$ and $\Omega_j$, respectively. Note that the surface modal expansion is truncated to a finite order $K_S$, for which the optimal choice is discussed in Sec.~\ref{sec:surface_modes_number_choice}.  Equation~\eqref{eq:surface_dynamical_system} governs the dynamics of the $\xi$-averaged acoustic potential $\overline{\varphi}$ and normal velocity $\overline{u}_{\xi}$ in the control volume $\mathscr{D}$, subjected to both the surface forcing from the adjacent subdomains $\Omega_i$ and $\Omega_j$, and the hydrodynamical interaction responsible for the complex conductivity. In the low frequency limit, or equivalently for very small $L_{\mathscr{D}}$, the first line of Eq.~\eqref{eq:surface_dynamical_system} enforces the flux continuity between $\Omega_i$ and $\Omega_j$, while the second line imposes a pressure jump related to $K_R$.\par

Finally, the surface modal expansions of Eq.~\eqref{eq:surface_modal_expansion}
allow the surface integrals in Eq.~\eqref{eq:subdomain_dynamical_system} to be rewritten as:
\begin{align}
\label{eq:surface_integral_projection}
\begin{aligned}
& \iint_{S_{ci}^{(m)}} \rho_0 c_0^2  \varphi^{S_{ci}^{(m)}} (\vec{x}_s,t)  \left[ \mathbf{\Lambda}^{-1}  \boldsymbol{\nabla_s} \pmb{\phi} (\vec{x}_s) \right]_n d^2 \vec{x}_s  = \rho_0 c_0^2 \left[ \mathbf{\Lambda}^{-1}   \left( \boldsymbol{\nabla_s} \pmb{\phi} | \ {}^t \mathscrbf{K} \right) \right]_{n,*} \boldsymbol{\nu} (t)  \\
& \iint_{S_{ci}^{(m)}} \rho_0 c_0^2   u_s^{S_{ci}^{(m)}} (\vec{x}_s,t) \left[ \mathbf{\Lambda}^{-1} \pmb{\phi} (\vec{x}_s) \right]_n  d^2 \vec{x}_s = \mathpzc{s}_i \rho_0 c_0^2 \left[ \mathbf{\Lambda}^{-1}   \left( \pmb{ \phi} | \ {}^t \mathscrbf{K} \right) \right]_{n,*} \boldsymbol{\mu} (t)
\end{aligned}
\end{align}
where $[\mathbf{M}]_{n,*}$ denotes the entire $n$\textsuperscript{th} row of a matrix $\mathbf{M}$. Equation~\eqref{eq:surface_integral_projection} shows that the surface source terms in Eq.~\eqref{eq:subdomain_dynamical_system} are now evaluated thanks to the surface modal projections $\left( \pmb{\nabla_s \phi} | \ {}^t \mathscrbf{K} \right)$ and $\left( \pmb{ \phi} | \ {}^t \mathscrbf{K} \right)$ rather than through simple piece-wise approximations as in Eq.~\eqref{eq:surf_int_approximation}. This yields a formulation robust with respect to both the frame ill-conditioning and the numerical noise resulting from its construction. Equations~\eqref{eq:surface_dynamical_system}-\eqref{eq:surface_integral_projection} show that the original state-space realization for the subdomain $\Omega_i$ derived in Chapter~\ref{chap:FRAME} and in Appendix~\ref{sec:ss_realization_subdomain} needs to be adapted. The reformulated state-space matrices are detailed in Appendix~\ref{sec:ss_realization_subdomain_with_complex_boundary}. In addition, these relations are used to define a state-space representation for the acoustics on the complex boundary $S_{ci}^{(m)}$, which is given in Appendix~\ref{sec:ss_realization_complex_boundary}.

\subsection{Selection algorithm for constructing the surface modal basis} \label{sec:surface_modes_number_choice}

As explained earlier, the subdomain $\Omega_i$ pressure and velocity must be accurately represented on its boundary $S_{ci}^{(m)}$, and since those are computed thanks to the frame modal expansion of Eq.~\eqref{eq:pressure_expansion_reminder}, it is in turn crucial to correctly evaluate the frame modes restrictions $\phi_n (\vec{x}_s)$ and their gradients $\nabla_s \phi_n (\vec{x}_s)$ on this surface. Equation~\eqref{eq:surface_integral_projection} and Eq.~\eqref{eq:surface_dynamical_system} show that these frame modes restrictions are approximated by their projections on the surface modal basis $( \mathscr{K}_k)_{k \geq 1}$, which write:
\begin{align}
\label{eq:surface_mode_approx}
\pmb{\phi } (\vec{x}_s) \approx \left( \pmb{ \phi} | \ {}^t \mathscrbf{K} \right) \pmb{\lambda}^{-1} \mathscrbf{K} (\vec{x}_s) \ , \ \bm{\nabla_s} \pmb{\phi} (\vec{x}_s) \approx \left( \bm{\nabla_s} \pmb{\phi} | \ {}^t \mathscrbf{K} \right)  \pmb{\lambda}^{-1} \mathscrbf{K} (\vec{x}_s)
\end{align}
where $\pmb{\lambda}$ is the diagonal matrix with coefficients $\lambda_k$. The relations in Eq.~\eqref{eq:surface_mode_approx} are only approximations because the surface modal basis $( \mathscr{K}_k)_{k \geq 1}$ is truncated up to a finite order $K_S$, and this truncated set might not be sufficient to \textit{exactly} represent the frame modes and their gradients on the surface. On one hand, a value of $K_S$ too low yields inaccurate projections in Eq.~\eqref{eq:surface_mode_approx}. On the other hand, including too many surface modes not only results in a more costly LOM because of a large number of DoF, but it can also produce small-amplitude unphysical terms in the matrices $\left( \pmb{ \phi} | \ {}^t \mathscrbf{K} \right)$ and $\left( \bm{\nabla_s} \pmb{\phi} | \ {}^t \mathscrbf{K} \right)$, which can in turn be considerably amplified due to the frame ill-conditioning (see the multiplication by $\mathbf{\Lambda}^{-1}$ in Eq.~\eqref{eq:surface_integral_projection}). It is therefore crucial to use an appropriate size for the surface modal basis. In this matter, a procedure was designed to automatically and robustly construct an optimal surface modal basis $( \mathscr{K}_k)_{k \geq 1}$. It proceeds in 3 steps and is applied in a similar fashion to $\left( \pmb{ \phi} | \ {}^t \mathscrbf{K} \right)$ and to $\left( \bm{\nabla_s} \pmb{\phi} | \ {}^t \mathscrbf{K} \right)$:
\begin{enumerate}[wide, labelwidth=!, labelindent=0pt]
\item{Initially, the surface modal basis $( \mathscr{K}_k)_{k \geq 1}$ is chosen to contain a relatively large number of modes, of which only some will be retained. The SVD of ${}^t \! \left( \pmb{ \phi} | \ {}^t \! \mathscrbf{K} \right) \times  \left( \pmb{ \phi} | \ {}^t \! \mathscrbf{K} \right)$ is computed to determine the rank of $\left( \pmb{ \phi} | \ {}^t \mathscrbf{K} \right)$ (since these two matrices have the same rank; also see Sec.~\ref{sec:spurious_svd_attenuation} for more detail regarding the SVD):
\begin{align}
\label{eq:svd_surface_modes_number}
{}^t \! \left( \pmb{ \phi} | \ {}^t \! \mathscrbf{K} \right)  \left( \pmb{ \phi} | \ {}^t \! \mathscrbf{K} \right) = \mathcalbf{V} \ \mathbf{\Sigma} \ {}^t \! \mathcalbf{V} 
\end{align}
The rank $r$ is defined as the number of singular values $\sigma_1, \sigma_2, ..., \sigma_r$ larger than a given threshold $\varepsilon' = \varepsilon_{rk} \sigma_1$. The $r$ corresponding singular vectors $\mathcalbf{V}_{*,1}, ...,\mathcalbf{V}_{*,r}$ are isolated.}
\item{For each singular vector $\mathcalbf{V}_{*,i}$ ($1\leq i \leq r $), its components $\mathcalbf{V}_{k,i}$ on the surface modes $\mathscr{K}_k$ are sorted in descending order:  $|\mathcalbf{V}_{k1,i}| \geq |\mathcalbf{V}_{k2,i} | \ge ... $. These components are then added one by one until a significant part of the singular vector $\mathcalbf{V}_{*,i}$ is recovered, or more precisely: $\mathcalbf{V}_{k1,i}^2 + \mathcalbf{V}_{k2,i}^2 + ... + \mathcalbf{V}_{kq,i}^2 \geq (1-\varepsilon_{svn}) | \mathcalbf{V}_{*,i} |^2 $ ($q$ being the smallest integer such that this inequality is verified). The surface modes $\mathscr{K}_{k1}, ..., \mathscr{K}_{kq}$ are therefore the modes that are necessary to represent the singular vector $\mathcalbf{V}_{*,i}$ and they are added to the list of modes to retain in the surface modal basis. This process is repeated for all the singular vectors $\mathcalbf{V}_{*,1}, ...,\mathcalbf{V}_{*,r}$, and the list of surface modes to retain is iteratively incremented.}
\item{Surface modes that are not included in the list previously computed are discarded and a new surface modal basis is constructed. Note that the $K_S$ modes retained in the surface basis are not necessarily the first $K_S$ eigenmodes solutions of the curvilinear Helmholtz eigenproblem~ of Eq.~\eqref{eq:surface_eigenproblem}.}
\end{enumerate}
This procedure is systematically applied to construct the surface modal basis: the size $K_S$ is therefore not a input required by the method, but is instead a parameter implicitly determined as soon as the subdomain frames are provided. The thresholds $\varepsilon_{rk}$ and $\varepsilon_{svn}$ are fixed to $\varepsilon_{rk} = 10^{-2}$ and $\varepsilon_{svn} = 0.2$. They do not require to be tuned for the examples presented below.

\paragraph{Size and elements of the surface modal basis} \mbox{}\\[-5mm]

\begin{figure*}[h!] 
\centering
\includegraphics[width=0.99\textwidth]{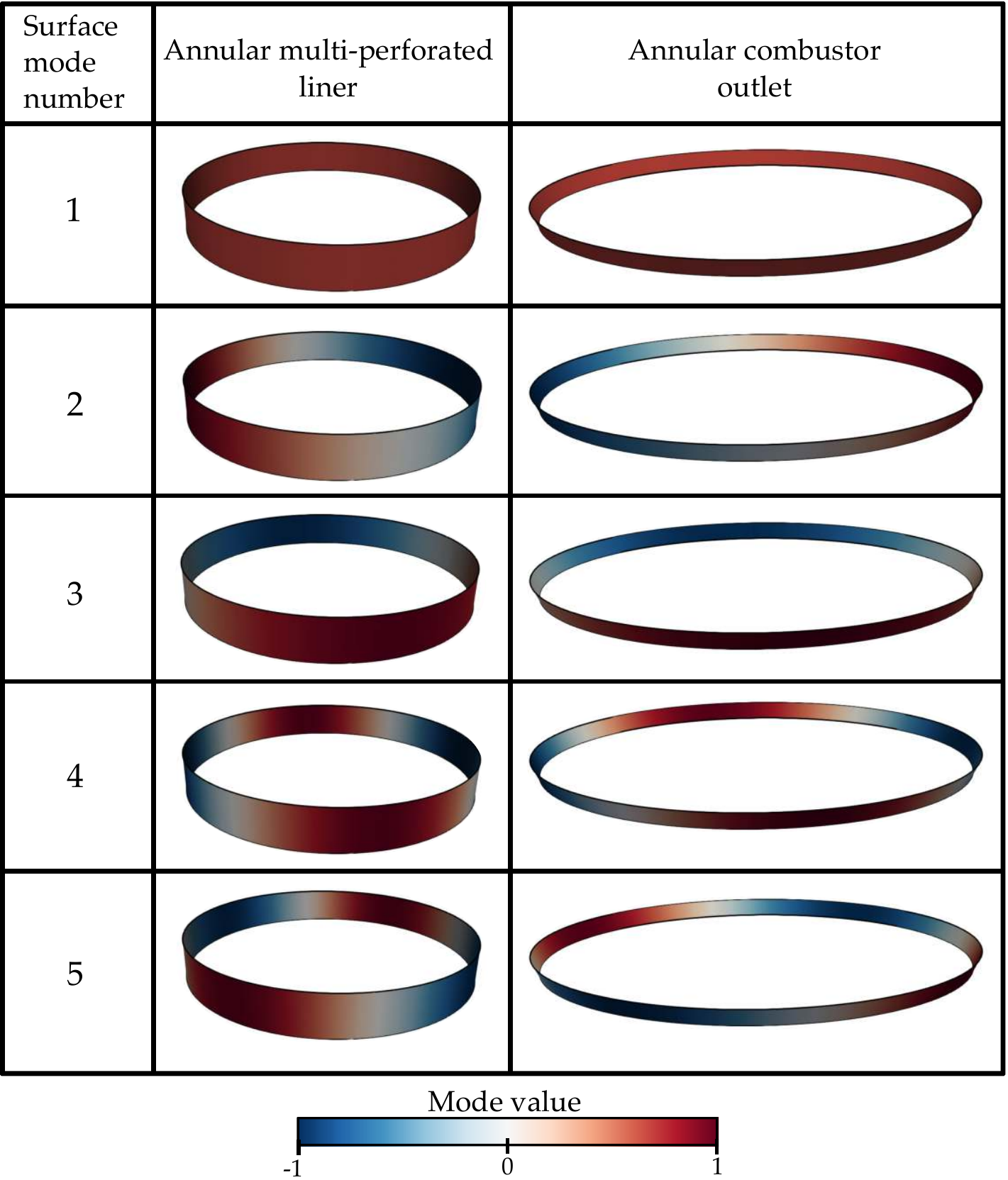}
\caption{Spatial shape of the first 5 elements $\pmb{\mathscr{K}_k} \bm{(\vec{x}_s)}$ contained in the surface modal bases used to model the annular liner (first column) and the conical outlet (second column). These modes are selected by the algorithm introduced above.}
\label{fig:surface_modes_shapes}
\end{figure*}

\begin{figure*}[h!] 
\centering
\includegraphics[width=0.99\textwidth]{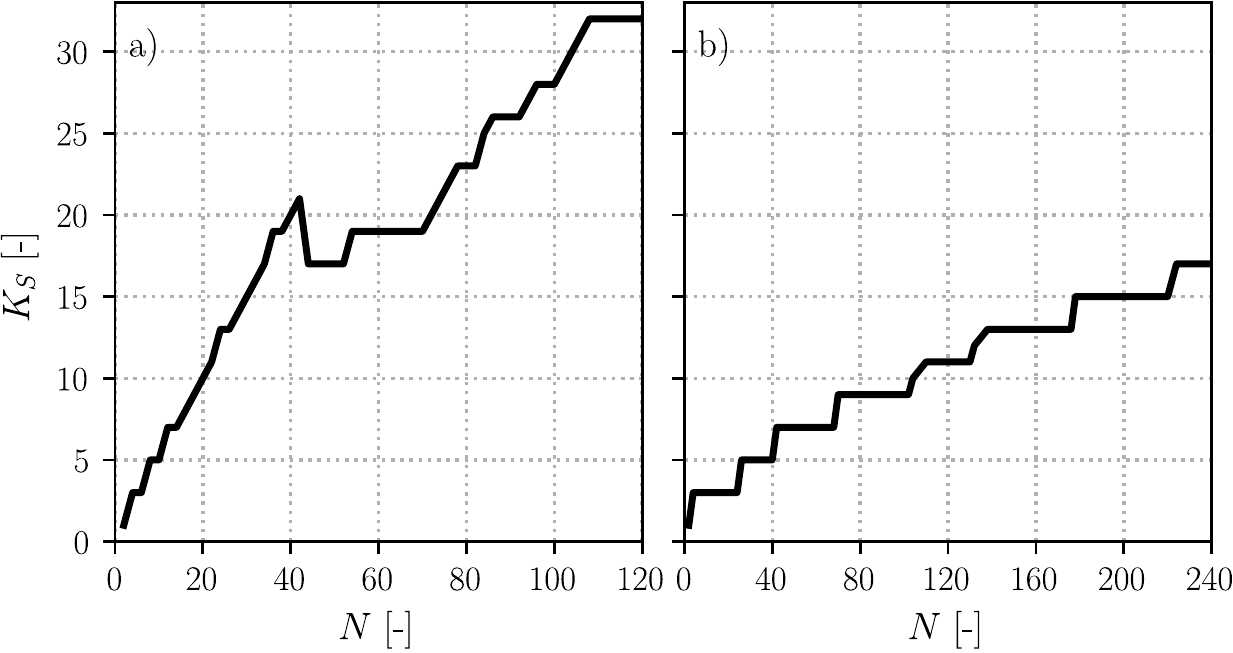}
\caption{(a) The surface modal basis size $\bm{K_S}$ for the annular liner, determined by the selection algorithm, as a function of the frames size $\bm{N}$ in the inner cylindrical subdomain and the outer annular subdomain. (b) Same, as a function of the frame size $\bm{N}$ in the combustor subdomain.}
\label{fig:surface_modes_numbers}
\end{figure*}

\noindent Two examples of geometrically complex boundaries will be presented in the following sections, namely an annular liner in Sec.~\ref{sec:surface_modal_expansion_convergence} and a conical outlet in Sec.~\ref{sec:surface_modal_expansion_annular_example}. In both cases, the selection algorithm described above is used to determine the optimal size $K_S$ of the truncated surface modal basis, as well as the modes that should be retained in its construction. The selected surface modes are only a function of the frames size $N$ in the adjacent subdomains, and are independent of the conductivity or impedance value on the boundary. The first few modes composing the surface modal bases for both cases are displayed in Fig.~\ref{fig:surface_modes_shapes}.
In both cases, the first surface mode is uniform, while the following ones come as pairs of orthogonal azimuthal modes. The sizes $K_S$ of the truncated surface modal bases are shown in Fig.~\ref{fig:surface_modes_numbers} as a function of the adjacent subdomains frames size $N$.
The evolution of $K_S(N)$ depends on a multitude of geometrical features inherent to both the complex boundary itself and to the connected subdomains. The trend is simpler for the combustor outlet (Fig.~\ref{fig:surface_modes_numbers}-(b)), since the boundary is in this case adjacent to a single subdomain: $K_S$ increases continuously with $N$, and scales as $K_S (N) \sim N/15$. For the annular multi-perforated liner (Fig.~\ref{fig:surface_modes_numbers}-(a)), since $S_c^{(1)}$ has two neighboring subdomains of different geometries, the dependence of $K_S$ with $N$ $( N = N^{(1)} = N^{(2)})$ is more complicated. It roughly scales as $K_S (N) \sim 4 N/15$, but presents a first linear increase from $N = 0 $ to $N=40$. It then reaches a plateau for $40 \leq N \leq 70$, and continues with a second linear increase for $N \geq 70$. Further analysis of the selection algorithm showed that in the initial linear portion, the value of $K_S$ set by the selection algorithm is constrained by the outer annulus frame, while the inner cylinder frame requires fewer surface modes to be accurately represented on  $S_c^{(1)}$. Conversely, in the second linear portion $K_S$ is constrained by the inner cylinder frame, whereas the outer annulus frame necessitates a smaller surface modal basis.

\section{Accuracy and convergence assessment} \label{sec:surface_modal_expansion_convergence}

\begin{figure}[h!] 
\centering
\includegraphics[width=67mm]{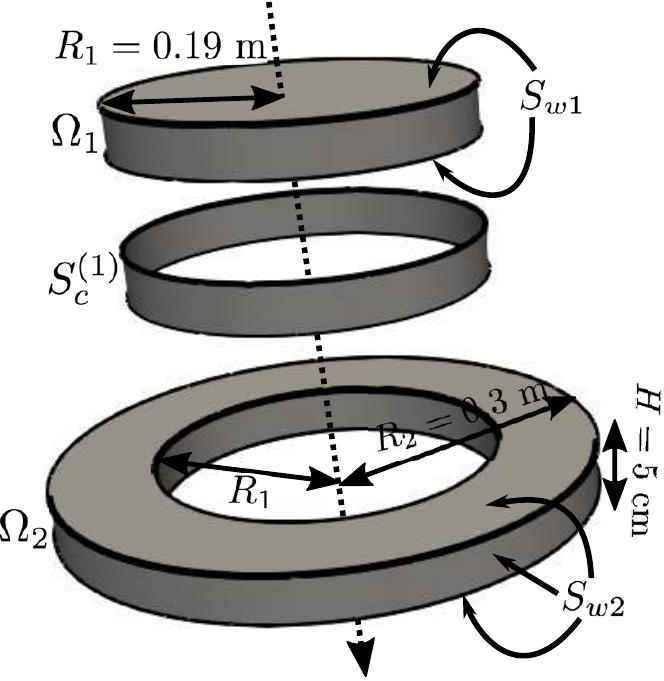}
\caption{Exploded view of the acoustic network consisting of 3 subsystems: 2 volume subdomains (an inner cylinder $\bm{\Omega_1}$ and an outer annulus $\bm{\Omega_2}$) that share a common connection boundary $\bm{S_c^{(1)}}$. The top, bottom, and lateral boundaries are rigid-walls. The mean fields are homogeneous within both subdomains, with $\bm{c_0 = 347.2}$~m/s and $\bm{\rho_0 = 1.176}$~kg/m\textsuperscript{3}.}
\label{fig:cylinder_geometry}
\end{figure}

The aim of this section is to evaluate the precision of the surface modal expansion method, and to provide an empirical verification of its convergence on a canonical case with analytically tractable reference solutions. The system of interest, representative of a simplified combustion chamber enclosed in its casing, is shown in Fig.~\ref{fig:cylinder_geometry}. It consists of a cylindrical geometry of radius $R_2$ and height $H$, delimited by rigid-walls and comprising an annular ribbon-like acoustic liner of radius $R_1$ characterized by its Rayleigh conductivity $K_R (j \omega)$.
The system is split into an acoustic network of 2 subdomains, and a complex boundary $S_c^{(1)}$ corresponding to the annular liner. In order to be able to satisfy any specified jump relation at this common boundary, modal frames $(\phi_n^{(1)} (\vec{x}))$ and $(\phi_n^{(2)} (\vec{x}))$ are built for each one of the subdomains. They are obtained numerically by using the FE Helmholtz solver AVSP~\cite{nicoud2007} separately on $\Omega_1$ and $\Omega_2$, that are meshed with uniform tetrahedral cells of dimension $\Delta x = H/10$. Since $H$ is much shorter than other dimensions, longitudinal modes are not included in these frames and will therefore not be discussed in the following. In addition, the number of modes in $(\phi_n^{(1)} (\vec{x}))$ and $(\phi_n^{(2)} (\vec{x}))$ is fixed to a same value $N = N^{(1)} = N^{(2)}$. The surface modal basis $(\mathscr{K}_k (\vec{x}_s))$ of $S_c^{(1)}$ is not computed by directly solving the curvilinear eigen-problem of Eq.~\eqref{eq:surface_eigenproblem}. The annular strip is instead replaced with a thin annulus where AVSP is used to resolve the classical three-dimensional homogeneous Helmholtz equation. Solutions of Eq.~\eqref{eq:surface_eigenproblem} are then obtained by interpolating back the obtained eigenmodes on the surface $S_c^{(1)}$.\par

After assembling every state-space representation of the network, the eigenvalues and eigenvectors of the whole system dynamics matrix $\mathbf{A}^f$ are solved. This yields, for every eigenmode $n$, its frequency $f_n$ and growth rate $\sigma_n$, its pressure mode shape $\Upsilon_{p,n} (\vec{x})$, and its velocity mode shapes $\Upsilon_{{u_r},n} (\vec{x})$ and $\Upsilon_{u_{\theta},n} (\vec{x})$. In the following, LOM solutions are compared to reference solutions (denoted with a superscript ${}^\mathcal{R}$) thanks to metrics defined for any scalar $s$ or field $g(\vec{x})$:
\begin{align}
\label{eq:errors_definitions}
E_{s} = \dfrac{| s^{\mathcal{R}} - s |}{s^{\mathcal{R}}} \ , \ E_{g} = \dfrac{|| g^{\mathcal{R}} (\vec{x}) - g (\vec{x}) ||}{|| g^{\mathcal{R}} (\vec{x}) ||} \ , \ \mathcal{E} \left\{ g  \right\} (\vec{x}) =  \dfrac{ \vert g^{\mathcal{R}} (\vec{x})   - g(\vec{x}) | }{\max \vert g^{\mathcal{R}} (\vec{x}) \vert } 
\end{align}
where $E_s$ is the relative error on the scalar $s$, $E_g$ is the L-2 norm relative error for the field $g (\vec{x})$, and $\mathcal{E} \left\{ g  \right\}$ is the local relative error for $g (\vec{x})$.

\subsection{Case of a subdomain simple interface ($K_R = \infty$)} \label{subsec:no_liner_example}

The limit case of an infinitely large conductivity $K_R$ is first inspected. In this situation, the term $Y_{K_R}$ in Eq.~\eqref{eq:curvilinear_Helmholtz_full} and Eq.~\eqref{eq:surface_dynamical_system} vanishes, such that at low-frequencies the acoustics dynamics in $S_{c}^{(1)}$ reduces to the quasi-static pressure continuity $\left( p^{\Omega_1} \vert \mathscrbf{K} \right) = \left( p^{\Omega_2} \vert \mathscrbf{K} \right)$ and velocity continuity $\left( u_s^{\Omega_1} \vert \mathscrbf{K} \right) = - \left(u_s^{\Omega_2} \vert \mathscrbf{K} \right)$. The liner then behaves as a simple interface between the cylinder and the annulus, and the reference solutions are therefore the eigenmodes of the cylinder of radius $R_2$ given by:
\begin{align}
\label{eq:analytical_solutions_cylinder}
\Upsilon_{p,mn}^{\mathcal{R}} (r,\theta) = J_n \left(\pi \beta_{mn} \dfrac{r}{R_2} \right) \sin  \left(  n (\theta - \theta_0) \right) \ , \ f_{mn}^{\mathcal{R}} = \dfrac{c_0 \beta_{mn}}{2 R_2}
\end{align}
where $J_n$ is the n\textsuperscript{th}-order Bessel function of the first kind, $\beta_{mn}$ are the roots of $J_n' (\pi \beta_{mn}) = 0$, and $\theta_0$ is an arbitrary constant. The LOM convergence is assessed by progressively increasing the number of modes $N$ in the frame expansions and by computing the errors defined in Eq.~\eqref{eq:errors_definitions} in function of $N$.\par

\begin{figure}[h!]
\centering
\includegraphics[width=0.9\textwidth]{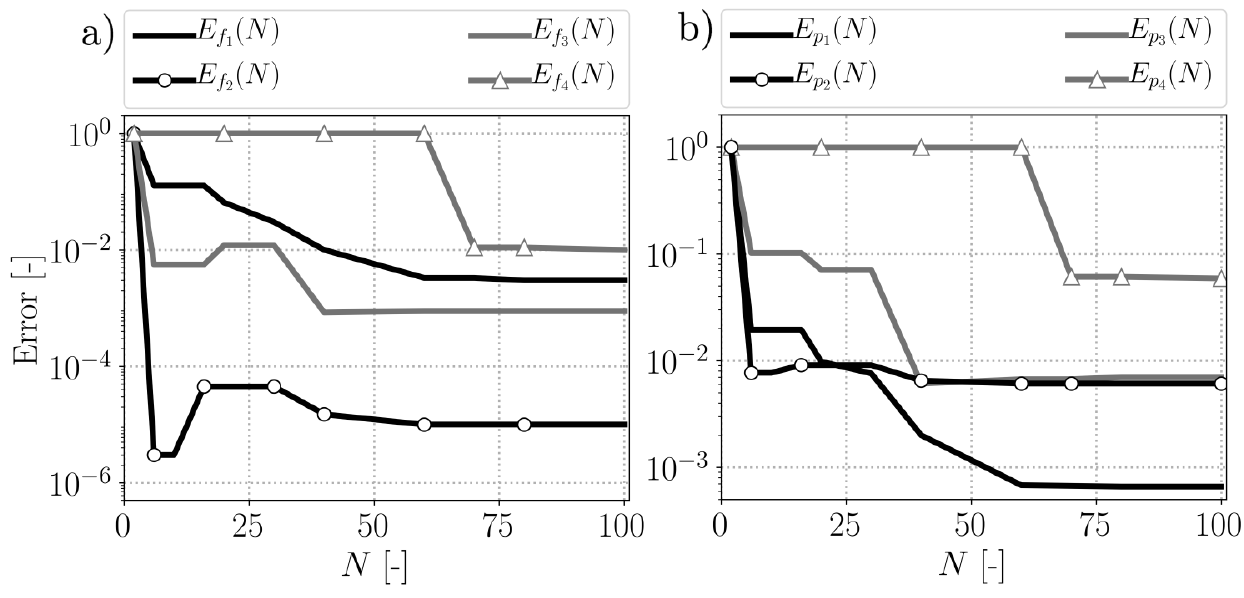}
\caption{(a) Frequency relative errors for mode 1 to mode 4, as a function of the modal frame size $\bm{N}$. (b) Pressure mode shape relative errors with respect to the L-2 norm, for mode 1 to 4, as a function of the modal frame size $\bm{N}$. Note the error logarithmic scale.}
\label{fig:cylinder_simple_errors}
\end{figure}

\begin{figure}[h!]
\centering
\includegraphics[width=0.99\textwidth]{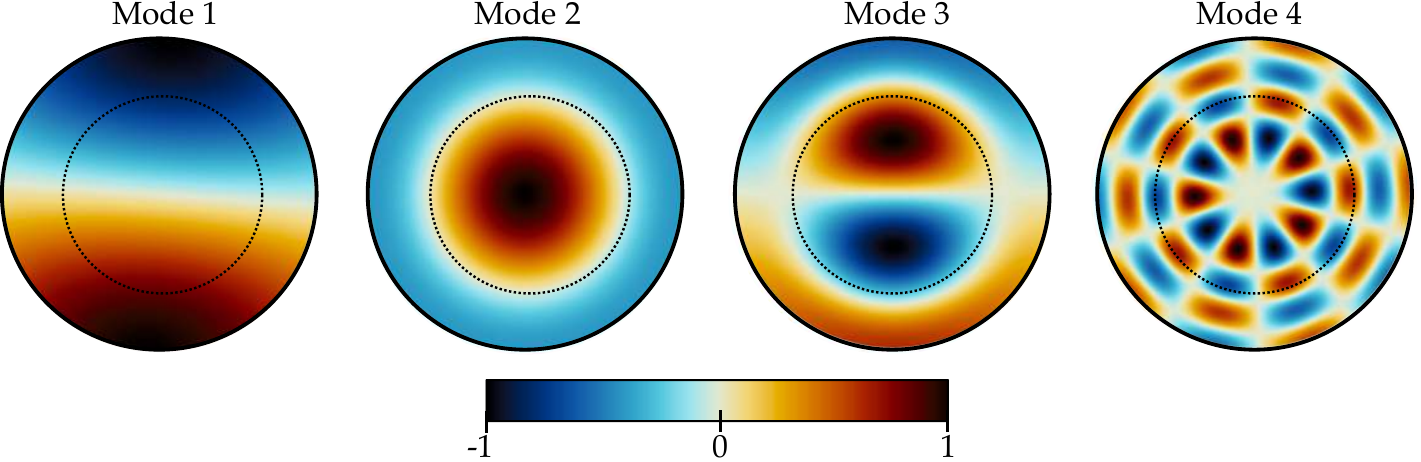}
\caption{Pressure shape of modes 1 to 4, for a modal frame size $\bm{N = 70}$. The dashed line represents the location of the connection surface $\bm{S_c^{(1)}}$ characterized by $\bm{K_R = \infty}$.}
\label{fig:cylinder_simple_mode_shapes}
\end{figure}

Figure~\ref{fig:cylinder_simple_errors} displays these results for a few select modes, namely mode 1 which is the first azimuthal mode ($n=1,m=0$), mode 2 designating the first radial mode ($n=0,m=1$), mode 3 referring to the first mixed-mode ($n=1,m=1$), and mode 4 which is a higher-order mixed mode ($n=3,m=5$).
Figure~\ref{fig:cylinder_simple_errors}-(a) shows that the frequencies computed with the LOM for modes 2 and 3 rapidly converges towards their reference values, with errors below 1\% for $N$ as low as 10. Further increasing the frame size results in a transitory stagnation, and even a slight deterioration, of the numerical solutions for these two modes. This is also observed on the mode shape errors in Fig~\ref{fig:cylinder_simple_errors}-(b). This behavior is attributed to the inclusion in the frame of modes that may not be accurately resolved due to numerical approximations in the FE solver used to generate them. This deterioration in the LOM results remains however limited, and vanishes for larger frames. The convergence of mode 1 is slower yet more monotonic: both its frequency and mode shape errors drop below 1\% at $N=40$ and reach $0.1$\% for $N$ greater than 60. Since mode 4 is higher-order ($f_4^{\mathcal{R}} = 3189$~Hz), it requires a larger frame to be accurately captured. Thus, the surface modal expansion method combined with the frame expansion achieves an overall satisfactory accuracy with relatively few DoF. For instance, with $N=70$ all the modes considered are resolved with an error below 1\% for both the frequency and the mode shape (except mode 4 which presents a 5\% error on its mode shape). This is confirmed in Fig.~\ref{fig:cylinder_simple_mode_shapes}, where the mode shapes of all 4 modes are shown to be accurately captured by the LOM.
Note in particular that the connection surface $S_c^{(1)}$ does not affect the modes shapes, and acts as expected as a simple interface between the two subdomains. Increasing the number of frame modes beyond $N=70$ does not improve the results, as the minimal achievable error is once again limited by the numerical approximations affecting the frame generation.

\subsection{Case of a multi-perforated liner} \label{subsec:liner_example}

After verification of its convergence in the elementary case of a simple interface between the cylinder and the annulus, the surface modal expansion performance is now evaluated in the more involved situation where the surface $S_c^{(1)}$ is a multi-perforated acoustic liner. Its Rayleigh conductivity is defined through a generalization of the classical Howe's model~\cite{howe1979} accounting for the plate thickness $h$~\cite{sun2002}. However, since the mathematical expression of this model comprises complex frequency-dependent Bessel functions, it cannot be directly translated into a state-space realization $\left\{  \mathbf{A_{K_R}},  \mathbf{B_{K_R}}, \mathbf{C_{K_R}} \right\}$. Instead, the original model is replaced with its 2\textsuperscript{nd}-order polynomial expansion given by:
\begin{align}
\label{eq:howe_approximation}
K_R (j \omega) = -K_R^A j \omega + K_R^B \omega^2 , \ \textrm{with} \ K_R^A = - \dfrac{\pi a^2}{2 U} \ , \ K_R^B = \dfrac{2 a^3}{3 U^2} + \dfrac{\pi a^2 h}{4 U^2}
\end{align}
where $a$ is the aperture radius and $U$ the bias flow speed. This expansion accurately approximates the original model (with discrepancies less than 1\%) for $a \omega / U \leq 0.2$, a range that encompass most multi-perforated liners encountered in practical applications. The conductivity state-space realization derived from Eq.~\eqref{eq:howe_approximation} is given in Appendix~\ref{sec:ss_realization_conductivity}. The procedure to obtain reference solutions is fully detailed in~\cite{gullaud2012} and reminded in Appendix~\ref{Appendix:B}. It relies on a separation of variables that leads to the resolution of a dispersion relation $F(j \omega) = 0$ in the complex plane. This last step cannot be performed analytically, and Powell's minimization algorithm~\cite{powell1964,press1992} is used to compute the roots of this equation.\par

\begin{figure}[h!]
\centering
\includegraphics[width=0.9\textwidth]{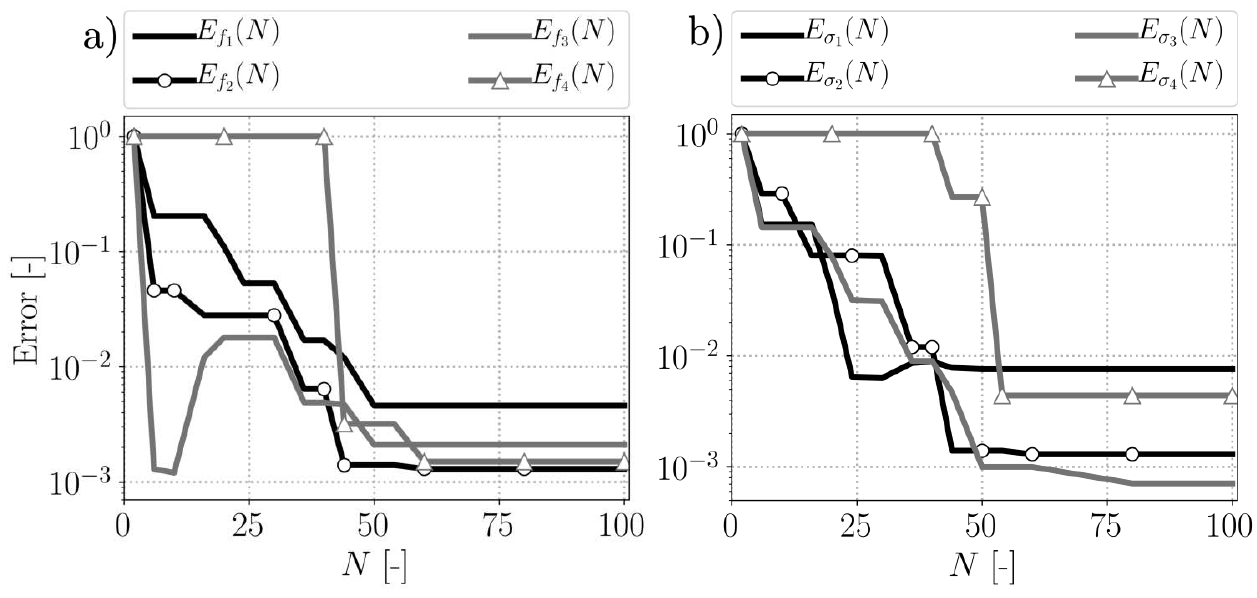}
\caption{(a) Frequency relative errors as a function of the modal frame size $\bm{N}$, for modes 1 to 4. (b) Growth-rate relative errors in function of $\bm{N}$ for modes 1 to 4. Errors are plotted with a logarithmic scale.}
\label{fig:cylinder_mlpf_errors}
\end{figure}

\begin{figure}[h!]
\centering
\includegraphics[width=0.99\textwidth]{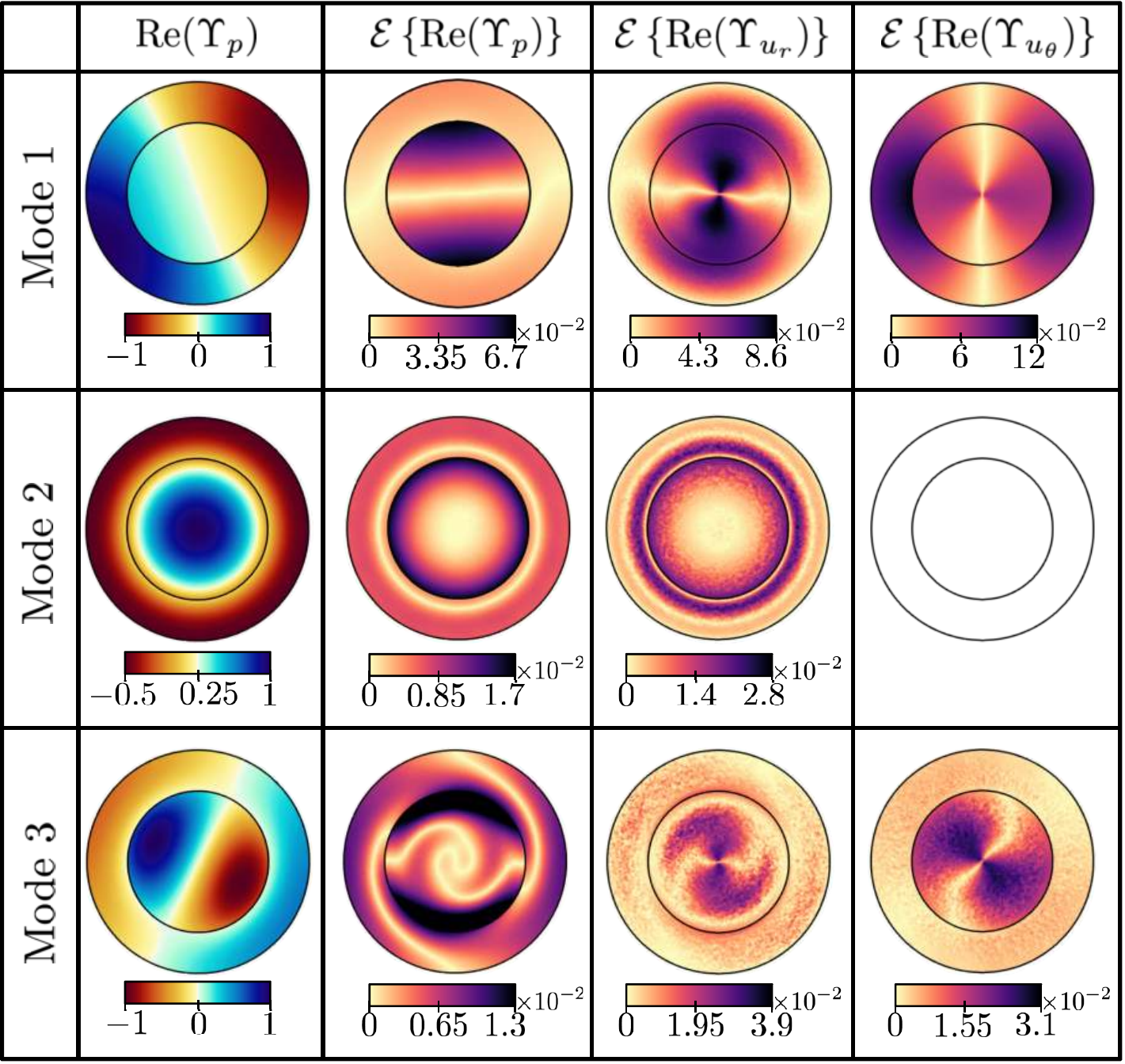}
\caption{Comparison between the mode shapes computed with the LOM ($\bm{N=40}$) and the reference solutions, for modes 1 to 3. First column: real part of the pressure mode shape computed with the LOM. Second column: local relative error on the pressure mode shape real part. Third column: local relative error on the radial velocity real part. Fourth column:  local relative error on the azimuthal velocity real part. The azimuthal velocity for mode 2 is not displayed since it is zero. All the fields $\pmb{\Re} \bm{ (\Upsilon_p) (\vec{x})}$, $\pmb{\Re} \bm{(\Upsilon_{u_r}) (\vec{x})}$ and, $\pmb{\Re} \bm{(\Upsilon_{u_{\theta}}) (\vec{x})}$ were normalized beforehand to fit in the range $\bm{[-1,1]}$.}
\label{fig:cylinder_mlpf_mode_shapes}
\end{figure}

The Rayleigh conductivity parameters are chosen to induce significant acoustic losses ($a = 0.5$~mm, $d = 5$~mm, $U = 10$~m/s, $h = 1$~mm, $\rho_u = \rho_0$). Low-order  results are then compared to reference solutions for the first azimuthal mode (mode 1, with $f_1^{\mathcal{R}} = 309.4$~Hz, $\sigma_1^{\mathcal{R}} = -62.6 \ \textrm{s}^{-1}$), the first radial mode (mode 2, with $f_2^{\mathcal{R}} = 604.4$~Hz, $\sigma_2^{\mathcal{R}} = -372.8 \ \textrm{s}^{-1}$), the first mixed mode (mode 3, with $f_3^{\mathcal{R}} = 806.8$~Hz, $\sigma_3^{\mathcal{R}} = -257.5 \ \textrm{s}^{-1}$), and a higher-order 3-3 mixed mode (mode 4, with $f_4^{\mathcal{R}} = 2553$~Hz, $\sigma_4^{\mathcal{R}} = -85 \ \textrm{s}^{-1}$). Figure~\ref{fig:cylinder_mlpf_errors} shows the frequency and growth rates relative errors, as a function of the number of modes $N$ in the frames $(\phi_n^{(1)} (\vec{x}))$ and $(\phi_n^{(2)} (\vec{x}))$.
The convergence trends are similar to the no-liner case of Sec.~\ref{subsec:no_liner_example}: for modes 1 to 3, the errors on both the frequency and the growth rate rapidly drops to only a few percents for $N$ lower than 20. The frequency of mode 3 presents a transitory deterioration starting at $N=20$ and vanishing for $N \geq 30$. Unsurprisingly, the higher-order mode 4 necessitates a larger frame to be accurately captured. For $N \geq 50$, both the frequencies and the growth rates of all the modes considered are accurately resolved with errors ranging from $0.1$~\% to $0.8$~\%. In Fig.~\ref{fig:cylinder_mlpf_mode_shapes} the errors committed by the LOM on the mode shapes are displayed for a fixed frame size ($N=40$).
The pressure mode shapes are accurately resolved by the LOM, with a maximum relative error not exceeding 7\% for mode 1, and 2\% for modes 2 and 3. Predictably, the maximum errors in these three cases  are reached in the vicinity of the multi-perforated liner where an important pressure discontinuity occurs. The radial and azimuthal velocity mode shapes are equally well captured, with relative errors locally peaking at a few percents only. These observations indicate that the combination of the surface and frame modal expansions not only enables the precise resolution of frequencies and growth rates in the presence of significant acoustic losses due to a liner, but also has the ability to capture the pressure and velocity fields, including the large pressure jump through the multi-perforated plate.

\section{Application to the partially reflecting outlet of a gas turbine annular combustor}
\label{sec:surface_modal_expansion_annular_example}

\begin{figure*}[h!]
\centering
\includegraphics[width=75mm]{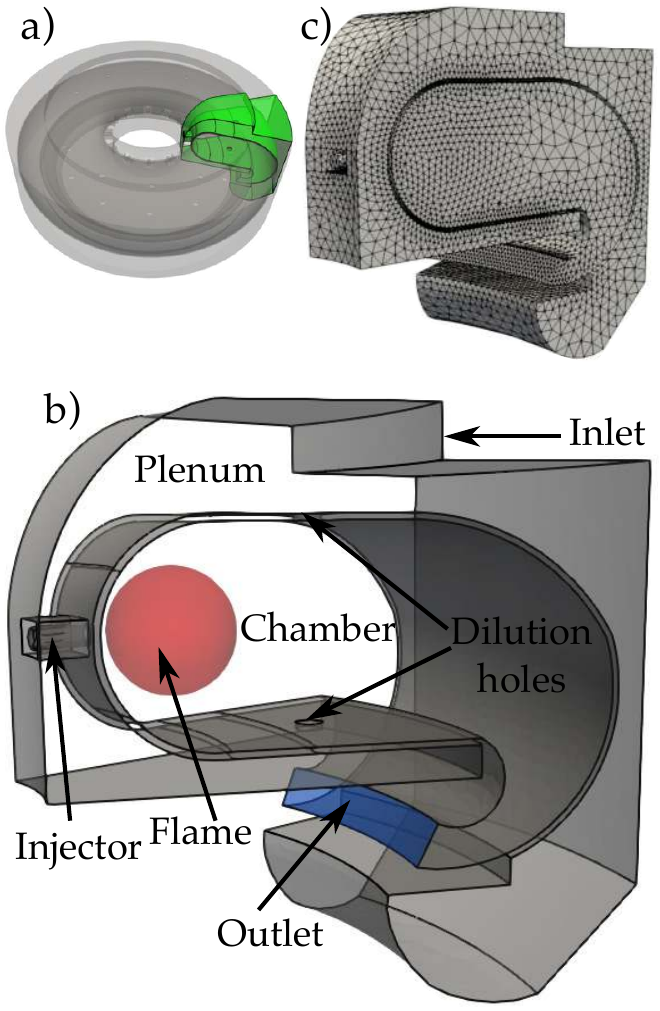}
\caption{(a) The annular combustor studied in this section, with one of its 12 sectors highlighted in green. (b) Closeup view of a sector. The red sphere indicates the spatial distribution of heat-release fluctuations, and the blue surface represents the partially reflecting chamber outlet. (c) Unstructured FEM mesh for one of the sectors. The mesh for the annular configuration, containing $\bm{2 \times 10^{6}}$ tetrahedral cells, is obtained by rotation and duplication of a single sector mesh.  For simplicity, the sound speed and density fields are assumed uniform in the entire domain, with $\bm{c_0 = 448.2}$~m/s and $\bm{\rho_0 = 0.706}$~kg/m\textsuperscript{3}.}
\label{fig:sme_combustor_mesh}
\end{figure*}

This section aims at assessing the ability of the surface modal expansion method to model complex-shaped impedance boundaries in realistic combustors where thermoacoustic instabilities may exist. The system of interest, represented in Fig.~\ref{fig:sme_combustor_mesh}, is an annular combustion chamber characteristic of those found in helicopter engine gas turbines.
It comprises 12 identical sectors, in each of which a $3$~cm radius spherical flame lies $4$~cm from the injector exit. The plenum and chamber are linked by dilution holes pierced through the wall. Acoustic losses due to hydrodynamic interactions that may exist at these holes are neglected. The focus is here on the chamber outlet, which may be chocked and give to a turbine. It usually cannot be considered as a rigid-wall or a pressure-release boundary, but is rather characterized by a finite impedance $Z$ (or equivalently a reflection coefficient $R = (Z-1)/(Z+1)$). In the seminal work of Marble and Candel~\cite{marble1977acoustic}, a compact chocked outlet has a real-valued impedance $Z = 2/((\gamma - 1) M_2)$, with $M_2$ the Mach number at the outlet. This expression shows that such boundary is only partially reflecting, which may induce acoustic losses and damp the unstable thermoacoustic modes of the combustor. The present approach is more generic: instead of using the model of~\cite{marble1977acoustic}, the impedance is varied from $Z = + \infty$ (\textit{i.e.} perfectly reflecting rigid-wall, or $R=1$) to $Z = 1$ (\textit{i.e.} non-reflecting boundary, or $R = 0$). From a design point of view, this strategy also allows the identification of the optimal outlet impedance that maximizes the damping of unstable thermoacoustic modes.\par

\begin{figure*}[h!]
\centering
\includegraphics[width=75mm]{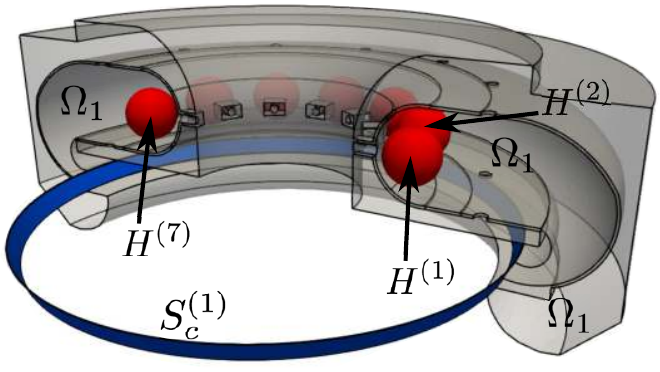}
\caption{A 180$\bm{{}^{\circ}}$ view of the acoustic network used in the LOM. The network contains 14 subsystems. The chamber, plenum, injectors, and dilution holes are gathered into a single subdomain $\bm{\Omega_1}$, which contains 12 heat sources $\bm{H^{(l)}}$ (in red). It is linked to a complex boundary $\bm{S_c^{(1)}}$ (in blue) representing the chamber outlet, and characterized by an impedance $\bm{Z}$. All the other boundaries are rigid-walls.}
\label{fig:sme_combustor_network}
\end{figure*}

The low-order acoustic network modeling the annular combustor is shown in Fig.~\ref{fig:sme_combustor_network}.
A single subdomain $\Omega_1$ comprising the entire combustor volume is used. It would also be possible to further split $\Omega_1$ into a collection of smaller subdomains (\textit{e.g.} with the plenum, chamber, and injectors as distinct subdomains), but as the aim is here to concentrate on the modeling of the chamber outlet, such decomposition is not necessary. The state-space representation for this latter is obtained by slightly adapting the dynamical system of Eq.~\eqref{eq:surface_dynamical_system} (since here $S_c^{(1)}$ is connected to only one subdomain). The expansion frame of $\Omega_1$, as well as the surface modal basis of $S_c^{(1)}$, are once again generated thanks to the 3D-FEM solver AVSP~\cite{nicoud2007}, for which the computational mesh is shown in Fig.~\ref{fig:sme_combustor_mesh}-(c). Note that the geometry of the complex boundary $S_c^{(1)}$ differs from that of the annular liner of Sec.~\ref{sec:surface_modal_expansion_convergence}, as its radius of curvature is not uniform but depends on the axial coordinate. The algorithm of Sec.~\ref{sec:surface_modes_number_choice} is used to select the $K_S (N)$ elements that are retained in the construction of the surface modal basis: $K_S (N)$ grows continuously, from $K_S (N = 20) = 3$ to $K_S (N = 240) = 17$. The response of a given flame $H^{(l)}$ to acoustic fluctuations is modeled by a constant $n-\tau$ FTF:
\begin{align}
\label{eq:FTF_n_tau}
\hat{Q}_l (\omega) = n_f \ e^{-j \omega \tau_f} \ \hat{u}_{f,l} (\vec{x}_{f,l}, \omega)
\end{align}
where $n_f = 2000$~J/m is the flame gain, $\tau_f = 1.667$~ms is the time-delay, and $\hat{u}_{f,l}$ is the acoustic velocity at the location $\vec{x}_{f,l}$. This reference point is located at the middle of each one of the 12 injectors. Similarly to previous chapters, the frequency-domain FTF of Eq.~\eqref{eq:FTF_n_tau} is converted into a state-space realization thanks to a Pole Base Function expansion of the time-delay term:
\begin{align}
\label{eq:PBF_expansion}
e^{-j \omega \tau_f} \approx \sum_{q = 1}^{M_{P}} \dfrac{-2 a_q j \omega}{\omega^2 + 2 c_q j \omega - \omega_{0q}^2}
\end{align} 
where the coefficients $a_q, c_q, \omega_{0q}$ are fitted with a specialized optimization algorithm proposed in~\cite{douasbin2018}.\par

\begin{figure*}[h!]
\centering
\includegraphics[width=0.99\textwidth]{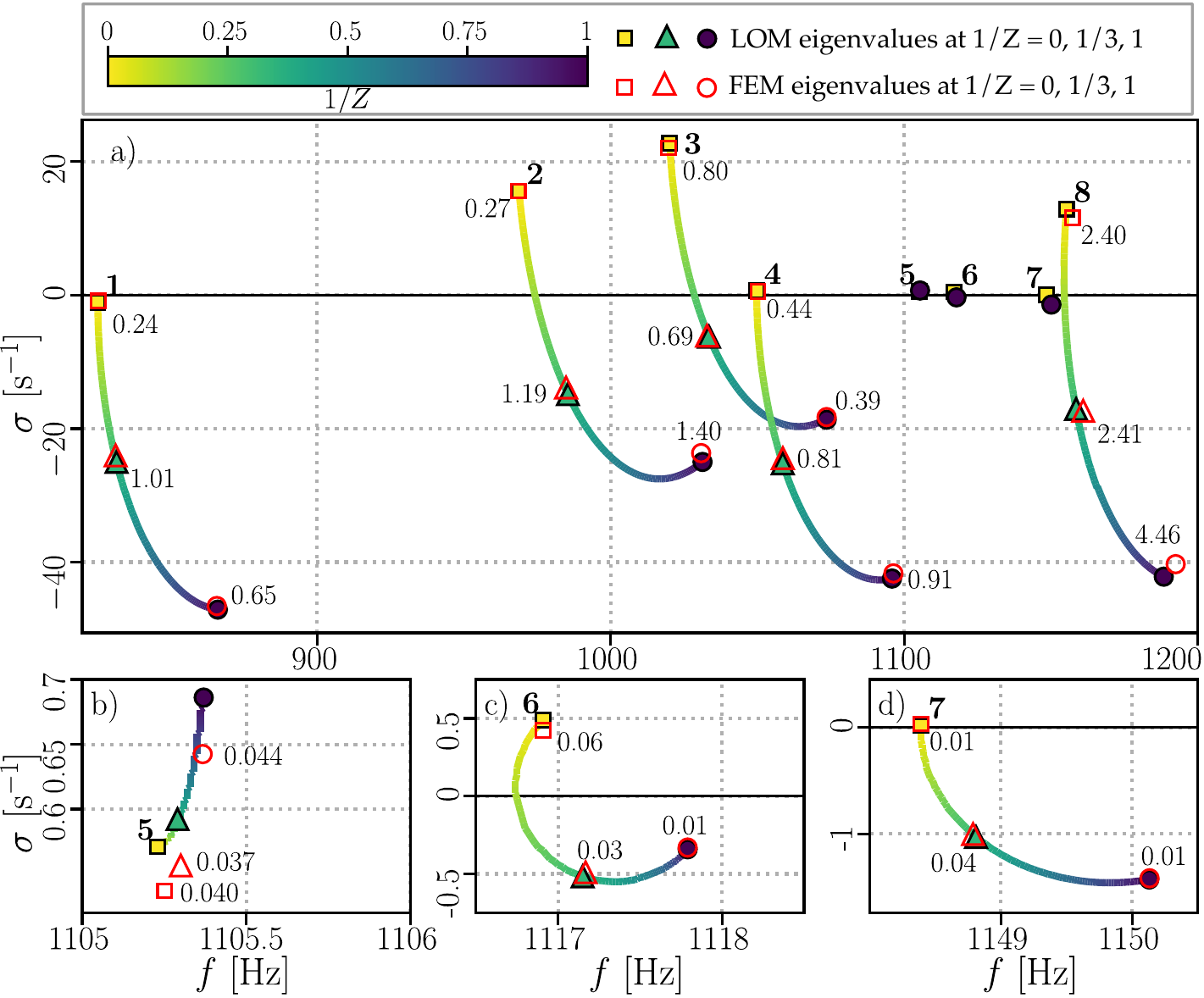}
\caption{(a) Trajectories in the complex-frequency plane $\bm{f - \sigma}$ of the 8 modes computed by the LOM in the range $\bm{[800 \textrm{ Hz} , \ 1200 \textrm{ Hz}]}$, as the outlet impedance is decreased from $\bm{Z = + \infty}$ to $\bm{Z = 1}$. Dark colored symbols indicate the complex eigenfrequencies computed by the LOM for $\bm{Z = + \infty}$, $\bm{Z = 3}$, and $\bm{Z = 1}$, while the red symbols show the corresponding FEM reference solutions. The numbers are the absolute errors (in Hz) committed by the LOM at these points. (b), (c), (d) Closeup views on the trajectories of modes 5, 6, and 7, respectively.}
\label{fig:sme_combustor_tracking_Rz}
\end{figure*}

\begin{figure*}[h!]
\centering
\includegraphics[width=136mm]{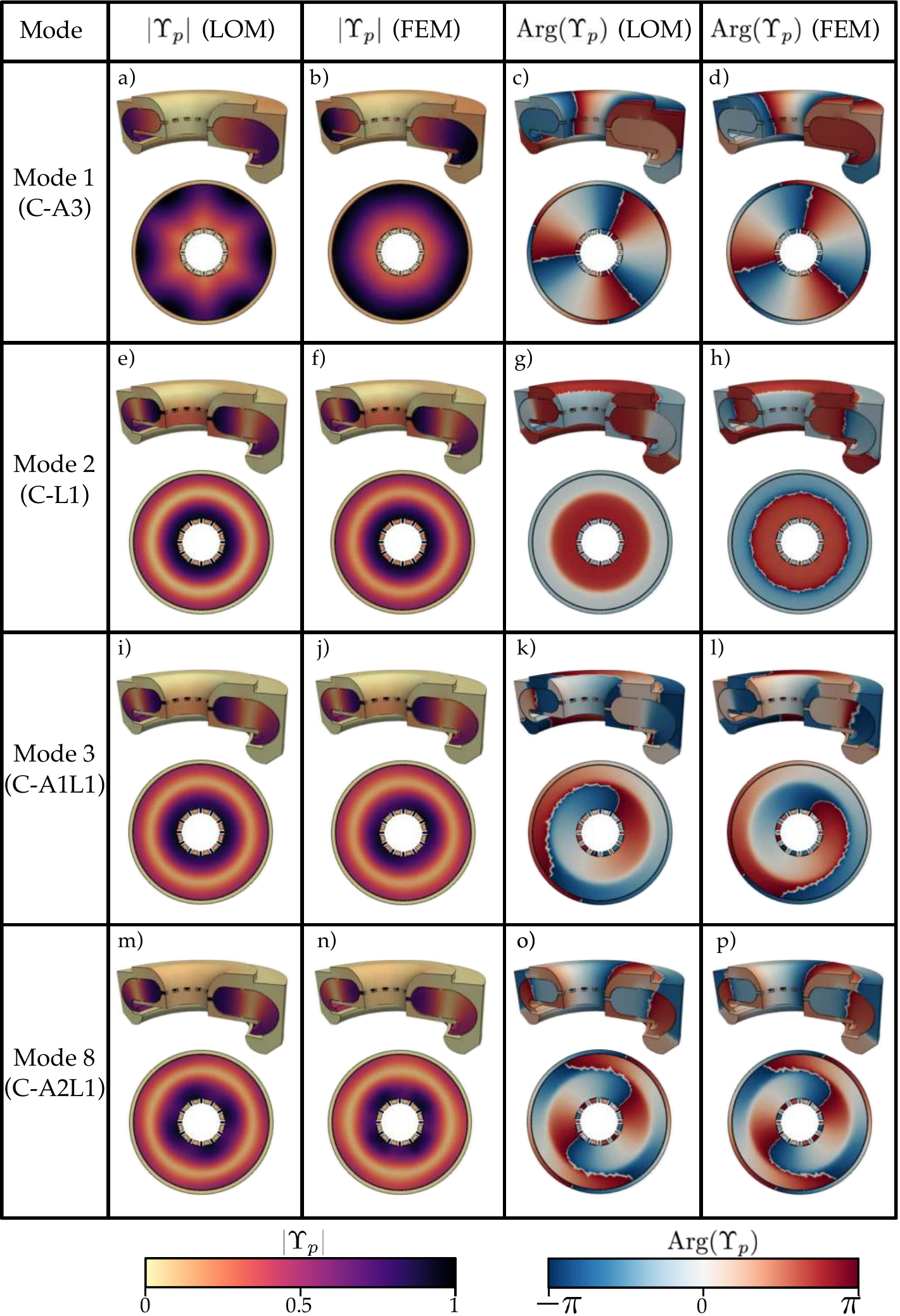}
\caption{Pressure mode shape $\bm{\Upsilon_p (\vec{x})}$ at $\bm{Z=1}$ of modes 1, 2, 3, and 8, computed with the LOM and compared to the FEM reference solutions. "C" stands for "Chamber", and AxLy denotes a mode of order x in the azimuthal direction and order y in the longitudinal one.}
\label{fig:sme_combustor_modes_shapes}
\end{figure*}

All the unstable thermoacoustic modes of the combustor are observed to lie between $800$~Hz and $1200$~Hz. A series of 100 LOM simulations are performed to continuously decrease the outlet impedance from $Z = + \infty$ to $Z = 1$. The frame size is fixed to $N = 120$, such that the modal surface basis contains $K_S = 11$ elements. All 8 modes comprised in the range $[800 \textrm{ Hz} , \ 1200 \textrm{ Hz}]$ are computed, and the trajectories they follow in the complex-frequency plane as $Z$ varies are shown in Fig.~\ref{fig:sme_combustor_tracking_Rz}. For validation purpose, AVSP is used to resolve the corresponding reference solutions for a few select impedance values, namely $Z = + \infty$ (perfectly reflecting rigid-wall), $Z = 3$ (partially reflecting), and $Z = 1$ (anechoic).
In addition, the spatial shapes of a few modes at $Z = 1$ are compared in Fig.~\ref{fig:sme_combustor_modes_shapes}.
An overall excellent agreement is found between the LOM and the FEM reference solutions for both frequencies, growth rates, and spatial shapes. Modes 1, 2, 3, 4, and 8 are chamber modes and are therefore significantly affected by the decrease of the outlet impedance (Fig.~\ref{fig:sme_combustor_tracking_Rz}-(a)). More particularly, modes 2, 3 and 8 are the most unstable thermoacoustic modes of the combustor for a perfectly reflecting outlet, but decreasing the impedance to $Z=4$ is sufficient to damp and stabilize them. This trend is well captured by the LOM, with errors that do not exceed $5$~Hz for mode 8, and stay below $1.5$~Hz for others. Modes 5, 6, and 7 are plenum modes, and the effect of the chamber outlet reflection coefficient on them is therefore more subtle. They remain marginally stable/unstable (\textit{i.e.} with growth rates close to zero), even under anechoic condition. Figures~\ref{fig:sme_combustor_tracking_Rz}-(b,c,d) evidence the capability of the surface modal expansion method to accurately represent even these minute growth rate and frequency variations. Counterintuitively, a perfectly anechoic outlet does not maximize the damping of modes 2 and 6: their optimal damping is rather reached for a larger reflection coefficient roughly equal to $0.18$. This trend is even more remarkable for mode 5 (1\textsuperscript{st} plenum longitudinal mode) as its growth rate steadily increases when $R$ decreases. Only a few minor differences are observed between the LOM spatial mode shapes and the associated FEM reference solutions, most noticeably on the modulus of mode 1 (Fig.~\ref{fig:sme_combustor_modes_shapes}-(a,b)), which nonetheless vanishes for larger $N$, and on the phase of mode 3 (Fig.~\ref{fig:sme_combustor_modes_shapes}-(k,l)). Note, however, that this latter is simply due to the spinning azimuthal component, which can either be clockwise or anticlockwise, and therefore does not indicate an error committed by the LOM.\par

\begin{figure*}[h!]
\centering
\includegraphics[width=144mm]{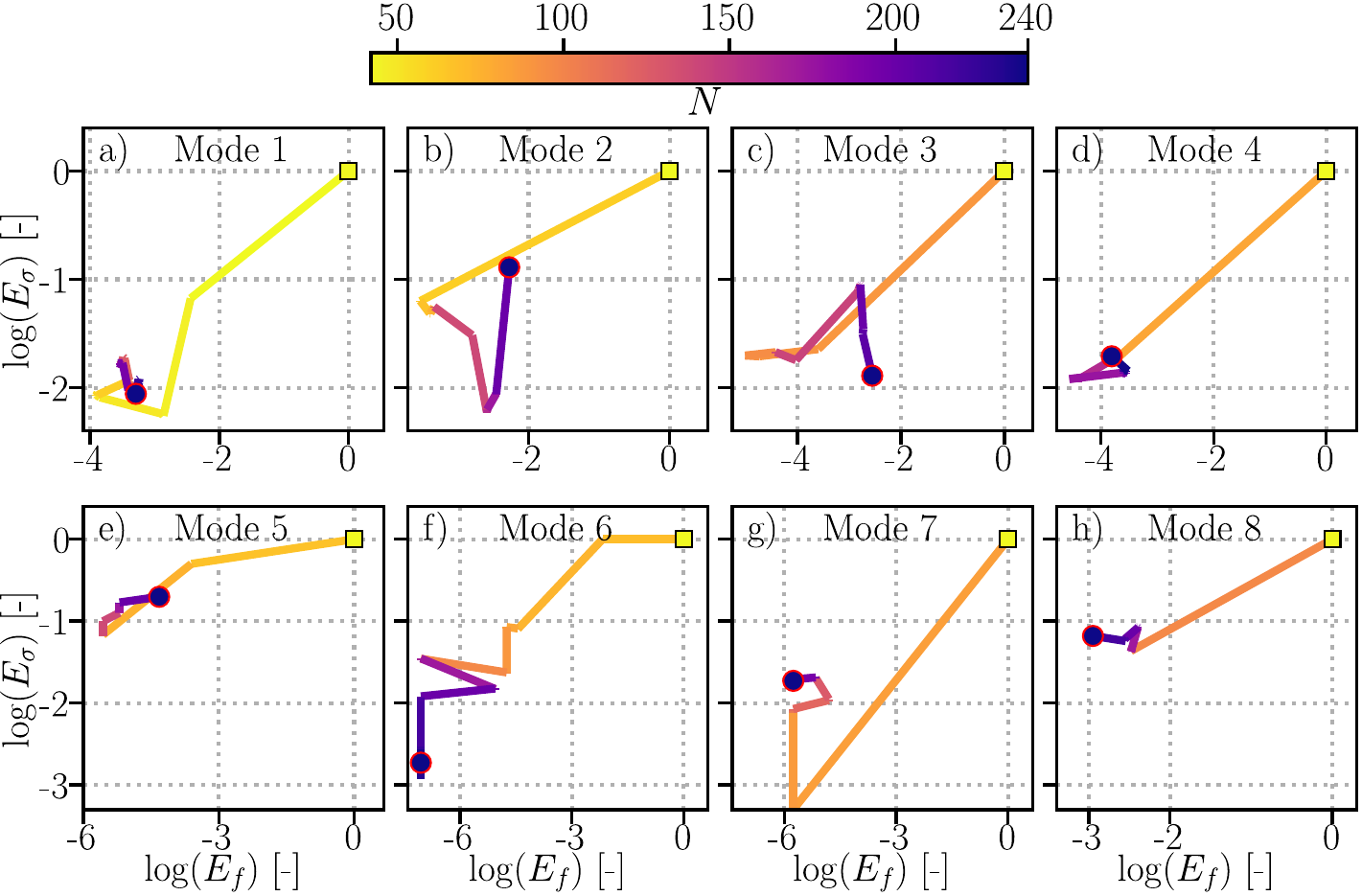}
\caption{Trajectories in the complex plane $\bm{E_{f} - E_{\sigma}}$ of the 8 modes of interest, as the frame size increases from $\bm{N = 40}$ to $\bm{N = 240}$. Note the logarithmic scale on both axes.}
\label{fig:combustor_convergence_N}
\end{figure*}

A second series of computations is performed to evaluate the accuracy of the LOM with respect to $N$ and $K_S(N)$. In this matter, the outlet condition is set to anechoic ($Z = 1$), and $N$ is progressively increased from $40$ to $240$. Equation~\eqref{eq:errors_definitions} is used to calculate the relative frequency error $E_f$ and growth rate error $E_{\sigma}$ between the LOM and the FEM reference solutions for each one of the 8 modes of interest; the trajectories they follow in the complex plane as $N$ varies are displayed in Fig.~\ref{fig:combustor_convergence_N}.
The convergence behavior is comparable to that observed in Sec.~\ref{subsec:no_liner_example} and Sec.~\ref{subsec:liner_example}: after an initial rapid convergence, a further increase in the frame size results in a stagnation of the errors, and even a slight deterioration for some modes. This latter remains however moderate, as all the modes are converged with frequency errors below 1\% and grow rate errors in the range 1\%-10\% for $N \geq 110$ (at the exception of mode 5, for which $E_{\sigma}$ reaches 20\%). This slight degradation of the results is once again attributed to the frame condition number that increases with $N$. Note that similarly to the liner case of Sec.~\ref{sec:surface_modal_expansion_convergence}, lower-order modes are converged with smaller frame and surface basis: for instance $48$ and $70$ frame elements are necessary to yield $E_f < 1$\% and $E_{\sigma} < 10$\% on mode 1 and mode 2, respectively. Conversely, higher-order modes necessitate larger expansions to be accurately resolved (\textit{e.g.} $110$ frame elements are required to achieve $E_f < 1$\% and $E_{\sigma} < 10$\% on mode 8). The overall LOM convergence is deemed satisfactory, as Fig.~\ref{fig:combustor_convergence_N} shows that the errors stay bounded within acceptable limits. This confirms the ability of the surface modal expansion method to accurately model the geometrical complexity of realistic generic outlet impedance in industrial combustors.

 \section{Conclusion}  \label{sec:frame_modal_expansion_conclusion}

The representation of geometrically complex boundaries, such as multi-perforated liners or chamber inlets and outlets, is out of reach for the majority of existing thermoacoustic LOMs. This restriction is a major obstacle preventing these models from dealing with realistic industrial combustors, and the literature on the subject is nearly non-existent. In the purpose of circumventing this difficulty and of improving state-of-the-art thermoacoustic LOMs, we introduced the surface modal expansion method to model these topologically complex boundaries. The approach relies on two principles: on one hand it takes advantage of the frame modal expansion proposed in Chapter~\ref{chap:FRAME} to accurately represent the acoustic velocity and pressure fields near non-rigid-wall boundaries. On the other hand, it exploits a reformulation of the Acousto-Elastic Method, used in vibroacoustics, to model the complex-shaped frontiers as two-dimensional manifolds where acoustic variables are expanded onto an orthogonal basis of modes solutions of a curvilinear Helmholtz problem.\par

The governing equations resulting from the surface modal expansion method are formulated in the state-space framework used in this work, and the corresponding state-space realizations are implemented in STORM. The potential of the surface modal expansion method for the low-order modeling of complex boundaries was estimated on a non-reactive case comprising an annular multi-perforated liner in a cylindrical geometry. This example evidenced a satisfactory convergence behavior, similar to that of the frame modal expansion presented in Chapter~\ref{chap:FRAME}. The ability of the surface modal expansion method to accurately represent the partially reflecting of an industrial gas turbine combustor in the presence of unstable thermoacoustic modes was then shown. However, further validations on more complex cases, combining dilution holes on multi-perforated liners and inlet and outlet boundaries, are still required to fully assess the potential of the surface modal expansion method. Note that the generation of the surface modal bases should preferably be performed with a curvilinear Helmholtz solver. Due to the unavailability of such solver at CERFACS, it was decided to generate the surface modal bases with the 3D FE solver AVSP. The preliminary construction of the surface modal bases is performed through a pre-processing tool implemented in STORM API.\par

\extraPartText{ \hfill \\[5ex]
\textbf{This part led to the following publications:} \\[5mm]
\textit{\textbf{1. Laurent, C.}, Esclapez, L., Maestro, D., Staffelbach, G., Cuenot, B., Selle, L., Schmitt, T., Duchaine, F., Poinsot, T. (2019). Flame wall interaction effects on the flame root stabilization mechanisms of a doubly-transcritical LO2/LCH4 cryogenic flame.} \textbf{Proceedings of the Combustion Institute}, 37(4), 5147-5154. \\[4mm]
\textit{\textbf{2. Laurent, C.}, Staffelbach, G., Nicoud, F., Poinsot, T. (2020). Heat-release dynamics in a doubly-transcritical LO\textsubscript{2}/LCH\textsubscript{4} cryogenic coaxial jet flame subjected to fuel inflow acoustic modulation.} (\textbf{Proceedings of the Combustion Institute}, Accepted) \\[4mm]
\textit{\textbf{3. Laurent, C.}, Poinsot, T. (2020). Nonlinear dynamics in the response of a doubly-transcritical  LO\textsubscript{2}/LCH\textsubscript{4} coaxial jet flame to fuel inflow acoustic perturbations.} (\textbf{Combustion and Flame}, Submitted)}
\part{Dynamics of a doubly-transcritical LO\textsubscript{2}/LCH\textsubscript{4} liquid rocket engine flame} \label{part:RG}
				
\chapter{Numerical simulation of real-gas flows} \label{chap:LES_RG}
\minitoc				

\begin{chapabstract}
This chapter is an introduction to flame dynamics in LRE conditions and to the numerical methods that are used throughout this work to simulate those. It starts by emphasizing the specificity of combustion in LRE conditions, which lies in the dominant non-ideal thermodynamics effects due to the injection of cryogenic reactants in a high-pressure environment. Previous studies interested in real-gas mixing and in high-pressure cryogenic flames are briefly reviewed. In particular, the widely documented Mascotte test rig (operated by ONERA, France) is described in some detail, as it is in this geometry that the present numerical simulations are performed. More specifically, this work is interested in the doubly-transcritical regime, where both oxygen and methane are injected as cryogenic liquids. The motivations behind this choice of geometry and operation conditions are explained. Then, the real-gas version of the solver AVBP, that is used here to perform LES and DNS of LRE flames is introduced. The governing equations it resolves, as well as the models and numerical methods it employs are depicted. Finally, an important contribution of this work is presented: it consists in the derivation of a chemical kinetic mechanism to model CH\textsubscript{4} combustion in LRE conditions. This kinetic scheme is used in the subsequent numerical simulations that are the object of the following chapters.
\end{chapabstract}

\section{Introduction to real-gas flows}   \label{sec:les_rg_intro}

In typical Liquid-Rocket Engines, the reactants are injected into the combustion chamber through hundreds of coaxial injectors. The oxidizer is usually pure oxygen, while the fuel can be pure hydrogen, methane, or rocket-grade kerosene. The combustion process that then occurs strongly differs from that encountered in gas turbines and is remarkable from a thermodynamical point of view for a number of reasons: (1) the high chamber pressure that can range from $50$~bars to more than $100$~bars, (2) the cold cryogenic reactants whose temperature can be lower than $80$~K, and (3) the high temperatures reached in the flame region, sometimes exceeding $4000$~K. The reactants mixing, the combustion process, and ultimately the flame dynamics strongly depend on these particular thermodynamic conditions, which therefore require a thorough modeling.

\subsection{Real-gas thermodynamics} \label{sec:thermo_RG}

\begin{figure}[h!]
\centering
\includegraphics[width=0.99\textwidth]{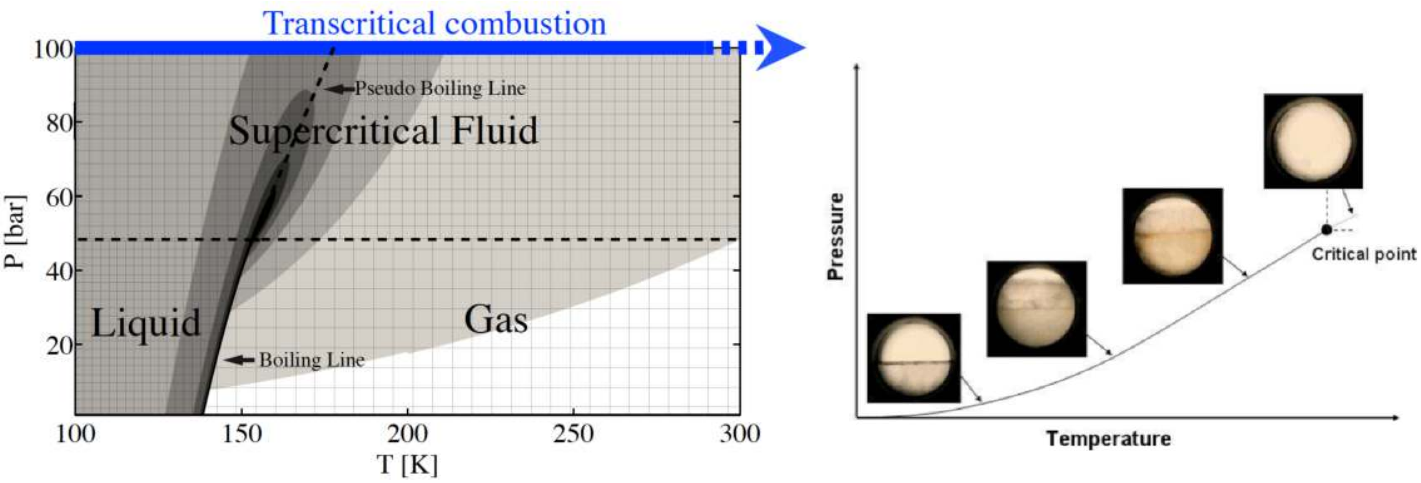}
\caption{Left: $\bm{P \! - \! T}$ diagram for pure O\textsubscript{2}, with contours of heat capacity $\bm{C_p}$ indicated in gray scale (white:  10\textsuperscript{3}~J/K.kg, black: 10\textsuperscript{4}~J/K.kg). Extracted from~\cite{Ruiz:2012phd}. Right: illustration of the vanishing liquid-gas interface in a $\bm{P \! - \! T}$ diagram. Extracted from~\cite{Poling:2001}.}
\label{fig:PT_diagram}
\end{figure}

The high ambient pressure that exists in a LRE combustion chambers enhances the interactions between the fluid molecules. These interactions result in a modification of the fluid thermodynamic properties, which does not behave as an ideal gas anymore, but rather as a so-called \textit{real-gas}. The most remarkable of these non-ideal behaviors occurs when the ambient pressure $P$ exceeds the fluid critical pressure $P_C$. As illustrated in the pressure-temperature diagram in Fig.~\ref{fig:PT_diagram} (left), when a liquid is introduced into a \textit{subcritical} environment at $P < P_C$ and is heated-up (for instance through chemical reactions), it undergoes a sudden vaporization to become an ideal-gas when its temperature meets the boiling line. On the contrary, when a liquid is injected in a \textit{supercritical} environment at $P > P_C$, a clear boiling line does not exist anymore. Instead, when its temperature reaches the pseudo-boiling region, the liquid undergoes a progressive and continuous transition to become a supercritical fluid. This pseudo-vaporization is called a \textit{transcritical} transformation, and the liquid in its initial injection state is sometimes referred as a transcritical fluid. The point that separates these two thermodynamic transition regimes is called the critical point and is located at $(P_C,T_C)$ in the $P \! - \! T$ diagram. 
Figure~\ref{fig:PT_diagram} (right) shows an interesting manifestation of a transcritical transformation: when the pressure is increased in a liquid-vapor system at equilibrium, the clear interface that exists between the liquid and the gas progressively vanishes. When the pressure exceeds $P_C$ this interface is blurred and nearly completely disappears.\par

\begin{figure}[h!]
\centering
\includegraphics[width=0.99\textwidth]{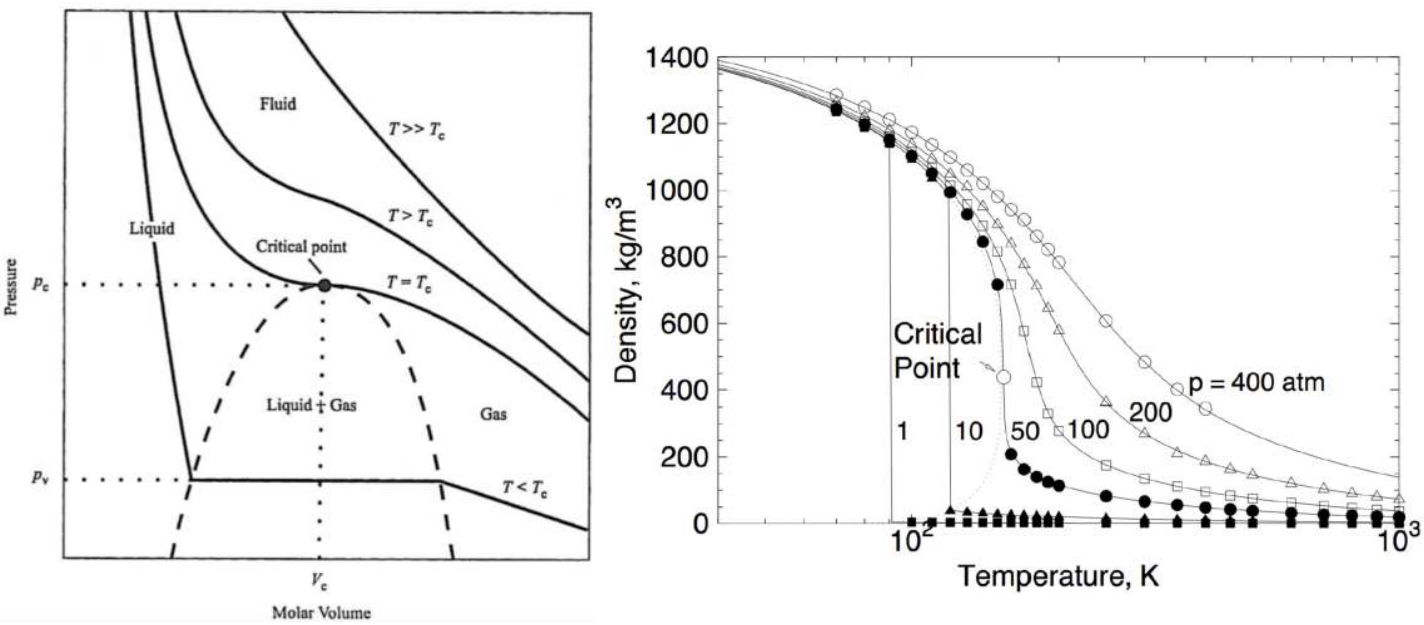}
\caption{Left: a typical $\bm{P \! - \! V}$ diagram for a pure fluid, with the critical point, and some isothermal lines. Right: oxygen density as a function of temperature, for increasing pressures. Extracted from~\cite{Mari:2015phd}.}
\label{fig:PV_diagram}
\end{figure}

Other key properties of fluids in elevated pressure environments are illustrated in Fig.~\ref{fig:PV_diagram} (left). The isothermal line $T = T_C$ is tangential to both the dew and the bubble curves at the critical point $P = P_C$, thus indicating that near its critical point, a fluid is infinitely compressible and infinitely dilatable. In other words, near the critical point, a small variation of pressure or temperature produces a very large variation of volume (or equivalently density).
This strongly nonlinear behavior is further evidenced in Fig.~\ref{fig:PV_diagram} (right). For subcritical or near-critical pressures, the density presents a sharp discontinuity at the boiling temperature, marking the frontier between the liquid and gas states. However, for supercritical pressures this discontinuity is smoothed: the higher the pressure, the larger the pseudo-boiling region. Note that for $P \gg P_C$ and $T \gg T_C$ the isothermal lines and the density evolution nearly match that of an ideal gas, which suggests that under these conditions a supercritical fluid does not strongly differ from an ideal one.\par

One of the simplest approaches to model the complex nonlinear thermodynamics observed at supercritical pressure consists in using a cubic Equation of State (EoS). Those are usually based on the Principle of Corresponding Sates~\cite{Hirschfelder:1954}, and also include correction terms accounting for non-spherical molecules interactions based on the acentric factor~\cite{Pitzer:1955}. In this work, the Soave-Redlich-Kwong (SRK) Equation of State~\cite{soave1972} is employed. Its expression writes:
\begin{align}
\label{eq:SRK}
P = \dfrac{\rho \bar{r} T}{1 - \rho b_m} \ - \ \dfrac{\rho^2 a_m(T)}{1+\rho b_m}
\end{align}
where $\bar{r}$ is the mixture gas constant, and $b_m$ and $a_m(T)$ are calculated as in~\cite{Yang2000}. Note that in the limits of small density or large temperature, it reduces to the ideal-gas law $P = \rho \bar{r} T$. This EoS is here selected for its relative simplicity, as well as for its accuracy in the prediction of the density at $T < T_C$. It is however known to commit larger errors near the critical point. Most importantly, it is crucial to notice that here a "\textit{dense-gas}" approach is used, such that only a single phase can be resolved and subcritical conditions ($P < P_C$, $T < T_C$) cannot be accounted for. Modeling multi-components phase-separation necessitates additional thermodynamic equations, such as the resolution of a Vapor-Liquid-Equilibrium. This strategy was employed in a few recent studies~\cite{Zips:2018,Traxinger:2018,Traxinger:2019}, but these advanced thermodynamic behaviors are not considered in the present work.

\subsection{Real-gas non-reactive flows}  \label{sec:RG_flows}

\begin{figure}[h!]
\centering
\includegraphics[width=0.99\textwidth]{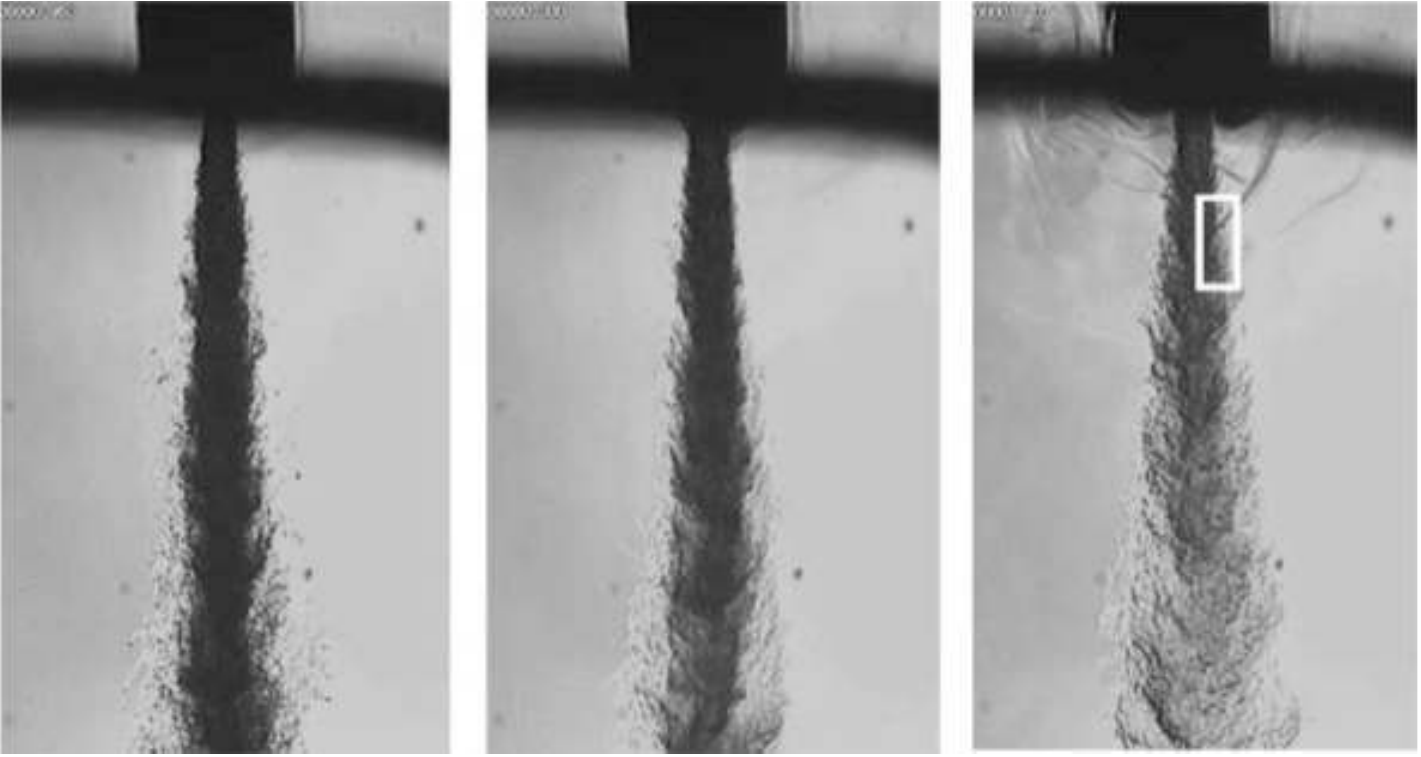}
\caption{Injection of a non-reactive cryogenic N\textsubscript{2} into a quiescent N\textsubscript{2} environment at $\bm{P}$. Left: subcritical ($\bm{P = 0.91 P_C}$). Center: supercritical ($\bm{P = 1.22 P_C}$). Right: supercritical ($\bm{P = 2.71 P_C}$). Extracted from~\cite{chehroudi2002}.}
\label{fig:jets_supercritical}
\end{figure}

Non-ideal thermodynamics in LRE high-pressure conditions have a first order effect on the fluid transport. This was first evidenced through the study of canonical non-reactive flows such as single droplet evaporation, or mono-component mixing layers~\citep{Yang2000,Bellan2000}. At subcritical pressure, the mixing of cryogenic reactants was found to be controlled by atomization and droplets evaporation. On the contrary, at supercritical pressure, the surface tension vanishes and the fluid transport is governed by turbulent mixing and mass diffusion. Later studies were interested in more complex problems typical of LRE configurations, such as the injection of non-reactive cryogenic round jets into high-pressure environments. A number of studies dealing with this subject were conducted by the DLR~\cite{mayer1996,mayer2001,mayer2003,oschwald2006} and the AFRL~\cite{chehroudi2002,chehroudi2012}. One of those is shown in Fig.~\ref{fig:jets_supercritical}; it confirms earlier observations asserting that jet breakup and atomization do not occur at supercritical pressure, as the mixing process is instead due to pseudo-boiling and turbulent mass transfer manifested through finger-like structures at the jet edge.
A number of theoretical works were also interested in identifying essential parameters that characterize the coaxial mixing. It was for instance evidenced that the length of the central dark core of a liquid jet injected into a gaseous medium is determined by the momentum flux ratio $J = \rho_g v_g^2 / \rho_l v_l^2$~\cite{villermaux2000,candel2011}.\par

The interaction of cryogenic round jets with acoustic instabilities were also the object of intense research efforts. Most remarkably, the response of a jet to an imposed  transverse acoustic modulation was proved to depend on its thermodynamical regime. Indeed, a subcritical jet response is rather strong, and increases even further as the pressure gets closer to $P_C$. It is maximum near the fluid critical point, but then sharply weakens at supercritical pressure~\cite{davis2007}.

\subsection{Real-gas flames} \label{sec:RG_flames}

\begin{figure}[h!]
\centering
\includegraphics[width=0.99\textwidth]{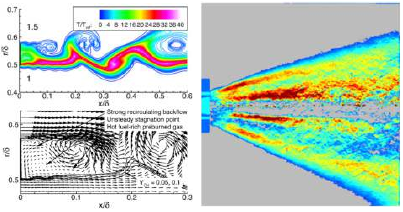}
\caption{Left: near-field of a LO\textsubscript{2}/GH\textsubscript{2} reactive mixing layer stabilized behind a splitter plate at $\bm{P = 101}$~bar. Extracted from~\cite{oefelein2005}. Right: Abel transform of time-averaged CH* emission images, in the case of a doubly-transcritical LO\textsubscript{2}/LCH\textsubscript{4} coaxial jet-flame at $\bm{P = 54}$~bar. Extracted from~\cite{singla2005}.}
\label{fig:flame_snapshots}
\end{figure}

The study of real-gas reactive flows started with the analysis of simplified systems, including the combustion of LO\textsubscript{2}/GH\textsubscript{2} and LO\textsubscript{2}/GCH\textsubscript{4} mixing layers behind splitter plates. This fundamental problem, representative of a small portion of a coaxial injector, was investigated through both experiments~\cite{CANDEL2006,juniper2003,juniper2003a,singla2006,singla2007} and high-fidelity DNS~\cite{Zong2007,Yang2000,oefelein1998,oefelein2005,oefelein2006,okongo2002dns}. In~\cite{oefelein2005} (see Fig.~\ref{fig:flame_snapshots}, left), it was for instance found that the flame root, dominated by diffusion effects due to large thermophysical properties gradients, is stabilized in a region of high shear behind the injector lip. In addition, the flame was not lifted, but was rather anchored in a strong recirculation backflow located directly in the vicinity of the splitter plate.
Larger scale shear coaxial jet flames were also the subject of intense research efforts both experimentally~\cite{candel1998,Juniper2000,juniper2003,singla2005} and numerically thanks to LES~\cite{Ruiz:2011,schmitt2009,Schmitt2011,schmitt2020}. A large variety of such flames were investigated: the most often the reactant is liquid oxygen and the fuel is gaseous hydrogen or methane. Remarkably, Singla \textit{et al.}~\cite{singla2005} were the only ones to characterize the so-called doubly-transcritical regime on a LO\textsubscript{2}/LCH\textsubscript{4} jet-flame. Some of their results, for instance in Fig.~\ref{fig:flame_snapshots} (right), evidenced a specific flame structure composed of doubly-conical reactive layers.\par

\begin{figure}[h!]
\centering
\includegraphics[width=0.99\textwidth]{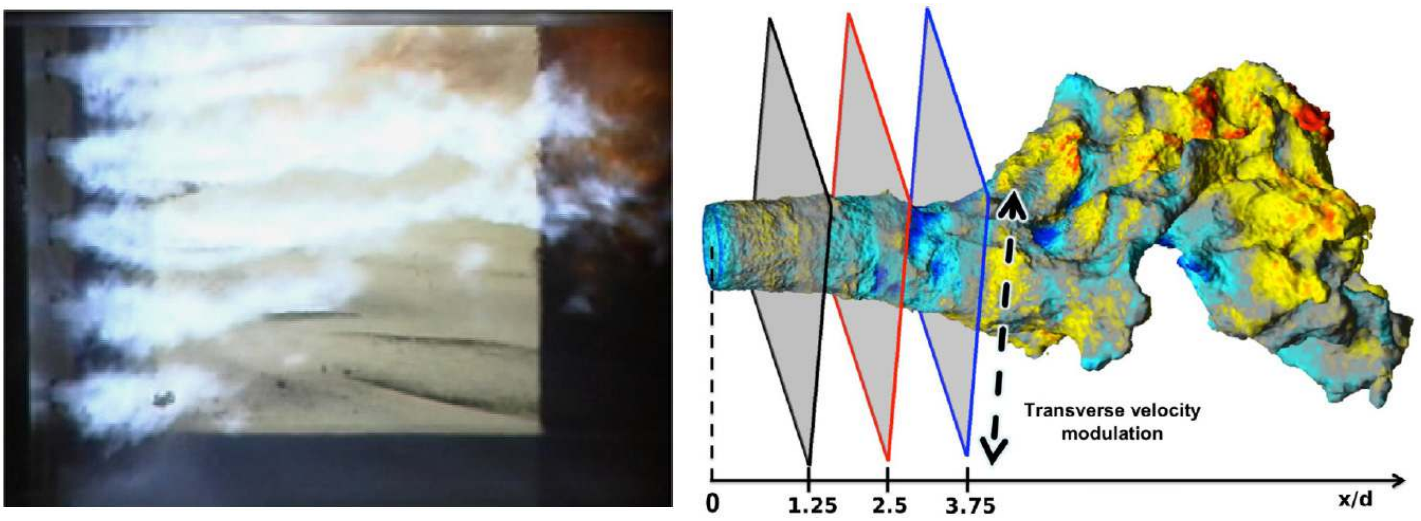}
\caption{Left: view of 5 cryogenic LO\textsubscript{2}/GCH\textsubscript{4} coaxial jet flames at $\bm{P = 26}$~bar, subjected to a strong transverse acoustic mode at $\bm{f = 3200}$~Hz. Extracted from~\cite{Mery2013}. Right: temperature iso-surface ($\bm{T = 2000}$~K) colored by axial velocity, for a single LO\textsubscript{2}/GCH\textsubscript{4} coaxial jet flame at $\bm{P = 114}$~bar subjected to a strong transverse mode at $\bm{f = 2000}$~Hz. Extracted from~\cite{Hakim2015}}.
\label{fig:flame_instab}
\end{figure}

The dynamics of LRE coaxial jet flames subjected to acoustic waves proved to be a challenging problem requiring state-of-the-art experimental facilities or numerical solvers. As a result, supercritical flame dynamics could be thoroughly investigated only very recently, for instance in a testbed comprising a few coaxial injectors~\cite{Richecoeur2006_bis,Mery2013} (see Fig.~\ref{fig:flame_instab}, left). LO\textsubscript{2}/GCH\textsubscript{4} flames were subjected to a large amplitude transverse acoustic mode, that noticeably enhanced cryogenic O\textsubscript{2} pseudo-boiling and mixing, resulting in flattened, shorter dark cores, as well as more compact and intense heat-release regions. LES of comparable mono-injector~\cite{Hakim2015} and multi-injector~\cite{Hakim2015_2,Schmitt2014} configurations led to identical observations. For example, in~\cite{Hakim2015} it was noticed that the flame oscillation patterns were strongly dependent on the acoustic forcing frequency: at high frequency, the flame and its oxidizer dark core were shortened and flattened and displayed rigid-body-like vertical oscillations, while at lower frequency the flame undergoes a flag-like flapping motion (see in Fig.~\ref{fig:flame_instab}, right).
LES was also employed to explore triggering, as well as LO\textsubscript{2}/GH\textsubscript{2} flame dynamics during thermoacoustic limit-cycles in the BKD, a 42-injectors rocket combustor~\cite{Urbano2016,Urbano2017,groning2016}. It was proved that the flames located near velocity nodes have the largest contribution to the instability, and the gain and phase of the heat release response at the unstable frequencies were computed. In another recent study, the particular effect of the fuel injection temperature on the BKD unstable modes was evaluated~\cite{schmitt2017bkd}.\par

\subsection{The Mascotte test rig} \label{sec:intro_Mascotte}

Only a very few experiential facilities are designed to sustain the extreme thermodynamic conditions encountered during a LRE combustion instability. The previously mentioned BKD, a 42-injectors LO\textsubscript{2}/GH\textsubscript{2} LRE operated by DLR Lampoldshausen is one of them~\cite{groning2016}. It can sustain supercritical pressures up to $100$~bar, and was used to study thermoacoustic insatbilities with frequencies ranging from $10$~kHz to $25$~kHz and of amplitude exceeding $30$~bar~\cite{groning2017}. However, due to its complexity and the presence of a large number of injectors, this configuration does not allow for a detailed characterization of cryogenic flame response to acoustic oscillations. Another experimental facility used to investigate thermoacoustic instabilities in LRE conditions is the Mascotte test rig, operated by ONERA. Its development started in the early 90s with the design of a single-injector chamber~\cite{habiballah1996,vingert2000}, that was for instance used in the work of Singla \textit{et al.}~\cite{singla2005} to study the structure of an unperturbed doubly-transcritical LO\textsubscript{2}/LCH\textsubscript{4} flame. Later, a multi-injector configuration was designed and a Very High Amplitude Modulator (VHAM) based on a system of spinning wheels was implemented to impose transverse acoustic waves at frequencies that can be selected~\cite{Richecoeur2006_bis,Mery2013}. Thus, the Mascotte test rig has been the subject of a large number of experimental and numerical studies, and it advantageously combines the characteristics of a realistic LRE to that of a well-characterized academic combusor. It therefore constitutes a natural choice to perform the present numerical simulations.\par

As the objective of the present study is to characterize the dynamics of a single cryogenic jet flame under imposed acoustic oscillations, the original mono-injector configuration of Mascotte is selected. In addition, it has to be noticed that most previous research efforts focused on supercritical flames where the oxygen is injected in a dense transcritical state, and the fuel (either H\textsubscript{2} or CH\textsubscript{4}) as a light supercritical fluid. Very few studies tackled the problem of a doubly transcritical regime, where both reactants are injected in a dense transcritical state. The experimental work of Singla \textit{et al.}~\cite{singla2005} is one of the only analysis reporting characteristics of such flames in a realistic shear-coaxial configuration. As this combustion regime is likely to occur in future LREs, a better understanding of governing flame dynamics phenomena in these specific conditions is highly needed. This study is therefore intended to deal with this specific doubly-transcritical regime, where both oxygen and methane are injected as cryogenic liquids.\par

\begin{figure}[h!]
\centering
\includegraphics[width=0.99\textwidth]{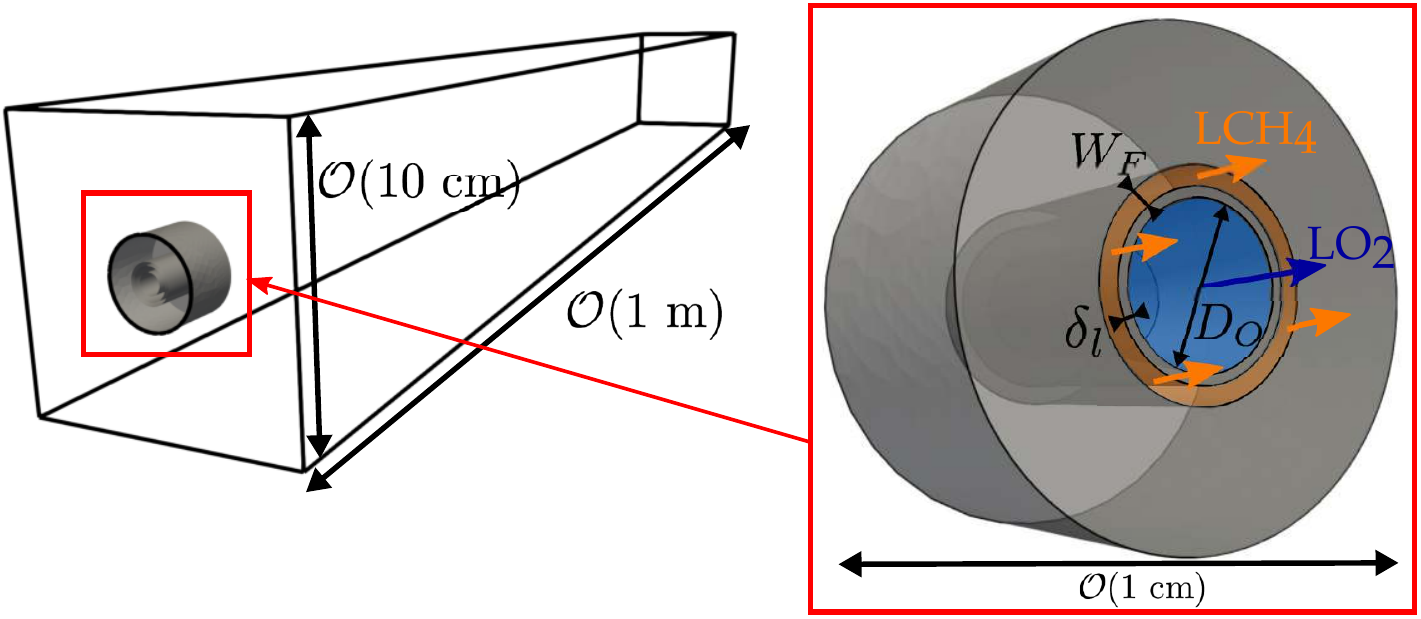}
\caption{Left: 3D view of the Mascotte configuration simulated in this work. Right: closeup view on the coaxial injector, with the central injection of LO\textsubscript{2} surrounded by the annular injection of LCH\textsubscript{4}. Other dimensions are: $\bm{\delta_l = \mathcal{O} (0.1 \ \textbf{mm})}$ and $\bm{D_O, W_F = \mathcal{O} (1 \ \textbf{mm})}$.}
\label{fig:geom_intro}
\end{figure}

The Mascotte geometry studied throughout this work (Fig.~\ref{fig:geom_intro}) is a parallelepipedic chamber with a coaxial injector at its backplate that comprises a central round injection of liquid oxygen (diameter $D_O$), surrounded by an annular injection of liquid methane (width $W_F$). Those are separated by a tapered lip of thickness $\delta_l$ at its tip.
The doubly-transcritical injection conditions (Tab.~\ref{tab:properties_mascotte}) are similar to the operation point T1 in~\cite{singla2005}.
\begin{table}[h!]
\centering
\begin{tabular}{c c c c c c}
\multicolumn{1}{c}{ } \T\B &  \multicolumn{3}{|c|}{$\dot{m}_{inj}/ \dot{m}_{inj}^{O2}$~~~~~~~$T_R^{inj}$~~~~~~~$P_R^{inj}$ } & $T_C$ & $P_C$\\
 \cline{1-6}
 \multicolumn{1}{c}{O\textsubscript{2} }&  \multicolumn{3}{|c|}{~~~~~$1.0$~~~~~~~~~~~$0.55$~~~~~$1.49$} & $155$~K & $50.4$~bar\\
 \cline{1-6}
 \multicolumn{1}{c}{CH\textsubscript{4} } &  \multicolumn{3}{|c|}{~~~~~$3.59$~~~~~~~~~$0.62$~~~~~~$1.64$} & $190$~K & $46$~bar\\
 \cline{1-6}
\end{tabular}
\caption{Injection conditions (mass flow rate, reduced pressures and temperatures) in Mascotte. The subscript \textsubscript{R} denotes the reduced temperature and pressure, defined as $\bm{T_R = T/T_C}$ and $\bm{P_R = P/P_C}$, respectively. The momentum flux ratio is $\bm{J = (\rho_{CH4} u_{CH4}^2)/(\rho_{O2} u_{O2}^2) = 33.3}$, and the global equivalence ratio is $\bm{\phi_g = 14.3}$.}\label{tab:properties_mascotte}
\end{table}
The only difference is that the chamber pressure has been increased to 75~bar (instead of 54~bar in the experimental study), giving higher reduced pressures. This discrepancy with the experimental conditions of~\cite{singla2005} is expected to strongly affect the flow features, and direct comparison with experimental data of~\cite{singla2005} is therefore ruled out. However, the aim of the present work is not to quantitatively reproduce experimental results, but rather to qualitatively describe flame dynamics driving mechanisms occurring in LREs conditions. The computation of this higher pressure operation point was motivated by two elements:
\begin{itemize}
\item{A pressure of 75~bar is still relevant to LREs conditions, and it therefore allows worthy qualitative and quantitative analyses. In addition, it could potentially be reproduced in an academic test rig such as Mascotte, which can sustain chamber pressures up to 100 bar.}
\item{Simulations of the experimental operating conditions T1 in~\cite{singla2005} were attempted during this PhD, but are not reported here. Indeed, at the experimental pressure (54~bar), both reactants pass near their respective critical points as they are heated up, resulting in high fluctuations of pressure and density, that are due to the high compressibility and dilatability combined with the numerical resolution of extremely large density gradients. These numerical oscillations destabilize the computation, unless a significant artificial viscosity is added, which considerably degrades the simulation quality. On the contrary, at 75~bar these numerical oscillations are weaker and do not require the addition of a large artificial viscosity, thus yielding a higher-fidelity computation.}
\end{itemize}

\section{Numerical simulation of reactive real-gas flows} \label{sec:les_equation}

Only a few Computational Fluid Dynamics (CFD) solvers have the ability to account for both non-ideal thermodynamics and reactive flows that are encountered in LREs. In this work, the real-gas (RG) version of the AVBP solver developed at CERFACS is used. This so-called \textit{AVBP-RG} is built upon the classical solver AVBP used for ideal-gas simulations~\cite{moureau2005}. Its development was initiated during the PhD of Schmitt, who validated it on reactive and non-reactive cases~\cite{schmitt2009,schmitt2010}, some of which in the Mascotte test rig~\cite{Schmitt2011,schmitt2020}. AVBP-RG was further improved later~\cite{Ruiz:2012phd,Mari:2015phd}, and was used to study LRE thermoacoustic instabilities in a number of complex cases, generally yielding a good agreement with experimental observations~\cite{Urbano2016,Hakim2015_2,schmitt2017bkd}.

\subsection{Navier-Stokes equations} \label{sec:NS_complete}

Similarly to ideal-gas CFD solvers, AVBP-RG resolves the compressible, reactive, multi-components Navier-Stokes equations, which write with the Einstein's rule of summation as:
 \begin{align}
& \frac{\partial  \rho }{\partial t}              + \frac{\partial}{\partial x_j}         ( \rho \: u_j ) = 0  \hspace{10pt} \textrm{(Mass conservation)}  \label{eq:continuity_ns} \\
& \frac{\partial  \rho \: u_i}{\partial t}        + \frac{\partial}{\partial x_j}         ( \rho \: u_i \: u_j )   =  - \: \frac{\partial}{\partial x_j} [ p \: \delta_{ij} - \tau_{ij} ]   \hspace{10pt} \textrm{(Momentum conservation)} \label{eq:momentum_ijk_ns}\\
& \frac{\partial \rho \: E}{\partial t}           + \frac{\partial}{\partial x_j}         ( \rho \: E \: u_j)              =  - \: \frac{\partial}{\partial x_j} [ u_i \: ( p \: \delta_{ij} - \tau_{ij} ) + q_j ] + \dot{\omega}_T    \hspace{5pt} \textrm{(Energy conservation)} \label{eq:energy_ijk_ns} \\
& \frac{\partial \rho_k}{\partial t}              + \frac{\partial}{\partial x_j}         ( \rho_k  \: u_j)                =  - \: \frac{\partial }{ \partial x_j} [ J_{j,k} ] + \dot{\omega}_k. \hspace{10pt} \textrm{(Species k conservation)} \label{eq:species_ijk_ns}
\end{align}
where $\rho = \sum_k \rho_k$ is the density, $u_j$ is the j\textsuperscript{th} component of the velocity, $p$ is the static pressure, $E$ is the energy, $Y_k$ is the mass fraction of species k, $\tau_{ij}$ is the viscous tensor, $q_j$ is the j\textsuperscript{th} component of the energy flux, $J_{j,k}$ is the j\textsuperscript{th} component of the of the diffusive flux of species k, $\dot{\omega}_T$ is the heat-release due to combustion, and $ \dot{\omega}_k$ is the reaction rate of species k. These latter terms are linked by the relation:
\begin{align}
\label{eq:definition_omega_T}
\dot\omega_T = - \sum_{k=1}^N \Delta h^0_{f,k} \dot\omega_k
\end{align}
where $\Delta h^0_{f,k}$ is the enthalpy of formation of species $k$ at temperature $T_0$. The species diffusive fluxes are obtained thanks to 1\textsuperscript{st}-order approximations~\cite{Hirschfelder:1969} and they read:
\begin{align}
J_{i,k} = - \rho \left( D_k \frac{W_{k}}{{W}} \ddi{X_k} -
 Y_k V_i^c \right) \label{eq:flux_species}
\end{align}
where the diffusion coefficient $D_k$ is computed by assuming constant Schmidt numbers $Sc_k$: $D_k = \mu / Sc_k$, with $\mu$ the dynamic viscosity. In Eq.~\eqref{eq:flux_species}, $V_i^c$ is a correction diffusion velocity ensuring global conservation of mass, that is given by:
\begin{align}
\label{eq:diff_velocity}
V^c_i = \sum_{k=1}^N D_k \frac{W_{k}}{{W}}\ddi{X_k} 
\end{align}
In multi-species flows, this correction term affects the heat-flux that has to account for heat transport due to species diffusion. It now writes:
\begin{align}
 q_i = \underbrace{- {\lambda} \, \ddi{T}}_{\mbox{\rm Heat
conduction}}  \quad
       \underbrace{- \rho \sum_{k=1}^N \left(D_k \frac{W_{k}}{{W}}
       \ddi{X_k} - Y_k V_i^c \right) {h}_k }_{\mbox{\rm Heat flux 
       through species diffusion}} 
       \label{eq:heat_flux}
\end{align}
where $\lambda$ is the mixture heat conductivity, and $h_k$ the mass enthalpy of species k. The viscous stress tensor in Eq.~\ref{eq:momentum_ijk_ns} is defined as:
\begin{align}
\tau_{ij} = 2 \mu \left( S_{ij}-\frac{1}{3} \delta_{ij}S_{ll} \right) \label{eq:stress_tensor}
\end{align}
where $S_{ij}$ is the strain rate tensor:
\begin{align}
S_{ij} = \frac{1}{2} \left( \frac{\partial u_i}{\partial x_j} + \frac{\partial u_j}{\partial x_i} \right)\label{eq:rate_of_strain_tens}
\end{align}
Finally, this set of equations should be completed by a thermodynamic EoS, that is here the SRK equation defined in Eq.~\eqref{eq:SRK}.\par

At this stage, the entirety of the terms in the previous equations and their computation have been detailed, with the exception of:
\begin{itemize}
\item{The species reaction rates $\dot\omega_k$, whose modeling is the object of Sec.~\ref{sec:kinetic_mechanism}.}
\item{The mass enthalpy $h_k$, and other thermodynamic variables, that directly depend on the EoS. The method to compute those is given in Sec.~\ref{sec:adaptation_RG}.}
\item{The transport coefficients $\lambda$ and $\mu$ that are also detailed in Sec.~\ref{sec:adaptation_RG}.}
\end{itemize}
In Chapter~\ref{chap:fwi_rg}, AVBP-RG is used to perform a DNS, that consists in resolving the set of equations described above, without a need for further modeling. In Chapters~\ref{chap:mascotte_linear}~and~\ref{chap:nonlinear_mascotte} AVBP-RG is used to carry out a series of LES, whose methodology is explained below.

\subsection{Filtered LES equations and subgrid-scale modeling} \label{sec:LES_intro}

\begin{figure}[h!]
\centering
\includegraphics[width=0.6\textwidth]{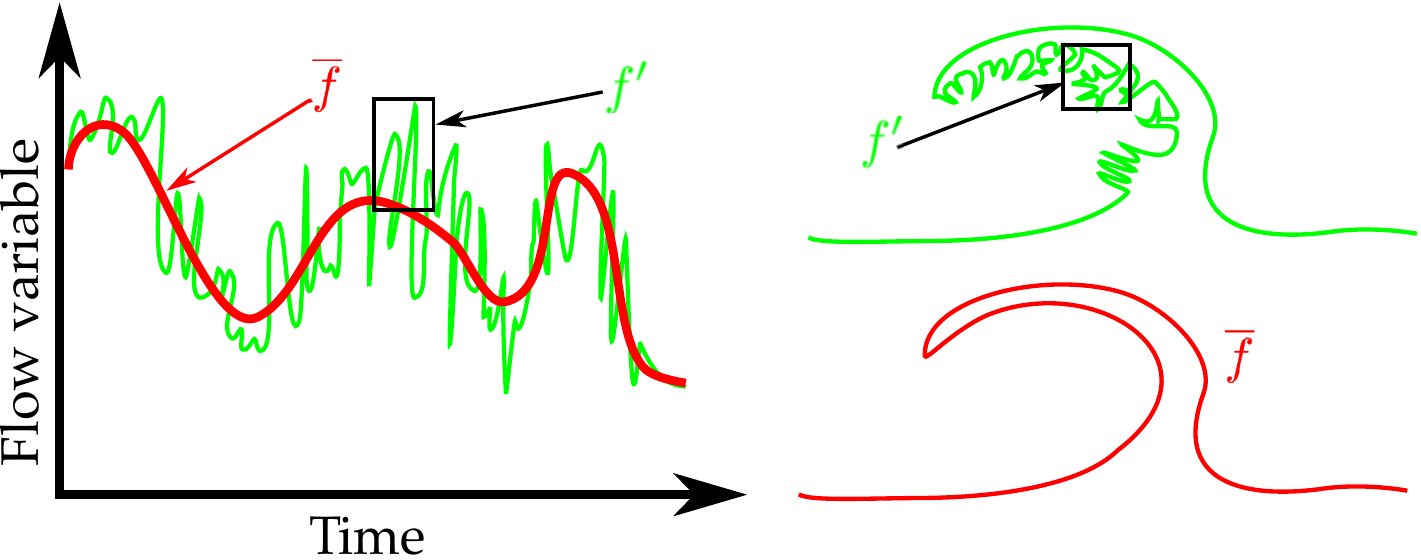}
\caption{Left: schematic representation of the effect due to the LES filtering on the temporal evolution of a flow variable $\bm{f}$ at a given point in the computational domain. Right: schematic representation of the effect due to the LES filtering on the spatial pattern of a flow variable $\bm{f}$.}
\label{fig:les_illustr}
\end{figure}

The LES concept lies in a phenomenological interpretation of the Kolmogorov cascade, asserting that in turbulent flows, the largest scales that contain the most energy dissipate this latter by breaking-down into finer and finer scales containing less and less energy, down to the smallest possible scale called the Kolmogorov length. Pragmatically, it therefore appears natural to fully resolve the energy-containing scales, that is the largest ones, and to replace the small ones by cheaper models. More formally, this idea leads to decomposing any flow variable $f$ as $f=\overline{f} + f'$, where $\overline{f}$ is the resolved part, and $f'$ is the unresolved or \textit{subgrid-scale} (SGS) part. Mathematically, $\overline{f}$ corresponds to a spatial low-pass filtering of $f$. These considerations are illustrated in Fig~\ref{fig:les_illustr}.
 In addition, dealing with variable density flows requires to introduce the mass-weighted Favre filtering of $f$ as $ \widetilde{f} =  \overline{\rho f} / \overline{\rho}$. Applying the Favre filtering to the set of equations Eq.~\eqref{eq:continuity_ns}-\eqref{eq:species_ijk_ns} yields the conservation equations for LES:
\begin{align}
&  {\partial  \overline{\rho} \over \partial t} +  {\partial \over \partial x_i} ( \overline{\rho} \: \widetilde{u_i} ) = 0 
\label{eq:mass_les_ijk} \\
& {\partial  \overline{\rho} \: \widetilde{u_i} \over \partial t}  + {\partial \over \partial x_j} ( \overline{\rho} \: \widetilde{u_i} \: \widetilde{u_j} ) = - \: {\partial \over \partial x_j} [ \overline{P} \: \delta_{ij} - \overline{\tau_{ij}} - \overline{\tau_{ij}}^t ]
\label{eq:momentum_les_ijk} \\
& {\partial \overline{\rho} \: \widetilde{E} \over \partial t} + {\partial \over \partial x_j} ( \overline{\rho} \: \widetilde{E} \: \widetilde{u_j}) = - \: {\partial \over \partial x_j} [ \overline{ u_j \: ( P \: \delta_{ij} - \tau_{ij} ) } + \overline{q_j} + \overline{q_j}^t ] + \overline{\dot{\omega}_T}
\label{eq:energy_les_ijk} \\
& {\partial \overline{\rho} \: \widetilde{Y_k}  \over \partial t} + {\partial \over \partial x_j} ( \overline{\rho} \: \widetilde{Y_k}  \: \widetilde{u_j} ) = - \: {\partial \over \partial x_j} [ \overline{J_{j,k}} + \overline{J_{j,k}}^t ] + \overline{\dot{\omega}_k}
\label{eq:species_les_ijk}
\end{align}
where the subscript ${}^t$ denotes turbulent quantities that require a modeling, as detailed below. Additionally, any nonlinear quantity of the form $\overline{f g}$ or $\widetilde{f g}$ is \textit{a priori} unknown and needs to be evaluated through some approximation. Note that these models are here identical to that of an ideal-gas LES. For some of these terms it is indeed quantitatively justified to retain the same SGS models as in real-gas flows~\cite{selle2007}, on the contrary for others this choice is dictated by the absence of reliable models specifically derived for real-gas conditions, which is still the subject of current research efforts~\cite{unnikrishnan2017}.

\paragraph{The Reynolds tensor} \mbox{}\\[-5mm]

\noindent The first turbulent term requiring modeling is the Reynolds stress tensor $\overline{\tau_{ij}}^t$; this is achieved thanks to the turbulent viscosity, or Boussinesq's hypothesis~\cite{Boussinesq:1877a}. This leads to: 
\begin{align}
\label{eq:reynolds_stress_closure}
\overline{\tau_{ij}}^t = 2 \: \overline{\rho} \: \nu_t \left( \widetilde{S}_{ij}
   - \frac{1}{3} \delta_{ij} \widetilde{S}_{ll} \right)
\end{align}
In this work, the turbulent SGS viscosity $\nu_t$ is obtained thanks to the SIGMA model~\cite{nicoud2011}:
\begin{align}
\label{eq:def_sigma_SGS}
\nu_t = (C_{\sigma} \Delta)^2 \dfrac{\sigma_3(\sigma_1 - \sigma_2) (\sigma_2 - \sigma_3)}{\sigma_1^2}
\end{align}
where $C_{\sigma} = 1.35$, $\Delta$ is the cube-root of the cell volume, and the $\sigma_i$ are the singular values of the tensor $G$ built from the resolved velocity gradient $g$: $G = {}^t \! g g $.

\paragraph{The SGS diffusive species flux vector} \mbox{}\\[-5mm]

\noindent The second turbulent term that requires to be modeled to close the problem is the SGS diffusive species flux vector $\overline{J_{i,k}}^t$ defined as:
\begin{align}
\label{eq:def_SGS_species_flux}
\overline{J_{i,k}}^t = \overline{\rho} \:
\left(\widetilde{u_i Y_k} - \widetilde{u}_i \widetilde{Y}_k \right)
\end{align}
This vector is modeled by a gradient-diffusion hypothesis:
\begin{align}
\label{eq:model_SGS_species_flux}
\left( D_k^t \frac{W_{k}}{\overline{W}}
 {\partial \widetilde{X}_k \over \partial x_i} - \widetilde{{Y_k}} {\widetilde{V_i}}^{c,t} \right) \ , \ \textrm{with} \ D_k^t = \frac{\nu_t}{S_{c,k}^t}
\end{align}
where the turbulent Schmidt number $S_{c,k}^t = 0.7$ is the same for all species. The turbulent correction velocity $\widetilde{V}_i^{c,t}$ is given by:
\begin{align}
\label{eq:model_SGS_turbu_correction_velocity}
\widetilde{V}_i^{c,t} = \sum_{k=1}^N \frac{\mu_t}{\overline{\rho} S_{c,k}^t  }  \frac{W_{k}}{\overline{W}} {\partial \widetilde{X}_k \over \partial x_i}
\end{align}

\paragraph{The SGS heat flux vector} \mbox{}\\[-5mm]

\noindent The third turbulent term that needs modeling is the SGS heat flux vector $\overline{q_i}^t$ defined as :
\begin{align}
\label{eq:def_SGS_heat_flux}
\overline{q_i}^t = \overline{\rho}
     ( \widetilde{u_i E} - \widetilde{u}_i \widetilde{E} )
\end{align}
This term is also modeled thanks to a gradient-diffusion assumption which gives:
\begin{align}
\label{eq:model_SGS_heat_flux}
\overline{q_i}^t = - \lambda_t
\, {\partial \widetilde{T} \over \partial x_i}
 + \sum_{k=1}^N \overline{J_{i,k}}^t \: \widetilde{\bar{h}_k} \ , \ \textrm{with} \ \lambda_t = \frac{{\mu_t} \overline{C_p}}{P_r^t}
\end{align}
where the turbulent Prandtl  number is fixed to $P_r^t  = 0.7$.

\paragraph{The filtered viscous terms} \mbox{}\\[-5mm]

\noindent A few additional filtered quantities appear in the set of LES conservation equations (Eq.~\eqref{eq:momentum_les_ijk} to Eq.~\eqref{eq:energy_les_ijk}) and need to be calculated. The laminar filtered stress tensor $\overline{\tau_{ij}}$ is computed through simple 1\textsuperscript{st}-order approximations~\cite{poinsotBook2005}:
\begin{align}
\label{eq:approx_filter_stress}
\overline{\tau_{ij}}  \approx 2 \overline{\mu}
(\widetilde{S}_{ij}-\frac{1}{3} \delta_{ij} \widetilde{S}_{ll}) \ , \ \textrm{with} \  \widetilde{S}_{ij} = \frac{1}{2}\left(\frac{\partial \widetilde{u}_i}{\partial x_j} + \dfrac{\partial \widetilde{u}_j}{\partial x_i}\right)
\end{align}
where the filtered viscosity $\overline{\mu}$ is approximated as $\overline{\mu( T)}  \approx \mu (\widetilde{T})$. The filtered diffusive species flux vector $\overline{J_{i,k}}$ writes:
\begin{align}
\label{eq:filtered_species_diff_flux}
\overline{J_{i,k}} \approx  - \overline{\rho}
 \left( \overline{D}_k \frac{W_{k}}{\overline{W}}
 {\partial \widetilde{X}_k \over \partial x_i} - \widetilde{{Y_k}} {\widetilde{V_i}}^c \right)
\end{align}
And the filtered heat-flux is given by:
\begin{align}
\label{eq:filtered_heat_flux}
\overline{q_i} \approx - { { \widetilde{\lambda} {\partial \widetilde{T} \over \partial x_i}} }
 + \sum_{k=1}^N {\overline{J_{i,k}}} \: {\widetilde{\overline{h}_k}}
\end{align}
These equations are obtained by assuming that the spatial variations of molecular diffusion are negligible and can be modeled thanks to simple gradient assumptions.

\paragraph{The filtered inviscid terms} \mbox{}\\[-5mm]

\noindent A single filtered inviscid term appearing in Eq.~\eqref{eq:energy_les_ijk} needs further developments to be computed. Similarly to the filtered viscous terms, it is evaluated based on 1\textsuperscript{st}-order approximation:
\begin{align}
\label{eq:filtered_press_vel}
\overline{u_j P} \delta_{ij} \approx \widetilde{ u_j } \overline{P } \delta_{ij}
\end{align}\par

Note that no combustion model is used in this work, or in other words the filtered reaction rates $\overline{\dot{\omega}_k}$ are directly computed from the available filtered variables following the Arrhenius law without further assumptions. This choice is based on the observation that diffusion flames, which include the LRE flames studied in this work, \textit{self-adapt} to the mesh that is provided. It is therefore not necessary to use combustion models such as the thickened flame employed for premixed flames LES~\cite{poinsotBook2005}. This approach, however, ignores the SGS effects, which may be significant and alter the flame length. In the present simulations, the grid is considered sufficiently resolved to limit these SGS effects.\par

Finally, the set of governing equations composed of  Eq.~\eqref{eq:continuity_ns} to Eq.~\eqref{eq:species_ijk_ns} in the case of a DNS, or Eq.~\eqref{eq:mass_les_ijk} to Eq.~\eqref{eq:species_les_ijk} in the case of a LES, is integrated thanks to numerical schemes. In this work, the schemes that are used belong to the class of Two-step Taylor-Galerkin finite element schemes~\cite{colin2000}: in Chapter~\ref{chap:mascotte_linear} and Chapter~\ref{chap:nonlinear_mascotte} the TTG4A scheme (3\textsuperscript{rd}-order in space and 4\textsuperscript{th}-order in time) is used, while in Chapter~\ref{chap:fwi_rg} the TTGC scheme (3\textsuperscript{rd}-order in space and time) is employed.

\subsection{Adaptations for real-gas flows} \label{sec:adaptation_RG}

In the previous paragraph, the only apparent difference due to real-gas flows is the use of the SRK EoS of Eq.~\eqref{eq:SRK}. A number of additional adaptations are nonetheless necessary to deal with transcritical and supercritical conditions. Those are listed below.

\paragraph{Computation of thermodynamic variables} \mbox{}\\[-5mm]

\noindent In the DNS or LES conservation equations presented above, a number of thermodynamic variables appear: for instance the heat capacity $C_p$, the species mass enthalpy $h_k$, the sensible energy $e_s$, etc. For a low-pressure ideal-gas, such thermodynamic variable $F$ only depends on the temperature and the mixture: $F = F^0(T,Y_k)$, in which case $F^0$ is computed thanks to the tabulation of coefficients over the temperature range of interest. In a high-pressure real-gas,  $F$ depends also on the density, which is accounted for thanks to a departure function $\Delta F (\rho,T,Y_k)$~\cite{Poling:2001}:
\begin{align}
\label{eq:def_departure_function}
F (\rho,T,Y_k) = F^0(T,Y_k) + \Delta F (\rho,T,Y_k)
\end{align}
An analytical expression for the departure function $\Delta F (\rho,T,Y_k)$ is obtained by integrating the differential $d F$ along a thermodynamic path, and by using the fact that any integral over a closed contour is zero.

\paragraph{Computation of transport coefficients} \mbox{}\\[-5mm]

\noindent For a low-pressure ideal-gas, the mixture dynamic viscosity $\mu$ only depends on the temperature such that $\mu = \mu^0 (T)$, and $\mu^0$ is computed thanks to a power-law:
\begin{align}
\label{eq:visco_power_law}
\mu^0 (T) = \mu_{ref} \left( \dfrac{T}{T_{ref}} \right)^a
\end{align}
where $\mu_{ref}$ is a reference viscosity value at a reference temperature $T_{ref}$. The conductivity $\lambda = \lambda^0 (T)$ is computed thanks to the constant Prandtl number ($Pr$) assumption: $\lambda^0 = \mu^0 C_p /Pr$. For a high-pressure real-gas, $ \mu$ and $\lambda$ depend on the density and their computations are then performed thank to Chung method~\cite{Chung:1984,chung1988}, which writes:
\begin{align}
\label{eq:Chung_method}
& \mu = \mu^0 (T) \ F_1 (T, \rho, T_C, \rho_C , \omega) \\
& \lambda = \lambda^0 (T) \ F_2 (T, \rho, T_C, \rho_C , \omega)
\end{align}
where $F_1$ and $F_2$ are semi-empirical correction functions accounting for high pressure deviation. For a multi-species fluid, the critical parameters (denoted with subscripts \textsubscript{C}) are replaced by their \textit{pseudo-critical} counterparts, calculated according to simple mixture-averaging rules~\cite{reid1987}.

\paragraph{Boundary conditions} \mbox{}\\[-5mm]

Many boundary conditions in AVBP-RG are based on the Navier Stokes Characteristic Boundary Condition (NSCBC)~\cite{poinsot1992} formulation. This formalism requires to convert the conservative variables $\rho$, $\rho u_i$, $\rho E$, $\rho Y_k$ into the characteristic variables (\textit{i.e.} waves). This conversion is highly dependent on the thermodynamics and needs to be adapted for a real-gas (see~\cite{schmitt2020,schmitt2009} for more details).

\paragraph{Numerical scheme}  \mbox{}\\[-5mm]

The TTGC and TTG4A numerical schemes used to integrate the conservation equations require the computation of the convective flux Jacobian. This matrix involves thermodynamic variables, and therefore needs to be adapted for real-gas flows (further details can be found in~\cite{schmitt2020,schmitt2009}).

\paragraph{Artificial viscosity} \mbox{}\\[-5mm]

The centered schemes implemented in AVBP are known to produce spurious oscillations, called wiggles, in the vicinity of large gradients. Is is therefore necessary to selectively damp these fluctuations, which is achieved thanks to the addition of an artificial viscosity. In AVBP, two type of artificial viscosities are used: a 2\textsuperscript{nd}-order viscosity that is based on a sensor intended to identify the regions of large gradients, and a background 4\textsuperscript{th}-order hyper-viscosity applied uniformly in the domain. For typical gaseous flames, the viscosity sensor is usually designed to identify the flame region, since it is where the largest gradients are located. On the contrary, in real-gas flows large density gradients may also exist in the pseudo-vaporization region, which necessitates a reformulation of the viscosity sensor. In addition, the real-gas artificial viscosity cannot be applied in the same fashion as the ideal-gas one: indeed, a variation of thermodynamic variable caused by the artificial viscosity may trigger even larger oscillations of other variables, due to the strongly nonlinear EoS. These observations led to the development of a Localized Artificial Diffusivity (LAD)~\cite{schmitt2020}, whose sensor $S_{\rho}$ is based on the density:
\begin{align}
\label{eq:LAD}
S_{\rho} = \dfrac{\vert \vec{u} . \vec{n} \vert \Delta t}{\Delta x} \left| \dfrac{\hat{\overline{\rho}} - \overline{\rho}}{0.01 \overline{\rho}} - \xi \right|
\end{align}
where $\Delta t$ is the time-step, $\Delta x$ the characteristic cell size, $\xi$ an activation threshold, and $\hat{f} = K * f$ is a spatial filtering of the variable $f$. The selective filter $K$ is built from the composition of a reference filter $G$ with its approximate deconvolution, as in~\cite{mathew2003}. The effective 2\textsuperscript{nd}-order viscosity is then applied to the conservative variables proportionally to the sensor $S_{\rho}$ (with the addition of some limiters for species mass fractions). In the case where the fluid is highly non-ideal (\textit{i.e.} the compressibility factor $Z_{th} = p/( \rho \overline{r} T)$
 is low: $Z_{th} < 0.9$) the artificial viscosity is applied in a non-conservative way to the pressure $p$, rather than to the energy $\rho E$.

\section{Derivation of a kinetic mechanism for CH\textsubscript{4} combustion in LRE conditions} \label{sec:kinetic_mechanism}

A point that was not discussed in the previous sections is the computation of the reaction rates $\dot{\omega}_k$ in Eq.~\eqref{eq:species_ijk_ns}. Most previous studies dealing with LRE combustion employed global schemes, or infinitely fast chemistry based on tabulated equilibrium~\cite{schmitt2020,Schmitt2011}. However, these simplified chemistry models are known to produce large errors and to induce serious limitations, for instance due to their inability to accurately account for flame quenching. This limitation is even more restrictive in the study of flame dynamics, where the reaction front may be submitted to wildly fluctuating strain rates. Unfortunately, fully detailed kinetic mechanisms are still too costly for LES, and very few reduced mechanisms have been derived and validated for CH\textsubscript{4}/O\textsubscript{2} combustion in conditions relevant to those found in LREs. One of the first goals of this work therefore consists in deriving an Analytically Reduced Chemical (ARC) scheme that: (1) yields an acceptable accuracy in comparison to detailed chemistry, and (2) is compact enough to be utilized in costly LES computations.\par

The reduction of a mechanism from a detailed one is based on the Directed Relation Graph Error Propagation (DRGEP) algorithm proposed by Pepiot \textit{et al.}~\cite{pepiot2008}. It proceeds in four stages:
\begin{enumerate}
\item{One or several reduction cases are defined. Here, these cases consist of two one-dimensional counterflow CH\textsubscript{4}/O\textsubscript{2} diffusion flames at $P = 54$~bar and $P = 90$~bar, respectively. Other parameters are the same for both flames: the global equivalence ratio is $\phi_g = 4$, the injection temperature is $T_{CH4}^{inj} = T_{O2}^{inj} = 200$~K, and the global strain rate is $a = 1000$~s\textsuperscript{-1}. These parameters are deemed representative of combustion in LRE conditions, and the two distinct pressures are intended to yield a reduced mechanism valid over a large pressure range. For each reduction case, a set of reduction targets is also defined. Here, the reduction targets are the integrated heat-release, the maximum temperature, and the spatial profiles of a few species of interest, including the reactants and key products (H\textsubscript{2}O, CO, CO\textsubscript{2}).}
\item{Species are progressively removed from the mechanism, by identifying species removals that do not contribute to large errors on the reduction targets. Once the committed error reaches a user-defined threshold, the species removal process is stopped.}
\item{The number of reactions in the mechanism is decreased following the same strategy. }
\item{A number of additional \textit{quasi-steady-state} (QSS) species are identified, essentially based on their fast reaction rates. These species are not transported in the subsequent LES or DNS, but are still used in the computation of the chemical reaction rates.}
\end{enumerate}
A large number of reduction strategies were attempted by modifying the reduction cases as well as the targets of interest. Only the one that provided the best overall results is discussed here. Note that the counterflow diffusion flame computations required by the reduction process, as well as other verifications presented below, are performed in ideal-gas conditions. This simplification is based on previous flame structure analyses~\cite{pons2009,Lacaze2012} showing that cryogenic reactants undergo a pseudo-vaporization \textit{before} entering the reaction zone. Thus, chemical reactions occur in a light supercritical fluid, whose thermodynamic properties do not strongly differ from that of an ideal-gas.\par

In order to maximize the chances of obtaining an accurate yet compact ARC mechanism, the reduction procedure is attempted on several detailed schemes, that were validated beforehand for high-pressure methane combustion. Those include the widely known GRI3.0~\cite{smith2000} specifically designed for methane combustion, the Jerzembeck mechanism~\cite{jerzembeck2009} valid for high-pressure high-temperature combustion of alcanes ranging from methane to heptane, the 17 species Lu skeletal mechanism~\cite{lu2008}, the RAMEC mechanism~\cite{petersen2007} purposely developed for high pressure methane kinetics in ramjet combustors, and the Zhukov mechanism~\cite{zhukov2009}. A preliminary chemical equilibrium computed at $T = 200$~K, $P = 54$~bar and $\phi = 1$ with the GRI3.0 shows the existence of 10 species in non-negligible proportions; it is therefore reasonable to target a reduced mechanism comprising 10 transported species. Any larger ARC scheme is then systematically deemed too expensive to be used in LES computations. For instance, the reductions of the RAMEC and the Zhukov detailed schemes yield reduced mechanisms comprising at least 15 species, which are then discarded in the following. Three high-pressure (HP) ARC schemes are therefore retained: the J\textsubscript{HP} (10 transported species, 39 reactions, 3 QSS species) reduced from the Jerzembeck detailed model, the GRI\textsubscript{HP} (9 transported species, 82 reactions, 7 QSS species) reduced from the GRI3, and the LU\textsubscript{HP} (9 species, 51 reactions, 3 QSS species). Discarding the mechanism reduced from the RAMEC may appear disputable: indeed, the RAMEC was validated against shock tube experimental data in a large pressure range comprised between 40 bars to 240 bars, and for mixtures with relatively low dilution level (about 70\%). On the contrary, high-pressure experimental validations of the GRI3.0 are rather modest: it was for instance evaluated on a 84 bars shock tube auto-ignition for a CH\textsubscript{4}/O\textsubscript{2} mixture diluted in 98\% of Argon~\cite{petersen1996}. Thus, the choice of discarding the mechanism reduced from the RAMEC while retaining the GRI\textsubscript{HP} is intended to reduce as much as possible the cost of the following LES and DNS. \par

 In the purpose of selecting the best ARC mechanism, in terms of accuracy and computational cost, a series of \textit{a posteriori} evaluations is conducted by comparing the reduced mechanisms to the GRI3.0 that serves as a reference. These comparisons are carried out on counterflow CH\textsubscript{4}/O\textsubscript{2} diffusion flames, as well as isobaric reactors, over a large pressure range, and in thermodynamic conditions representative of LRE combustion. Some of these comparisons are presented below. For information, they also include the LU13~\cite{lu2008}, a popular ARC mechanism commonly used for atmospheric CH\textsubscript{4}/Air flames. Computations are performed with the Cantera software~\cite{goodwin2009}, and are based on the ideal-gas EoS. Note that the validation cases considered here do not include any premixed flames, as LRE combustion is usually dominated by the non-premixed regime.\par

\begin{figure}[h!]
\centering
\includegraphics[width=0.99\textwidth]{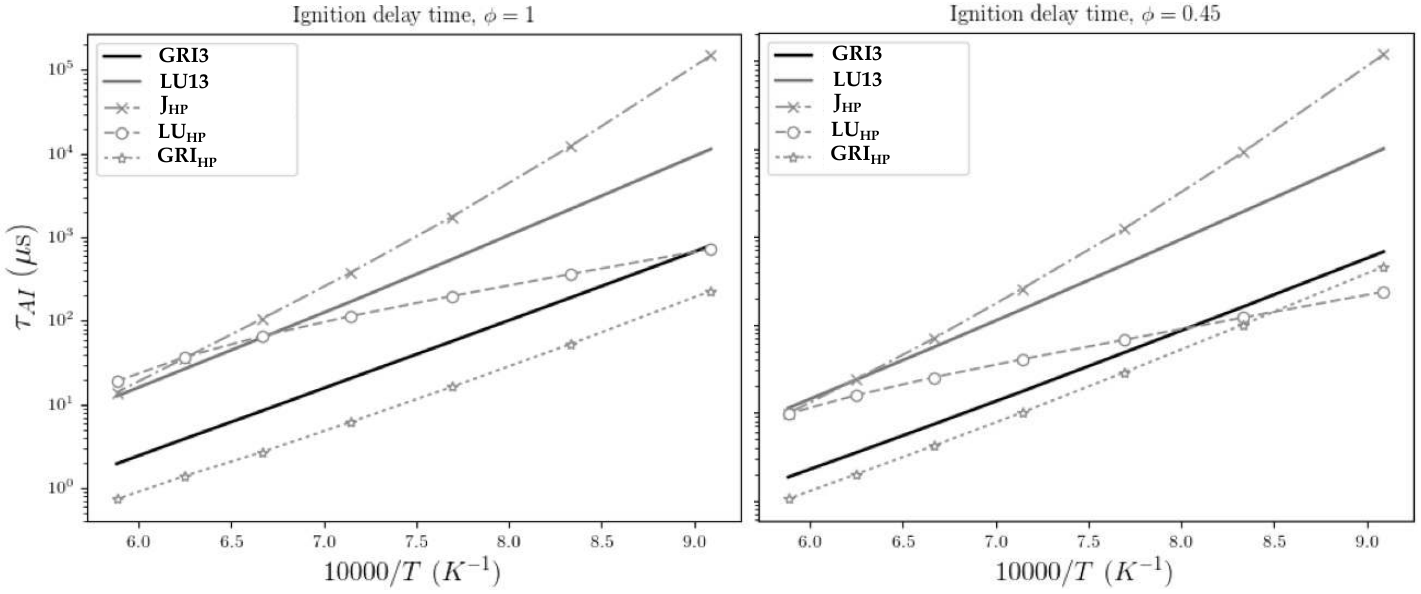}
\caption{Comparison of the ignition delay time $\bm{\tau_{AI}}$ in an isobaric CH\textsubscript{4}/O\textsubscript{2} reactor at $\bm{P = 54}$~bar, as a function of the initial temperature, for the five kinetic schemes considered. Left: $\bm{\phi = 1}$. Right: $\bm{\phi = 0.45}$.}
\label{fig:arc_tau_ai}
\end{figure}

\begin{figure}[h!]
\centering
\includegraphics[width=0.99\textwidth]{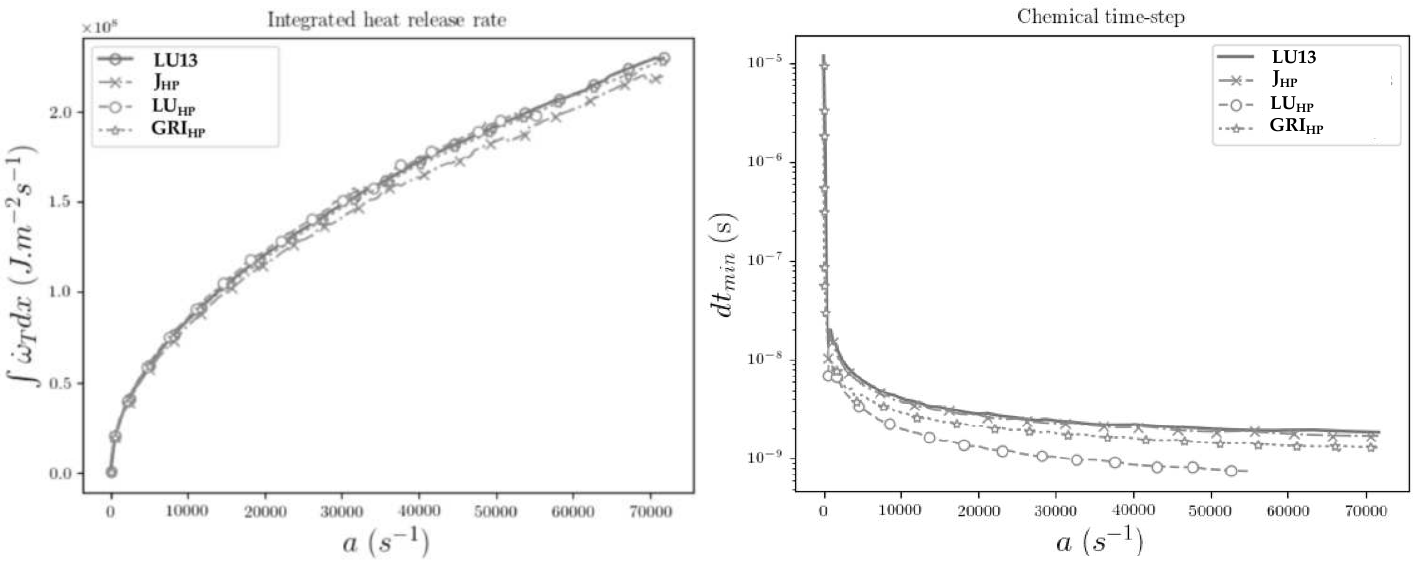}
\caption{Left: comparison of the integrated heat-release rate between 4 of the kinetic schemes considered, for a series of CH\textsubscript{4}/O\textsubscript{2} counterflow diffusion flames at increasing strain rate $\bm{a}$ (other parameters are: $\bm{P = 54}$~bar, $\bm{T_{CH4}^{inj} = T_{O2}^{inj} = 200}$, $\bm{\phi_g = 1}$). Right: comparison between the smallest chemical time scale $\bm{dt_{min}}$ of four of the kinetic schemes considered, for the same set of counterflow diffusion flames.}
\label{fig:arc_hr_dt}
\end{figure}

Figure~\ref{fig:arc_tau_ai} shows comparisons between the auto-ignition delay times $\tau_{AI}$ in an isobaric reactor at $P = 54$~bar, for each one of the mechanisms under consideration. Two conditions are considered: a stoichiometric mixture ($\phi = 1$) and a mixture at $\phi = 0.45$, which is the equivalence ratio minimizing $\tau_{AI}$.
The LU13 scheme results in a difference of roughly one order of magnitude in comparison to the GRI3.0 reference, which indicates that this ARC mechanism is limited to atmospheric CH\textsubscript{4}/Air flames and is not applicable for high-pressure oxycombustion. This point justifies the need to specifically derive a reduced mechanism for LRE conditions. The reduced J\textsubscript{HP} scheme produces even larger discrepancies than the LU13, with an ignition delay time over-estimated by 3 orders of magnitude at low temperatures. This poor accuracy disqualifies it for the subsequent multi-dimensional CFD simulations. On the contrary, the LU\textsubscript{HP} and the GRI\textsubscript{HP} yield acceptable agreements with the GRI3.0 reference. As LRE coaxial jet flames are usually highly strained, it is necessary to assess the behavior of the considered mechanisms over a wide range of strain rates. This is illustrated in Fig.~\ref{fig:arc_hr_dt}, where a series of CH\textsubscript{4}/O\textsubscript{2} counterflow diffusion flames at $P = 54$~bar are computed for strain rates ranging from $a = 1$~s\textsuperscript{-1} to $a = 7 \times 10^{4}$~s\textsuperscript{-1}. 
The total heat-release rate integrated over the one-dimensional domain is nearly identical for all four schemes considered, which suggests that these kinetic models respond in a similar fashion to a strain rate increase. A key characteristic of a mechanism is the order of magnitude of its smallest chemical time-scale $dt_{min}$, as this parameter directly influences the time-step that must be used in an unsteady CFD simulation. In Fig.~\ref{fig:arc_tau_ai} (right) it is shown that the chemical time-scale (computed as $dt_{min} = \min_{k} \min_x |Y_k(x) / \dot{\omega}_k(x)| $) of the LU\textsubscript{HP} is several times smaller than that of other reduced mechanisms. Integrating the chemical source terms in a DNS or a LES thanks to the LU\textsubscript{HP} would therefore require a much smaller time-step, which would in turn significantly increase the computational cost.\par

\begin{figure}[h!]
\centering
\includegraphics[width=0.99\textwidth]{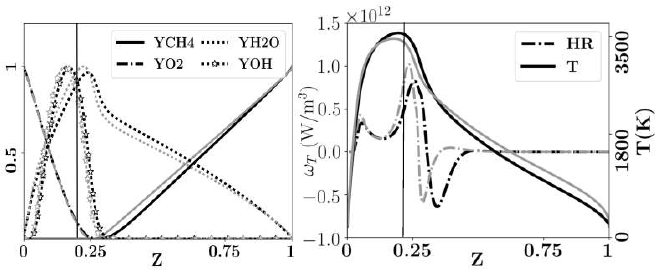}
\caption{Comparison between flame structures obtained with GRI3.0 (dark lines), and the reduced GRI\textsubscript{HP} scheme (light lines), for a CH\textsubscript{4}/O\textsubscript{2} counterflow diffusion flame at $\bm{P=75}$~bar, $\bm{\phi_g = 4}$, injection temperatures $\bm{T_{inj} = 200}$~K, and strain rate $\bm{a = 1000}$~s\textsuperscript{-1}. Normalized mass fraction profiles of selected species (left), temperature and heat release (right) are presented in mixture fraction space (computed with Bilger's definition~\cite{bilger1990}). Vertical lines represent the stoichiometry $\bm{Z=0.20}$.}
\label{fig:counterflow_1}
\end{figure}

In the light of these observations, the GRI\textsubscript{HP} ARC mechanism appears as achieving an acceptable trade-off between accuracy and cost: it is therefore implemented in AVBP and used in the multi-dimensional CFD computations presented in the next chapters. Further validation of this reduced kinetic model are provided in Fig.~\ref{fig:counterflow_1}, where it is compared to the GRI3.0 on the structure of a CH\textsubscript{4}/O\textsubscript{2} counterflow diffusion flame at $P = 75$~bar and $a = 1000$~s\textsuperscript{-1}. The heat release profiles obtained with both mechanisms are quite similar, presenting three reaction peaks: a primary peak in the rich side of the flame ($Z = 0.25$), a secondary weaker peak in the lean side ($Z=0.07$), and an endothermic zone in the rich side ($Z=0.3$). In both cases, the thickness of the primary reaction peak in the physical space is $\rm 125 \ \mu m$. Although the reduced mechanism tends to underestimate the maximum temperature (-2.5\%), and to overestimate the primary reaction peak (+20\%), an overall satisfactory agreement is observed.
The flame displayed in Fig.~\ref{fig:counterflow_1} is expected to be roughly representative of combustion modes encountered in the following numerical simulations. It is in particular used in Chapter~\ref{chap:fwi_rg} as a reference for comparison. Note that the average volumetric heat release rate is  $\dot{\Omega}_0 =  3.5 \times 10^{11} \rm \ W/m^{3}$, and the heat release per flame surface area is $\Phi_0 = 40.4 \ \rm MW/m^{2}$ (obtained by integration over the flame thickness).

\section{Conclusion} \label{sec:les_rg_conclusion}

The characterization of cryogenic flame response to acoustic oscillations is a notoriously challenging problem, which explains why flame dynamics in LRE conditions are not as thoroughly understood as in gas turbines. On one hand, the extreme conditions encountered in LRE combustion chambers render any experiment costly and limit the availability of advanced diagnostics. On the other hand, the non-ideal thermodynamics represent a major difficulty for CFD solvers, which must not only account for modified transport properties, but also employ specific numerical procedures to deal with the resolution of large thermophysical gradients. The real-gas version of AVBP is one of the few LES solvers that has these characteristics and that has been validated on a sufficient number of relevant cases. Similarly, the Mascotte test rig is arguably one of the most widely documented in the literature related to LRE thermoacoustic instabilities. The essence of the present work therefore consists in combining these two well-known tools to investigate flame dynamics in the doubly-transcritical regime, a set of thermodynamic conditions that has been left largely unexplored, but that is nonetheless likely to occur in future LREs.\par

Interestingly, anterior research efforts mostly studied the dynamics of supercritical coaxial jet flames subjected to large amplitude transverse acoustic waves, for only a very few forcing frequencies. Strong acoustic perturbations have the merit of generating clearly visible dynamics, thus permitting macroscopic analyses of the dark core or the flame surface evolution, as in~\cite{Mery2013,Hakim2015,Hakim2015_2,Urbano2017}. However, intense excitations may trigger nonlinear phenomena, such as flame response saturation or higher-harmonics generation, that are not characteristic of the small fluctuations that exist at the onset of an instability. As a result, previous works were mostly restricted to qualitative examinations: they did not compute a Flame Transfer Function (FTF) relating the fluctuations of heat-release to the acoustic forcing over a wide range of frequencies, nor did they provide sufficient quantitative information to guide the development of analytical flame response models. Thorough theoretical investigation of diffusion jet flames dynamics were conducted~\cite{Magina2013,Magina2016,tang2019}, but only for gaseous laminar systems which are too distant from LRE combustion. It is also worth mentioning that the vast majority of previous computational studies related to cryogenic flame were performed with simplified chemical kinetic models, as well as adiabatic boundaries.\par

To complete earlier research, considerable computational resources are employed in the present work (about 40 million cumulated CPU hours). These significant assets allows for the exploration of a number of ambitious directions, with the intent of pushing the frontiers of existing knowledge related to flame dynamics in LRE conditions. These research efforts include:
\begin{itemize}
\item{The first LES of a doubly-transcritical LO\textsubscript{2}/LCH\textsubscript{4} coaxial jet flame, based on a detailed chemical scheme.}
\item{One of the first attempts to compute and analyze a complete FTF of a cryogenic coaxial jet flame over a wide frequency range.}
\item{The clear identification of nonlinear flame response mechanisms caused by large amplitude acoustic oscillations.}
\item{One of the first attempts to study the effects of flame-wall interaction on the anchoring mechanism of a LO\textsubscript{2}/LCH\textsubscript{4} flame thanks to a DNS with complex chemistry combined with an unsteady conjugate heat transfer problem.}
\end{itemize}

These numerical simulations and their results are the object of the next chapters.

				
\chapter{Forced linear dynamics in a doubly-transcritical LO\textsubscript{2}/LCH\textsubscript{4} coaxial jet-flame} \label{chap:mascotte_linear}
\minitoc				
 
\begin{chapabstract}
In this chapter, LES are used to study the linear response of a doubly-transcritical LO\textsubscript{2}/LCH\textsubscript{4} coaxial jet flame subjected to fuel inflow acoustic harmonic oscillations. The geometry and the operating conditions are that of the academic test rig Mascotte, introduced in the previous chapter. The simulations are performed with the real-gas version of the AVBP solver, and use the complex kinetic scheme for CH\textsubscript{4} high-pressure oxycombustion that was derived earlier. After a brief description of the numerical setup, unforced simulations are used to analyze the overall flame topology, its principal features, as well as its intrinsic dynamics. The existence of three self-sustained hydrodynamic modes is evidenced. Then a series of forced LES, where the methane injector is modulated by small-amplitude harmonic acoustic waves, provides a thorough insight into the flame linear response for a wide range of forcing frequencies, spanning from approximately 1kHz to 20kHz. Local Flame Transfer Functions (FTF) are computed and analyzed: regions of preferential heat-release response are observed to be highly dependent on the forcing frequency. The interaction between the forced and the intrinsic flame dynamics is also assessed. Then, an analysis based on a flame sheet assumption is conducted to distinguish the main sources of heat release fluctuations: the primary contribution comes from the species diffusivity oscillations, while the density variations have a negligible effect. The second largest contributor is either the mixture fraction gradient or the flame surface area, depending on the forcing frequency. The FTFs are expected to be useful for thermoacoustic Low-Order Models or Helmholtz solvers, and the subsequent analysis has the potential to guide future development of analytical models for flame dynamics in LREs.
 \end{chapabstract}

\section{Numerical setup} \label{sec:linear_rg_setup}

The geometry of interest if the academic Mascotte test bench introduced in the previous Chapter (Fig.~\ref{fig:geom_intro}), that is the subject of an extensive literature related to LRE combustion. Its operating conditions, given in Tab.~\ref{tab:properties_mascotte}, are similar to those of the point T1 in~\cite{singla2005} at the exception of the pressure that is increased to 75 bar. Most importantly, both reactants (O\textsubscript{2} and CH\textsubscript{4}) are injected as dense cryogenic fluids, which characterizes the doubly-transcritical regime. The real-gas version of the LES solver AVBP (see Sec.~\ref{sec:les_equation}) is employed to solve the three-dimensional reactive, multi-species, and compressible Navier-Stokes equations, with non-ideal thermodynamics modeled thanks to the Soave-Redlich-Kwong (SRK) equation of state of Eq.~\eqref{eq:SRK}. The TTG4A finite-element scheme~\cite{colin2000} (3\textsuperscript{rd} order in space, 4\textsuperscript{th} in time) adapted for real-gas conditions is used. The $\sigma$-model~\cite{nicoud2011} closes the subgrid stress tensor. Reaction kinetics are based on the complex mechanism for high-pressure CH\textsubscript{4} oxycombustion that was reduced from the GRI3.0 mechanism in Sec.~\ref{sec:kinetic_mechanism}. The LAD density-based selective sensor is used to damp spurious numerical oscillations, by adding an artificial viscosity in the regions of large thermophysical gradients.\par

\begin{figure}[h!]
\centering
\includegraphics[width=0.9\textwidth]{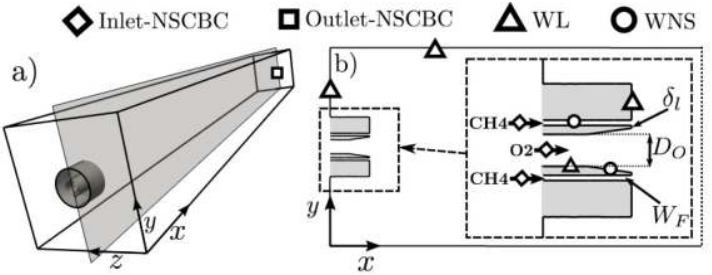}
\caption{(a) Three-dimensional view of the computational domain corresponding to the Mascotte test rig. (b) Closeup view of the near injector region. The symbols detail the boundary conditions in the simulation (WL: Law of the Wall, WNS: Resolved No Slip Wall).}
\label{fig:schematic_intro}
\end{figure}

The computational domain, recalled in Fig.~\ref{fig:schematic_intro} for convenience, is a parallelepipedic chamber with a coaxial injector at its backplate that comprises a central round injection of liquid oxygen (diameter $D_O$), surrounded by an annular injection of liquid methane (width $W_F$). Those are separated by a tapered lip of thickness $\delta_l$ at its tip.
The boundary conditions for the simulation are shown in Fig.~\ref{fig:schematic_intro}-(b). One-seventh power-law velocity profiles and synthetic turbulence~\cite{smirnov2000} are imposed at both fuel and oxidizer inlets through NSCBC boundary conditions~\cite{poinsot1992}. The mean bulk injection velocities of the fuel and oxidizer are respectively denoted $u_{CH4}^0$ and $u_{O2}^0$, or $u_{F}^0$ and $u_{O}^0$ depending on the circumstances. Note that the O\textsubscript{2} inflow is roughly one order of magnitude slower than the  CH\textsubscript{4} stream. When necessary, acoustic wave modulation at the fuel inlet is applied with the method proposed in~\cite{Kaufmann2002}. All wall boundaries are assumed adiabatic. The walls of both O\textsubscript{2} and CH\textsubscript{4} injection channels that are located near the injector exit plane are resolved, and no-slip conditions are applied. The tip of the injector lip is also resolved. On the contrary, the upstream portions of the injection lines are not resolved and are rather modeled with a classical Law of the Wall. It is \textit{a posteriori} verified that the wall coordinate $y^{+}$ is between 2 and 7 at the lip tip, and between 5 and 10 in the reactants injection channels. Counterintuitively, the region near the lip of the injector is not the most constraining regarding the near-wall mesh resolution: indeed, the temperature is high in this region due to the presence of the flame, which yields a larger viscosity and thus a lower $y^{+}$. Conversely, the very low viscosity in the cryogenic reactants injection lines gives larger $y^{+}$.

\begin{figure}[h!]
\centering
\includegraphics[width=0.99\textwidth]{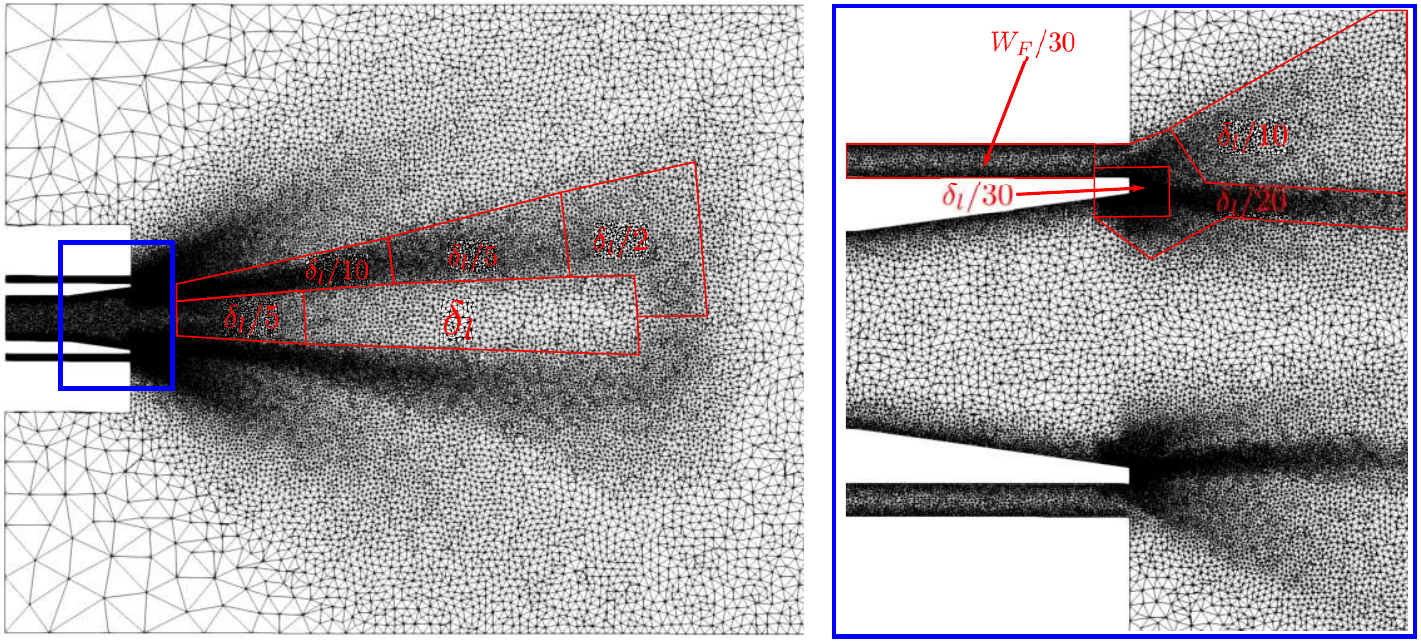}
\caption{Visualization of the 40 million cells mesh used in the LES, with characteristic cell sizes indicated. Left: flame region. Right: closeup view of the near-injector region.}
\label{fig:mesh_visu}
\end{figure}

Computations are performed on a 3D unstructured mesh of 40 million tetrahedral cells, shown in Fig.~\ref{fig:mesh_visu}.
This mesh is generated thanks to the following strategy:
\begin{itemize}
\item{A first preliminary LES is performed on a coarse 10 million cells mesh, until the flame is stabilized.}
\item{Physically relevant metrics are computed based on the coarse averaged solution. Here, these metrics are the averaged temperature field, indicating the flame brush, as well as the LIKE criterion~\cite{daviller2017} which marks the kinetic energy dissipation in the CH\textsubscript{4} annular jet.}
\item{The coarse grid is then remeshed thanks to the adaptive mesh refinement algorithm contained in the open-source library MMG~\cite{dobrzynski2008,dapogny2014}, which selectively adds cells in regions identified by the physical metrics previously computed.}
\end{itemize}
The mesh resolution is $\delta_l/30$ in the near-injector region, which corresponds to cells of a few micro-meters. Note that this cell size is several times smaller than the thickness of the one-dimensional counterflow flame displayed in Fig.~\ref{fig:counterflow_1}, for which the strain rate ($a = 1000 \ \textrm{s}^{-1}$) is deemed representative of that in the present simulations. The resolution is progressively coarsened downstream.

\section{Flame topology and intrinsic dynamics} \label{sec:linear_rg_flame_topology_and_instrinsic_dynamics}

In order to provide a better understanding of the flame response to acoustic perturbations, this section first describes the unforced flame topology, as well as its intrinsic dynamics. This simulation was run and averaged over 25 ms, which is roughly 10 convective times necessary for a particle of fluid in the slow O\textsubscript{2} jet to reach the end of the flame.

\subsection{Global flame topology} \label{sec:linear_rg_flame_topology}

\begin{figure}[h!]
\centering
\includegraphics[width=0.7\textwidth]{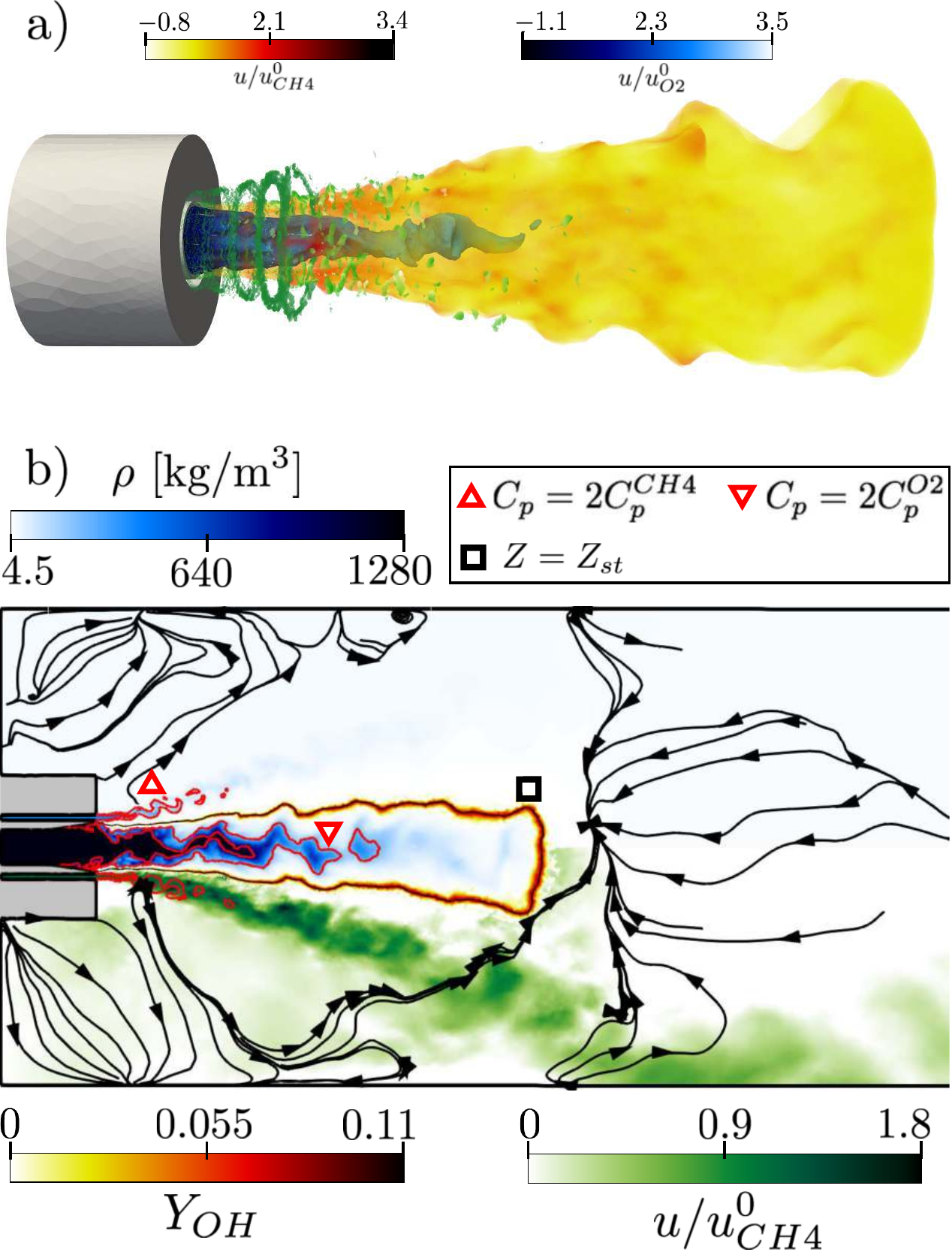}
\caption{(a) Yellow: instantaneous isosurface of temperature $\bm{T = 2000}$K colored by axial velocity, blue: isosurface of density $\bm{\rho = 0.5 \rho_{O2}^0}$ colored by axial velocity, green: isosurface of Q-criterion $\bm{Q = (u_{CH4}^0/W_F)^2}$. (b) Two-dimensional instantaneous fields of density (blue), axial velocity (green, only in the lower-half), and OH mass fraction (yellow). Red lines are contours of heat capacity $\bm{C_p}$, the dark line marked by a square is the contour of stoichiometric mixture fraction $\bm{Z=Z_{st}=0.2}$. Lines with arrows are the flow streamlines. $\bm{C_p^{CH4}}$ and $\bm{C_p^{O2}}$ are the reactants heat capacities in their respective injection conditions.}
\label{fig:instant_map}
\end{figure}

\begin{figure}[h!]
\centering
\includegraphics[width=0.7\textwidth]{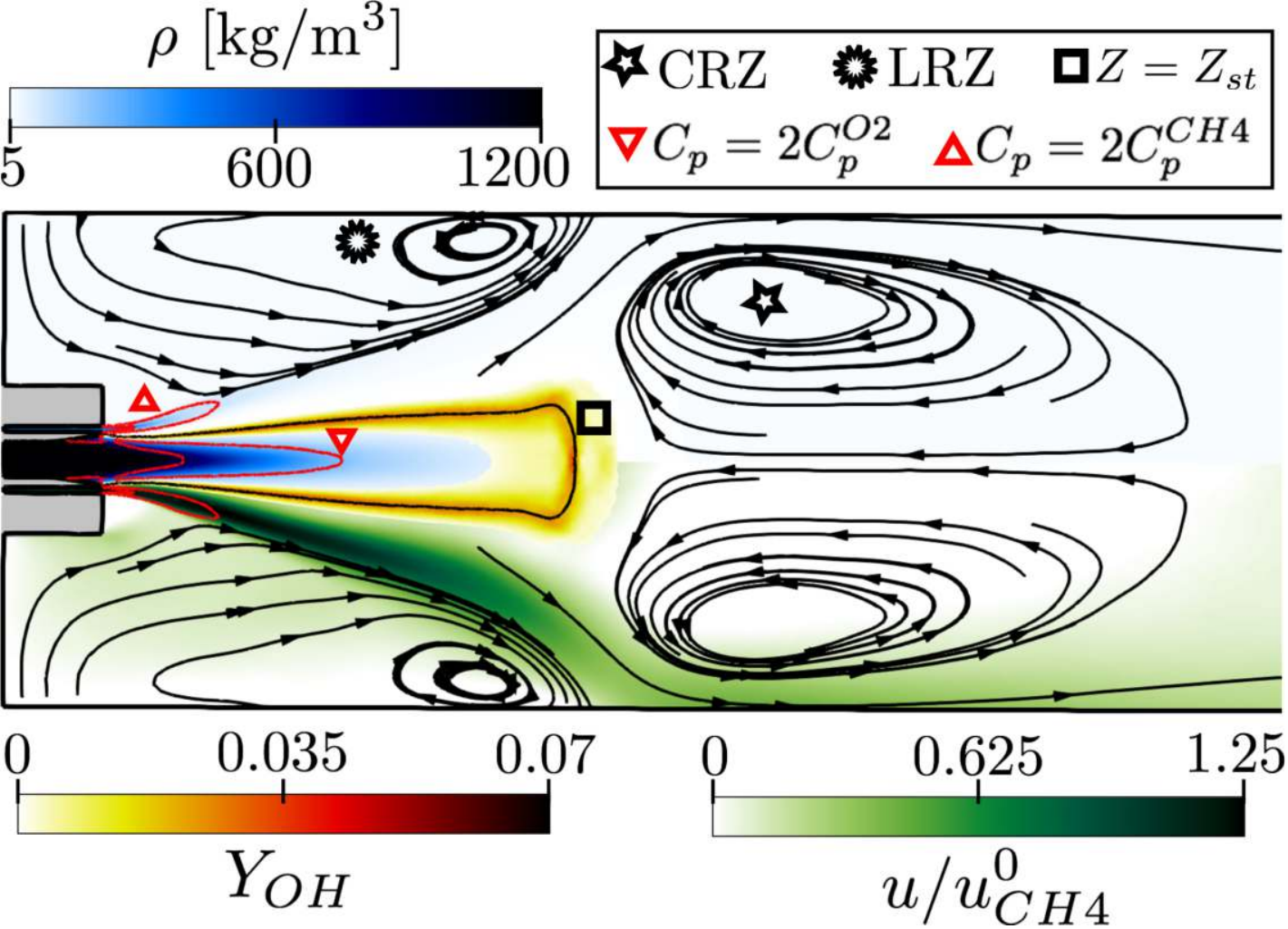}
\caption{Averaged fields of density (blue), axial velocity (green, only in the lower-half), and OH mass fraction (yellow). Red lines are contours of heat capacity $\bm{C_p}$, the dark line marked by a square is the contour of stoichiometric mixture fraction $\bm{Z=Z_{st}=0.2}$. Lines with arrows are the flow streamlines. CRZ: Central Recirculation Zone, LRZ: Lateral Recirculation Zone.}
\label{fig:ave_map}
\end{figure}

Instantaneous solutions are presented in Fig.~\ref{fig:instant_map}, and averaged fields of density, axial velocity, and OH radical mass fraction are shown in Fig.~\ref{fig:ave_map}.
Remarkable features characteristic of confined cryogenic coaxial jet flames are observed. The dense O\textsubscript{2} core extends over a considerable length in the chamber, before being completely vaporized in a region of intense pseudo-boiling located between $x=1.5 D_O$ and $x=4D_0$ from the injector exit ($C_p$ contour in Fig.~\ref{fig:ave_map}). Contrarily, the annular CH\textsubscript{4} jet experiences an earlier pseudo-boiling between $x=0.5D_O$ and $x=2D_O$, such that downstream of this location the fuel stream is in a light gaseous state. The dark core vaporization induces a significant dilatation of the oxidizer flow, which in turn provides radial momentum to the surrounding flame and light methane jet. As a result, this one mixes with hot combustion products, is pushed outwards, and the streamlines deviate from their original trajectory to impinge on the chamber wall. The impact of this fast and light fluid on the solid boundary generates two large recirculation zones: the Central Recirculation Zone (CRZ) and the Lateral Recirculation Zone (LRZ). Upstream of this divergence, the CH\textsubscript{4} annular jet has a constant opening angle of $17^{\circ}$, while the flame (mixture fraction contour in Fig.~\ref{fig:ave_map}) is narrower with a nearly constant opening angle of $5^{\circ}$. The flame lies in a region of low axial velocity, and its thin flame brush indicates that it is relatively stable with limited displacements. Its length is $L_f = 13.3 D_O$, and it terminates abruptly in a region where it encounters the strong adverse flow of the CRZ.\par

Figure~\ref{fig:instant_map} reveals rich dynamics in the physical processes described above. Jet breakup causes detachment of pockets of cryogenic O\textsubscript{2} from the dark core, which results in intermittent vaporization and mixing. Isosurfaces of Q-criterion evidence vortex rings that are formed at the methane injector exit and convected downstream in the annular jet with a clearly identifiable wavelength. The flame surface is also visibly affected. These coherent wave-like patterns suggest the existence of strong intrinsic hydrodynamic instabilities, which are the object of Sec.~\ref{sec:intrinsic_dynamics}.

\paragraph{Mesh sensitivity assessment} \mbox{}\\[-5mm]

\begin{figure}[h!]
\centering
\includegraphics[width=0.99\textwidth]{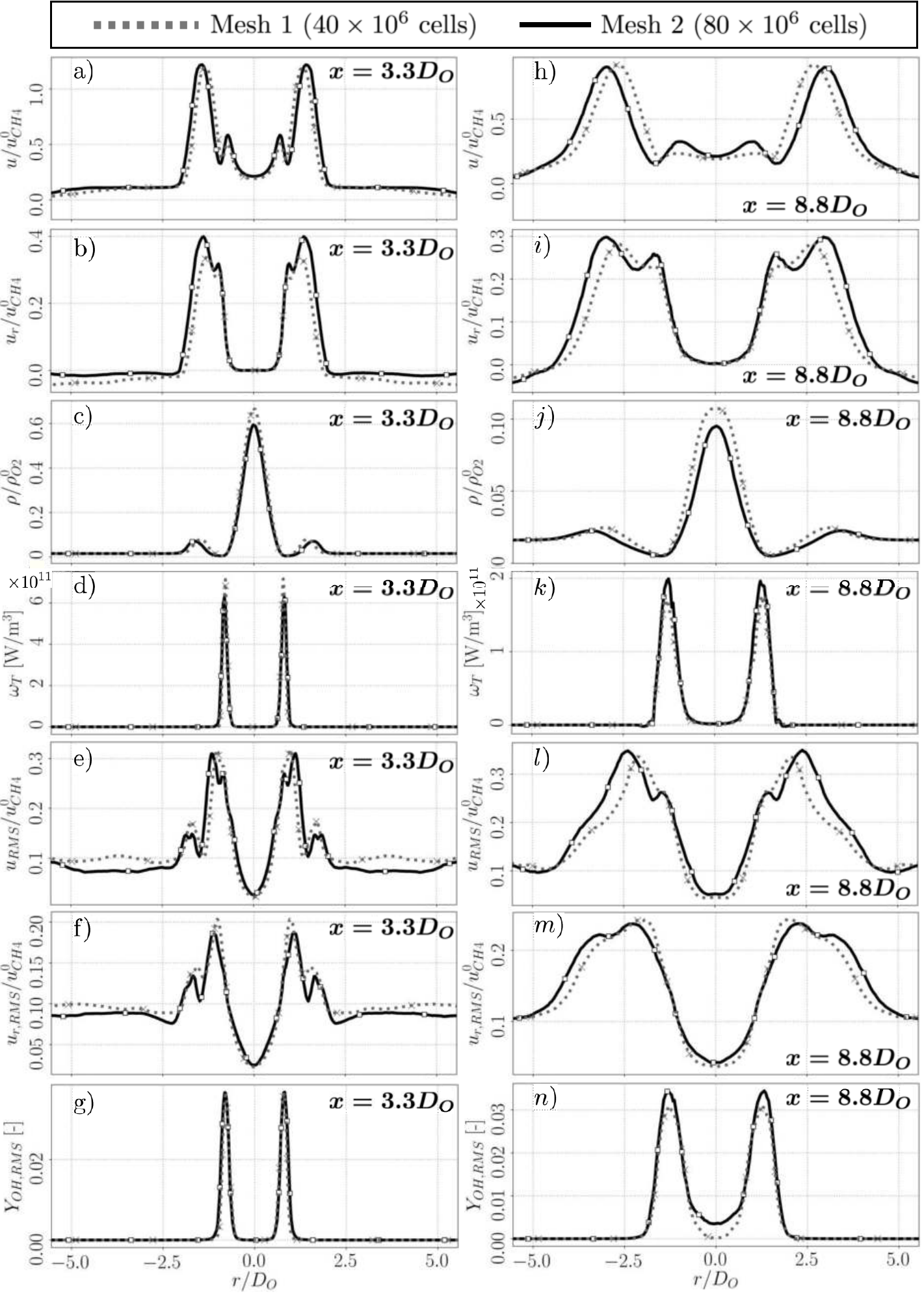}
\caption{(a)-(g) Comparison between averaged axial velocity $\bm{u}$, radial velocity $\bm{u_r}$, density $\bm{\rho}$, heat-release rate $\bm{\omega_T}$, axial velocity RMS $\bm{u_{RMS}}$, radial velocity RMS $\bm{u_{r,RMS}}$, and OH mass fraction RMS $\bm{Y_{OH,RMS}}$ computed on both meshes, at the axial location $\bm{x=3.3D_O}$. (h)-(n) Same, at the axial location $\bm{x = 8.8 D_O}$.}
\label{fig:mesh_sensitivity}
\end{figure}

This unforced LES is also used to evaluate the computation sensitivity to the mesh resolution. Two simulations are performed on distinct meshes: Mesh 1 with 40 million tetrahedral cells (used for most results in this work), and the second one with 80 million tetrahedral cells. This latter grid was generated by applying the adaptive refinement strategy of Sec.~\ref{sec:linear_rg_setup} to the former one. Averaged and RMS radial profiles are displayed in Fig.~\ref{fig:mesh_sensitivity}.
An overall satisfactory agreement is observed, even on usually highly sensitive quantities such as the heat-release rate (Fig.~\ref{fig:mesh_sensitivity}-(d),(k)) or the RMS of OH mass fraction (Fig.~\ref{fig:mesh_sensitivity}-(g),(n)). Most importantly, the flame mean position and thickness are well captured on Mesh 1. The only noticeable difference is a slightly larger opening angle of the methane annular jet on Mesh 2 ($18.5^{\circ}$ instead of $17^{\circ}$ on Mesh 1). The wider annular jet is related to a slightly shorter O\textsubscript{2} dark core on Mesh 2 (Fig.~\ref{fig:mesh_sensitivity}-(c),(j)), attesting an earlier oxidizer pseudo-boiling. However, these differences are minor and are not expected to significantly affect the flame forced dynamics.

\subsection{Near-injector region} \label{sec:near_injector} \label{sec:linear_rg_near_injector}

\begin{figure}[h!]
\centering
\includegraphics[width=0.99\textwidth]{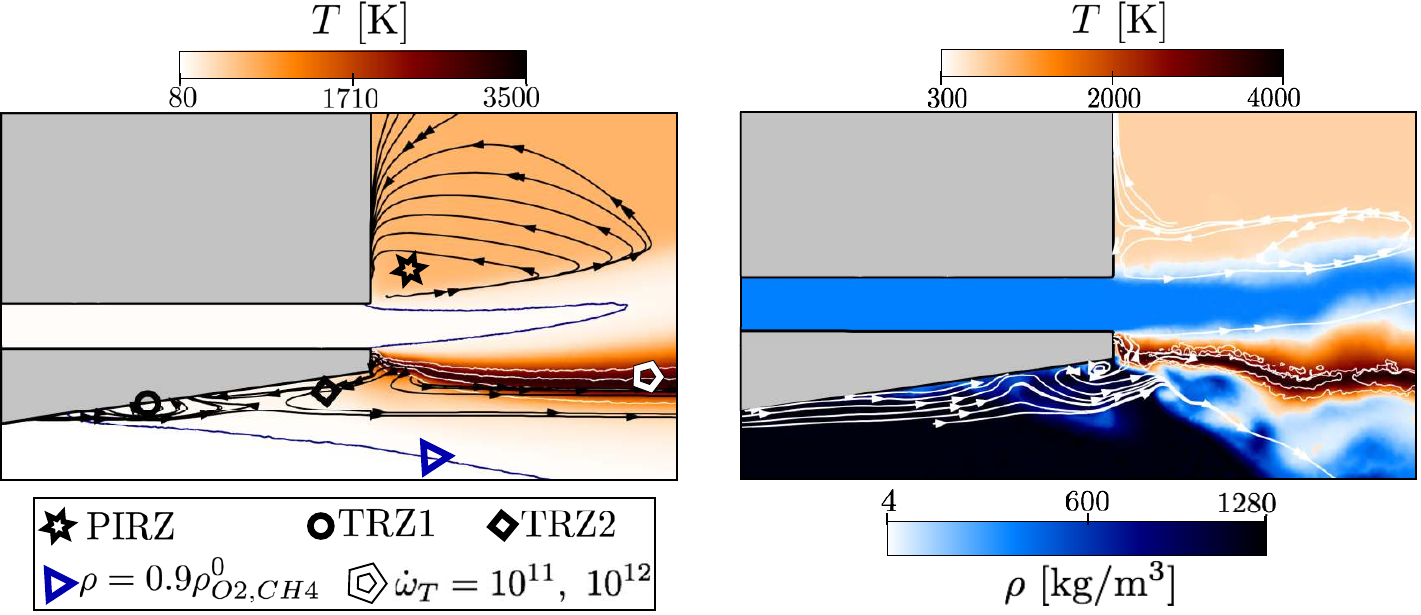}
\caption{Left: near-injector averaged field of temperature. The blue lines near the O\textsubscript{2} and the CH\textsubscript{4} streams are density isolines at $\bm{\rho = 0.9 \rho_{O2}^0}$ and $\bm{0.9 \rho_{CH4}^0}$, respectively. White lines are heat-release rate $\bm{\dot{\omega}_T}$ isolines. Dark lines with arrows are streamlines. PIRZ: Post-Injector Recirculation Zone. TRZ: Tapering Recirculation Zone. Right: near-injector instantaneous fields of temperature (red-dark) and density (blue-dark). White lines with arrows are flow streamlines.}
\label{fig:near_injector}
\end{figure}

In LRE cryogenic flames, the anchoring of the flame root at the coaxial injector lip is known to have first-order effects on the overall flame topology, as well as on its response to acoustic perturbations. The unforced LES solutions are therefore used to examine the stabilization mechanisms of the flame root. Figure~\ref{fig:near_injector} shows time-averaged and instantaneous near-injector fields. 
The flame anchoring is strongly influenced by the use of adiabatic boundary conditions: the reactive layer is indeed directly attached onto the injector lip, with instantaneous wall temperatures reaching $4000$~K, and as a result no triple-flame is observed. When it exits the injector, the high-momentum dense CH\textsubscript{4} jet not only creates a Post-Injector Recirculation Zone (PIRZ), but also pushes the flame towards the O\textsubscript{2} stream. This backflow induces a lip Tapering Recirculation Zone (TRZ2) located beneath the lip near its tip. A second oppositely rotating TRZ (TRZ1) is generated upstream by the oxidizer flow detachment from the injector wall. This flow detachment, induced by the lip tapering angle, directly influences the O\textsubscript{2} pseudo-boiling, as it coincides with a sharp density decrease that marks the frontier of the oxidizer dense core ($\rho = 0.9 \rho_{O2}^0$ in Fig.~\ref{fig:near_injector}, left).\par

The two counter-rotating recirculation zones TRZ1 and TRZ2 play a fundamental role on the strong anchoring of the flame root, as they induce a low-velocity region confined between the dense O\textsubscript{2} core and the injector lip. This region acts as a pre-burning gas reservoir enhancing the O\textsubscript{2} pseudo-vaporization, that directly supplies oxidizer to the reactive layer stabilized in the wake of the lip. Note that the formation of this pocket of light fluid is also very dependent on the thermal boundary condition, since it may loose heat through the walls. This anchoring mechanism, and in particular the effect of Flame-Wall Interaction, is the object of further investigations in Chapter~\ref{chap:fwi_rg}.

\subsection{Intrinsic flame dynamics} \label{sec:intrinsic_dynamics}

\begin{figure}[h!]
\centering
\includegraphics[width=0.8\textwidth]{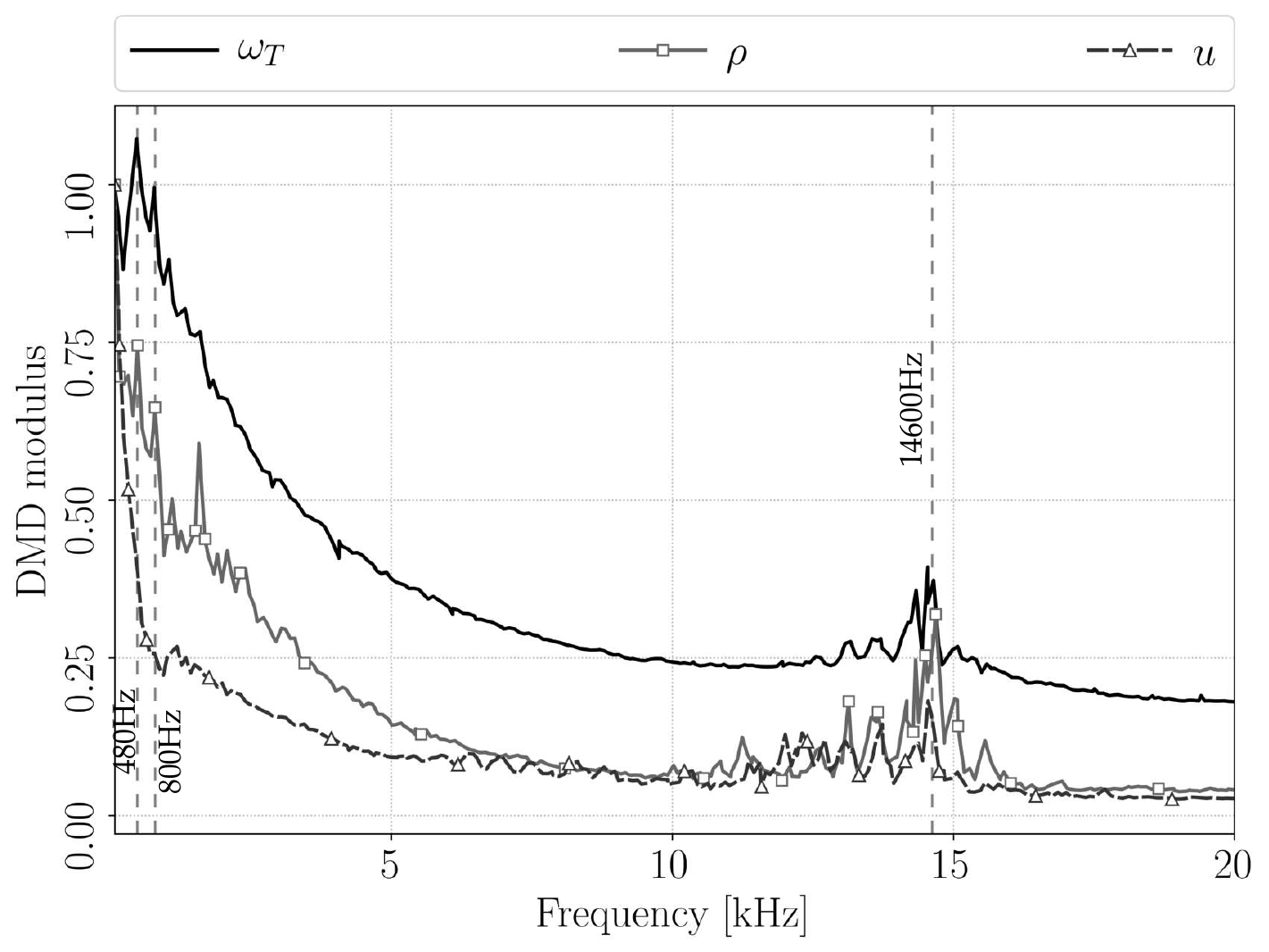}
\caption{Three-dimensional Dynamic Mode Decomposition spectra of the heat-release rate $\bm{\dot{\omega}_T}$, the density $\bm{\rho}$, and the axial velocity $\bm{u}$. For all variables, the mean fields are subtracted before computing the DMD. The resulting modes are then normalized such that their respective moduli at the first frequency point are unity.}
\label{fig:dmd_spectrum}
\end{figure}

The examination of the unforced flame topology in Sec.~\ref{sec:linear_rg_flame_topology} suggested the existence of elaborate intrinsic dynamics. Since these self-sustained oscillations may interact with the flame response to imposed acoustic perturbations, it is necessary to thoroughly characterize them beforehand. In this matter, Fourier analyses of a set of time-signals recorded at different locations in the computational domain (not shown here) evidence the occurrence of clearly identifiable frequencies on several flow variables. In order to gain a better insight into the spatio-temporal features of these self-sustained oscillations, three-dimensional Dynamic Mode Decomposition~\cite{schmid2010} (DMD) of the heat-release rate $\dot{\omega}_T$, the density $\rho$, and the axial velocity $u$ are computed. The three resulting DMD spectra (\textit{i.e.} the L-2 norm of the modes) are shown in Fig.~\ref{fig:dmd_spectrum}.
Three distinct modes are expressed on at least two of the flow variables of interest: two low-frequency modes at $f = 480$~Hz and $f = 800$~Hz are visible on the density and the heat-release spectra. A significantly higher-frequency band of activity is centered around a third mode at $f = 14600$~Hz, and is observed on all three variables. Note the presence of another peak in the density spectrum at roughly $1600$~Hz; however, since this mode is not expressed on other variables, it is not discussed in the following.\par

\begin{figure}[h!]
\centering
\includegraphics[width=0.95\textwidth]{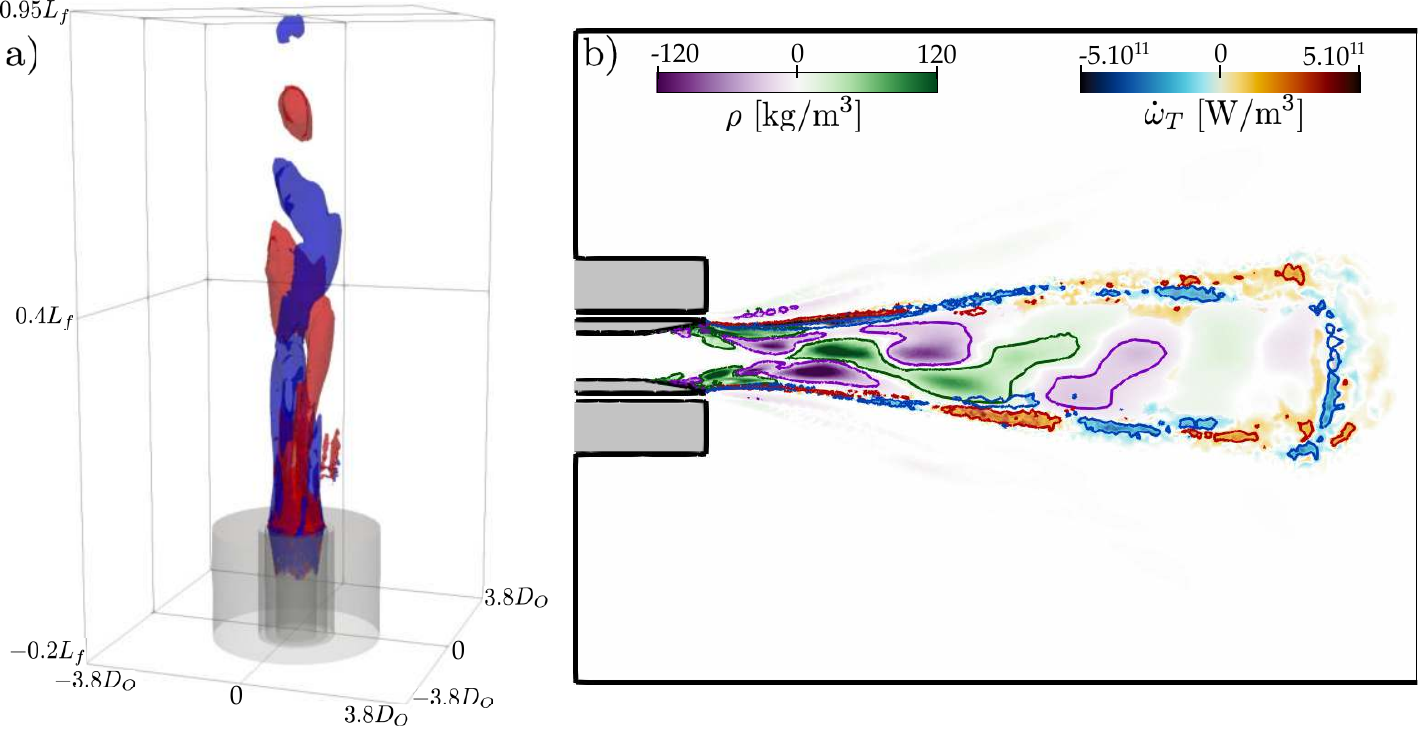}
\caption{Real part of the first DMD mode at $\bm{f = 480}$~Hz. Left: isosurface $\bm{\rho = 20}$~kg/m\textsuperscript{3} (red) and $\bm{\rho = -20}$~kg/m\textsuperscript{3} (blue) of the density mode. Right: two-dimensional cut of the first density mode (purple-green) superimposed on the first heat-release mode  (blue-red).}
\label{fig:dmd_mode480}
\end{figure}

\begin{figure}[h!]
\centering
\includegraphics[width=0.95\textwidth]{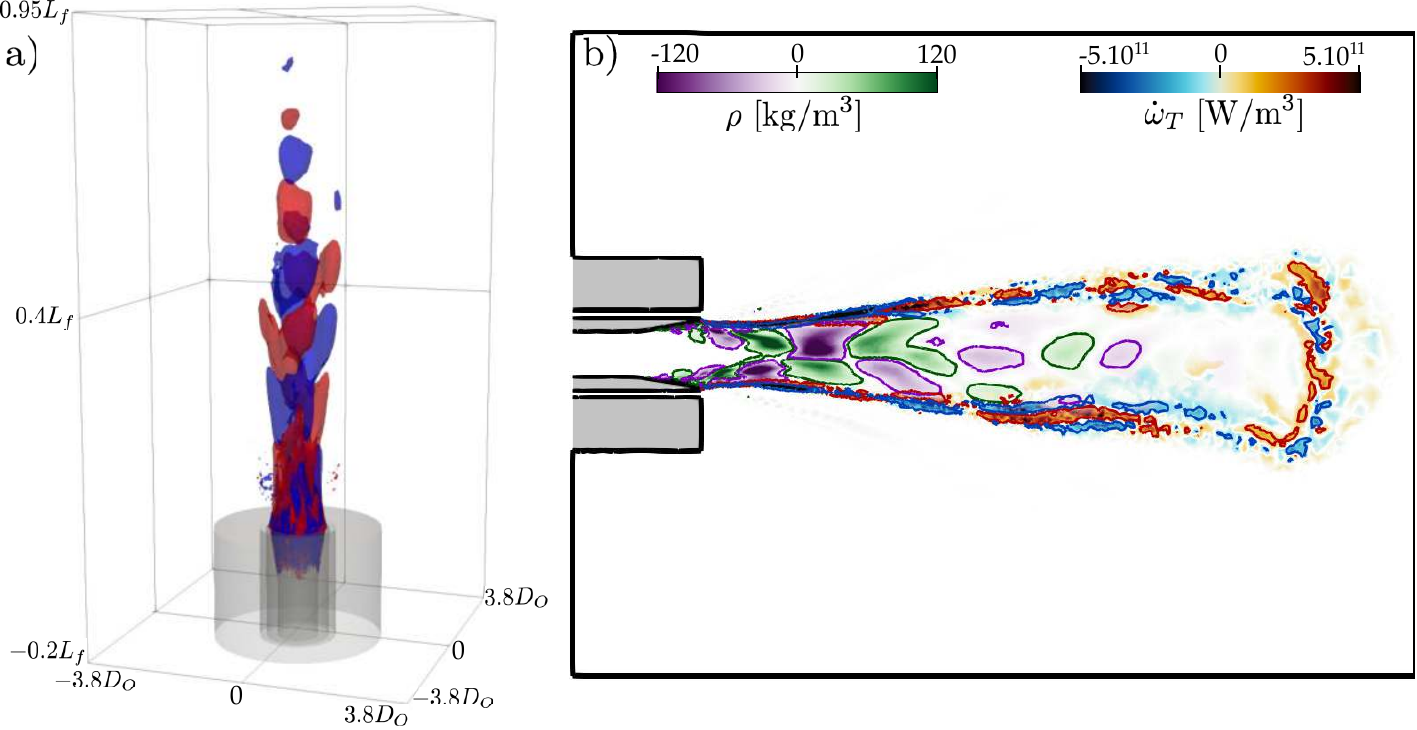}
\caption{Real part of the second DMD mode at $\bm{f = 800}$~Hz. Left: isosurface $\bm{\rho = 15}$~kg/m\textsuperscript{3}  (red) and $\bm{\rho = -15}$~kg/m\textsuperscript{3} (blue) of the density mode. Right: two-dimensional cut of the second density mode (purple-green) superimposed on the second heat-release mode  (blue-red).}
\label{fig:dmd_mode800}
\end{figure}

The spatial structure of the first mode at $f = 480$~Hz is displayed in Fig.~\ref{fig:dmd_mode480}.
It comprises significant density variations in the O\textsubscript{2} core, that originate from the tapered portion of the lip, a region characterized by an intense oxidizer pseudo-boiling. At the exit of the injector, the density mode structure consists of a few straight segments, which then combine into two main branches. These branches then start to roll-up from $x = 2 D_O$, which corresponds to the limit of the dense  O\textsubscript{2} core (see Fig.~\ref{fig:ave_map}), and ultimately disintegrate downstream. These observations suggest that this mode corresponds to a jet break-up hydrodynamic instability occurring in the cryogenic central stream. Its longitudinal wave-length is roughly $L_f/3$ (see Fig.~\ref{fig:dmd_mode480}, right), which corresponds to an approximate frequency $f_{app} \approx 3 \overline{u}_{O2}/L_f$, where $\overline{u}_{O2}$ is the center-line velocity. Using the velocity value from Fig.~\ref{fig:mesh_sensitivity}-(a,h) yields $f_{app} \approx 507$~Hz, which is relatively close to the DMD frequency. The jet-breakup also generates local heat-release fluctuations, by pushing denser oxidizer pockets towards the flame front, that is in turn displaced outwards. The second mode at $f = 800$~Hz, shown in Fig.~\ref{fig:dmd_mode800}, has a very similar structure to the first one.
It comprises a roll-up pattern in the oxidizer stream characteristic of a jet-breakup. Its wave-length appears to be approximately $L_f/6$, which suggests that it is the secondary jet-breakup instability, or equivalently the first harmonic of the mode at $480$~Hz.\par

\begin{figure}[h!]
\centering
\includegraphics[width=0.95\textwidth]{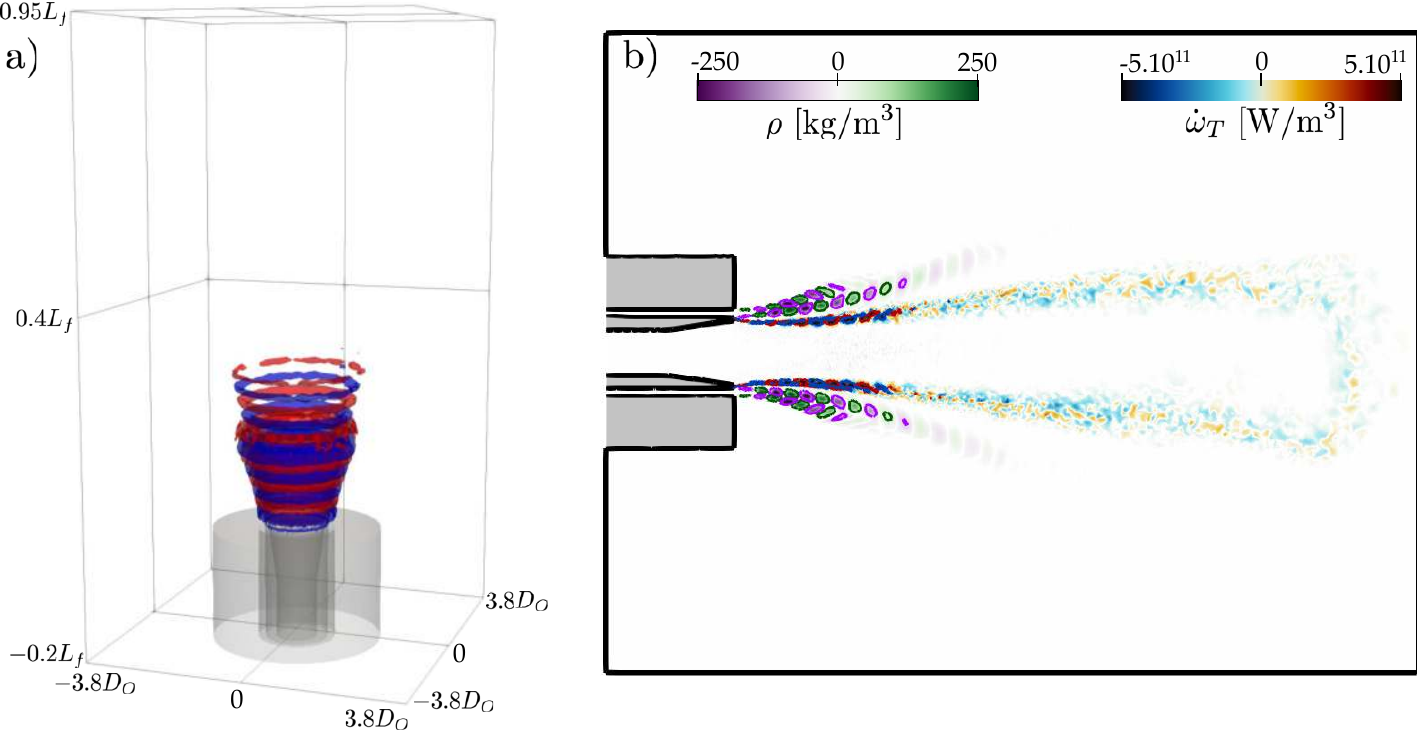}
\caption{Real part of the third DMD mode at $\bm{f = 14600}$~Hz. Left: isosurface $\bm{\rho = 60}$~kg/m\textsuperscript{3}  (red) and $\bm{\rho = -60}$~kg/m\textsuperscript{3} (blue) of the density mode. Right: two-dimensional cut of the third density mode (purple-green) superimposed on the third heat-release mode (blue-red).}
\label{fig:dmd_mode14600}
\end{figure}

\begin{figure}[h!]
\centering
\includegraphics[width=0.6\textwidth]{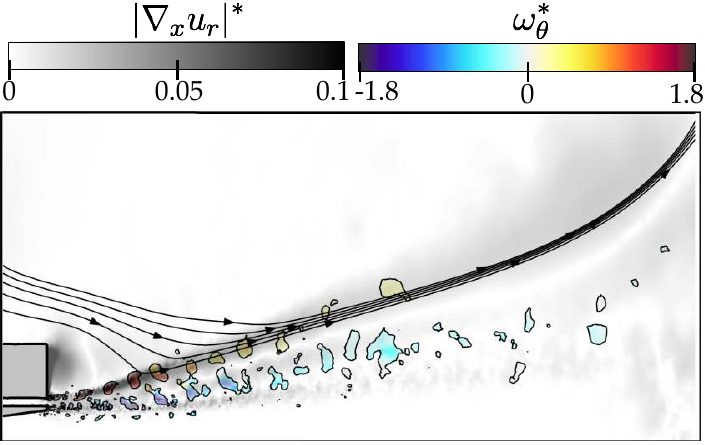}
\caption{Two-dimensional cut of superimposed instantaneous and time-averaged solutions. The thin dark contours are instantaneous isolines of $\bm{Q}$-criterion, that are colored by the normalized azimuthal vorticity $\bm{\omega_{\theta}^{*} = ( \omega_{\theta} W_F ) / u_{CH4}^0 }$ (blue-red). The gray colormap represents the normalized mean flow strain rate $\bm{| \nabla_x \overline{ u_r } |^{*} = ( | \nabla_x \overline{ u_r } | W_F) / u_{CH4}^0}$, and the dark lines with arrows are mean flow streamlines.}
\label{fig:vortices_mode14600}
\end{figure}

The higher-frequency mode at $f = 14600$~Hz, shown in Fig.~\ref{fig:dmd_mode14600} presents a remarkably different structure.
Unlike the previous modes, it is expressed on all three flow variables considered, in a region close to the annular CH\textsubscript{4} jet exiting the injector. It generates variations of both density and velocity, in phase with fluctuations of heat-release rate. This structure suggests that it consists of a double-layer of vortex rings that are generated at the fuel inlet and are convected downstream along the flow streamlines. This pattern is characteristic of a Kelvin-Helmholtz hydrodynamic instability. A Strouhal number for this vortex shedding mode can be computed as $St = f W_F / u_{F}^0  \approx 0.26$, which is comparable to values usually observed for simple laminar bluff-body flows. When vortices pass near the reactive layer, they perturb it, thus resulting in heat-release oscillations. Interestingly, the inner layer of vortex rings is convected over a longer length ($\approx 2D_O$) than the outter layer ($\approx 1D_O$). This specific pattern can be explained by an examination of the mean flow strain rate presented in Fig.~\ref{fig:vortices_mode14600}.
Note the nondimensionalization introduced here and used throughout this chapter: the axial and radial length scales are normalized by the fuel injector width $W_F$ (\textit{i.e.} roughly the vortex core dimension at the injector exit), while the velocities are normalized by that of the fuel injection $u_{CH4}^0$. The outer vortex rings layer lies at the frontier of the LRZ, and is therefore subjected to a large tangential strain rate, which in turn results in a faster dissipation induced by the intense shear stress.\par

These intrinsic unsteady phenomena may participate in the flame response to imposed acoustic perturbations, and it is therefore necessary to consider their potential interaction with the imposed excitation to thoroughly describe the flame dynamics. The interaction between the forced flame dynamics and the high-frequency vortex-shedding mode at $f = 14600$~Hz is discussed in Sec.~\ref{sec:linear_rg_forced_interaction}. On the contrary, as the first two jet breakup modes have frequencies that are significantly lower than that of usual LREs thermoacoustic instabilities, they will be mostly ignored in the following. More pragmatically, simulating the interaction between these low-frequency self-sustained instabilities, and a high-frequency acoustic forcing would require to capture a sufficiently large number of oscillation periods, which would result in a tremendous computational cost. This task is deemed unfeasible with the available resources, and it is therefore not considered below. Note however that in the subsequent analysis, the time-signals are systematically detrended before computing a Fourier transform, such that the effect of the low-frequency oscillations is attenuated as much as possible.\par

\section{Linear flame response to fuel inflow acoustic modulation} \label{sec:linear_rg_flame_response}

Distinct LES are performed to impose acoustic harmonic perturbations at the fuel inlet, for 16 forcing frequencies $f$ comprised in the range $\mathcal{O}(1 \textrm{kHz})$-$\mathcal{O}(20 \textrm{kHz})$. The targeted modulated inlet velocity reads:
\begin{align}
\label{eq:target_modulation_velocity}
u'_{CH4} = A u_{CH4}^0 \sin (2 \pi f t)
\end{align}
This pulsation is imposed through the method proposed in~\cite{Kaufmann2002}: the actual fluctuating inflow velocity may then slightly differ from Eq.~\eqref{eq:target_modulation_velocity} in order to avoid the reflection of acoustic waves at the fuel inlet. Since the goal is to evaluate the flame \textit{linear} response to these modulations, the excitation amplitude $Au_{CH4}^0$ is low ($A = 0.025$), such that it is characteristic of the onset of an instability. This linearity is \textit{a posteriori} verified below. Each simulation is run for at least 20 periods to ensure the dissipation of transients and the convergence of Fourier statistics. Modulations at the oxidizer inlet are also attempted, but no noticeable flame response is observed. Since the bulk inflow velocity $u_{O2}^0$ is roughly one order of magnitude lower than $u_{CH4}^0$, such forcing produces very small absolute value perturbations, which are difficult to detect using Fourier analysis due to the ambient turbulent noise. As a result, it is not guaranteed that this apparent absence of flame response to acoustic excitation of the oxidizer inflow is stemming from a physical phenomenon, or simply the consequence of signals that are too weak to be clearly quantified.\par

Due to the high-frequency of LRE thermoacoustic instabilities and the length of coaxial jet flames, the heat-release region is not assumed acoustically compact (a usual hypothesis for gas-turbine combustion): therefore all quantities depend on both the forcing frequency $f$ and the axial location $x$. In the following,  normalized quantities based on the flame length $L_f$ and the fuel bulk inflow velocity $u_{CH4}$ are introduced as: $f^{*} = fL_f/u_{CH4}^0$ and $x^{*} = x/L_f$. In addition, for any variable $v$ the Fourier coefficient at the forcing frequency $f^{*}$ is denoted $\hat{v}$.

\subsection{Flame Transfer Function} \label{sec:linear_rg_FTF}

\begin{figure}[h!]
\centering
\includegraphics[width=0.6\textwidth]{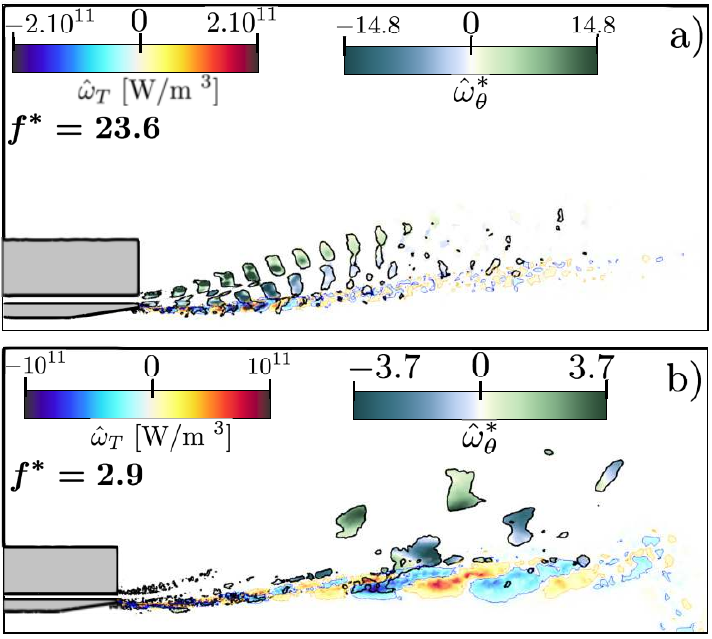}
\caption{(a) Maps of the Fourier coefficients real parts for a high-frequency forcing ($\bm{f^{*}=23.6}$). Dark lines: isoline of $\bm{\hat{Q}}$-criterion, colored by the normalized azimuthal vorticity $\bm{\hat{\omega}_{\theta}^{*} = (\hat{\omega}_{\theta} W_F)/(Au_{CH4}^0)}$. Blue-red: Fourier coefficient of the heat-release rate $\bm{\hat{\omega}_T}$. (b) Same, for a low-frequency forcing ($\bm{f^{*}=2.9}$). }
\label{fig:FFT_vortices}
\end{figure}

The imposed acoustic perturbations produce remarkably different responses, depending on the forcing frequency $f^{*}$. This is evidenced in Fig.~\ref{fig:FFT_vortices}, where the Fourier transforms of azimuthal vorticity and heat-release are mapped for a low-frequency forcing ($f^{*}=2.9$ in Fig.~\ref{fig:FFT_vortices}-(b)), and a high-frequency one ($f^{*}=23.6$ in Fig.~\ref{fig:FFT_vortices}-(a)).
When they exit the fuel injector, acoustic waves generate shear, which in turn produces pairs of counter-rotating vortex rings that are convected downstream in the annular jet. At high-frequency these vortices are generated directly at the injector nozzle and are convected over a length of roughly $0.4 L_f$ before dissipating. Conversely, at low-frequency the larger vortex rings are engendered more downstream (at approximately $x=0.5L_f$) and are transported to the end of the flame. When vortices travel in the vicinity of the flame, they generate a variable strain rate that induces flame fluctuations: there exists therefore a strong correlation between these dinstictive vortex dynamics and the resulting heat-release oscillations. At $f^{*}=23.6$, the heat-release response concentrates in a thin layer extending from the injector exit to $x=0.3L_f$ in the wake of the lip.  In contrast, for $f^{*}=2.9$,  $\hat{\omega}_T$ has a lower intensity, but is distributed over a longer and wider area mostly lying in the second half of the flame. The broader heat-release brush suggests in this case large flame displacements.\par

\begin{figure*}[h!]
\centering
\includegraphics[width=0.99\textwidth]{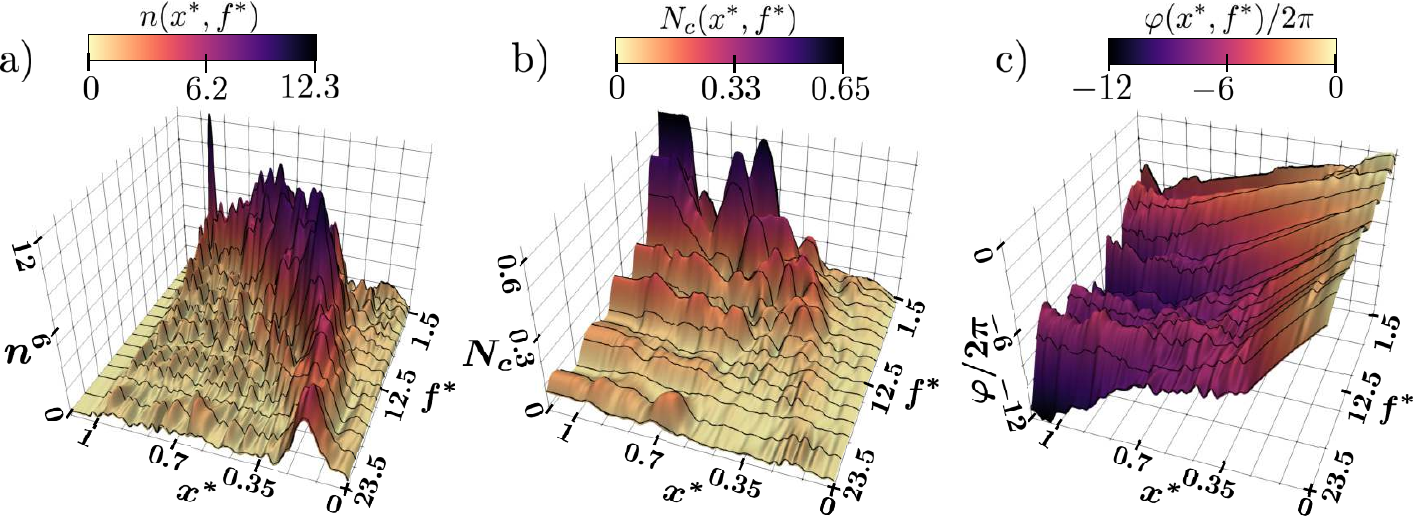}
\caption{(a) Local FTF gain $\bm{n(\st{x},\st{f})}$. (b) Cumulative FTF gain $\bm{N_c(\st{x},\st{f})}$. (c) Local FTF phase $\bm{\varphi(\st{x},\st{f})}$. The dark lines are points directly computed from the 16 forced frequencies. For vizualization purposes the surfaces were obtained by cubic spline interpolation along the $\bm{\st{f}}$ axis.}
\label{fig:gain_phase_3d}
\end{figure*}

To provide a quantitative evaluation of the flame forced dynamics, a FTF relating the heat-release fluctuations to the acoustic excitation is defined in the frequency domain:
\begin{align}
\label{eq:HRfluct_def}
\dfrac{\left( \hat{Q}/Q_0 \right)}{\left( \hat{u}/u_0 \right)} = \dfrac{L_f }{Q_0 \left( \hat{u}/u_0 \right)} \int_{0}^{1} \hat{q} (x^{*}) dx^{*}
\end{align}
where $Q_0$ and $\hat{Q}$ are the averaged and fluctuating global flame power, $u_0$ and $\hat{u}$ are the averaged and fluctuating bulk velocity in the fuel injector near its outlet, and $\hat{q} (x^{*})$ is the fluctuating heat-release per unit length. The local FTF gain $n(\st{x},\st{f})$ and phase $\varphi(\st{x},\st{f})$ are defined as:
\begin{align}
\label{eq:FTF_def}
\dfrac{L_f \hat{q} (x^{*})}{Q_0 \left( \hat{u}/u_0 \right)} = n(\st{x},\st{f}) e^{j \varphi(\st{x},\st{f})}
\end{align}
A cumulative FTF gain $N_c(\st{x},\st{f})$ is also defined:
\begin{align}
N_c(\st{x},\st{f}) = \left| \dfrac{L_f }{Q_0 \left( \hat{u}/u_0 \right)} \int_{0}^{\st{x}} \hat{q} (x^{*})  d \st{x'}  \right|
\end{align}
Note that $N_c(\st{x} = 1,\st{f})$ corresponds to the global gain of the entire flame, and that if the flame were compact then the local gain $n (\st{x},\st{f})$ would be uniformly equal to the global gain. The local FTF is displayed in Fig.~\ref{fig:gain_phase_3d}.\par
The local FTF gain map (Fig.~\ref{fig:gain_phase_3d}-(a)) corroborates observations already made above: at low-frequency forcing the region of strongest heat-release response is long and spans over the entire second half of the flame. As the excitation frequency increases, this region shortens and shifts upstream, such that for high-frequency forcing it is localized at the injector exit. This evolution of the preferential response region is a direct consequence of the vortex rings dynamics described earlier. The local FTF phase (Fig.~\ref{fig:gain_phase_3d}-(c)) behaves similarly: for low $\st{f}$ it is a linear function of $\st{x}$ over the entire flame, which indicates a constant propagation speed $u_{\omega_T}$ of heat-release disturbances along the axial direction. For higher $\st{f}$, the portion of constant propagation speed shrinks: it only extends from $\st{x}=0$ to $\st{x}=0.3$ at $\st{f}=23.6$. Downstream of this location, the local phase reaches a plateau where it stagnates (\textit{i.e.}~the heat-release presents in-phase bulk oscillations). The shape of this plateau corresponds to the upper boundary of the strongest response region in Fig.~\ref{fig:gain_phase_3d}-(a). In addition, $u_{\omega_T}$ increases with frequency (see later in Fig.~\ref{fig:propagation_speeds}): for $\st{f}=1.5$, $u_{\omega_T} = 0.35 u_{CH4}^0$, while $u_{\omega_T} = 1.1 u_{CH4}^0$ for $\st{f}=23.5$. In its linear portion, the phase evolves rapidly with $\st{x}$, which results in heat-release fluctuations that are spatially out of phase. Conversely, points within the plateau oscillate in phase. This directly affects the cumulative gain (Fig.~\ref{fig:gain_phase_3d}-(b)), that shows that the second half of the flame has always the largest contribution to the overall heat-release response. Indeed, even though responses are relatively small in the plateau, they oscillate synchronously, and therefore produce a strong overall contribution. On the contrary, heat-release fluctuations in the preferential response region are not synchronized due to a fast propagation, and thus cancel each other out. Finally, note that the local gain largely exceeds those typical of gas-turbine flames, whereas the global gain has more usual values.\par

\begin{figure}[h!]
\centering
\includegraphics[width=0.5\textwidth]{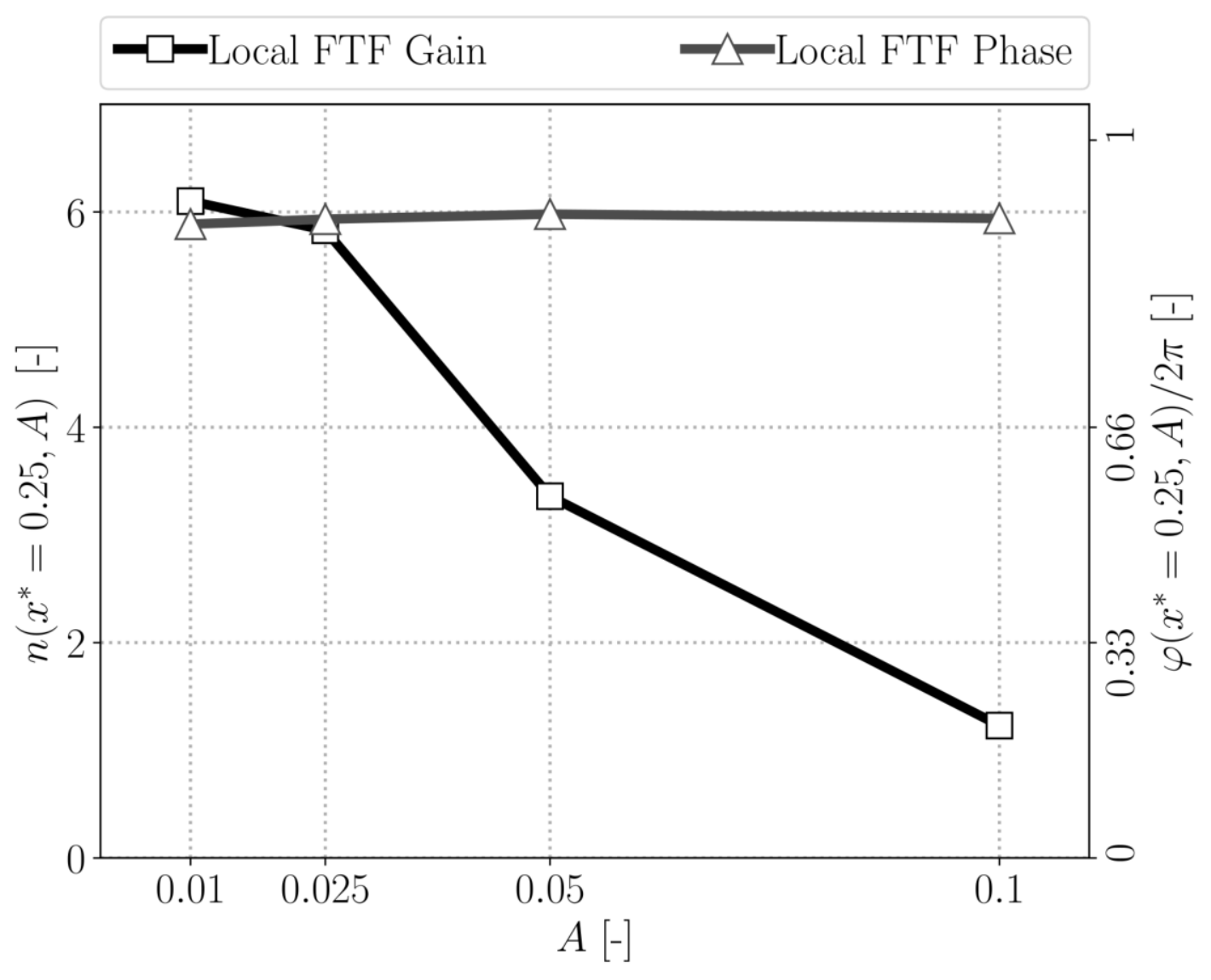}
\caption{Local FTF gain $\bm{n}$ and phase $\bm{\varphi}$ at the location $\bm{x^{*} = 0.25}$, and at the forcing frequency $\bm{f^{*} = 14.8}$, as a function of the modulation amplitude $\bm{A}$.}
\label{fig:linearity_check}
\end{figure}

The linear dependence of the flame response with respect to the modulation amplitude $A$, usually occurring at the onset of a thermoacoustic instability, is verified by performing additional simulations at $A = 0.01$, $A = 0.05$, and $A = 0.1$. For a forcing at $\st{f} = 14.8$, the preferential heat-release response region presents a peak at $\st{x} = 0.25$. The local FTF gain and phase at this location are shown in Fig.~\ref{fig:linearity_check}.
The phase appears to be independent of the excitation magnitude $A$. On the contrary, the local FTF gain decreases with $A$, indicating a response saturation, a nonlinear phenomenon often observed in forced flame dynamics. However, the values of $n(A)$ for $A = 0.01$ and $A = 0.025$ do not strongly differ ($6.1$ \textit{vs.} $5.9$), evidencing a nearly-linear dependence of the heat-release fluctuations with respect to $A$ in this range. Thus, these observations confirm that the acoustic modulations performed in this section trigger a linear response of the flame, and that nonlinear phenomena are negligible. This point is however not true for larger forcing amplitudes, which will be discussed in Chapter~\ref{chap:nonlinear_mascotte}.

\subsection{Analysis of heat-release dynamics contributions} \label{sec:linear_rg_hr_analysis}

\begin{figure}[h!]
\centering
\includegraphics[width=0.99\textwidth]{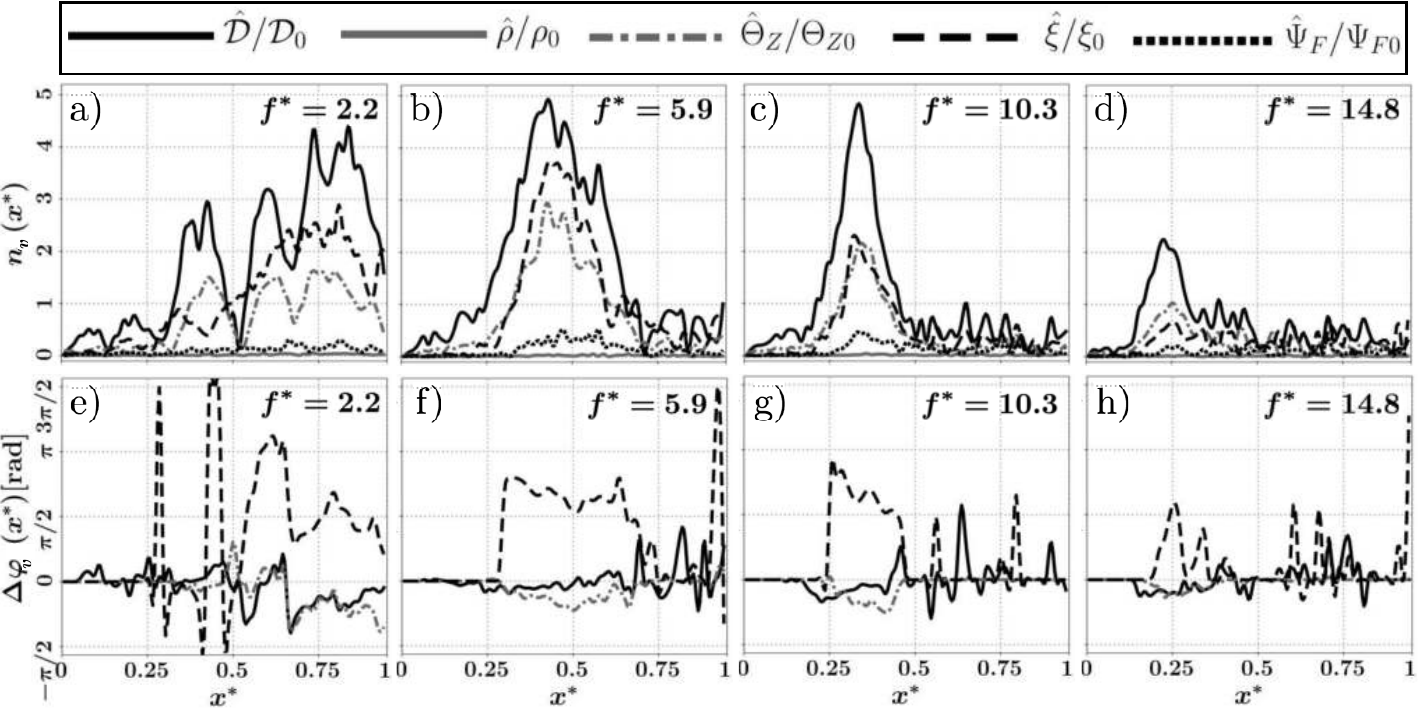}
\caption{(a)-(d) CTF gains $\bm{n_v}$ along the flame axis, at 4 forcing frequencies. (e)-(h) Same, for the CTF phase shift $\bm{\Delta \varphi_v}$ along the flame axis. Phase-shift values between $\bm{-\pi/2}$ and $\bm{\pi/2}$ indicate positive contributions to the heat-release response, while values between $\bm{\pi/2}$ and $\bm{3 \pi/2}$ are negative contributions.}
\label{fig:axial_contrib}
\end{figure}

The distinct physical contributions to the forced heat-release dynamics are here evaluated through a comprehensive analysis. The reaction layer is assumed to be the infinitely thin sheet $S_f$ corresponding to the mixture-fraction isosurface $Z=Z_{st}$. Following the analytical developments of~\cite{Magina2013}, the global flame power writes:
\begin{align}
\label{eq:global_flame_power_fluct}
Q(t) = \iint_{S_f} \dot{m}_R h_R dS = \int_{0}^{1} \dot{m}_R  \xi h_R \ d \st{x}
\end{align}
where $\dot{m}_R$ is the reactants mass burning rate per unit area, $h_R$ is the heat of reaction per unit mass of reactants burnt (assumed constant), and $\xi / 2 \pi$ is the local mean flame radius (\textit{i.e.}~the flame surface per unit length). For a non-premixed flame, the reactants mass burning rate per unit area reads:
\begin{align}
\label{eq:mass_reactants_burning_rate}
\dot{m}_R = \dot{m}_{F} + \dot{m}_{O} = \rho \mathcal{D} \left| \pdv{Y_F}{n} \right| + \rho \mathcal{D} \left| \pdv{Y_O}{n} \right| = \rho \mathcal{D} \left| \dfrac{\partial Z}{\partial n} \right|   \left(  \left| \dfrac{\partial Y_O}{\partial Z} \right| + \left| \dfrac{\partial Y_F}{\partial Z} \right|  \right)_{Z = Z_{st}}
\end{align}
where $\mathcal{D}$ is the species diffusivity coefficient (assumed equal to the mixture fraction diffusivity), $Y_0$ and $Y_F$ are the oxidizer and fuel mass fractions, and $\partial / \partial n$ is the gradient in the direction normal to the flame sheet. Injecting Eq.~\eqref{eq:mass_reactants_burning_rate} into Eq.~\eqref{eq:global_flame_power_fluct} yields the the local heat-release per unit length:
\begin{align}
\label{eq:contrib_full}
q(\st{x},t) = \bigg[ \rho \mathcal{D} \underbrace{\left| \dfrac{\partial Z}{\partial n} \right| }_{\Theta_Z} \underbrace{ \left(  \left| \dfrac{\partial Y_O}{\partial Z} \right| + \left| \dfrac{\partial Y_F}{\partial Z} \right|  \right) }_{\Psi_F} \xi h_R\bigg]_{Z=Z_{st}} 
\end{align}
After splitting each variable $v$ into its mean ($v_0$) and fluctuating ($v'$) parts, a linearization of Eq.~\eqref{eq:contrib_full} yields:
\begin{align}
\label{eq:contrib_split}
\dfrac{q'}{q_0 } = \left[ \dfrac{\rho'}{\rho_0} + \dfrac{\mathcal{D}'}{\mathcal{D}_0} + \dfrac{\Theta_Z'}{\Theta_{Z0}} + \dfrac{\Psi_F'}{\Psi_{F0}} + \dfrac{\xi'}{\xi_0} \right]_{Z=Z_{st}}
\end{align}
Note that this decomposition implicitly neglects the influence of turbulent fluctuations. Accounting for these latter would require a triple decomposition $v = v_0 + v' + v''$, where $v''$ are non-coherent, broadband turbulent oscillations~\cite{reynolds1972}. Applying the Fourier transform at the forcing frequency $\st{f}$ to Eq.~\eqref{eq:contrib_split} gives:
\begin{align}
\label{eq:contrib_split_Fourier}
\dfrac{\hat{q}}{q_0 } = \left[ \dfrac{\hat{\rho}}{\rho_0} + \dfrac{\hat{\mathcal{D}}}{\mathcal{D}_0} + \dfrac{\hat{\Theta}_Z}{\Theta_{Z0}} + \dfrac{\hat{\Psi}_F}{\Psi_{F0}} + \dfrac{\hat{\xi}}{\xi_0} \right]_{Z=Z_{st}}
\end{align}
In Eq.~\eqref{eq:contrib_split_Fourier}, the local-heat release fluctuations are decomposed into 5 distinct contributions $v$. Similarly to Eq.~\eqref{eq:FTF_def}, a Contribution Transfer Function (CTF) is defined for each one of those:
\begin{align}
\label{eq:contribs_FTF}
\dfrac{L_f q_0 \hat{v} (x^{*})}{Q_0 v_0 \left( \hat{u}/u_0 \right)} = n_v(\st{x},\st{f}) e^{j \varphi_v(\st{x},\st{f})} \ , \ \Delta \varphi_v = \varphi_v - \varphi
\end{align}
where $\Delta \varphi_v (\st{x},\st{f}) $ is the phase shift of the fluctuations of $v$, with respect to those of the local heat-release per unit-length $q (\st{x})$. The species diffusivity $\mathcal{D}$ is computed as the mixture mass-average of CH\textsubscript{4} and O\textsubscript{2}, and includes both molecular and SGS turbulent diffusion. Unsteady flame sheets are extracted from the simulations, and the individual CTFs are displayed in Fig.~\ref{fig:axial_contrib}.\par
The CTF gains (Fig.~\ref{fig:axial_contrib}(a)-(d)) behave similarly as those of the FTF of Sec.~\ref{sec:linear_rg_FTF}: for high-frequency forcing, all contributions are  highest near the injector, while at lower frequencies they prevail in a wider region located more downstream. Density fluctuations are remarkably low, which could seem surprising in a flow where dense cryogenic fluids coexist with light burnt gases. It is however consistent with~\cite{Lacaze2012}, where the structure of transcritical counterflow diffusion flames was shown to consist of a vaporization zone without any reaction occurring, followed by a gaseous reactive layer. Consequently, density fluctuations in the perturbed cryogenic methane jet vanish before reaching the flame surface, under the effect of the intense pseudo-boiling. Internal flame structure variations (represented by the term $\hat{\Psi}_F$) participate to the heat-release response in a larger extent than density, but are still roughly one order of magnitude lower than the others. The vortices-induced strain rate is significantly smaller than its extinction limit, and is therefore not sufficient to disturb the flame structure from its \textit{flamelet} regime. Diffusivity fluctuations are the overall major contributor: the heat-release response is mostly driven by the faster or slower diffusion of reactants towards the flame front, which affects the local fluxes and results in variations of the mass burning rate. These diffusivity fluctuations may originate from both SGS mixing variations, caused by the vortices strain-rate, and molecular diffusion oscillations engendered either by pressure waves that accompanied these vortex-rings, or by the modification of the local mixture composition. The secondary contributor is frequency dependent: at high $\st{f}$ it is the mixture fraction gradient fluctuations $\hat{\Theta}_Z$ in the near injector region, in agreement with previous studies in laminar gaseous diffusion jet flames~\cite{Magina2013,Magina2016,tang2019}. At $\st{f} = 10.3$ (Fig.~\ref{fig:axial_contrib}-(c)) a balance is reached between $\hat{\Theta}_Z$ and the flame surface area contribution $\hat{\xi}$. At low $\st{f}$ the trend inverts and $\hat{\xi}$ prevails in the second half of the flame. The relative participations of $\hat{\Theta}_Z$ and $\hat{\xi}$ can be explained by the flame topology (Fig.~\ref{fig:ave_map}). Near the injector the flame is strongly stabilized by a strong shear and is confined between the two dense cryogenic jets: it is not easily displaced by the vortex-induced strain-rate, which rather affects the mixing efficiency through the mixture fraction gradient (see the thin heat-release response layer in Fig.~\ref{fig:FFT_vortices}-(a)). Conversely, when vortices are generated downstream at lower frequencies, they strain a portion of the flame lying in a light gaseous flow with lower shear, where the flame is only weakly stabilized and can therefore be easily displaced (see the broader heat-release response layer in Fig.~\ref{fig:FFT_vortices}-(b)). Finally, the examination of CTF phase shifts in Fig.~\ref{fig:axial_contrib}(e)-(h) shows that both diffusivity and mixture fraction gradient oscillate nearly in phase with the local-heat release, and have therefore a positive contribution. On the contrary, flame surface area mostly fluctuates in phase quadrature ($\Delta \varphi_{\xi} = \pi/2$) or phase opposition ($\Delta \varphi_{\xi} =\pi$) with $q'(x)$, and has hence an adverse effect that tends to limit the local response strength. The vortex-rings dynamics are once again responsible for this behavior: as a vortex travels in the vicinity of the flame it enhances the local burning rate ($q'(x)>0$) through higher-strain rate ($\Theta_Z '>0$) and faster diffusion ($\mathcal{D}'>0$), but simultaneously contracts the flame inwards ($\xi'<0$).\par

\begin{figure}[h!]
\centering
\includegraphics[width=0.99\textwidth]{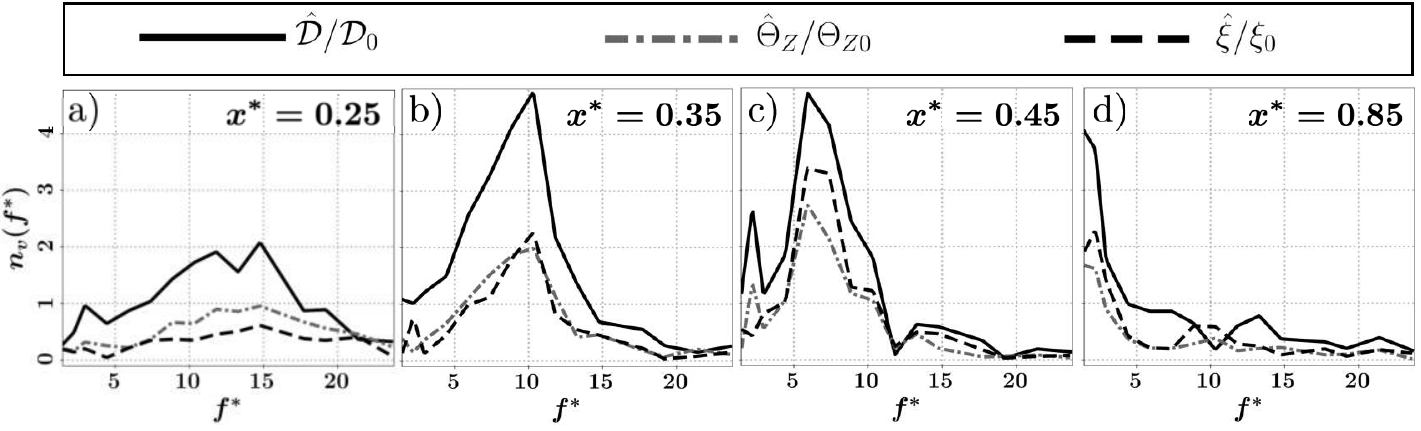}
\caption{CTF gains of the three dominant contributions $\mathcalbf{D}$, $\bm{\Theta_Z}$, and $\bm{\xi}$, at 4 distinct locations along the flame axis, as a function of the forcing frequency $\bm{\st{f}}$.}
\label{fig:contrib_fixed_locations}
\end{figure}

The frequency-dependence of the contributions to the heat-release response is further emphasized in Fig.~\ref{fig:contrib_fixed_locations}, which shows the CTF gains over the entire range of forcing frequencies, at a few fixed locations $\st{x}$.
Each one of the dominant contributions $\hat{D}$, $\hat{\Theta}_Z$, and $\hat{\xi}$ behaves as a band-pass filter in the first half of the flame. In the near-injector region (Fig.~\ref{fig:contrib_fixed_locations}-(a)), the three CTFs share a common peak at $\st{f} \approx 15$, their passbands are wide, but their respective magnitudes are rather low. For downstream locations (Fig.~\ref{fig:contrib_fixed_locations}-(b,c)), the CTFs peak frequency decreases as $\st{x}$ increases, their passbands narrow, and their magnitudes increase. This evolution leads to CTFs near the flame tip (Fig.~\ref{fig:contrib_fixed_locations}-(d)) that behave as narrow-band low-pass filters. The relative prevalence of either the flame surface fluctuations $\hat{\xi}$ or  the strain-rate fluctuations $\hat{\Theta}_Z$ does not appear to depend on the frequency: near the injector, strain-rate oscillations dominate from $\st{f} = 6$ up to $\st{f} = 22$, whereas in the second half of the flame surface oscillations are preponderant over the entire frequency range. Section~\ref{sec:linear_rg_vortex_dyn} and Sec.~\ref{sec:linear_rg_forced_interaction} will show that the frequency dependence of the CTFs peak results from a constructive superposition of vortices-induced perturbations.

\section{Vortex dynamics} \label{sec:linear_rg_vortex_dyn}

The spatial heat-release response, as well as its distinct contributions, appear to be strongly correlated to the generation and convection of vortex-ring pairs in the fuel annular jet. This section aims at detailing the forced dynamics of these vortical waves, which ultimately control the flow disturbances to which the flame front is subjected.

\subsection{Vortex generation} \label{subsec:vortex_generation}

The frequency-dependent preferential heat-release response region (see Fig.~\ref{fig:gain_phase_3d}-a)) seems to correspond to the zone where the vortex-ring pairs are formed. In particular, at high modulation frequencies vortices are created near the injector exit, which triggers considerable flame fluctuations in this region (Fig.~\ref{fig:FFT_vortices}-(a)); conversely, for a low forcing frequency, vortices are engendered downstream which explains a preferential response region located in the second half of the flame (Fig.~\ref{fig:FFT_vortices}-(b)). This distinctive vortex generation can be explained by a \textit{loading and firing} mechanism introduced by Matveev and Culick~\cite{matveev2003b}. In this low-order vortex-shedding model, that was for instance recently employed to study lock-in phenomena in combustion instabilities~\cite{britto2019,singaravelu2016}, the vortex generation is governed by two successive processes:
\begin{itemize}
\item{The\textit{ loading} phase, during which the circulation $\Gamma$ at the corners of the fuel injector exit increases nonlinearly, following the law:
\begin{align}
\label{eq:circulation_evolution}
\dv{\Gamma}{t} = \dfrac{1}{2} \left[ u_{CH4}^0 +  A u_{CH4}^0  \sin \left( 2 \pi f (t - \tau_{ac}) \right)  \right]^2
\end{align}
where $\tau_{ac}$ is the acoustic propagation time between the computational domain inlet, and the fuel injector exit. Equation~\eqref{eq:circulation_evolution} is a quasi-static generalization of the vortex-shedding model of Clements~\cite{clements1973}.
}
\item{The \textit{firing} phase: when the corner circulation $\Gamma$ reaches a critical value $\Gamma_{sep}$, the nascent vortex abruptly detaches from the solid boundary. As it starts to be convected downstream, its initial circulation is $\Gamma_{sep}$. In the meanwhile, the corner circulation $\Gamma$ drops to zero, and the loading phase of the next vortex starts. The separation circulation $\Gamma_{sep}$ is given by:
\begin{align}
\label{eq:circulation_critical}
\Gamma_{sep} (t) = \dfrac{u_{CH4}^0}{2 f_{St,0}} \left[ u_{CH4}^0 +  A u_{CH4}^0  \sin \left( 2 \pi f (t - \tau_{ac}) \right)  \right]
\end{align}
where $f_{St,0}$ is the natural vortex-shedding frequency in the absence of any modulation. In the present case, it can be assimilated to the frequency of the intrinsic Kelvin-Helmholtz mode presented in Sec.~\ref{sec:intrinsic_dynamics} ($f_{St,0} = 14600$~Hz or $St = 0.26$).}
\end{itemize}
After separating from the corners, the vorticity sheets roll-up to form the vortex-rings that are then convected downstream.\par

\begin{figure}[h!]
\centering
\includegraphics[width=0.70\textwidth]{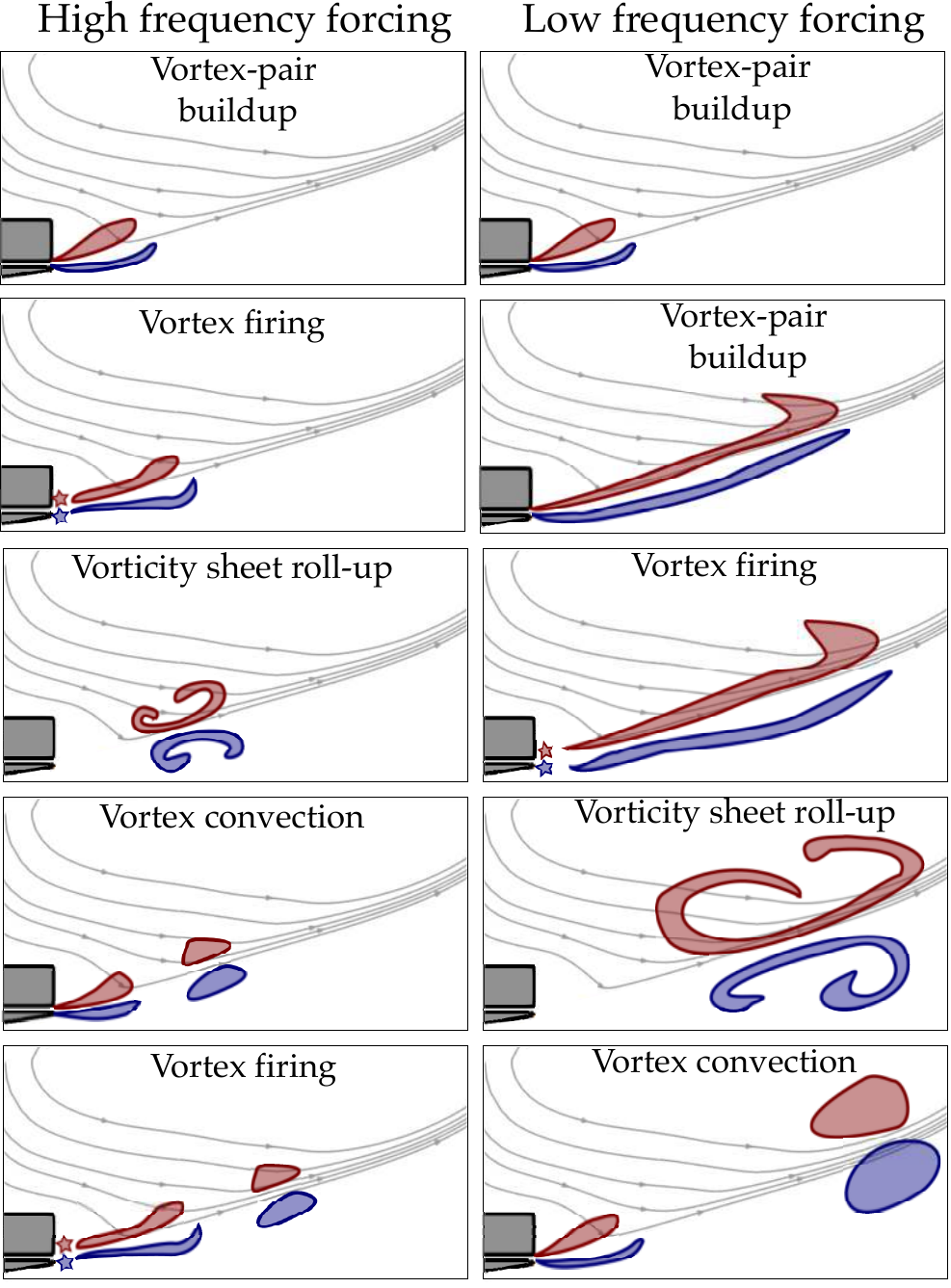}
\caption{Schematic of vortex build-up and firing mechanisms, for a high-frequency forcing (left) and a low-frequency forcing (right).}
\label{fig:schematic_vortex}
\end{figure}

As illustrated in Fig.~\ref{fig:schematic_vortex}, this mechanism governing the vortex growth and detachment yields significantly different vortical structures, depending on the forcing frequency $f$.
During the loading stage, a pair of vorticity sheets progressively expands at the corners of the fuel injector exit. For a high-frequency modulation, the firing occurs when these sheets are still located relatively close to the injector. The sheets then roll-up and yield a vortex-ring pair in the vicinity of the nozzle. Conversely, for a  low-frequency pulsation, the vorticity sheets grow further into the flow before the detachment occurs, which results in larger vortex-rings that are spawned downstream. Note that since the amplitude $A$ is small, first-order approximations of Eq.~\eqref{eq:circulation_evolution} and Eq.~\eqref{eq:circulation_critical} yield a separation frequency close to the forcing frequency: $f_{sep} \approx f$. Consequently, the vortex generation region location scales as the inverse of $f$: $x_{sep} \sim u_{CH4}^{0} / f$. This frequency-dependence of the vortex generation region is consistent with the shape of the preferential heat-release response region observed in Fig.~\ref{fig:gain_phase_3d}.\par

\subsection{Vortex convection} \label{subsec:vortex_convection}

\begin{figure}[h!]
\centering
\includegraphics[width=0.80\textwidth]{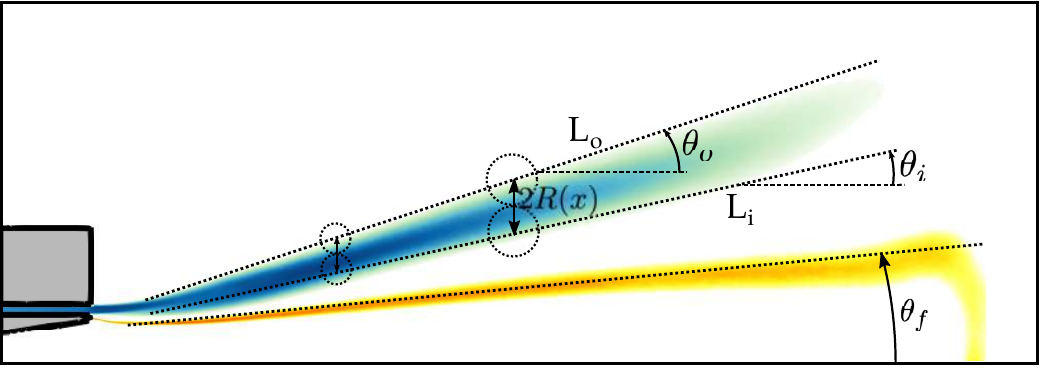}
\caption{Schematic of the inner and outer vortex rings convection paths, respectively denoted $\bm{L_i}$ and $\bm{L_o}$. Their opening angles $\bm{\theta_i}$ and $\bm{\theta_o}$, as well as the flame opening angle $\bm{\theta_f}$ are indicated. The typical vortex radius scale $\bm{R(x)}$ is also shown. Blue: mean axial velocity, yellow: averaged temperature.}
\label{fig:schematic_convection_vortices}
\end{figure}

\begin{figure}[h!]
\centering
\includegraphics[width=0.6\textwidth]{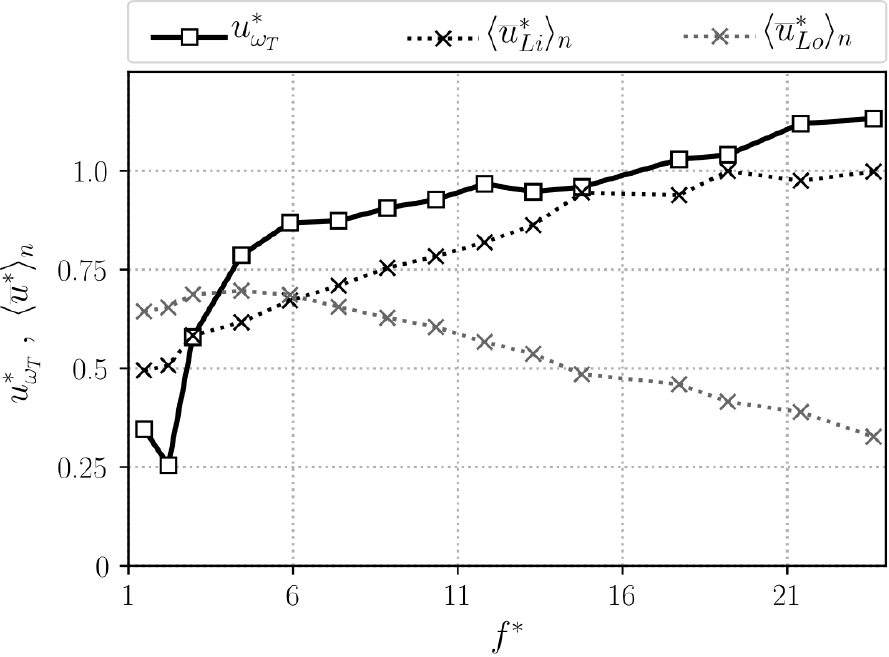}
\caption{Comparison between the normalized heat-release propagation speed $\bm{u_{\omega_T}^{*} = u_{\omega_T}/u_{CH4}^0}$, and the equivalent mean flow velocities $\bm{\langle u_{L_i}^{*} \rangle_n}$ and $\bm{\langle u_{L_o}^{*} \rangle_n}$ on the inner and outer convection lines, as defined by Eq.~\eqref{eq:equivalent_mean_flow_velocity}.}
\label{fig:propagation_speeds}
\end{figure}

Analyzing the vortex-ring pairs convection paths is the second key ingredient necessary to comprehend the disturbances that they may induce on the flame front. An examination of their trajectories shows that they travel along straight trajectories located in the shear-layers at the edges of the fuel annular jet. More precisely, and as illustrated in Fig.~\ref{fig:schematic_convection_vortices}, the outer vortex-rings are convected along the line $L_o$ forming an opening angle $\theta_o = 19^{\circ}$, while the inner vortex-rings propagate along $L_i$ lying roughly midway between the flame front and $L_o$, with an opening angle $\theta_i \approx (\theta_o + \theta_f)/2 \approx 12^{\circ} $.
Note that the convection paths $L_o$ and $L_i$ are aligned with the mean flow streamlines (see in Fig.~\ref{fig:ave_map} and Fig.~\ref{fig:vortices_mode14600}). The correlation between the heat-release fluctuations and the influence of the vortex rings pairs can be further evidenced by defining equivalent mean flow velocities in the lines $L_i$ and $L_o$:
\begin{align}
\label{eq:equivalent_mean_flow_velocity}
\langle u_{L_i}^{*} \rangle_n (\st{f}) = \dfrac{1}{\int_{0}^{1} n( \st{x}, \st{f}) d \! \st{x} } \int_{0}^{1} n(\st{x},\st{f}) \ \dfrac{\overline{u} \left( \st{x}, tan(\theta_i) \st{y} \right)}{u_{CH4}^0}  d \! \st{x} 
\end{align}
where $n$ is the local FTF gain defined in Eq.~\eqref{eq:FTF_def} and shown in Fig.~\ref{fig:gain_phase_3d}-(a), and $\overline{u}$ is the mean flow axial velocity in $L_i$. An analogous definition is used to compute an equivalent mean flow velocity $\langle u_{L_o}^{*} \rangle_n (\st{f})$ in $L_o$. The $n$-weighted spatial average in Eq.~\eqref{eq:equivalent_mean_flow_velocity} allows the definition of a unique equivalent mean flow velocity in each line $L_i$ and $L_o$, corresponding to the peak of the preferential heat-release response region. Figure~\ref{fig:propagation_speeds} compares these equivalent mean flow velocities to the propagation speed $u_{\omega_T}$ of the heat-release perturbations along the flames axis (computed thanks to the slope of the FTF phase in Fig.~\ref{fig:gain_phase_3d}-(c)).
The propagation speed of the heat-release disturbances appears to be strongly correlated to the equivalent mean flow velocity in the inner line $L_i$, as they both increase with $\st{f}$. This common growth is due to the preferential heat-release response region that is located closer to the injector exit at high $\st{f}$, a zone that is characterized by larger axial velocities on the inner line $L_i$. In contrast, the equivalent mean flow velocity on the outer line $L_o$ is of the same order as $\langle u_{L_i}^{*} \rangle_n$ at low-frequencies, but decreases with $\st{f}$, which is attributed to the fact that $L_o$ lies close to the low velocity field of the PIRZ near the injector (see Fig.~\ref{fig:near_injector}) and of the LRZ further downstream (Fig.~\ref{fig:ave_map}). In the light of these conversations, the heat-release fluctuations seem to mainly originate from the flow perturbations induced by the inner layer of vortex-rings, while the outer layer vortices seem to have little effect on the flame front (but may still interact with those in the inner layer).\par

Even though the propagation of heat-release perturbations seems related to the transport of the vortex-rings along the trajectory $L_i$, the analysis of mean velocities such as the one analyzed in Eq.~\eqref{eq:equivalent_mean_flow_velocity} and Fig.~\ref{fig:propagation_speeds} is not sufficient to accurately estimate the instantaneous convection speed of a vortex, which is a notoriously difficult task. Indeed, vortices generated in the wake of bluff bodies often travel at the limit between regions of significantly different mean velocities: for instance in the present case, the outer-layer vortices are convected at the frontier between the high-momentum CH\textsubscript{4} jet and the LRZ, while the inner-layer vortex-rings move between the CH\textsubscript{4} jet and a lower axial velocity region enclosing the flame. To account for the true convection speed that may differ from the primary flow velocity, Matveev and Culick~\cite{matveev2003b} used a simple model that writes:
\begin{align}
\label{eq:convection_speed_model_0}
u_{conv} (x,t) = \alpha \overline{u} (x) + u'(x,t)
\end{align}
where $u_{conv}$ is the instantaneous vortex convection speed, the coefficient $\alpha$ accounts for the difference between the convection speed and the local mean flow velocity $\overline{u}$, and $u'$ represents the effect of the local flow fluctuations. This latter term can be neglected in the present case, since the modulation amplitude is small. Dotson \textit{et al.}~\cite{dotson1997} found the coefficient $\alpha$ to be between $0.5$ and $0.6$ in solid rocket motors cavities. To estimate the true convection speeds in $L_i$ and $L_o$, azimuthal vorticity time-signals over both lines are extracted, and two distinct frequency-domain Vorticity Transfer Functions (VTF) are defined based on Fourier coefficients:
\begin{align}
\label{eq:vorticity_transfer_function_definition}
\vert \hat{\omega}_{\theta}^{*} \vert (\st{x}, \st{f}) \ e^{j \varphi_{\omega_{\theta}} (\st{x}, \st{f})} = \dfrac{R(\st{x})}{A u_{CH4}^0}  \hat{\omega}_{\theta} (\st{x}, \st{f})
\end{align}
where the normalization coefficient $R(\st{x})$ is the typical vortex radius scale at the axial location $\st{x}$, defined as the half-length separating the lines $L_i$ and $L_o$ (see in Fig.~\ref{fig:schematic_convection_vortices}). Since the difference between the opening angles $\theta_o$ and $\theta_i$ is small, it is expressed as $R(x) = (\theta_o - \theta_i) x /2$. The nondimensionalization of Eq.~\eqref{eq:vorticity_transfer_function_definition} allows the relative strengths of vortices of different sizes to be directly compared based on their associated VTF gains $\vert \hat{\omega}_{\theta}^{*} \vert$. Indeed, vortex-rings generated close to the injector are smaller, and their vorticity is then naturally higher than that of wider vortices engendered downstream. Thus, with the nondimensionalized Eq.~\eqref{eq:vorticity_transfer_function_definition}, two vortices with equal circulation but different sizes yield the same VTF gain $\vert \hat{\omega}_{\theta}^{*} \vert$. The propagation speed $v_{\omega_{\theta}}$ in the direction of the convection path ($L_i$ or $L_o$) is computed thanks to the VTF phase $\varphi_{\omega_{\theta}} (\st{x}, \st{f})$:
\begin{align}
\label{eq:convection_speed_computation}
v_{\omega_{\theta}}^{*} (\st{x}, \st{f}) = \dfrac{2 \pi}{u_{CH4}^0} \left( \pdv{\varphi_{\omega_{\theta}}}{x} \right)^{-1} 
\end{align}
Similarly to Eq.~\eqref{eq:equivalent_mean_flow_velocity}, an equivalent propagation speed is defined for each trajectory $L_i$ and $L_o$ as:
\begin{align}
\label{eq:equivalent_convection_speed}
\langle v_{\omega_{\theta}}^{*}  \rangle (\st{f}) = \dfrac{1}{\int_{0}^{1} \vert \hat{\omega}_{\theta}^{*} \vert (\st{x}, \st{f}) \ d \! \st{x}} \int_{0}^{1} \vert \hat{\omega}_{\theta}^{*} \vert (\st{x}, \st{f}) \ v_{\omega_{\theta}}^{*}  (\st{x}, \st{f}) \  d \! \st{x}
\end{align}
where the $\vert \hat{\omega}_{\theta}^{*} \vert$-weighted spatial average allows the computation of a unique equivalent propagation speed representative of the vorticity preferential response region. To estimate the coefficient $\alpha$ of Eq.~\eqref{eq:convection_speed_model_0}, a normalization of the  convection speed by the local mean flow is introduced as $\langle v_{\omega_{\theta}}  \rangle / \langle \overline{v}  \rangle \approx \alpha$, where  $\langle \overline{v}  \rangle$ is the equivalent mean velocity in the direction of the propagation trajectory, defined analogously to Eq.~\eqref{eq:equivalent_convection_speed}.

\begin{figure}[h!]
\centering
\includegraphics[width=0.90\textwidth]{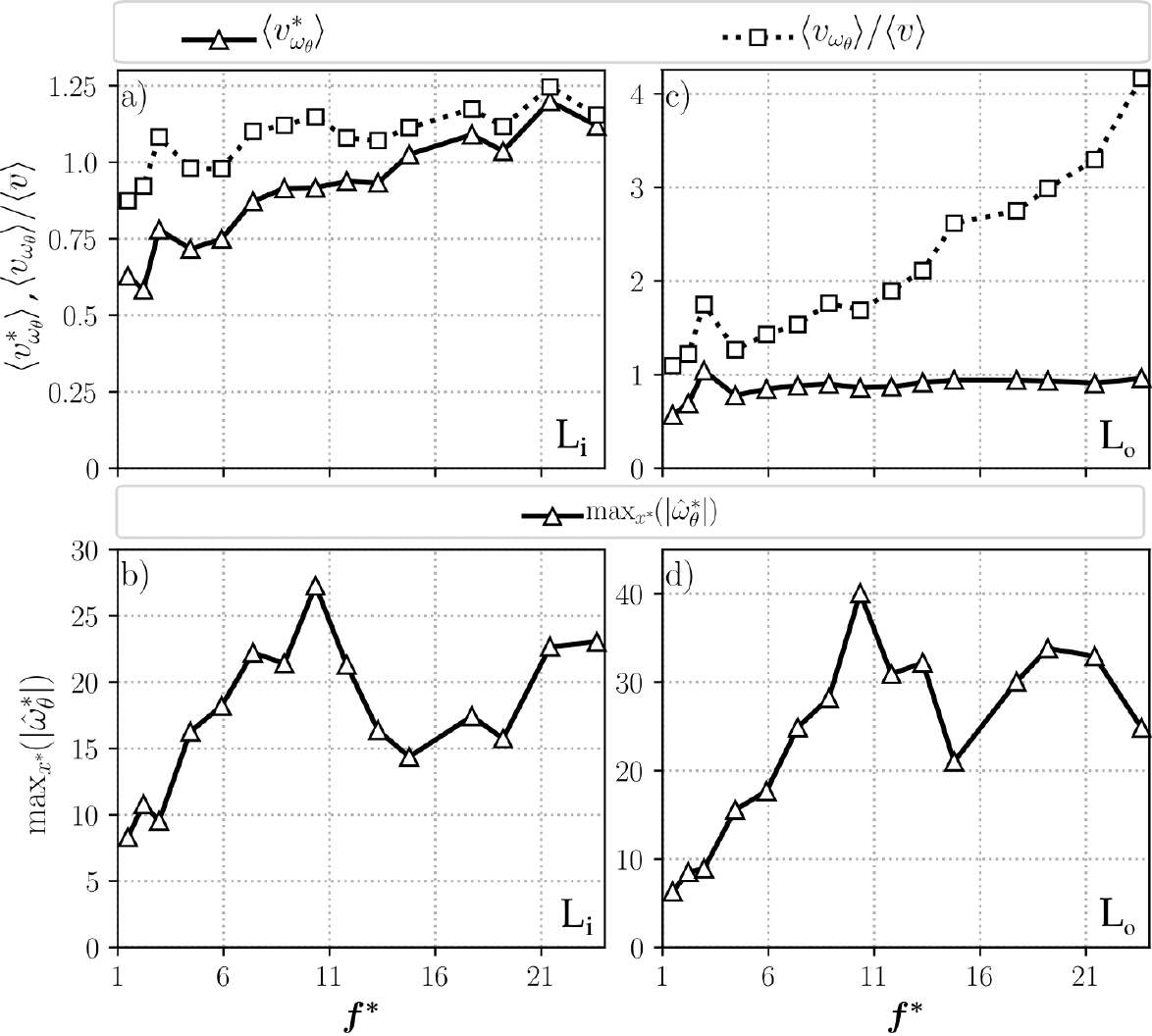}
\caption{Comparison of the vortex responses on both trajectories $\bm{L_i}$ (left column) and $\bm{L_o}$ (right column), as a function of the modulation frequency $\bm{\st{f}}$. (a) and (c): Comparison of the equivalent vortex convection speeds, defined by Eq.~\eqref{eq:equivalent_convection_speed}. (b) and (d): Comparison of the maximal VTF peaks magnitude for each forcing frequency.}
\label{fig:vortex_speed_strength}
\end{figure}

Figure~\ref{fig:vortex_speed_strength} shows comparisons between the equivalent convection speeds and magnitudes of the VTF gain on both trajectories $L_i$ and $L_o$.
On the inner layer $L_i$, the vortex-rings move at a speed similar to that of the local mean flow, yielding a coefficient $\alpha$ close to unity (Fig.~\ref{fig:vortex_speed_strength}-(a)). More precisely, $\alpha$ increases from $0.85$ at low frequencies, up to $1.15$ at high $\st{f}$. The convection coefficient in the outer layer $L_o$ also increases with $\st{f}$ , but in a considerably larger extent, with $\alpha \approx 1$ at low $\st{f}$ and $\alpha \approx 4$ at high $\st{f}$ (Fig.~\ref{fig:vortex_speed_strength}-(b)). However, this growth is not due to a faster convection of vortices, since the equivalent propagation speed $\langle v_{\omega_{\theta}}^{*}  \rangle$ nearly does not vary for $\st{f} > 5$, but rather to an equivalent mean flow velocity $\langle \overline{v}  \rangle$ that diminishes. Indeed, as the frequency increases the vorticity preferential response region progressively shifts upstream, to a low-velocity area under the increasingly stronger influence of the LRZ and the PIRZ adverse flows (see Fig.~\ref{fig:ave_map} and Fig.~\ref{fig:near_injector}). This decrease of the $L_o$ mean velocity is also visible in Fig.~\ref{fig:propagation_speeds}. In contrast, the inner-layer of vortex rings does not travel at the edge of any recirculation zone, which partly explains the more moderate variation of $\alpha$. Independently of their evolution, the values of $\alpha$ computed significantly exceed those usually encountered in gaseous flows. Even more remarkably, the outer vortex-rings travel at speeds close to $u_{CH4}^0$ over most of the frequency range, and independently of the local mean flow velocity in the region where they are emitted. This can be explained by the fact that these waves not only induce fluctuations of vorticity, but also of density, as vortices trap pockets of cryogenic liquid roughly 100 times denser that the surrounding fluid. When they are formed, outer vortex-rings then have a strong convective momentum due to a large density; they are therefore entrained by this inertia and move at a speed close to the bulk injection velocity, which largely exceeds that of the local mean flow. On the contrary, the inner vortex-rings motion is more sensitive to the local mean velocity since they live closer to the flame, and are therefore subjected to a more intense pseudo-boiling that quickly decreases their inertia. This loss of density is even more sudden at high $\st{f}$, when inner vortices are engendered close to the injector. In this case, they then accelerate under the influence of the pseudo-boiling, which may explain their propagation speed surpassing the injection velocity ($\langle v_{\omega_{\theta}}   \rangle \approx 1.15 u_{CH4}^0 $ at $\st{f} = 23.6$ in Fig.~\ref{fig:vortex_speed_strength}-a)).\par

The vortex-rings strength is measured by the peak magnitude in the VTF gains, shown in Fig.~\ref{fig:vortex_speed_strength}-(b,d). The vorticity responses in the outer and the inner convection paths have a similar shape, with a dominant peak at an intermediate frequency $\st{f} \approx  10$, and a secondary maximum at higher frequency near $\st{f} \approx 21$. The presence of these peaks suggests a resonance mechanism, that is further discussed in Sec.~\ref{sec:linear_rg_forced_interaction}. The vorticity response is stronger in the outer trajectory $L_o$ than in $L_i$, at the exception of low modulation frequencies ($\st{f} < 6$), where their intensities are comparable. This overall dominance of the outer vortex-rings is due to their distance to the flame front: they indeed propagate in a non-reactive, relatively cold, and less viscous environment, whereas the inner vortex-rings live close to the reaction front, such that their strength is mitigated by interactions with the flame and by the large viscosity of the hot gases. On the contrary, at low modulation frequencies both inner and outer vortex-rings travel further away from the flame front (see for instance in Fig.~\ref{fig:schematic_vortex} or in Fig.~\ref{fig:FFT_vortices}-(b)), which translates into an identical response strength. Note that the peaks at $\st{f} \approx  10$ and $\st{f} \approx 21$ on the VTF gain are not as noticeable on the FTF gain (Fig.~\ref{fig:gain_phase_3d}-(a)), or on the CTF gains (Fig.~\ref{fig:contrib_fixed_locations}), which suggests that there is no direct proportionality between the vortex and the heat-release fluctuations intensities.\par

To conclude, the analysis of the vortex-rings convection paths, especially thanks to the computation of the VTFs leads to three major observations:
\begin{itemize}
\item{The inner vortex-rings are sensitive to the local mean flow, as their propagation speed approaches the local mean velocity.}
\item{Because of their greater inertia, the outer vortex rings are less sensitive to the local mean velocity: they instead travel at a speed approximately equal to the injection velocity $u_{CH4}^0$, independently of their location in the flow.}
\item{The outer vortex-rings are globally stronger than the inner ones, as they live further away from the flame front. However, strong vorticity perturbations do not necessarily produce equally intense heat-release fluctuations.}
\end{itemize}

\section{Interaction between forced and intrinsic dynamics} \label{sec:linear_rg_forced_interaction}

As mentioned in Sec.~\ref{sec:intrinsic_dynamics}, the unperturbed flame dynamics are characterized by several self-sustained coherent oscillations, among which a dominant vortex-shedding mode affecting the CH\textsubscript{4} stream. Its spatial structure (Fig.~\ref{fig:dmd_mode14600}) is similar to that of the flow patterns resulting from the imposed acoustic modulations, and its frequency $\st{f}_i = 21.4$ lies in the forcing range considered, such that it may interact with the modulated flame dynamics.\par

A phenomenon commonly encountered in such situation is the occurrence of a resonance when the system is excited at its natural frequency (\textit{i.e.}~$\st{f} = \st{f}_i$), resulting in a more intense response. However, the local FTF gain shown in Fig.~\ref{fig:gain_phase_3d}-(a), as well as the CTF gains of Fig.~\ref{fig:contrib_fixed_locations}, do not display any remarkable peak around $\st{f}_i$, which rules out the existence of a resonant heat-release response. The vorticity waves convected along the inner and outer streamlines $L_i$ and $L_o$ nonetheless strongly resonate when they are modulated at the subharmonic frequency $\st{f} = \st{f}_i / 2$, with a secondary peak at the natural frequency $\st{f} = \st{f}_i$, as evidenced in Fig.~\ref{fig:vortex_speed_strength}-(b,d). Subharmonic resonance is often observed in flows involving vortex-shedding, and is usually associated with vortex pairing. Remarkably, in Fig.~\ref{fig:vortex_speed_strength}-(a,c), $\st{f} = \st{f}_i / 2$ is the only frequency where the vortex rings in the inner and the outer layers travel at the same speed: for $\st{f} > \st{f}_i / 2$ the inner layer vortical waves propagate more slowly than that in the outer layer, while for $\st{f} < \st{f}_i / 2$ vortical waves are convected faster in $L_i$ than in $L_o$. This observation suggests that the subharmonic resonance at $\st{f} = \st{f}_i / 2$ is due to a pairing between the inner and outer layers vortex rings, that synchronize to travel at the same speed. The local flow disturbances induced by both vortices are then in phase and add up, thus producing an amplified response. Note that the local FTF and its distinct contributions also present significant gains at the subharmonic frequency $\st{f} = \st{f}_i / 2$ (for instance Fig.~\ref{fig:axial_contrib}-(c) and Fig.~\ref{fig:contrib_fixed_locations}-(b)), even though their respective resonance peaks are not as evident as that of the vorticity response.\par

\begin{figure}[h!]
\centering
\includegraphics[width=0.99\textwidth]{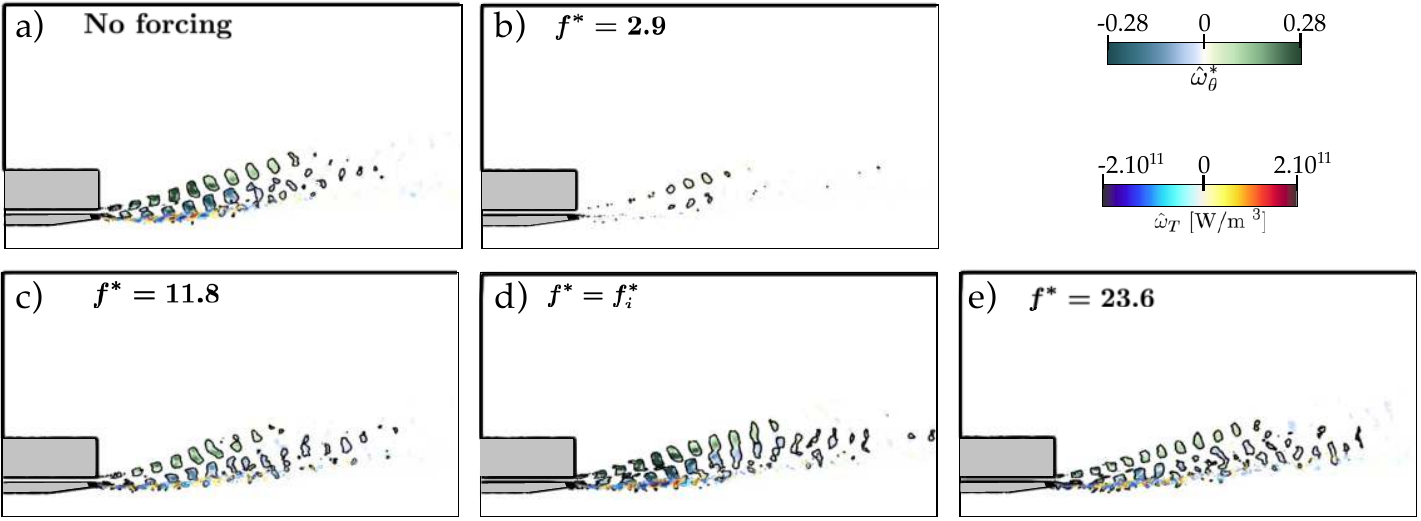}
\caption{Real parts of the Fourier coefficients computed at the self-sustained frequency $\bm{\st{f}_i}$, for different modulation frequencies $\bm{\st{f}}$. Blue-red colormap: heat-release Fourier coefficient $\bm{\hat{\omega}_T}$. Blue-green colormap: Fourier coefficient of the normalized azimuthal vorticity $\bm{\bm{\hat{\omega}_{\theta}^{*} = (\hat{\omega}_{\theta} W_F)/u_{CH4}^0 }}$. (a) Reference case without acoustic modulation. (b) to (e) Cases with increasing forcing frequency $\bm{\st{f}}$.}
\label{fig:interact_intrinsic}
\end{figure}

Another interaction that may occur between the flame intrinsic and forced dynamics is an alteration of its self-sustained vortex-shedding mode by the imposed modulation. This possibility is assessed by performing Fourier analyzes at the natural frequency $\st{f}_i$ for different forcing instances at $\st{f}$. The spatial structures of the vorticity and heat-release Fourier coefficients at $\st{f}_i$ are shown in Fig.~\ref{fig:interact_intrinsic}.
When forced at $\st{f} = \st{f}_i$ (Fig.~\ref{fig:interact_intrinsic}-(d)), the vortical and heat-release fluctuations patterns are identical to that of the intrinsic Kelvin-Helmholtz mode, with a slightly stronger response intensity. This implies that the self-sustained vortex-shedding mode induces relatively small-amplitude flow disturbances, triggering a linear flame response. When the system is excited at an intermediate to high frequency $\st{f}$ (Fig.~\ref{fig:interact_intrinsic}-(c) and (e)), the instrinsic mode structure is not affected, but its intensity is lessened. On the contrary, the intrinsic Kelvin-Helmholtz mode cannot coexist with a low-frequency forcing (Fig.~\ref{fig:interact_intrinsic}-(a)), as the flow response is in this case tuned on the modulation. These observations suggest that at low-frequency, the flame response is unimodal with dominant fluctuations at $\st{f}$. Conversely, for intermediate to high-frequency, the flame response is bimodal, with a primary component at $\st{f}$ and a secondary weaker component at $\st{f}_i$. This latter may be of first-order importance on the development of a thermoacoustic instability, as it may prevent the establishment of a stable limit-cycle or promote phenomena such as mode switching. A generalized type of transfer function would therefore need to be formulated to model this bimodal flame response. Note that additional fluctuations at $\st{f}-\st{f}_i$ and $\st{f}+\st{f}_i$ are also possible, but are nonetheless unlikely due to the small amplitudes of both the acoustic modulation and the self-sustained vortex-shedding.

\section{Conclusion}  \label{sec:linear_rg_conclusion}  

A series of LES have been performed to explore the forced dynamics of a doubly-transcritical LO\textsubscript{2}/LCH\textsubscript{4} LRE jet flame in the Mascotte academic test rig. 
The imposed excitations consist of small amplitude acoustic forcing at the fuel injector, that trigger a linear flame response over a wide frequency range relevant to thermoacoustic instabilities usually encountered in LREs. These computations were a numerical challenge that required 18.5 million CPU hours on a machine equipped with Intel Xeon 8168 processors (2.7GHz).\par

\begin{figure}[h!]
\centering
\includegraphics[width=0.99\textwidth]{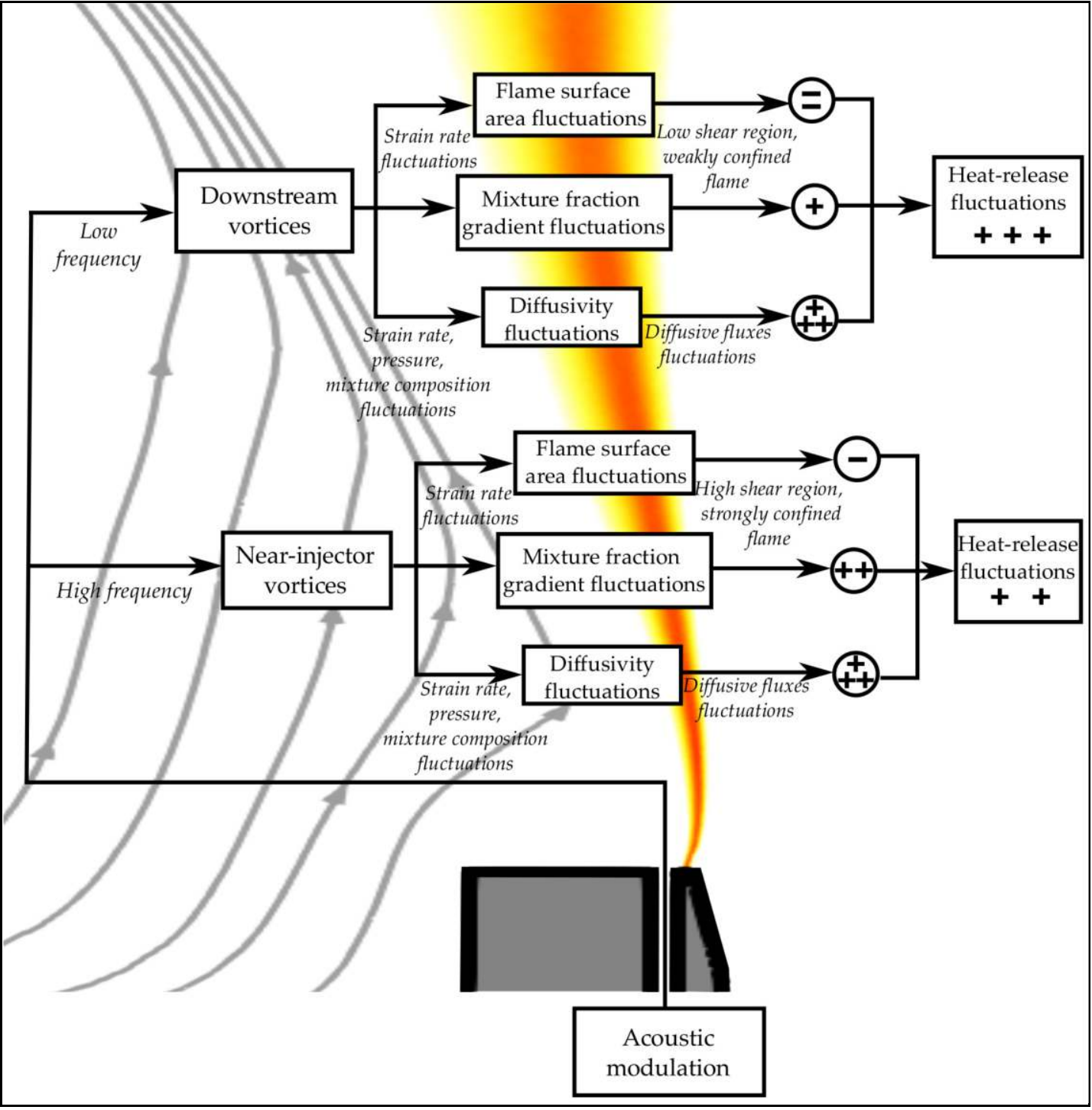}
\caption{Schematic of the different physical mechanisms that contribute to the heat-release linear response to acoustic modulation at the fuel inlet.}
\label{fig:schematic_contributions}
\end{figure}

The unforced flame first displayed elaborate intrinsic dynamics featuring break-up modes of the cryogenic oxidizer jet, as well as a vortex-shedding mode in the fuel annular stream characteristic of a Kelvin-Helmholtz instability. Then, the heat-release response to harmonic acoustic modulation appeared to be directly controlled by the dynamics of vortex-rings pairs traveling in the fuel annular jet. The computation of a non-compact FTF over the range $\mathcal{O}(1\textrm{kHz})$-$\mathcal{O}(20\textrm{kHz})$ revealed frequency-dependent preferential flame response regions, strongly correlated to the area of vortex generation and convection: the higher the excitation frequency, the closer to the injector both regions are localized. A comprehensive analysis of the heat-release dynamics evidenced three major contributions to the flame response: the variations of species diffusivity that globally dominate, those of mixture fraction gradient that are significant at high-frequency in the near-injector region, and the flame surface area fluctuations that become stronger at low-frequency in the second half of the flame. These contributions and their underlying mechanisms are summarized in Fig.~\ref{fig:schematic_contributions}.
This analysis has the potential to guide future development of theoretical models for LRE flame response, which are expected to account for the 3 major contributions identified. Additionally, the FTF data is available, and expected to be embedded in acoustic solvers for the prediction of thermoacoustic instabilities in multi-injectors LREs.

\chapter{Nonlinear dynamics in a doubly-transcritical LO\textsubscript{2}/LCH\textsubscript{4} coaxial jet flame} \label{chap:nonlinear_mascotte}
\minitoc				

\begin{chapabstract}
The previous chapter investigated the linear response of a LO\textsubscript{2}/LCH\textsubscript{4} LRE flame, subjected to small amplitude acoustic modulation imposed at the fuel inflow. This perturbation regime is typical of the onset of a thermoacoustic instability, but significantly differs from an established large-amplitude limit-cycle, which may trigger nonlinear phenomena in the flame response. The present chapter therefore extends the analysis initiated in the previous one, by assessing the nonlinear dynamics of the same doubly-transcritical flame in the Mascotte test-rig, subjected to strong fuel inflow acoustic modulation, over a wide frequency range spanning from $\mathcal{O} (1 \ \textrm{kHz})$ to $\mathcal{O} (20 \ \textrm{kHz})$. It starts with the identification of two key nonlinear features affecting the heat-release, namely the response intensity saturation and the generation of higher harmonics. A local Flame Describing Function is computed and commented. The vortex nonlinear dynamics, which are strongly correlated to the flame response, are then discussed. Finally, the physical mechanisms contributing to the heat-release response saturation are identified, and their respective nonlinear behaviors are compared.
\end{chapabstract}

\section{Introductory remarks} \label{sec:NL_intro}

The numerical setup is identical to the one of Chapter~\ref{chap:mascotte_linear}. The geometry is that of the single-injector Mascotte test rig (see in Fig.~\ref{fig:geom_intro} and Fig.~\ref{fig:schematic_intro}), operating in the doubly-transcritical regime. Both propellants (LO\textsubscript{2} and LCH\textsubscript{4}) are injected as dense cryogenic fluids into the combustion chamber at $P  = 75$~bar (see Tab.~\ref{tab:properties_mascotte}). A series of LES are performed to impose harmonic acoustic modulation of the fuel inflow, at 16 distinct frequencies $f$ ranging from $\mathcal{O} (1 \ \textrm{kHz})$ to $\mathcal{O} (20 \ \textrm{kHz})$. This frequency range is representative of thermoacoustic instabilities usually observed in LREs. The target fluctuating inlet velocity, imposed thanks to the methodology of~\cite{Kaufmann2002}, writes:
\begin{align}
u_{CH4}' = A u_{CH4}^0 \sin (2 \pi f t)
\end{align}
Nondimensionalized forcing frequency and axial location are defined based on the flame length and the fuel inlet bulk velocity: $\st{f} = f L_f/u_{CH4}^0$, $\st{x} = x/L_f$. Several other nondimensionalized quantities are introduced later and detailed when necessary. Simulations are run for at least 20 oscillations periods to ensure the dissipation of transients and the convergence of Fourier statistics. In addition, every time signal is systematically detrended before computing Fourier coefficients, in order to mitigate the effect of intrinsic low-frequency oscillations detailed in Sec.~\ref{sec:intrinsic_dynamics}. The notation $\hat{v}$ refers to the Fourier coefficient of the quantity $v$ at the forcing frequency $\st{f}$, unless specified otherwise.\par

At the onset of a thermoacoustic instability, the perturbations are small in comparison to the mean flow quantities, and they therefore induce a flame response that scales linearly with their amplitude. This linear regime was emulated in Chapter~\ref{chap:mascotte_linear} by imposing small-amplitude acoustic modulations ($A = 0.025$). During a combustion instability, the onset stage is often followed by a continuous growth of the fluctuations intensity, until a stable limit-cycle is reached. In this latter phase, the perturbations strength is significantly larger: for instance in~\cite{Urbano2017}, the bulk fuel inflow velocity was observed to oscillate from 10\% to 15\% of its mean value. These strong disturbances significantly alter the flame response, that then entails nonlinear features. Those can induce elaborate phenomena, such as chaotic oscillations, harmonics and subharmonics generation, or even flame blow-off and flash-back. In the present work, two common nonlinear flame response mechanisms are investigated:
\begin{itemize}
\item{The flame response saturation, or in other words a relative decrease in the strength of the heat-release fluctuations. This saturation is a key ingredient in  the establishment of a stable limit-cycle, as it slows down and ultimately stops the growth of the thermoacoustic oscillations when an equilibrium state is reached. The heat-release saturation was evidenced in~\ref{fig:linearity_check}, where the local FTF gain is shown to decrease with $A$.}
\item{The generation of higher harmonics, that is a known route to chaos in laminar combustion systems, through the period-doubling process that it induces~\cite{kashinath2014}. Subharmonics generation may also occur, but are not considered here due to the considerable associated computation cost.}
\end{itemize}
In the light of these preliminary remarks, imposing a 10\% amplitude modulation ($A = 0.1$) appears as a reasonable choice to trigger nonlinear flame response mechanisms representative of a LRE thermoacoustic limit-cycle. The following Sections provide an analysis of the LES results based on comparisons with the linear response computed in Chapter~\ref{chap:mascotte_linear} (where the forcing amplitude was 4 times smaller than that used here).

\section{Nonlinear heat-release response} \label{sec:nonlinear_hr}

\begin{figure}[h!]
\centering
\includegraphics[width=0.6\textwidth]{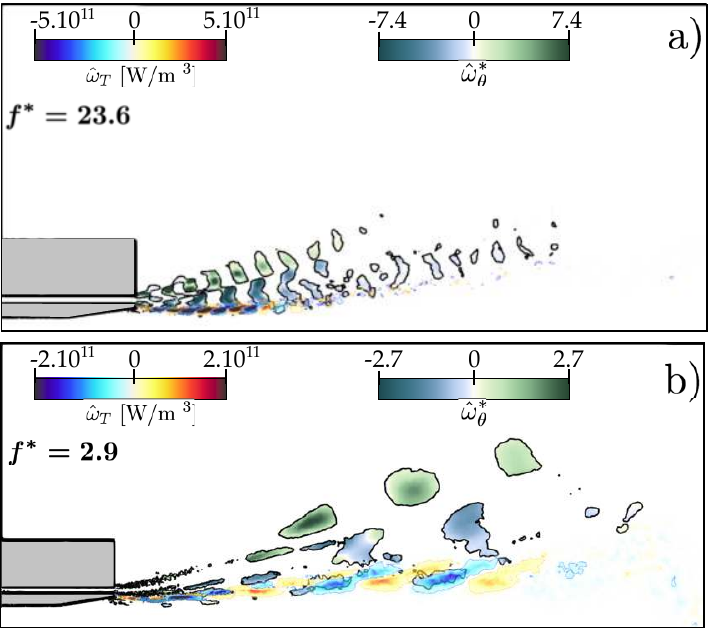}
\caption{(a) Maps of the Fourier coefficients real parts for a high-frequency forcing ($\bm{f^{*}=23.6}$). Dark lines: contour of $\bm{\hat{Q}}$-criterion, colored by the normalized azimuthal vorticity $\bm{\hat{\omega}_{\theta}^{*} = (\hat{\omega}_{\theta} W_F)/(Au_{CH4}^0)}$. Blue-red: Fourier coefficient of the heat-release rate $\bm{\hat{\omega}_T}$. (b) Same, for a low-frequency forcing ($\bm{f^{*}=2.9}$).}
\label{fig:NL_vortices_fft}
\end{figure}

The two-dimensional maps of the azimuthal vorticity and heat-release Fourier coefficients shown in Fig.~\ref{fig:NL_vortices_fft} reveal a flow response spatial pattern similar to the one observed with a low-amplitude forcing (see Fig.~\ref{fig:FFT_vortices} for comparison).
As they leave the fuel injector, acoustic waves generate a variable shear that in turn produces pairs of vortex-rings which are convected along the mean flow streamlines. At high frequency, these vortices are generated directly at the injector exit, while at low frequency they are spawned downstream in the second half of the flame. As they travel in the vicinity of the reaction front, these vortices perturb it, which produces in-phase heat-release fluctuations. The absolute local heat-release oscillations are naturally more intense for this large amplitude forcing than in the linear case ($\hat{\omega}_T (A=0.1) \sim 5 \times 10^{11}$ \textit{vs.} $\hat{\omega}_T (A=0.025) \sim 2 \times 10^{11}$  at $\st{f} = 23.6$, and $\hat{\omega}_T (A=0.1) \sim 2 \times 10^{11}$ \textit{vs.} $\hat{\omega}_T (A=0.025) \sim 1 \times 10^{11}$  at $\st{f} = 2.9$). The relative vortex response is however roughly diminished by half ($\hat{\omega}_{\theta}^{*} (A=0.1) \sim 7$ \textit{vs.} $\hat{\omega}_{\theta}^{*} (A=0.025) \sim 14$  at $\st{f} = 23.6$, and $\hat{\omega}_{\theta}^{*} (A=0.1) \sim 2$ \textit{vs.} $\hat{\omega}_{\theta}^{*} (A=0.025) \sim 4$  at $\st{f} = 2.9$).\par

\subsection{Local Flame Describing Function} \label{subsec:FDF}

The nonlinear dependence of the heat-release fluctuations $\hat{\omega}_T$ with respect to the forcing amplitude $A$ suggested in Fig.~\ref{fig:NL_vortices_fft} can be more precisely quantified through the computation of a frequency-domain Flame Describing Function (FDF), which is a generalization of the FTF concept to nonlinear systems~\cite{Noiray:2008,gelb1968}. Since the modulation frequencies are high, and since the flame is longer than in a usual gaseous case, it is not considered compact. Consequently, a local FDF depending on both the forcing frequency $\st{f}$ and the axial location $\st{x}$ is computed. This type of non-compact FDF was for instance computed by Li \textit{et al.}~\cite{li2017a} for a 1 meter long turbulent premixed gaseous flame. The local FDF gain and phase are defined in a similar fashion as in Sec.~\ref{sec:linear_rg_FTF}:
\begin{align}
\label{eq:FDF_def}
n(\st{x},\st{f}) e^{j \varphi(\st{x},\st{f})} = \dfrac{L_f \hat{q} (x^{*})}{Q_0 \left( \hat{u}/u_0 \right)}
\end{align}
where $\hat{q} (x^{*})$ is the Fourier coefficient of the fluctuating heat-release rate par unit length, $Q_0$ is the mean flame power, and $u_0$ is the mean bulk velocity in the fuel injection line. The associated cumulative FDF gain $N_c(\st{x},\st{f})$, which quantifies the overall heat-release oscillations, is still defined as:
\begin{align}
\label{eq:FDF_global_gain}
N_c(\st{x},\st{f}) = \left| \dfrac{L_f }{Q_0 \left( \hat{u}/u_0 \right)} \int_{0}^{\st{x}} \hat{q} (x^{*})  d \st{x'}  \right|
\end{align}
It is reminded that $N_c(\st{x} = 1,\st{f})$ is the global gain commonly used for compact flames. Note that the computation of a complete FDF would require to perform additional sets of simulations at different amplitude levels. In the present work, only a small portion of this FDF at a fixed amplitude $A=0.1$ is considered.\par

\begin{figure}[h!]
\centering
\includegraphics[width=0.99\textwidth]{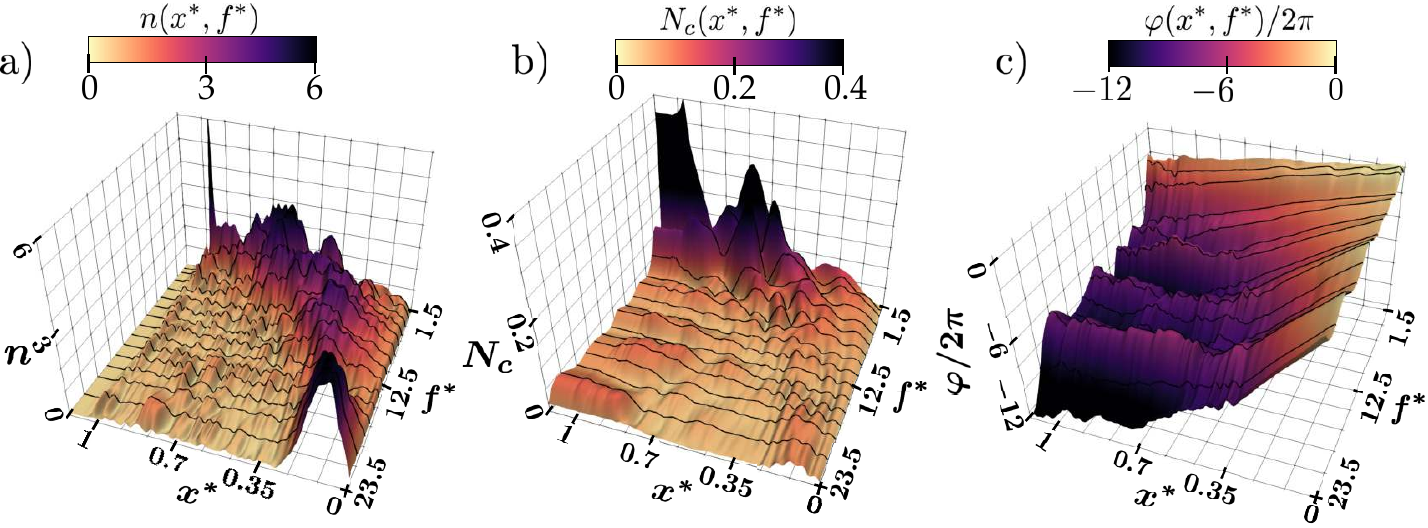}
\caption{Maps of the local FDF for a forcing at $\bm{A=0.1}$. (a) Local FDF gain $\bm{n(\st{x},\st{f})}$. (b) Cumulative FDF gain $\bm{N_c(\st{x},\st{f})}$. (c) Local FDF phase $\bm{\varphi(\st{x},\st{f})}$. The dark lines are points directly computed from the 16 forced frequencies. For visualization purposes the surfaces were obtained by cubic spline interpolation along the $\bm{\st{f}}$ axis.}
\label{fig:NL_3DFDF}
\end{figure}

\begin{figure}[h!]
\centering
\includegraphics[width=0.99\textwidth]{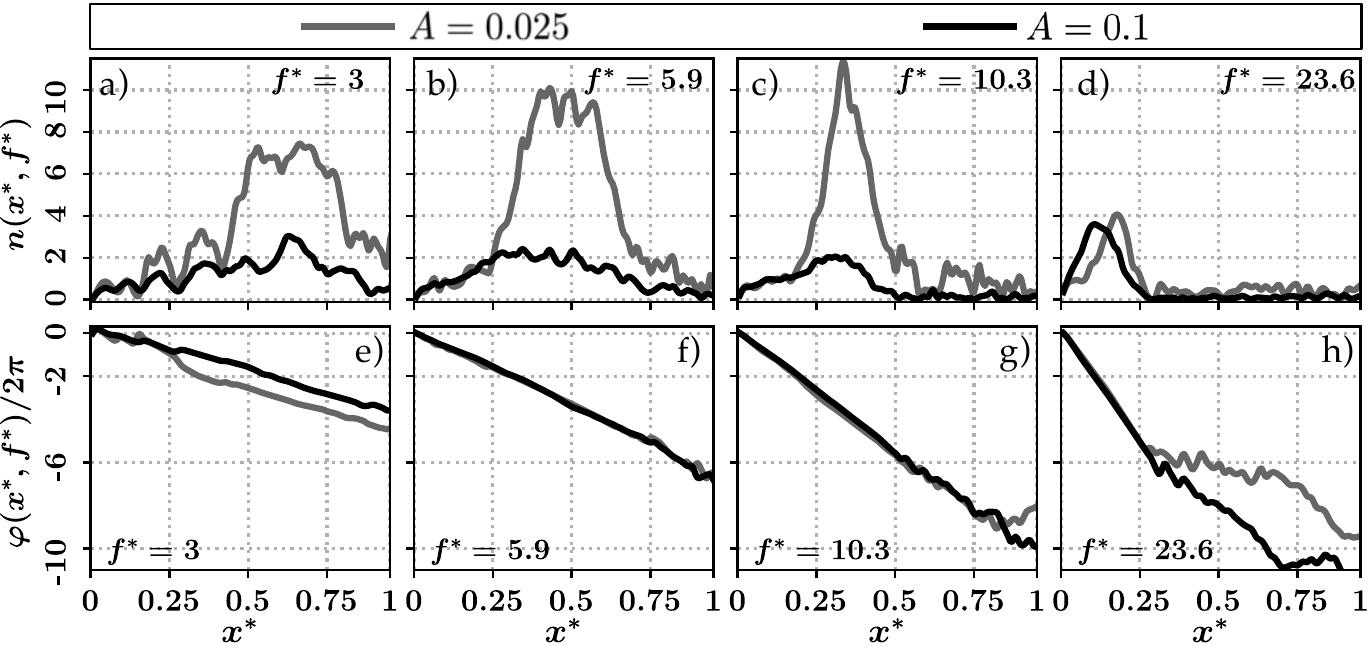}
\caption{Comparison between the local FDF (case $\bm{A =0.1}$) and the local FTF (case $\bm{A = 0.025}$), over the entire flame length, for a few select forcing frequencies $\bm{\st{f}}$. (a) to (d): local FDF/FTF gains $\bm{n (\st{x}, \st{f})}$. (e) to (h): local FDF/FTF phases $\bm{\varphi (\st{x}, \st{f})}$.}
\label{fig:NL_comparison_axial_FTF}
\end{figure}

The local FDF gain, phase, and cumulative gain are represented over the frequency range of interest and the flame length in Fig.~\ref{fig:NL_3DFDF}.
The local FDF gain (Fig.~\ref{fig:NL_3DFDF}-(a)) presents a preferential response region whose topology is similar to that observed in the linear case (see Fig.~\ref{fig:gain_phase_3d} for comparison). At high-frequency, the heat-release response concentrates in a small region at the injector exit, while for higher $\st{f}$ this region widens and is shifted downstream towards the flame tip. Differences are however observed on the magnitude of this preferential response region. In the linear regime, the FTF gain reaches maximal values of the order of $12$ near the half of the flame, for intermediate forcing frequencies ($ 8 < \st{f} < 11 $). In contrast, large amplitude modulations yield a FDF gain peaking at roughly $5$ in the near-injector region for high $\st{f}$.  This decrease of the relative heat-release oscillations intensity clearly evidences a nonlinear saturation of the flame response, which can be more quantitatively assessed through the direct comparisons between the local FDF and FTF shown in Fig.~\ref{fig:NL_comparison_axial_FTF}.
At low frequency, the FDF saturation is rather marked, as the local gain in the preferential region is roughly divided by a factor of 3 between the low-amplitude and the large-amplitude modulations (Fig.~\ref{fig:NL_comparison_axial_FTF}-(a)). It is even more pronounced at intermediate frequencies, where $n(A=0.1) \sim n(A=0.025)/5$ near the half of the flame (Fig.~\ref{fig:NL_comparison_axial_FTF}-(b,c)). On the contrary, the local heat-release response does not saturate for high-frequency forcing in the preferential response region located near the injector (Fig.~\ref{fig:NL_comparison_axial_FTF}-(d)). In this latter case, the only noticeable difference between the linear and the nonlinear regime is a slight displacement of the response peak upstream, towards the injector exit.\par

\begin{figure}[h!]
\centering
\includegraphics[width=0.70\textwidth]{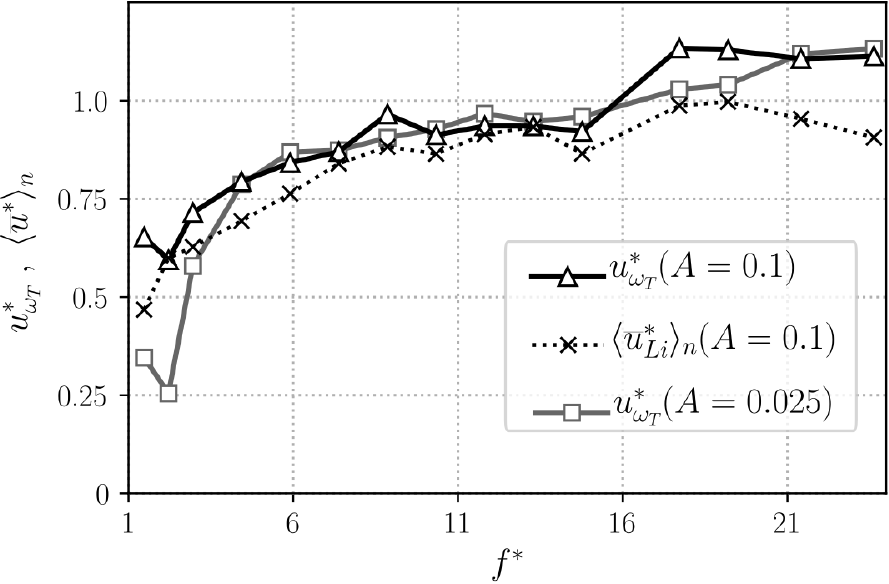}
\caption{Comparison between the propagation speeds $\bm{u_{\omega_T}^{*} = u_{\omega_T}/u_{CH4}^0}$ of the heat-release disturbances along the flame axis, as a function of the forcing frequency $\bm{\st{f}}$, for the large-amplitude and the small-amplitude modulations ($\bm{A = 0.1}$ and $\bm{A = 0.025}$, respectively). The equivalent mean flow velocity $\bm{\langle \overline{u}^{*}_{L_i} \rangle}$ along the inner streamline $\bm{L_i}$ (defined in Eq.~\eqref{eq:equivalent_mean_flow_velocity}) is also indicated for reference.}
\label{fig:NL_propagation_speeds}
\end{figure}

The FDF phase (Fig.~\ref{fig:NL_3DFDF}-(c) and Fig.~\ref{fig:NL_comparison_axial_FTF}-(e) to (h)) is also similar to that obtained in the low-amplitude case: it evolves linearly with $\st{x}$ from the exit of the injector to a downstream location where it reaches a plateau and stagnates. This pattern indicates a first region of constant propagation speed $u_{\omega_T}$ of the heat-release disturbances along the axial direction, followed by a second zone presenting bulk uniform heat-release oscillations. Note that the plateau does not exist at low $\st{f}$, suggesting a constant propagation speed $u_{\omega_T}$ over the entire flame length. Direct comparisons between the propagation speeds $u_{\omega_T}$ in the linear and the nonlinear regimes are provided in Fig.~\ref{fig:NL_propagation_speeds}.
Comparable values are obtained in both cases for frequencies $\st{f}  > 5$, thus indicating that the propagation of the  heat-release disturbances along the flame axis is not noticeably affected by nonlinear response mechanisms. Remarkable differences are however observed for $\st{f} < 5$, where the heat-release perturbations propagate twice as fast in the nonlinear regime as in the linear one. Interestingly, in the former case $u_{\omega_T}^{*}$ appears more strongly correlated to the equivalent mean flow velocity $\langle \overline{u}^{*}_{L_i} \rangle$ in the inner streamline $L_i$ (where vortex rings are convected), whereas in the linear regime $u_{\omega_T}^{*}$  is significantly lower than the equivalent mean flow velocity.\par

These dissimilarities in the propagation speeds $u_{\omega_T}^{*}$ have first-order effects on the saturation of the cumulative FDF gain $N_c$ (Fig.~\ref{fig:NL_3DFDF}-(b)). Because of the faster propagation at $\st{f} < 5$, the heat-release fluctuates more in-phase over the flame length in the nonlinear regime than in the linear one. This spatio-temporal coherence of the heat-release fluctuations directly results in a global flame response that is only slightly weakened in comparison to that obtained with small-amplitude modulations ($N_c (\st{x} = 1, A=0.1) \sim 0.4$ \textit{vs.} $N_c (\st{x} = 1, A=0.025) \sim 0.6$ at $\st{f} = 2.9$). On the contrary, even though the local saturation at high $\st{f}$ is not as pronounced as that at lower $\st{f}$ (see Fig.~\ref{fig:NL_comparison_axial_FTF}-(a) to (d)), the global response saturation is more marked in the former case. This is due to a propagation speed $u_{\omega_T}$ nearly identical to the one in the linear regime (Fig.~\ref{fig:NL_comparison_axial_FTF}-(f) to (h)), which produces out of phase heat-release oscillations. This weak spatio-temporal coherence yields global FDF gains for $\st{f} > 5$ significantly lower than that of the FTF.\par

To conclude, the computation and analysis of the local FDF leads to three major observations:
\begin{itemize}
\item{An significant saturation of the local heat-release response in the preferential response region is evidenced for low to intermediate forcing frequencies. Such local saturation does not exist for high-frequency modulation.}
\item{The FDF and the FTF phases are nearly identical, at the exception of a faster propagation of the heat-release perturbations along the flame axis in the nonlinear regime for $\st{f}<5$.}
\item{Counterintuitively, the global flame response saturation differs from the local one. The global gain saturation is more pronounced for high forcing frequencies, while it is only slightly marked for low to intermediate $\st{f}$, which directly results from the faster propagation speed at $\st{f}<5$.}
\end{itemize}

\subsection{Higher harmonics generation} \label{seubsec:harmonics}

\begin{figure}[h!]
\centering
\includegraphics[width=0.60\textwidth]{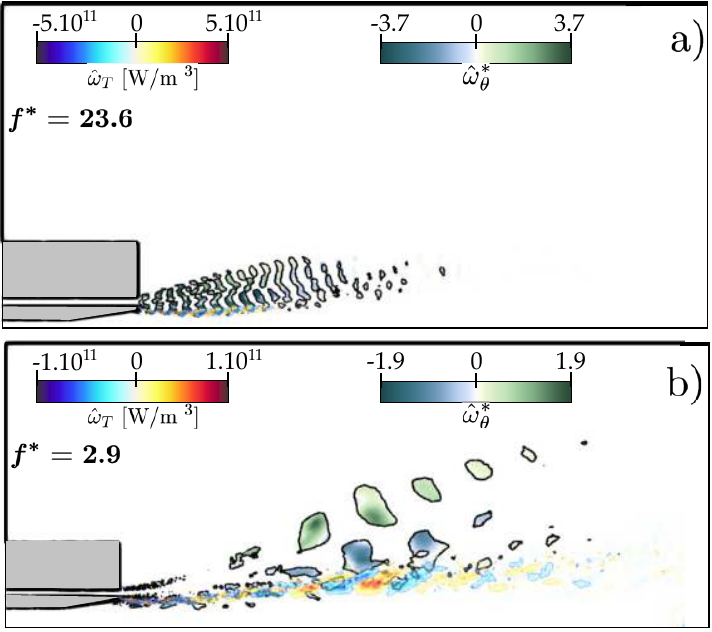}
\caption{(a) Maps of the Fourier coefficients real parts, computed at the 1\textsuperscript{st}-harmonic $\bm{2 \st{f}}$, for a high-frequency forcing ($\bm{f^{*}=23.6}$). Dark lines: contour of $\bm{\hat{Q}}$-criterion, colored by the normalized azimuthal vorticity $\bm{\hat{\omega}_{\theta}^{*} = (\hat{\omega}_{\theta} W_F)/(Au_{CH4}^0)}$. Blue-red: Fourier coefficient of the heat-release rate $\bm{\hat{\omega}_T}$. (b) Same, for a low-frequency forcing ($\bm{f^{*}=2.9}$).}
\label{fig:NL_harmonics_2Dmap_vortices}
\end{figure}

Another nonlinear response feature commonly encountered in flame subjected to strong acoustic perturbations is the generation of higher harmonics, at frequencies multiple of the fundamental forcing frequency $\st{f}$. Because of obvious limitations related to simulations cost and data storage, only the first harmonic at $ 2 \st{f}$ is considered in the present work. The possible occurrence of such harmonic response in the modulated LES is assessed by performing Fourier analyses at $2 \st{f}$. In this section (and only this one), the notation $\hat{v}$ therefore refers to the Fourier coefficient of the variable $v$ computed at the frequency $2 \st{f}$. Figure~\ref{fig:NL_harmonics_2Dmap_vortices} shows maps of the vorticity and heat-release harmonic responses for low-frequency and high-frequency modulations.
In comparison to the fundamental response at $\st{f}$ (see in Fig.~\ref{fig:NL_vortices_fft}), the 1\textsuperscript{st}-harmonic flow response displays a similar pattern comprising pairs of vortex rings in the fuel stream that perturb the reactive layer and therefore result in heat-release fluctuations at $2 \st{f}$. Identical frequency-dependent response regions are revealed, with a near-injector preferential response for a high-frequency modulation and a wider downstream preferential response for low-frequency excitation. Expectedly, in both cases the 1\textsuperscript{st}-harmonic flow structure presents shorter wavelengths that are halved in comparison to the fundamental, indicating the occurrence of a secondary vortex-shedding mode in the fuel stream. Note that the harmonic response relative amplitudes are lower than the fundamental ones: for instance at $\st{f} = 2.9$ the 1\textsuperscript{st}-harmonic vorticity response is roughly halved in comparison to the fundamental, while at $\st{f} = 23.6$ it is approximately only 50\% weaker.\par

\begin{figure}[h!]
\centering
\includegraphics[width=0.99\textwidth]{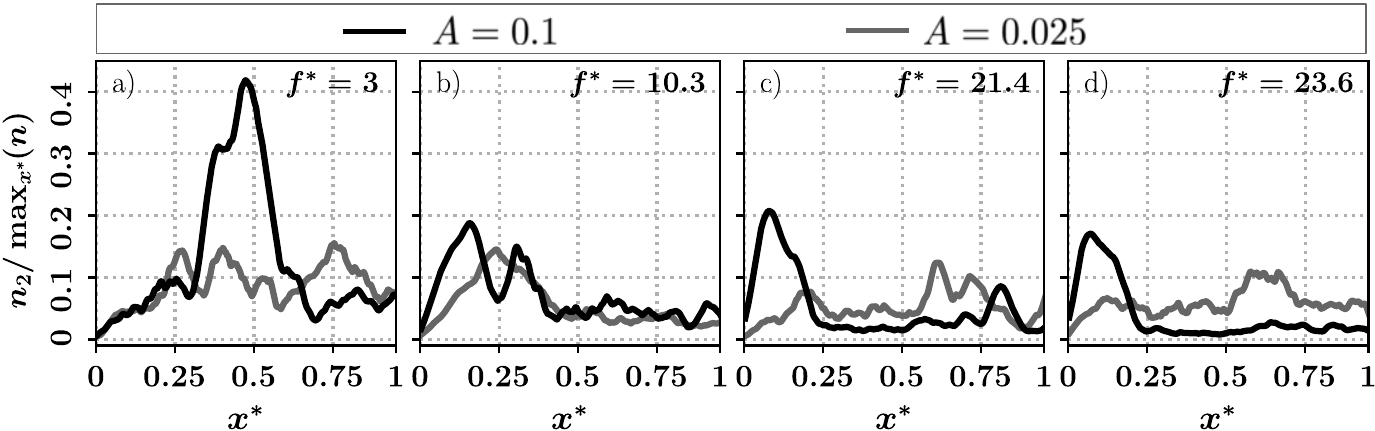}
\caption{Comparison of the 1\textsuperscript{st}-harmonic gain $\bm{n_2}$ with respect to the fundamental gain $\bm{n}$, between the nonlinear ($\bm{A = 0.1}$) and the linear ($\bm{A = 0.025}$) regimes, at a few select forcing frequencies $\bm{\st{f}}$.}
\label{fig:NL_harmonics_comparisons}
\end{figure}

More quantitatively, harmonics generation in the flame nonlinear response can be evaluated by a generalized frequency-domain transfer function. In this matter, Haeringer \textit{et al.}~\cite{haeringer2019} introduced the Extended Flame Describing Function (xFDF), which relies on the observation that if the flame response produces higher harmonics, the forcing velocity is then  likely to also contain those. Thus, the heat-release 1\textsuperscript{st}-harmonic $\hat{q}_2$ can originate from two distinct contributions: (1) the 1\textsuperscript{st}-harmonic FDF that transforms the fundamental of the velocity signal $\hat{u}_1$ into a heat-release oscillation at $2 \st{f}$, and (2) the fundamental FDF that converts the velocity harmonic $\hat{u}_2$ oscillating at $2 \st{f}$ into a heat-release fluctuation also at $2 \st{f}$. Formally, this yields the definition of the 1\textsuperscript{st}-harmonic FDF:
\begin{align}
\label{eq:xFDF_definition}
n_2 (\st{x},\st{f},A) e^{j \varphi_2 (\st{x},\st{f},A) } = \dfrac{L_f \hat{q}_2 (\st{x})}{ Q_0 (\hat{u}_1/ u_{0})} - n (\st{x},\st{f},A) e^{j \varphi (\st{x},\st{f},A) } \left( \dfrac{\hat{u}_2}{\hat{u}_1} \right)
\end{align}
where $n$ and $\varphi$ refer to the fundamental FDF defined in Eq.~\eqref{eq:xFDF_definition}, and $A$ is still the magnitude of the fundamental velocity $\hat{u}_1$ normalized by $u_{CH4}^0$. However in the present case, Fourier analyzes of velocity time-series in the fuel injection line did not reveal the appearance of any harmonic $\hat{u}_2$. The 1\textsuperscript{st}-harmonic FDF can therefore be computed by only retaining the first term in the right-hand side of Eq.~\eqref{eq:xFDF_definition}. In Fig.~\ref{fig:NL_harmonics_comparisons}, the strength of the harmonic flame response is assessed in the linear and nonlinear regimes by comparing the 1\textsuperscript{st}-harmonic gain $n_2 (\st{x}, \st{f})$ to the fundamental gain $n$ in the preferential response region (computed as $\max_{\st{x}} (n)$).
Unsurprisingly, large-amplitude modulations yield significantly stronger harmonic responses than the small-amplitude ones, over most of the frequency range. Note that this point is not true at $\st{f} = 10.3$, where the harmonic response magnitude of both cases are comparable (Fig.~\ref{fig:NL_harmonics_comparisons}-(b)). This is however not due to the generation of the harmonic by nonlinear mechanisms in the flame response, but rather to the presence of the self-sustained vortex-shedding mode at $\st{f}_i=21.4 \approx 2 \st{f}$, as discussed in Sec.~\ref{sec:intrinsic_dynamics} and Sec.~\ref{sec:linear_rg_forced_interaction}. In the nonlinear regime, the 1\textsuperscript{st}-harmonic response is the strongest at low forcing frequencies near the middle of the flame (Fig.~\ref{fig:NL_harmonics_comparisons}-(a)),  where $n_2$ reaches about 40\% of the fundamental response $n$. The 1\textsuperscript{st}-harmonic response strength is lower at higher $\st{f}$, but still of significant importance as it peaks at 20\% of the fundamental gain. The 1\textsuperscript{st}-harmonic propagation speed along the flame axis (computed from the FDF phase $\varphi_2 (\st{x}, \st{f})$) does not noticeably differ from the fundamental one, with values of $1.2 u_{CH4}^0$ at $\st{f} = 23.6$ and $0.7 u_{CH4}^0$ at $\st{f} = 1.5$.

\section{Nonlinear vortex dynamics} \label{sec:nonlinear_vortex}

Section~\ref{sec:linear_rg_vortex_dyn} showed that in the linear regime, the heat-release response is tightly related to the dynamics of vortex-rings pairs that are generated and convected in the CH\textsubscript{4} annular jet. Resemblances in the spatial response patterns of the flame and of the vorticity  observed in Fig.~\ref{fig:NL_vortices_fft} suggest a similar mechanism in the nonlinear regime. This section therefore aims at assessing nonlinear vortical waves dynamics in the fuel stream, with an emphasis on the relation between the respective saturations of the vorticity and flame responses, as well as on the convection speed of the vortex-rings. Note that the generation of vorticity higher harmonics is likely, as evidenced in Fig.~\ref{fig:NL_harmonics_2Dmap_vortices}, but is not discussed in this work. The convection trajectories of the vortices pairs are observed to be identical to those in the linear regime: vortex-rings formed at the outer corner of the CH\textsubscript{4} injector travel along a straight line $L_o$, while those engendered at the inner corner move along the path $L_i$, as illustrated in Fig.~\ref{fig:schematic_convection_vortices}. Nonlinear vortical waves dynamics are here quantified thanks to two frequency-domain Vorticity Describing Function (VDF), which are defined analogous to a FDF, as the generalization of the Vorticity Transfer Function of Eq.~\eqref{eq:vorticity_transfer_function_definition}:
\begin{align}
\label{eq:NL_VDF_definition}
\vert \hat{\omega}_{\theta}^{*} \vert (\st{x}, \st{f}) \ e^{j \varphi_{\omega_{\theta}} (\st{x}, \st{f})} = \dfrac{R(\st{x})}{A u_{CH4}^0}  \hat{\omega}_{\theta} (\st{x}, \st{f})
\end{align}
where the normalization by the typical vortex radius $R(\st{x})$ allows the direct comparison between the strengths associated to vortices of different sizes. A VDF is computed on each trajectory $L_i$ and $L_o$.

\subsection{Vorticity preferential response regions} \label{subsec:NL_vortex_response_region}

\begin{figure}[h!]
\centering
\includegraphics[width=0.70\textwidth]{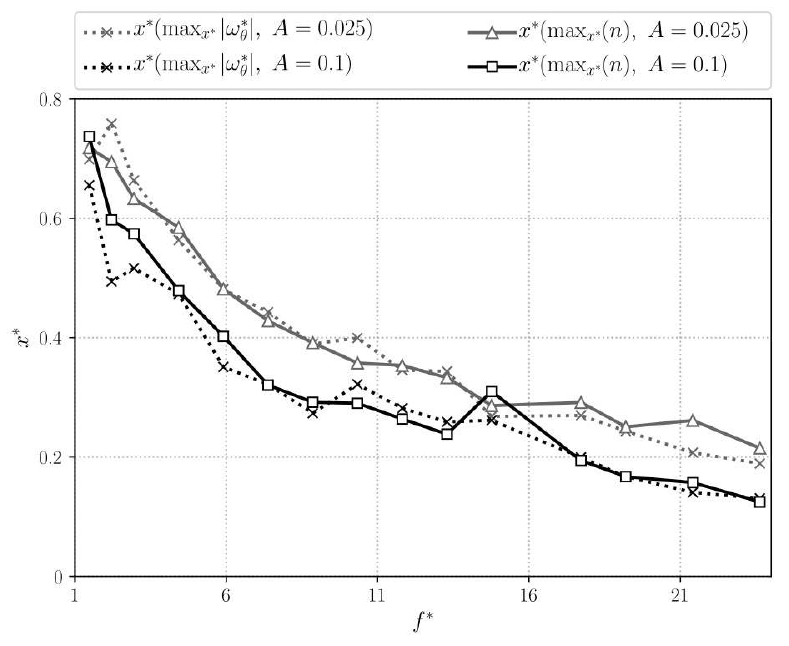}
\caption{Comparison of the axial locations of the heat-release and vorticity preferential response regions in the nonlinear and the linear regimes, as a function of the modulation frequency $\bm{\st{f}}$. The vorticity preferential region is the one of the inner layer $\bm{L_i}$.}
\label{fig:NL_response_region}
\end{figure}

The characteristic shape of the heat-release response region, observed in both linear and nonlinear regimes (Fig.~\ref{fig:NL_3DFDF}-(a)), was shown in Sec.\ref{subsec:vortex_convection} to correspond to a vorticity preferential response region. This latter results from a \textit{loading and firing} mechanism that governs the vortex-shedding dynamics, yielding a formation area located at $\st{x} \sim u_{CH4}^0 / \st{f}$, for small-amplitude modulations. Nonlinear phenomena altering the vortex generation are here estimated in Fig.~\ref{fig:NL_response_region} by comparing the locations of the heat-release and vorticity response regions between the linear and the nonlinear regimes. Those are computed as the locales of the peaks in the FDF gain and VDF gain, respectively.
In both linear and nonlinear cases, the vorticity and the heat-release response regions appear strongly correlated, which confirms the direct influence of the vortex-rings generation on the flame response, even at large modulation amplitudes. Similar vortex generation locations roughly scaling as $\st{x} \sim u_{CH4}^0 / \st{f}$ are observed, with the difference that the response regions in the nonlinear regime are slightly shifted upstream towards the injector. This deviation was already visible on the local FDF in Fig.~\ref{fig:NL_comparison_axial_FTF}. Note that the shifting length $\Delta x$ between the large and the low amplitude cases is the same for the vorticity and the heat-release response regions, and that it seems independent of the forcing frequency, with an approximate value $\Delta x \approx 0.1 L_f$. Such translation is not negligible, since it approaches one-third of the heat-release perturbation wavelength at $\st{f} = 2.9$, and more than two heat-release wavelengths at $\st{f} = 23.6$.

\subsection{Vortex convection}
\label{subsec:NL_vortex_convection}

The nonlinear convection of the vortex-rings along their respective trajectories is estimated thanks to the equivalent propagation speed $\langle v_{\omega_{\theta}}^{*} \rangle (\st{f})$ in the direction of the path $L_i$ or $L_o$. It is computed with the gradient of the VDF phase $\varphi_{\omega_{\theta}} (\st{x}, \st{f})$ based on the definition of Eq.~\eqref{eq:convection_speed_computation} and on the $\vert \hat{\omega}_{\theta}^{*}  \vert$-weighted spatial average detailed in Eq.~\eqref{eq:equivalent_convection_speed}. The coefficient $\alpha \approx \langle v_{\omega_{\theta}} \rangle / \langle \overline{v} \rangle$, defined in Eq.~\eqref{eq:convection_speed_model_0}, is of particular interest to quantify the difference between the true vortex convection speed and the local mean flow velocity. The local VDF gains $\vert \hat{\omega}_{\theta}^{*}  \vert$ are used to measure the vortices strength.\par

\begin{figure}[h!]
\centering
\includegraphics[width=0.90\textwidth]{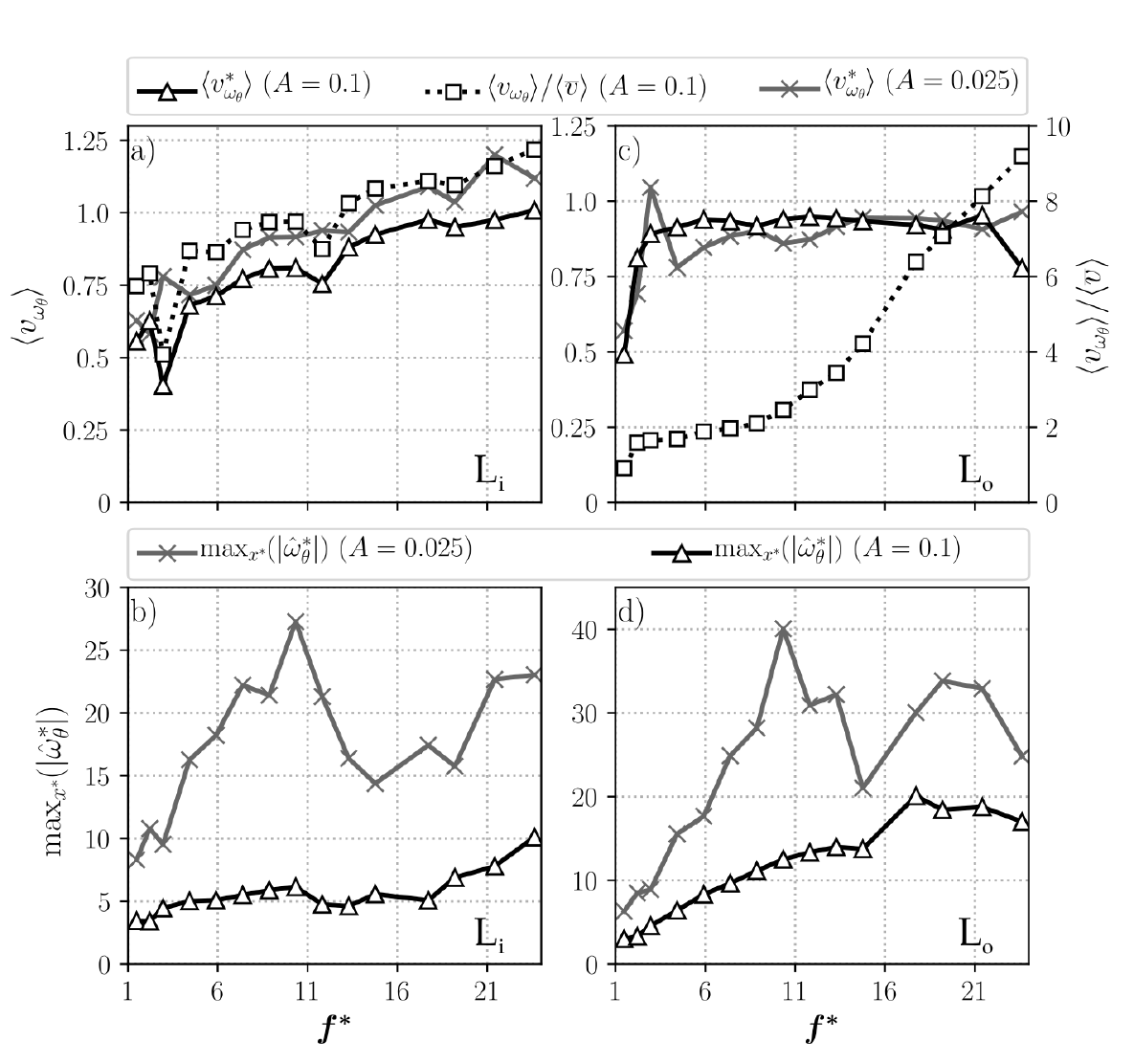}
\caption{Comparison of the vortex responses in the nonlinear and the linear regimes, on both trajectories $\bm{L_i}$ (left column) and $\bm{L_o}$ (right column), as a function of the modulation frequency $\bm{\st{f}}$. (a) and (c): Comparison of the equivalent vortex convection speeds, defined by Eq.~\eqref{eq:equivalent_convection_speed}. (b) and (d): Comparison of the maximal VDF and VTF peaks magnitudes for each forcing frequency.}
\label{fig:NL_vortex_speed_strength}
\end{figure}

Figure~\ref{fig:NL_vortex_speed_strength}-(a,c) compares the linear and nonlinear vortex-rings propagation speeds along $L_i$ and $L_o$.
The inner vortex propagation speed $\langle v_{\omega_{\theta}} \rangle$ presents a comparable growth  in the linear and the nonlinear cases, as it increases by approximately $0.5 u_{CH4}^0$ between $\st{f} = 1.5$ and $\st{f} = 23.6$ (Fig.~\ref{fig:NL_vortex_speed_strength}-(a)). It is however lower for large-amplitude than for small-amplitude modulations, with a deviation approaching $0.1 u_{CH4}^0$ over most of the frequency range. As a result, the nonlinear convection speed never exceeds the bulk injection velocity, whereas it reaches $1.15 u_{CH4}^0$ at high frequencies in the linear case. This slower vortex propagation in the nonlinear regime is likely to be responsible for the upstream translation of the preferential response regions towards the injector evidenced in Fig.~\ref{fig:NL_response_region}. The coefficient $\alpha$ was found to stay close to unity in the linear regime (varying from $0.9$ at low $\st{f}$ to $1.1$ at high $\st{f}$ in Fig.~\ref{fig:vortex_speed_strength}). This sensitivity of the convection speed with respect to the local mean flow was attributed to the vortex inertia that quickly decays after its shedding, under the influence of the intense pseudo-boiling due to the proximity of the flame. In contrast, at larger modulation amplitude the variation of $\alpha$ is more pronounced, since it grows from $0.75$ at $\st{f} = 1.5$ to $1.25$ at $\st{f} = 23.6$. This shows that in the nonlinear regime the vortex-rings motion in the inner layer $L_i$ is less sensitive to the local mean velocity. Such difference is imputed to the true convection speed that is affected not only by the local mean velocity, but also by the local velocity fluctuations induced by the vortex pairs themselves (see the term $u'(x,t)$ in Eq.~\eqref{eq:convection_speed_model_0}). In contrast, the outer vortices convection speed $\langle v_{\omega_{\theta}} \rangle$ is only slightly influenced by the larger amplitude forcing (Fig.~\ref{fig:NL_vortex_speed_strength}-(c)). It indeed remains close to $0.9 u_{CH4}^0$ over most of the frequency range, independently of the region where the vortices travel. Similarly to the linear regime, this low sensitivity to the local mean flow is due the large inertia of the outer vortex-rings, as they trap dense pockets of cryogenic fuel. This invariance of $\langle v_{\omega_{\theta}} \rangle$ combined with the translation of the preferential response region towards the injector results in an increase of $\alpha$ at high $\st{f}$ even more dramatic than in the linear case. Indeed, the outer vortex-rings are in this case forced through the PIRZ (Fig.~\ref{fig:near_injector}) where $\langle \overline{v} \rangle$ is low, which in turn amplifies $\alpha$.\par

The saturation of the vorticity response strength is revealed in Fig.~\ref{fig:NL_vortex_speed_strength}-(b,d), where the magnitude of the peaks in the VDF and VTF gains are compared. Both inner and outer vortex-rings responses are considerably weaker at $A = 0.1$. Remarkably, the resonant peaks that were observed around $\st{f} = 10$ in the linear regime have vanished, and the vorticity response strength only presents a slow, continuous growth with $\st{f}$. In Sec.~\ref{sec:linear_rg_forced_interaction} the apparition of these peaks was imputed to equal vortex convection speeds in $L_i$ and $L_o$ around $\st{f} = 10$, resulting in a synchronization of both vortices, that in turn produces stronger vorticity fluctuations. On the contrary, in the nonlinear regime this synchronization does not occur at $\st{f} \approx 10$ ($\langle v_{\omega_{\theta}} \rangle_{\st{f} = 10} = 0.78 u_{CH4}^0$ in $L_i$ \textit{vs.} $\langle v_{\omega_{\theta}} \rangle_{\st{f} = 10} = 0.9 u_{CH4}^0$ in $L_o$), but rather  near $\st{f} \gtrsim 18$ (Fig.~\ref{fig:NL_vortex_speed_strength}-(a,c)), which corroborates the shifting of the maximal VDF gain towards higher frequencies. Most importantly, this profound modification of the vorticity response intensity is directly related to the heat-release response saturation (Fig.~\ref{fig:NL_comparison_axial_FTF}-(a) to (d)). At low to intermediate frequencies, the heat-release and inner vortex-rings saturations are similar. At $\st{f}=3$, the FDF and VDF gains are roughly halved in comparison to the linear regime. The maximum saturation occurs around  $\st{f}=10$, where both FDF and VDF gains are divided by a factor 5. Some dissimilitudes are however visible at high frequency, since there is no noticeable saturation of the heat-release, whereas the vorticity response on $L_i$ displays a saturation factor of 2 to 3, and that on $L_o$ a saturation factor of 1.5.\par

To conclude, the analysis of the Vorticity Describing Functions computed on the trajectories $L_i$ and $L_o$ leads to three major observations:
\begin{itemize}
\item{The propagation of vortex-rings along $L_i$ is less sensitive to the local mean flow velocity in the nonlinear regime than in the linear one. The true vortex convection speed is believed to be reduced under the effect of the local velocity fluctuations induced by the vortices themselves. On the contrary, the propagation of the outer vortex-rings is not altered.}
\item{A direct consequence of the previous point is a translation of the heat-release and vorticity preferential response regions towards the injector. }
\item{The saturation of the vorticity response is sufficient to explain that of the heat-release for low to intermediate modulation frequencies. Because of the inner-vortex rings lower convection speed, the minimal vorticity saturation occurring when both inner and outer vortices synchronize, is now observed at high frequencies. It is however still noticeable, which cannot explain the absence of saturation of the flame response at high $\st{f}$.}
\end{itemize}

\section{Contributions to HR nonlinear dynamics} \label{sec:nonlinear_contributions}

In this section, the nonlinear dynamics of the distinct contributions to the heat-release response are analyzed to identify potential saturation mechanisms. The reasoning of Sec.~\ref{sec:linear_rg_hr_analysis} based on the flame sheet assumption is repeated. The heat-release fluctuations are decomposed into five distinct contributions, which are recalled here for convenience:
\begin{align}
\label{eq:NL_contrib_split_Fourier}
\dfrac{\hat{q}}{q_0 } = \left[ \dfrac{\hat{\rho}}{\rho_0} + \dfrac{\hat{\mathcal{D}}}{\mathcal{D}_0} + \dfrac{\hat{\Theta}_Z}{\Theta_{Z0}} + \dfrac{\hat{\Psi}_F}{\Psi_{F0}} + \dfrac{\hat{\xi}}{\xi_0} \right]_{Z=Z_{st}}
\end{align}
where $\mathcal{D}$ is the species diffusion coefficient, $\Theta_Z$ represents the mixture fraction gradient, $\Psi_F$ is related to the internal flame structure, and $\xi$ is the flame surface area per unit length (\textit{i.e.} proportional to the local mean flame radius). Once again, Eq.~\eqref{eq:NL_contrib_split_Fourier} is used to define 5 Contribution Describing Functions (CDFs), as amplitude-dependent generalizations of the Contribution Transfer Functions (CTFs) detailed in Eq.~\eqref{eq:contribs_FTF}. For each contribution $v$ of Eq.~\eqref{eq:NL_contrib_split_Fourier}, the associated CDF writes:
\begin{align}
\label{eq:contribs_CDF}
n_v(\st{x},\st{f}) e^{j \varphi_v(\st{x},\st{f})} = \dfrac{L_f q_0 \hat{v} (x^{*})}{Q_0 v_0 \left( \hat{u}/u_0 \right)}
\end{align}
The reader is reminded that this equation is obtained through a linearization, followed by a Fourier transform at the modulation frequency $\st{f}$. This approach may be justified for small-amplitude fluctuations, but not in a nonlinear regime where higher-order terms $\widehat{v_i v_j}$ may be of primary importance. Note however that such terms are likely to produce higher-harmonic oscillations, and since the present section only aims at assessing the saturation mechanisms, Eq.~\eqref{eq:contribs_CDF} is still a reasonable approximation.\par

\begin{figure}[h!]
\centering
\includegraphics[width=0.99\textwidth]{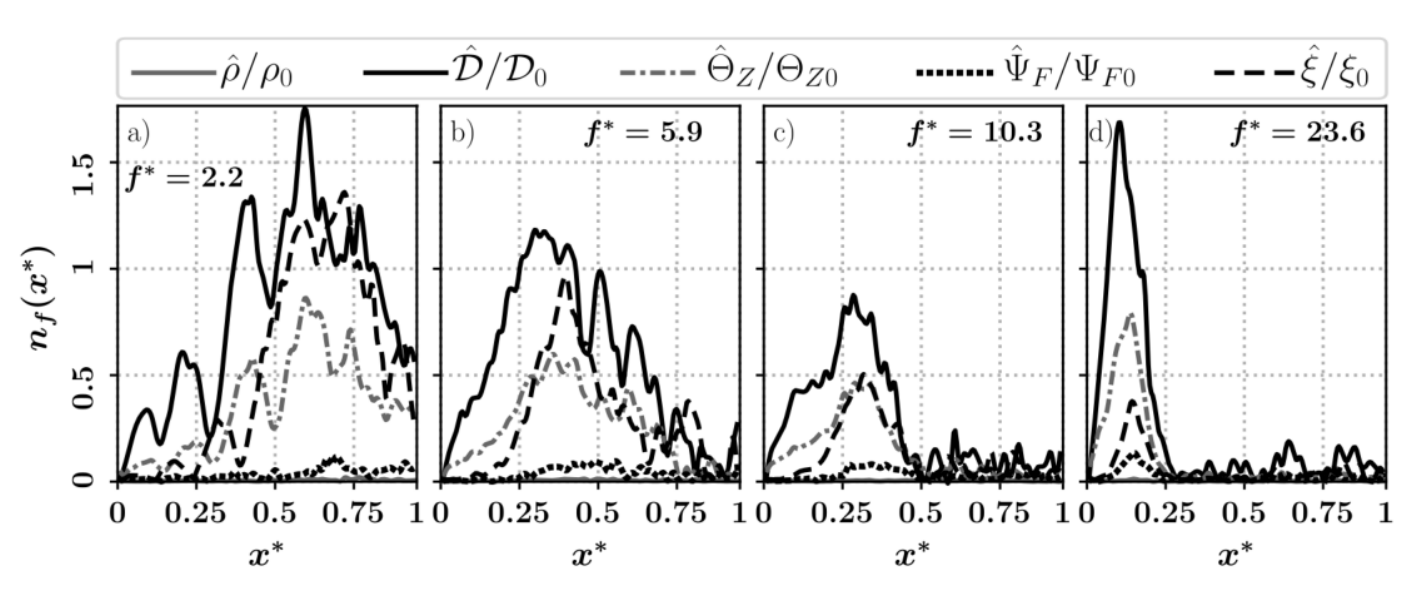}
\caption{CDF gains $\bm{n_v}$ along the flame axis, at 4 forcing frequencies.}
\label{fig:NL_contributions}
\end{figure}

Figure~\ref{fig:NL_contributions} compares the 5 CDF gains along the flame axis for a few forcing frequencies.
The relative weights of the contributions are highly similar to those obtained in the linear regime (see Fig.~\ref{fig:axial_contrib} for comparison). The diffusion coefficient fluctuations are the overall major contributor to the heat-release response. The second most dominant contribution stems from the mixture fraction gradient fluctuations in the near-injector region at high frequency, or from the flame surface area in the second half of the flame at lower frequency. The internal flame structure and the density do not vary sufficiently to significantly contribute to the flame response. The clear similitudes observed between the linear and the nonlinear regimes suggest that the physical mechanisms responsible for these relative contributions are identical to those already described in Sec.~\ref{sec:linear_rg_hr_analysis}.\par

\begin{figure}[h!]
\centering
\includegraphics[width=0.99\textwidth]{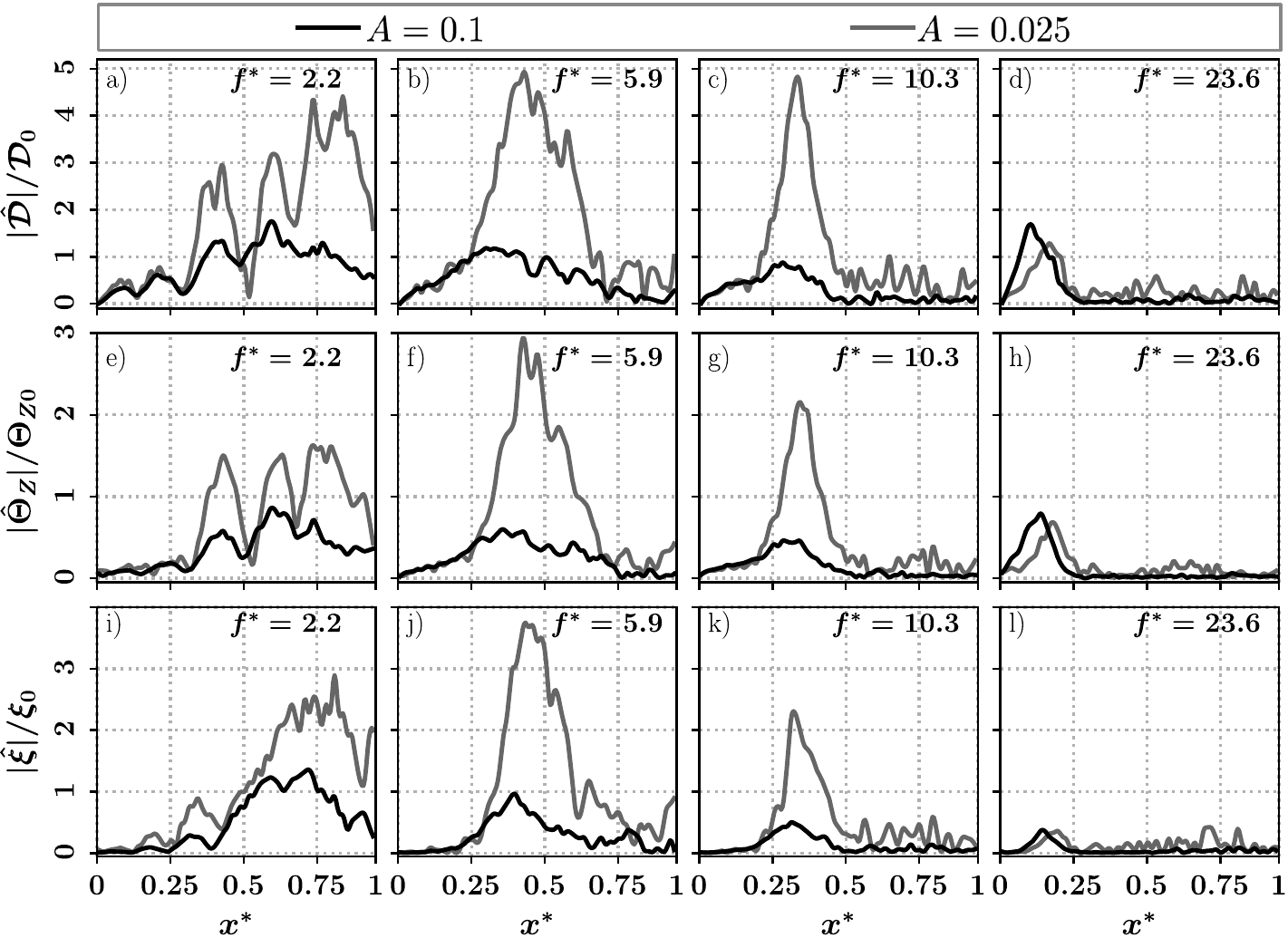}
\caption{Comparison between the CDF and the CTF gains along the flame axis for a few select modulation frequencies $\bm{\st{f}}$. First row: species diffusion coefficient contribution. Second row: mixture fraction gradient contribution. Third row: flame surface area contribution.}
\label{fig:NL_contributions_comparison}
\end{figure}

In order to quantify the saturation of each one of the three dominant contributions ($\mathcal{D}$, $\Theta_Z$, and $\xi$), their respective CDF gains are compared to the CTF gains in Fig.~\ref{fig:NL_contributions_comparison}.
At low frequency (Fig.~\ref{fig:NL_contributions_comparison}-(a,e,i)), the contributions' responses are approximately half as strong in the nonlinear case than in the linear one, which also corresponds to the saturation observed on the FDF gain (Fig.~\ref{fig:NL_comparison_axial_FTF}-(a)). However, the flame saturation is not in this case due to the contributions' saturation, since the vorticity (Fig.~\ref{fig:NL_vortex_speed_strength}-(a)), considered as the origin of the flow disturbances affecting the reaction front, is also twice smaller for $A = 0.1$ than for $A = 0.025$. The same remark is true at intermediate frequency (second and third columns in Fig.~\ref{fig:NL_contributions_comparison}), where the heat-release, its contributions, and the vorticity (Fig.~\ref{fig:NL_vortex_speed_strength}-(b,d)) display a similar saturation factor comprised between 4 and 5.  In contrast, the contributions responses are remarkably different at high frequency (Fig.~\ref{fig:NL_contributions_comparison}-(d,h,l)): the diffusion and the mixture-fraction fluctuations are respectively 50\% and 20\% more intense in the nonlinear than in the linear regime. This trend goes against that visible on the vorticity response, which still saturates at high $\st{f}$. The enhanced contributions of $\mathcal{D}$ and $\Theta_Z$ to the heat-release fluctuations may counterbalance the vortex-rings saturation, and would therefore explain the only slight flame saturation observed at high frequencies in Fig.~\ref{fig:NL_comparison_axial_FTF}-(d).\par

To conclude, the examination  of the Contribution Describing Functions leads to three major observations:
\begin{itemize}
\item{The contributions to the heat-release fluctuations, their relative weights, and their governing mechanisms appear to be similar in the nonlinear and linear regimes.}
\item{The dominant contributors are not responsible for the flame saturation for low to intermediate modulation frequencies, since their respective responses are in line with that of the vorticity.}
\item{Most remarkably, at high frequencies the species diffusion and mixture-fraction gradient variations are more intense in the nonlinear than in the linear regime. They therefore offset the vorticity response saturation, and result in the absence of heat-release saturation.}
\end{itemize}

\section{Conclusion}  \label{sec:nonlinear_mascotte_conclusion}  

A series of LES was performed to assess the nonlinear dynamical response of a doubly-transcritical LO\textsubscript{2}/LCH\textsubscript{4} cryogenic flame to harmonic acoustic modulation imposed at the fuel injector. Particular emphasis was put on extending the linear analysis of Chapter~\ref{chap:mascotte_linear} to the case of large-amplitude forcing. Two nonlinear mechanisms were especially targeted, namely the flame response saturation, and the generation of higher-harmonics. These computations necessitated considerable computational resources, including 15 million CPU hours on a machine equipped with Intel Xeon 8168 processors (2.7GHz).\par

A portion of a non-compact Flame Describing Function was computed at a fixed modulation amplitude, equal to 10\% of the bulk injection velocity, and over a wide frequency range spanning from $\mathcal{O} ( 1 \ \mathrm{kHz}) $ to $\mathcal{O} ( 20 \ \mathrm{kHz}) $. The examination of this FDF evidenced a more pronounced heat-release response saturation for low to intermediate forcing frequencies, in the downstream half of the flame. Conversely, at high frequencies the heat-release preferential response region, located near the injector exit, does not noticeably saturate, but is shifted upstream towards the injector. An analysis of the nonlinear vortex dynamics and of the contributions to the heat-release fluctuations led to a physical interpretation of these results. More precisely, the flame saturation at low to moderate frequencies was attributed to that of the vortex-rings pairs traveling at the edges of the CH\textsubscript{4} annular jet. The pronounced vorticity response saturation at moderate frequencies was in turn imputed to the lower propagation speed of the inner vortices in comparison to the linear regime, thus resulting in a desynchronization of the vortex-rings pair and in a weaker response. The absence of heat-release saturation at high frequency is believed to originate from a moderate vorticity saturation, counterbalanced by enhanced fluctuations of the species diffusion coefficient and of the mixture-fraction gradient.\par

The generation of a first harmonic at twice the modulation frequency was observed in the heat-release and vorticity fields. A portion of a non-compact Extended Flame Describing Function was computed to estimate the strength of this harmonic in comparison to the fundamental. It was shown that the harmonic generation occurs mostly at low to intermediate frequencies near the middle of the flame, where the 1\textsuperscript{st}-harmonic magnitude approaches 40\% of the fundamental.\par

The proposed analyses are expected to be useful to guide future theoretical studies interested in the nonlinear response of LRE coaxial jet flames to acoustic perturbations. In particular, the strong influence of the complex vortex dynamics on the flame response was unveiled, which emphasizes the necessity to develop advanced shedding and convection models for vortex-ring pairs in cryogenic variable-density flows. Finally, the computed FDF and xFDF are made available and can be embedded in thermoacoustic low-order models of multi-injectors LREs.

				
\chapter{Flame-wall interaction effects on the flame root stabilization mechanisms of a doubly-transcritical LO\textsubscript{2}/LCH\textsubscript{4} coaxial jet-flame} \label{chap:fwi_rg}
\minitoc				

\begin{chapabstract}
In the previous chapters, as in the majority of existing studies related to LRE combustion, the boundary condition at the injector lip is supposed adiabatic. This assumption is highly disputable, considering the submillimetric lip thickness and the proximity of both cold cryogenic reactants and hot combustion products. The present chapter therefore aims at thoroughly assessing the effect of Flame-Wall Interaction (FWI) on the anchoring mechanism of a doubly-transcritical LO\textsubscript{2}/LCH\textsubscript{4} jet-flame in the geometry of the Mascotte test rig. It uses high-fidelity quasi-DNS with complex chemistry to study the flame root stabilization mechanisms, and the FWI effects are accounted for by considering the unsteady conjugate heat-transfer problem. It starts by introducing the multi-physics numerical framework employed to resolve such a strongly coupled problem. A simulation with adiabatic boundary conditions is also performed for comparison. Simulation results are then presented. An analysis of the flame root structure provides a detailed insight into its stabilization mechanisms and shows that the large wall heat losses at the lips of the coaxial injector are of primary importance. It is found that adiabatic walls simulations lead to enhanced cryogenic reactants vaporization and mixing, and to a quasi-steady flame, which anchors within the oxidizer stream. Then, the analysis of the non-adiabatic flame dynamics  evidences self-sustained oscillations of both lip temperature and flame root location at similar frequencies. The flame root moves from the CH\textsubscript{4} to the  O\textsubscript{2} streams at approximately $450$Hz, affecting the whole flame structure. This distinctive anchoring mechanism is expected to have first-order effects on the global flame topology and dynamics, including its response to acoustic perturbations. To the knowledge of the author, this is one of the first attempts to simulate a doubly-transcritical shear-coaxial flame including FWI and complex CH\textsubscript{4}/O\textsubscript{2} chemistry. 
\end{chapabstract}

\section{The conjugate heat-transfer problem} \label{sec:fwi_rg_code_coupling}

\begin{figure}[h!]
\centering
\includegraphics[width=0.85\textwidth]{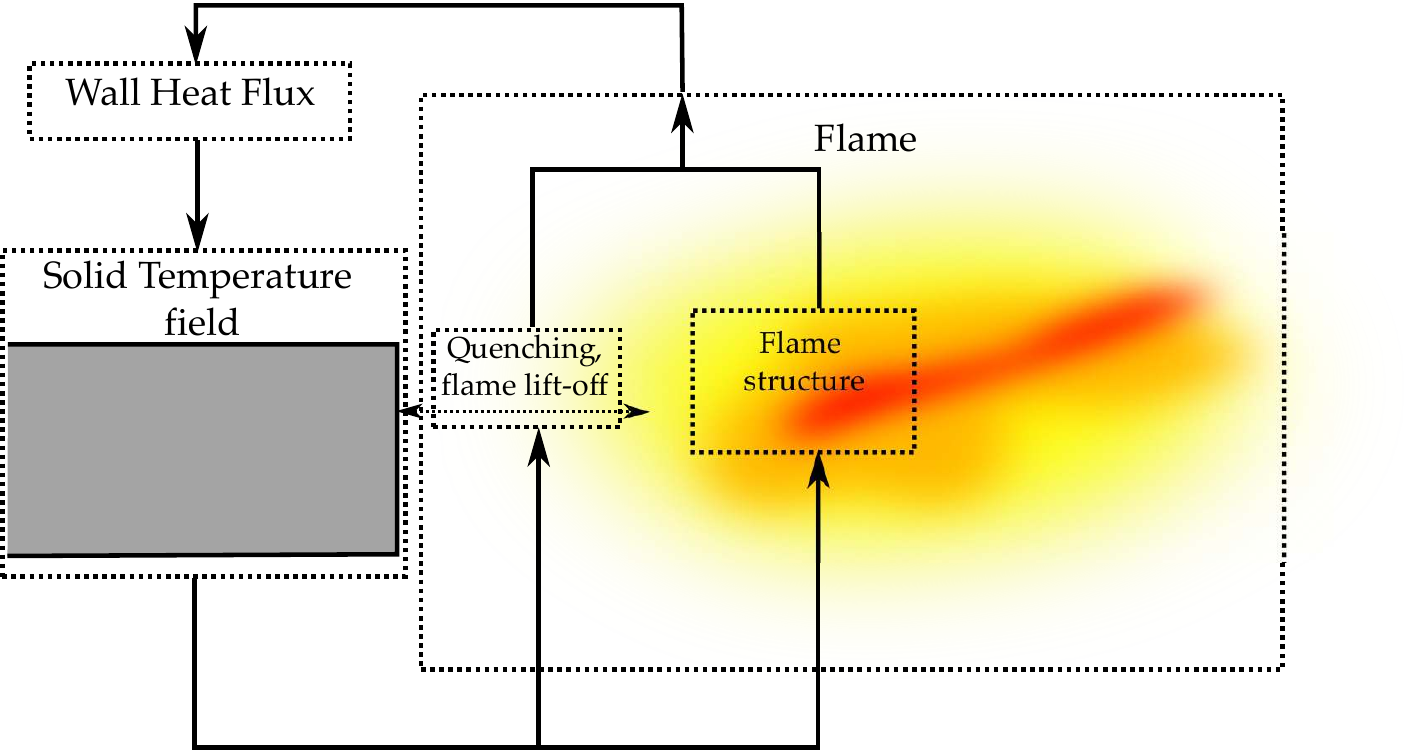}
\caption{Schematic of the conjugate heat transfer problem, or in other words the two-way coupling due to Flame-Wall Interaction.}
\label{fig:FWI_illustr}
\end{figure}

Flame-Wall Interaction (FWI) is known to strongly affect flame anchoring, for instance by inducing flame quenching and lift-off. These effects are in turn of prevalent importance on the flame global topology, on its intrinsic dynamics, as well as on its response to acoustic perturbations~\cite{Mejia:2018}.  Conversely, any modification of the flame features may result in a modification of the heat flux that the flame provides to solid boundaries, which may perturb the temperature field within the wall. This two-way coupling between reactive flow phenomena and heat conduction in the solid constitutes a \textit{conjugate heat transfer problem}, as illustrated in Fig.~\ref{fig:FWI_illustr}.
For most combustion applications, this coupling is rather loose: indeed, the heat diffusion time-scale in the solid can be as large as $\mathcal{O} (1 \ \textrm{s})$, while the flow characteristic time-scale is in $\mathcal{O} (1 \ \textrm{ms})$. The temperature within the wall is then insensitive to fast oscillations in the fluid (but still depends on the mean flow and on slower oscillations), such that only the average solid temperature field is involved in the conjugate heat transfer problem. However, as briefly mentioned in~\cite{oefelein2005}, the dynamic flame-wall coupling is expected to be stronger in the case of a LRE coaxial injector, due to the combination of very high heat fluxes and thin lips. This effect is expected to be even more pronounced in the case of doubly transcritical combustion, since both reactants are injected in a cryogenic state, which will be confirmed later. Note that the numerical modeling of FWI usually requires complex chemistry, due to the relatively cold temperature in the vicinity of the solid boundary. Thus, most previous numerical studies of LRE ccombustion overlooked the FWI problem at the coaxial injector lips. They instead relied on two crude simplifications: (1) adiabatic boundary conditions were assumed at the coaxial injector lip~\cite{oefelein2006,oefelein2005}, and (2) simplified chemistry, tabulated~\cite{Lacaze2012,Schmitt2011} or based on two-step reduced schemes~\cite{schmitt2009}, was employed. In the light of these observations, the present work assesses the FWI influence on the anchoring mechanisms of the doubly-transcritical LO\textsubscript{2}/LCH\textsubscript{4} flame studied in the previous chapters. Most importantly, it considers an unsteady, synchronized conjugate heat transfer problem, where the flame and the lip temperature field may be strongly coupled.\par

\subsection{Numerical setup} \label{sec:numerical_setup}

\begin{figure}[h!]
\centering
\includegraphics[width=0.8\textwidth]{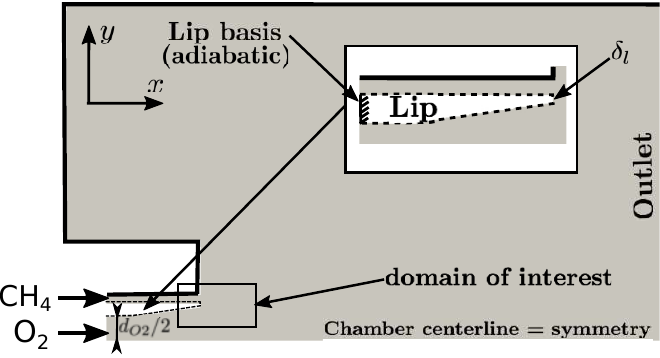}
\caption{The two-dimensional computational domain. Thick dark lines represent no-slip adiabatic walls for the fluid domain. Dashed lines are non-adiabatic no-slip walls were fluid and solid domains are coupled. The basis of the lip (hashed line) is an adiabatic boundary condition for the solid domain.}
\label{fig:geometry_2D}
\end{figure}

As the present study focuses on flame root stabilization, computations are performed on a two-dimensional slice of the Mascotte geometry (see Fig.~\ref{fig:geom_intro}). The computation domain, shown in Fig.~\ref{fig:geometry_2D} is also truncated in the axial direction.
The area of interest, of dimensions $2 d_{O2}\times 1 d_{O2}$, is located near the injector lip of thickness $\delta_{l}$, and consists of cells of dimension 0.03$\delta_{l}$ (\textit{i.e.}~a few $\mu \rm m$). The diffusion flame reactive thickness (of order $100 \ \mu$m for the strain rate levels considered here) implies a high resolution of the front with at least 10-16 cells across the reactive thickness. Injection conditions (recalled in Tab.~\ref{tab:properties_fwi} for convenience) are the same as those employed in the previous chapters: they are similar to the operation point T1 in~\cite{singla2005}. The only difference is that the chamber pressure has been increased to 75~bar (instead of 54~bar in the experimental study), giving higher reduced pressures.
\begin{table}[h]
\centering
\begin{tabular}{c c c c c c}
\multicolumn{1}{c}{ }&\multicolumn{3}{|c||}{ Injection conditions} & \multicolumn{2}{c}{\makecell{Thermal \\ properties}}\\
 \cline{2-6}
\multicolumn{1}{c}{ } \T\B &  \multicolumn{3}{|c||}{$\dot{m}_{inj}/ \dot{m}_{inj}^{O2}$~~~~~~~$T_R^{inj}$~~~~~~~$P_R^{inj}$ } & \multicolumn{2}{c}{$e_f$ (SI)~~~~~~~$\kappa_f$ }\\
 \cline{1-6}
 \multicolumn{1}{c}{O\textsubscript{2} }&  \multicolumn{3}{|c||}{~~~~~$1.0$~~~~~~~~~~~$0.55$~~~~~$1.49$} & \multicolumn{2}{c}{$260$~~~~~~~~~$75.5$ }\\
 \cline{1-6}
 \multicolumn{1}{c}{CH\textsubscript{4} } &  \multicolumn{3}{|c||}{~~~~~$3.59$~~~~~~~~~$0.62$~~~~~~$1.64$} & \multicolumn{2}{c}{$2200$~~~~~~~~~$8.9$ }\\
 \cline{1-6}
 \multicolumn{1}{c}{HG} &  \multicolumn{3}{|c||}{-} & \multicolumn{2}{c}{$80$~~~~~~~~~~~$245$ }\\
 \cline{1-6}
\end{tabular}
\caption{Injection conditions (mass flow rate, reduced pressures and temperatures), and fluid thermal properties (effusivity $\bm{e}$ (SI), and effusivity ratio $\bm{\kappa_f}$) for reactants and hot gases (HG). The global equivalence ratio is $\bm{\phi_g = 14.3}$, the momentum flux ratio is $\bm{J = (\rho_F u_F^2)/(\rho_{O} u_O^2) = 33.3 }$, and the wall effusivity is $\bm{e_w = 1.9 \times 10^{4}}$ (SI). }\label{tab:properties_fwi}
\end{table}
One-seventh power-law velocity profiles are imposed at both reactants inlets through NSCBC boundary conditions~\cite{poinsot1992} adapted for real-gas thermodynamics. All walls are resolved and a no-slip velocity condition is imposed. Only the temperature field in the lip of the central injector is coupled with the fluid domain, such that the only non-adiabatic solid boundaries are the interfaces between the splitter plate and the fluid domain. The strategy to compute thermal quantities at this non-adiabatic boundary is detailed in the following paragraphs.\par

At this point it is interesting to define some quantities relevant to conjugate heat transfer. The fluid effusivity is defined as $e_f = (\rho C_p \lambda)^{1/2}$, with $\lambda$ the thermal conductivity and $C_p$ the specific heat capacity. We also define the fluid effusivity ratio by $\kappa_f = e_w/e_f$. The effusivity allows to evaluate the ability of a body to exchange heat with its surrounding. Thus, for $\kappa_f \gg 1$, temperature at the solid/fluid interface is mostly determined by the wall temperature, and this one can be considered as isothermal. On the contrary, if $\kappa_f \ll 1$ the interface temperature is imposed by the fluid and the wall may be considered as adiabatic. Intermediate values of $\kappa_f$ correspond to boundaries that are neither adiabatic nor isothermal. Thermal properties (Tab.\ref{tab:properties_fwi}) show the dominant thermal influence of the cryogenic CH\textsubscript{4} stream on the lip. None of the interfaces can be considered as adiabatic. Most importantly, the wall conduction characteristic time-scale is $\tau_w = (\delta_{l}^2 \rho C_p/\lambda) = 0.6 \ \rm ms$, which is comparable to that of unsteady motions in the fluid. This very short time scale-scale, that differs from most usual FWI applications, justifies the need for a strongly coupled unsteady conjugate heat transfer problem, where both the solid and the fluid temperatures vary on similar time scales.\par

\subsection{The AVTP solver} \label{sec:AVTP}

Resolving the conjugate heat-transfer problem requires to solve for the unsteady temperature field within the lip (white domain in Fig.~\ref{fig:geometry_2D}). In this work, this is performed thanks to the solid thermal solver AVTP~\cite{Amaya:2010a,Amaya:2010}, developed at CERFACS. Its structure is based on AVBP, and it solves the heat equation on unstructured hybrid meshes:
\begin{align}
\label{eq:heat_equation}
\rho^S C_p^S \pdv{T^S}{t} = \pdv{}{xi} \left( \lambda^S \pdv{T^S}{xi} \right)
\end{align}
where superscripts ${}^S$ denote quantities related to the solid. In Eq.~\eqref{eq:heat_equation} the diffusion term is numerical integrated thanks to the $2 \Delta$ operator~\cite{Lamarque:2007}. The time-integration is here carried out with a 1\textsuperscript{st}-order forward Euler scheme:
\begin{align}
\label{eq:Euler_1st_order}
\pdv{T^S}{t} \approx \dfrac{T_{n+1}^S - T_{n}^S}{dt_S}
\end{align}
where $T_{n+1}^S$ and $T_{n+1}^S$ are the solid temperature fields at the instants $n+1$ and $n$, respectively, and $dt_S$ is the solid time-step. This scheme is stable if the Fourier number $Fo$ is small enough: $Fo = (\lambda^S dt_S ) / ( \Delta_x^2 C_p^S \rho^S ) < 0.5 $, with $\Delta_x$ the cube-root of the cell volume. In the present simulation, the solid time-step is fixed to a constant value, such that $Fo = 0.1$. Classical heat conduction boundary conditions are implemented in AVTP. Those include for instance the isothermal condition ($T_w^S = T_{ref}$), the adiabatic condition ($q_w^S = 0 $), and the heat-loss condition ($q_w^S = q_{ref}$), where $T_w^S$ and $q_w^S$ are the lip temperature and heat-flux at the fluid-solid interface.\par

In the present simulation, the mesh used by the thermal solver is uniform, composed of triangular cells of size $\delta_{l}/120$. A quick evaluation of a characteristic diffusion length in the lip shows that it is considerably shorter than the axial length of the solid domain considered here: the lip basis is therefore close to adiabatic conditions.\par

\subsection{Multi-physics code coupling} \label{sec:multiphysics}

\begin{figure}[h!]
\centering
\includegraphics[width=0.9\textwidth]{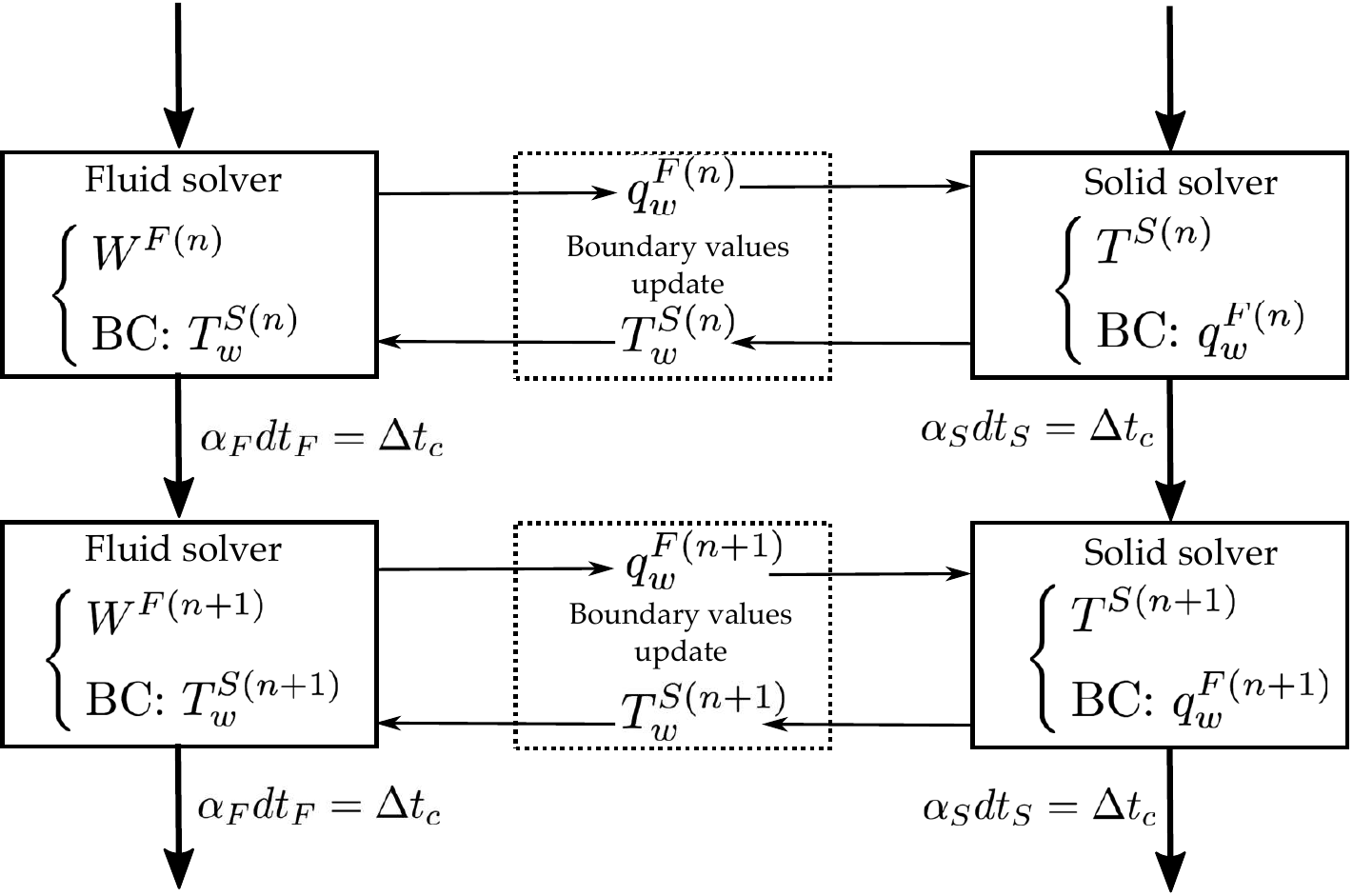}
\caption{Flowchart of the Parallel Coupling Strategy (PCS). Both fluid and solid solvers run in parallel on different set of processors. Boundary data ensuring the physical coupling are exchanged at the fluid/solid interface at meeting points, that are defined thanks to the exchange frequencies $\bm{\alpha_F}$ and $\bm{\alpha_S}$.}
\label{fig:PCS_illustr}
\end{figure}

The most crucial components in the resolution of the conjugate heat transfer problem is the coupling between the fluid solver and the solid thermal solver. In this work, the Parallel Coupling Strategy (PCS) designed by Duchaine \textit{et al.}~\cite{Duchaine:2013}, and based on the OpenPalm library\cite{duchaine2009} is employed. In this approach, depicted in Fig.~\ref{fig:PCS_illustr}, both solvers run in parallel on distinct set of processors. 
Some meeting points are defined and serve to schedule the exchange of boundary data values necessary to enforce a physical coupling. At each meeting point, both solvers update their respective thermal boundary conditions that are then used to continue the temporal integration. There exist different possibilities to define this exchange, however Giles \textit{et al.}~\cite{Giles:1997} proved that the conjugate heat transfer problem is stable only if an isothermal condition is imposed to the fluid domain, and a heat-loss (or Neumann) condition is imposed to the solid domain. Thus, at the coupling event $n$, AVBP-RG and AVTP respectively exchange their   wall heat flux $q_w^{F(n)}$ and their wall temperature $T_w^{S(n)}$. After this coupling event, AVBP-RG computes $\alpha_F$ iterations of constant time-step $dt_F$, with an isothermal boundary at the injector lip defined by $T_w^{S(n)}$. Meanwhile, AVTP computes $\alpha_S$ iterations at constant time-steps $dt_S$ with a heat-loss condition determined by $q_w^{F(n)}$. At the next coupling event $n+1$, both boundary values are updated from the fields that have been integrated, and so forth. Note that since an unsteady and strongly coupled conjugate heat transfer problem is considered here, both solvers need to be \textit{synchronized} in time, that is $\alpha_F dt_F = \alpha_S dt_S$. This is enforced by fixing $dt_S$ to a constant value $dt_S = 0.1 ( \Delta_x^2 C_p^S \rho^S )  / \lambda^S$, and by setting a large exchange frequency with $\alpha_S = 50$ and $\alpha_F=5$; $dt_F$ is then fixed to a constant value $dt_F = dt_S \alpha_S dt_S / \alpha_F$. This approach ensures a fast convergence of the solid solver and avoids any aliasing, but also results in a significant computational cost.

\section{Flame root stabilization mechansims} \label{sec:fwi_rg_stabilization_mechanisms}

Two simulations are performed, the first one (superscript \textsuperscript{C}) using conjugate heat transfer at the injector lip (Fig.~\ref{fig:geometry_2D}) thanks to the coupling strategy described above, and the second one (superscript \textsuperscript{A}) with adiabatic boundary conditions at the lip.
Characteristic quantities of the one-dimensional counterflow diffusion flame shown in Fig.~\ref{fig:counterflow_1} are used to nondimensionalize some of the results: the average volumetric heat release rate $\dot{\Omega}_0 =  3.5 \times 10^{11} \rm \ W/m^{3}$ is used to normalize the heat release computed in the numerical simulations (noted $HR^{*}$), and the heat release per flame surface area $\Phi_0 = 40.4 \ \rm MW/m^{2}$ (obtained by integration over the flame thickness) is used to normalize the wall heat flux (noted $\Phi_{w}^{*}$). Note also that the thickness of this one-dimensional flame in the physical space is $\rm 125 \ \mu m$, which is much larger than the smallest cell size used in the present simulations.\par

\subsection{Flame root structure} \label{sec:FR_structure}

\begin{figure*}[h!]
\centering
\includegraphics[width=0.99\textwidth]{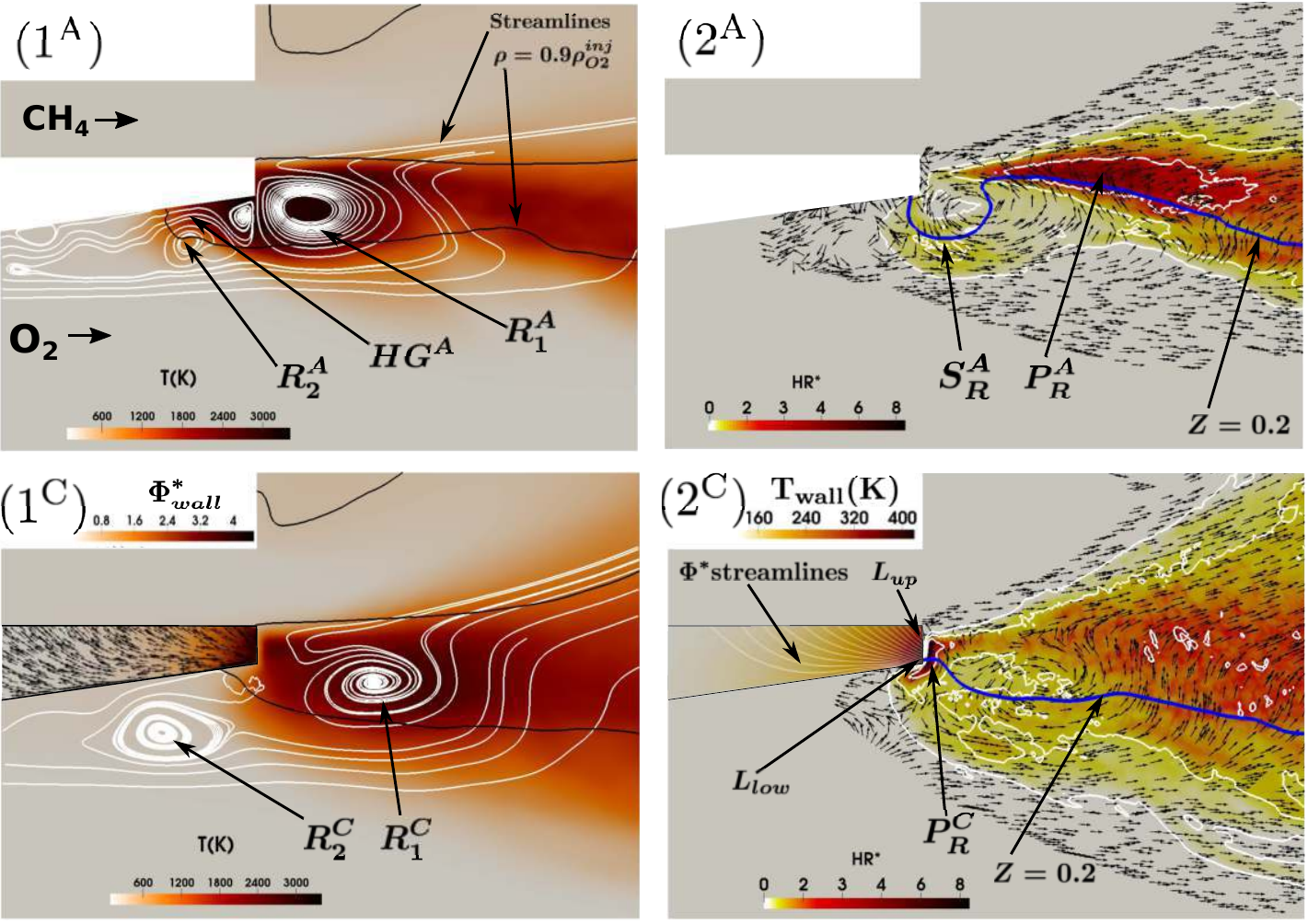}
\caption{(1) Averaged temperature field and velocity streamlines in the fluid. Averaged heat flux magnitude and vector field in the wall. (2) Averaged heat release rate and velocity vector field in the fluid. Averaged temperature and heat flux streamlines in the wall. Superscripts \textsuperscript{A} (resp.~\textsuperscript{C}) denote the simulation with adiabatic conditions at the lip (resp.~the conjugate heat transfer problem).}
\label{fig:ave_maps}
\end{figure*}

Both simulations were run and averaged over 12~ms, which roughly corresponds to 5 convective times relative to a fluid particle in the slow dense oxygen core, or equivalently to 20 characteristic diffusion times in the solid. Averaged fields of heat release rate, fluid temperature, wall temperature, and heat flux are shown in Fig.~\ref{fig:ave_maps}. 

The flame root structure obtained in the adiabatic case presents some distinguishable features. In Fig.~\ref{fig:ave_maps}-(1\textsuperscript{A}) two corotating recirculation zones are visible: a first one ($R_1^A$) is located in the lip's wake, while the second one ($R_2^A$) generated by upstream flow separation lies beneath the lip, in the oxygen injection stream. A smaller vortex appears to be trapped and stretched between $R_1^A$ and $R_2^A$, where a pocket of hot gases ($HG^A$) is trapped underneath the lip. The density isoline $\rho = 0.9\rho_{O2}^{inj}$ allows the visualization of the position of the dense oxygen core, and this one appears to retract under the influence of $HG^A$, which facilitates cryogenic O\textsubscript{2} pseudo-boiling. In Fig.~\ref{fig:ave_maps}-(2\textsuperscript{A}), two distinct reaction zones can be identified: an intense primary reaction zone ($P_R^A$) is located in the lip's wake, at a distance of approximately $1.5\delta_{l}$ from the lip's end, in a region of high shear between the two streams. This stabilization mechanism is similar to that described in~\cite{oefelein2005} for a LO\textsubscript{2}/GH\textsubscript{2} flame. However, unlike the LO\textsubscript{2}/GH\textsubscript{2} flame, the present case also displays a weaker secondary reaction zone ($S_R^A$), situated in the zone of high shear between $R_1^A$ and $R_2^A$. This secondary reaction zone, anchored onto the lip within the oxygen stream, is fueled through the pocket of hot gases $HG^A$, which facilitates cryogenic reactants vaporization, and therefore their mixing. This effect is noticeable as $S_R^A$ lies close to the stoichiometry line $Z=0.2$: gaseous CH\textsubscript{4} invades the dense O\textsubscript{2} stream and is responsible for the anchoring of the secondary flame front beneath the lip. Compared to the LO\textsubscript{2}/GH\textsubscript{2} flame studied in~\cite{oefelein2005}, this stabilization mechanism is probably due to a significantly higher momentum flux ratio $J = (\rho_F u_F^2)/(\rho_{O} u_O^2)$ ($J = 33.3$ here, while $J = 0.24$ in~\cite{oefelein2005}).  The stabilization mechanism in the adiabatic case seems extremely dependent on the formation and the expansion of the pocket of hot gases $HG^A$. The temperatures reached by the lip extremity for this adiabatic case range from approximately $300$~K to $3500$~K, and are not realistic.\par

On the contrary (see Fig.~\ref{fig:ave_maps}-(2\textsuperscript{C})), the fully coupled problem yields a relatively cold time-averaged lip temperature, which does not exceed $450$~K at its extremity. It is also worth noting the large temperature gradient across the lip thickness, varying from $450$~K at the lower corner of the lip ($L_{low}$) to $300$~K at the point $L_{up}$ near its upper corner. The cold lip induces significant heat losses at the wall (up to $\Phi_w^* = 4.5$, see Fig.~\ref{fig:ave_maps}-(1\textsuperscript{C})) and prevents the formation of the region of hot gases $HG^A$ seen for the adiabatic case, thereby changing the entire stabilization mechanism. The two main recirculation zones $R_1^C$ and $R_2^C$ still exist, but are slightly more separated than in the adiabatic case, so that no area of high shear is found at their common frontier. Mixing is also notably affected by wall heat losses: because of lower temperatures at the lip, vaporization of reactants is less efficient, and gaseous CH\textsubscript{4} is not transported within the oxygen stream, as in the adiabatic case. Only one primary region of reaction ($P_R^C$) is observed in the coupled simulation. It is located close to the lip extremity, and extends over a long narrow area spreading from the lower corner $L_{low}$ to the point  $L_{up}$ closer to the upper corner. This suggests periodic oscillations of the flame, as discussed below. A wider downstream flame brush also tends to indicate large amplitude fluttering of the flame front induced by oscillations of the flame root.\par

Detailed flame structures (Fig.~\ref{fig:ave_maps}-(3\textsuperscript{A}) and (3\textsuperscript{C})) in the mixture fraction space show that with adiabatic boundary conditions, the flame mostly burns in a pure diffusion mode, as its structure is very close to that of the one-dimensional flame presented in Fig.~\ref{fig:counterflow_1}.
\begin{figure*}[h!]
\centering
\includegraphics[width=0.99\textwidth]{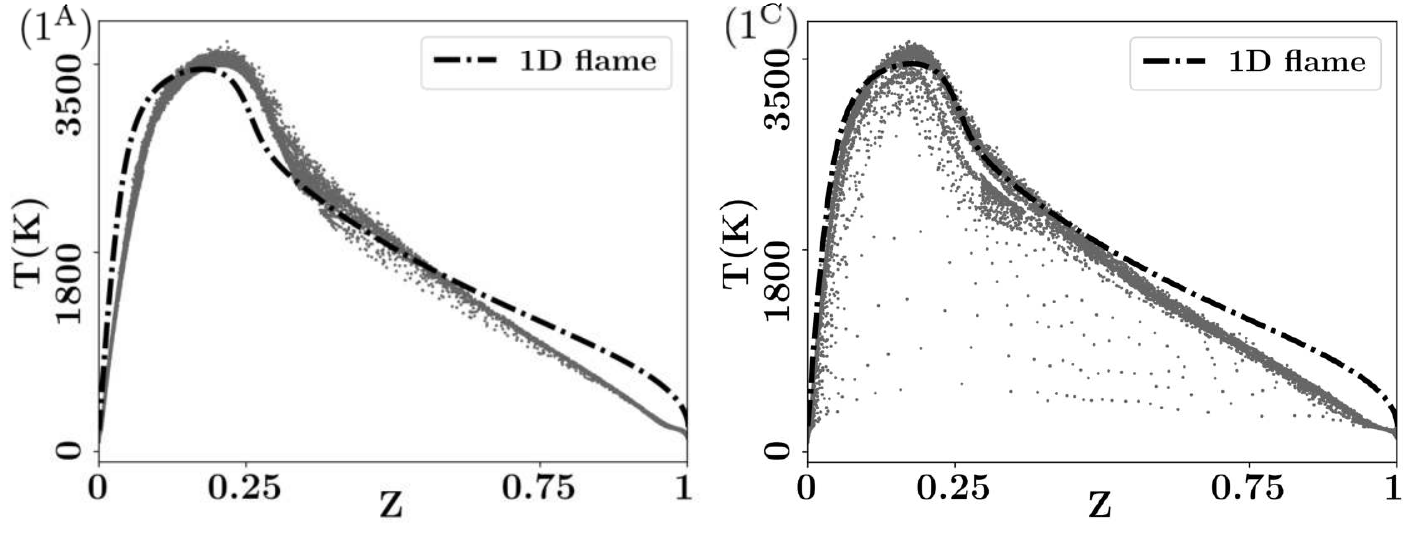}
\caption{Scatter plots of the flame structure in mixture fraction space, compared to the flame displayed in Fig.~\ref{fig:counterflow_1}. Superscripts \textsuperscript{A} (resp.~\textsuperscript{C}) denote the simulation with adiabatic conditions at the lip (resp.~the conjugate heat transfer problem).}
\label{fig:ave_scatter}
\end{figure*}
Conversely, the coupled problem gives a flame that burns in diffusion as well as partially-premixed modes: reactions do not occur close to the wall due to low temperatures, and reactants can mix before reacting.

\subsection{Flame root dynamics} \label{sec:FR_dynamics}

\begin{figure*}[h!]
\centering
\includegraphics[width=0.65\textwidth]{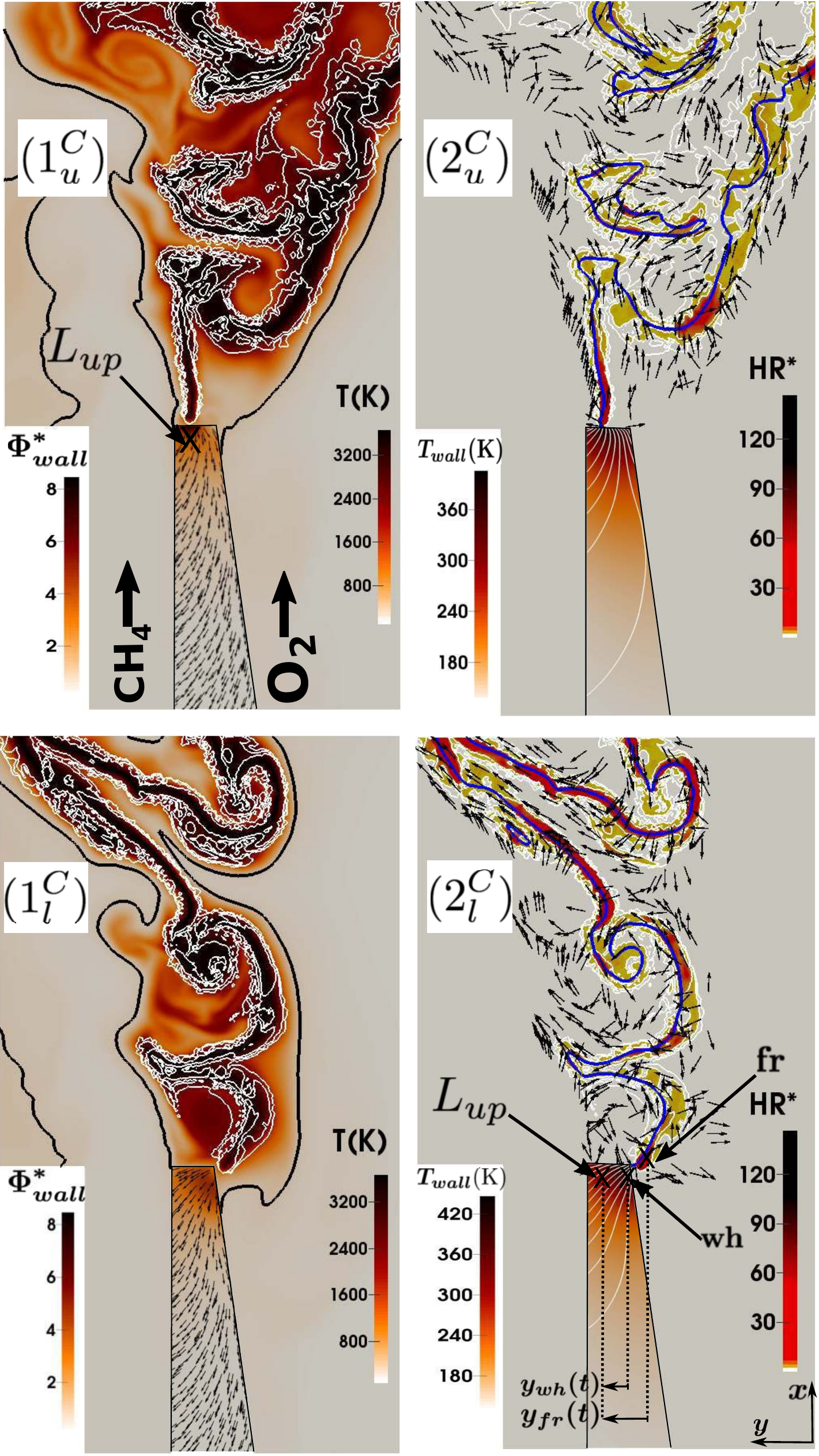}
\caption{(1) Instantaneous temperature field and heat release contours in the fluid. The dark line represents the $\bm{\rho=0.9\rho_{O2}^{inj}}$ isoline. Instantaneous heat flux magnitude and vector field in the wall. (2) Instantaneous heat release rate and velocity vector field in the fluid. The blue line represents the stoichiometry line $\bm{Z=0.2}$. Instantaneous temperature field with heat flux streamlines in the wall. Subscripts \textsubscript{u} (resp.~\textsubscript{l}) denote an instant when the flame root is located at the point $\bm{L_{up}}$ (resp.~ $\bm{L_{low}}$).}
\label{fig:solut_maps}
\end{figure*}

In the adiabatic case, the primary reaction zone $P_R^A$ does not exhibit significant oscillations, and similarly to the LO\textsubscript{2}/GH\textsubscript{2} flame of~\cite{oefelein2005}, it is stabilized in the highly sheared mixing layer between the two reactant streams. The weaker secondary reaction zone $S_R^A$, anchored beneath the lip is more sensitive to perturbations in the cryogenic oxidizer stream, and undergoes slight intermittentcy. The flame root obtained with FWI effects is more unstable and displays strong periodic self-sustained fluctuations. The primary reaction zone $P_R^C$ oscillates between two positions $L_{low}$ located near the lower lip corner, and $L_{up}$ closer to the upper lip corner. Meanwhile, the lip temperature field changes in phase with the flame root motion: at the lower corner $L_{low}$, temperature varies between $345$~K and $505$~K, while it varies between $278$~K and $355$~K at the upper corner. Snapshots of temperature, heat release, wall heat flux, and wall temperature fields at two instants $t_{up}$ and $t_{low}$, corresponding to the two distinct flame root locations, are provided in Fig.~\ref{fig:solut_maps}.
When $P_R^C$ is close to the point $L_{up}$,  the flame is subjected to a strong shear due to the proximity of the high momentum CH\textsubscript{4} stream. As a consequence, the flame root is a relatively straight and unperturbed segment, followed by large scale vortices downstream. At the same time, wall heat losses contribute to rapid heat-up of the upper corner of the lip ($\Phi_w^* = 8.5$), resulting in a relatively homogeneous temperature field across the lip thickness ($T_{Low} = 360$~K and $T_{up} = 340$~K). Conversely, when the main reaction zone $P_R^C$ is located near the lower corner $L_{low}$, shear is weaker due to low velocities in the oxygen stream. The flame root is stretched and rolled up by large scale vortical structures in the mixing layer, and it undergoes a flapping motion. In the meantime, the wall temperature gradient across the lip thickness reaches its maximal value ($T_{Low} = 490$~K and $T_{up} = 280$~K), due to the proximity of the reaction zone to the lip's lower corner.\par

\begin{figure}[h]
\centering
\includegraphics[width=0.999\textwidth]{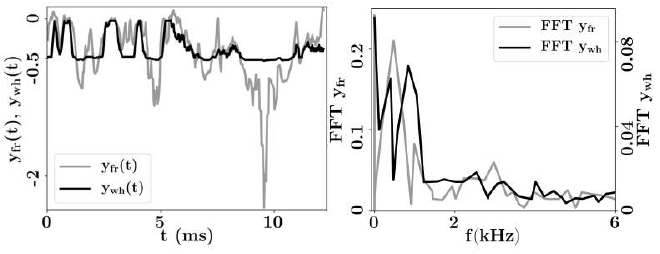}
\caption{Left: time-series of the flame root location $\bm{y_{fr}(t)}$ and the wall hotspot location $\bm{y_{wh}(t)}$. Right: Fourier spectra associated to these two signals. }
\label{fig:signals}
\end{figure}

\begin{figure}[h]
\centering
\includegraphics[width=0.75\textwidth]{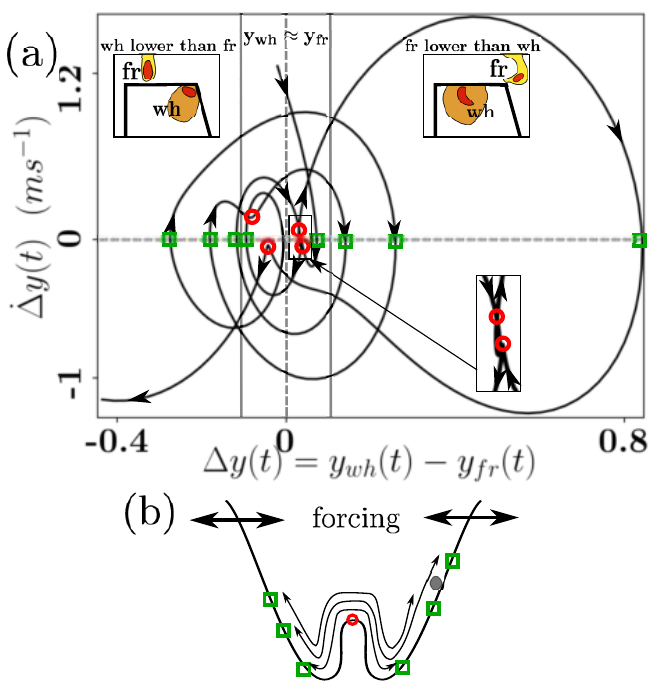}
\caption{(a) Phase-portrait of the coupled system in the plane ($\bm{\Delta y(t) ,\dot{\Delta} y(t)}$). Red circles indicate unstable equilibrium points at cusps in the trajectory. Green squares represent unstable attracting points. (b) Schematic view of an analogous forced double-well oscillator.}
\label{fig:phase_portrait}
\end{figure}

In order to further investigate coupling mechanisms responsible for the self-sustained oscillations affecting simultaneously the flame root and the wall temperature, two quantities of interest are recorded over time. The first one is the radial location $y_{fr}(t)$ of the flame root (noted fr), and the second one is the radial location $y_{wh}(t)$ of the wall hotspot (noted wh), defined as the point of maximum temperature in the lip (Fig.~\ref{fig:solut_maps} right). Both locations are normalized by the lip thickness $\delta_{l}$. Coordinates origin is taken at the point $L_{up}$ indicated in Fig.~\ref{fig:solut_maps}. Temporal evolution and Fourier spectra of the two signals are shown in Fig.~\ref{fig:signals}.
A strong correlation between flame root and wall hotspot fluctuations appears. Both are dominated by two modes, with a first fundamental mode at roughly $450$Hz and its harmonic at approximately $900$Hz. Thus, self-sustained oscillations affecting the flame root seem to result from a complex flame/heat transfer coupling at the wall. Pressure fluctuations were also recorded at several locations in the domain, and as resulting spectra did not exhibit any of the previous frequencies, the influence of acoustics was ruled out as a potential cause of the unsteady flame anchoring. The wall hotspot and flame root coupled dynamics can be further studied by considering the difference $\Delta y(t) = y_{wh}(t) - y_{fr}(t)$ (still normalized by the lip thickness $\delta_{l}$). One-dimensional Dynamic Mode Decomposition~\cite{schmid2010} is used to isolate the most energetic components of the signals $y_{fr}(t)$ and $y_{wh}(t)$, corresponding to the peaks in Fig.~\ref{fig:signals}. The resulting phase-portrait in the plane ($\Delta y(t) ,\dot{\Delta} y(t)$) is displayed in Fig.~\ref{fig:phase_portrait}.
The origin ($0,0$) in the phase-plane represents instants where wall hotspot and flame root face each other and move at the same speed. A quick examination of the phase-portrait allows to rule out the existence of any distinct closed limit-cycle, as the trajectory is chaotic and alternates between small orbits around the origin (\textit{i.e.}~fr and wh are close to each other and have similar motion), and larger orbits (\textit{i.e.}~fr and wh are far from each other). Most importantly, 4 cusps located near the origin are clearly visible on the trajectory showing that $y_{fr} \approx y_{wh}$ is an unstable equilibrium position. After reaching a cusp, the trajectory is sent back to a larger orbit. These dynamics are very similar to that of a particle in a forced double-well oscillator, the central bump representing the unstable equilibrium situation where wall hotspot and flame root face each other. The particle cannot reach exactly the top of the central bump, and slight overshoots in its displacement cause it to fall back on the other side  and oscillate on a larger orbit. Analogously, flame root and wall hotspot are never exactly in phase, and overshoots in the flame displacement cause this equilibrium situation to diverge to large values of $\Delta y$. Thus, the coupled dynamics consists of a combination of small-amplitude oscillations around the equilibrium state $y_{fr} \approx y_{wh}$, and larger-amplitude oscillations with the flame root position $y_{fr}$ and the wall hotspot position $y_{wh}$ decorrelated from each other. Physically, this unsteady coupling can be explained by unsteady heat convection phenomena occurring in the fluid. (1) When the flame is stabilized near the point $L_{up}$ large vortices downstream of the straight flame root bring hot gases close to the lower corner $L_{low}$ (see in Fig.~\ref{fig:solut_maps}). The lip hotspot then moves from the point $L_{up}$ to $L_{low}$, and the flame root follows its motion. (2) Conversely, when the flame is stabilized near $L_{low}$, vortices convect hot gases towards the upper corner, and the reverse mechanism takes place.

\section{Conclusion}  \label{sec:fwi_rg_conclusion}  

Adiabatic boundary conditions are often used in numerical simulations of flame dynamics, especially in LRE conditions where little experimental data is available. As a result, the effects of Flame-Wall Interactions on both the anchoring of the flame root, and on its forced and intrinsic dynamics is a problem that was mostly left unexplored. The present study circumvented the lack of experimentally measured wall temperatures by resolving an unsteady conjugate heat-transfer problem, where the temperature field within the injector lip was strongly coupled to a two-dimensional high-fidelity simulation of the doubly transcritical LO\textsubscript{2}/LCH\textsubscript{4} coaxial flame studied throughout this thesis. The use of the detailed kinetic scheme for CH\textsubscript{4} oxycombustion in LRE conditions, derived in Sec.~\ref{sec:kinetic_mechanism}, was also a key feature of this simulation.\par

Significantly different stabilization mechanisms are found between simulations using adiabatic walls and  the fully coupled computations. More precisely, in the former case, the flame is mainly stabilized in a sheared mixing layer between reactants streams, similarly to  LO\textsubscript{2}/GH\textsubscript{2} flames. However, adiabatic walls appear to facilitate cryogenic reactants vaporization and therefore their mixing. This effect results in the formation of a secondary reaction front within the O\textsubscript{2} injection stream. Conversely, the coupled problem exhibits no stable mixing layer and strong self-sustained oscillations of the flame root and of the temperature field in the lip. Both were found to be bimodal and to oscillate at similar frequencies, indicating a strong influence of heat transfer in the lip on the flame root location. As a result, successive flame root locations were observed (1) in the proximity of the CH\textsubscript{4} stream, where the flame is stabilized by high shear, (2) and in the proximity of the O\textsubscript{2} stream, where the flame root is stretched and rolled-up by vortical structures. The chaotic coupled dynamics of this system can be compared to that of a particle sliding in a forced double-well oscillator. These self-sustained oscillations are expected to be important for the flame response to external forcing, but also for the overall flame stabilization process. Thus, the three-dimensional flame dynamics LES performed in Chapter~\ref{chap:mascotte_linear}~and~\ref{chap:nonlinear_mascotte} would need to be recomputed by adding the resolution of a conjugate heat transfer problem. However, as this task requires immense computational resources, it was deemed unfeasible in the present thesis.\par

Finally, note that this study assumes perfectly smooth injector walls. Although this point was not investigated, flow features taking place at such small scale are expected to be highly sensitive to wall roughness, which could potentially alter the reported flame root stabilization mechanisms.

				

\chapter*{Conclusions and Perspectives} \label{chap:Conclusions}
\addcontentsline{toc}{chapter}{Conclusions and Perspectives}

This PhD thesis is dedicated to the development of methods and models intended to enable the numerical prediction of thermoacoustic instabilities in complex industrial systems, such as Liquid Rocket Engines or gas turbines. A wide variety of computational tools, ranging from cheap low-order models to costly high-fidelity simulations, were utilized and perfected. This work was carried out following two principal directions: the first one consisted in designing more accurate and efficient low-order models for the resolution of unstable thermoacoustic modes in complex combustors, while the second one was centered on the use of high-fidelity Large Eddy Simulations and Direct Numerical Simulations to explore the complex dynamics in an academic high-pressure cryogenic burner, representative of a Liquid Rocket Engine.

\paragraph{Low-Order Modeling of thermoacoustic instabilities in complex systems} \mbox{}\\[-5mm]

The first axis of this PhD thesis lies in the vein of previous studies conducted at CERFACS, most notably by Bauerheim \textit{et al.}~\cite{bauerheim2014,bauerheim2014bis,bauerheim2015,bauerheim2016}. It appeared that usual low-order modeling methods rely on drastic simplifications of the combustors' geometries, which significantly alter their applicability to industrial systems. The present work therefore aimed at improving these existing methods by designing a novel low-order strategy that would allow for the prediction of unstable thermoacoustic modes in combustion systems that are of particular interest for CERFACS industrial partners. This was achieved through the implementation of the code STORM, based on the following fundamental principles:
\begin{itemize}
\item{The number of Degrees of Freedom is reduced as mush as possible, such that a low computational cost is ensured. This was achieved by making use of modal expansions, which generally yield lower-order problems than other strategies commonly encountered.}
\item{The LOM is highly modular and flexible, which eases the straightforward modification of physical or geometrical parameters. In this matter, the LOM code STORM relies on a divide and conquer strategy, where complex combustors are split into acoustic networks of simpler elements, thanks to the state-space formalism.}
\item{Non-rigid-wall frontiers, such as complex impedance boundaries that are ubiquitous in industrial combustors, can be accurately modeled. This particular point was addressed thanks to the frame modal expansion, a generalization of the classical Galerkin expansion, that unlike the latter, is able to precisely resolve the acoustic fields in the vicinity of non-trivial boundaries.}
\item{Geometrically complex boundaries, which usually consist of wide and curved panels, can be accounted for. This is enabled by the combination of the frame expansion with the newly introduced surface modal expansion method. This latter approach is analogous to a Galerkin expansion, in the case of a curvilinear Helmholtz problem formulated in the vicinity of a topologically complex surface.}
\end{itemize}
The major contribution of this thesis to the low-order modeling of thermoacoustic instabilities is undoubtedly the frame expansion method, which constitutes the backbone of STORM. The modularity of flexibility of the approach was further enhanced thanks to the design of an Application Programming Interface (API) intended to generate and manage the complex acoustic networks that STORM can resolve. The implementation of this API was initiated during this PhD thesis, and will be continued in future ones. Even though this work did not present any application of STORM to actual industrial combustors, such results were nonetheless recently obtained on a Safran Aircraft Engines annular combustor. A strong emphasis was however put on empirically assessing the convergence and the precision of the proposed numerical methods. The author believes that the strategies introduced during this PhD have the potential to drastically improve state-of-the-art thermoacoustic low-order models. Future development at CERFACS will include the use of STORM for Uncertainty Quantification, adjoint optimization, robust design, as well as temporal nonlinear simulations of thermoacoustic limit-cycles and other chaotic phenomena.

\paragraph{High-fidelity simulations of a doubly-transcritical LO\textsubscript{2}/LCH\textsubscript{4} coaxial jet-flame} \mbox{}\\[-5mm]

The second research direction followed during this PhD was guided by the urgent need to characterize the doubly-transcritical combustion of methane and oxygen, a regime that is mostly unexplored, but that is nonetheless likely to occur in future reusable Liquid Rocket Engines. In this matter, several high-fidelity simulations of the Mascotte academic test rig, operating in conditions relevant to that of an industrial LRE, were performed. The specificities of these simulations lied in the non-ideal thermodynamics that are resolved, as well as in the use of a complex kinetic scheme that was purposely derived for high-pressure methane oxicombustion. The allocation of considerable computational resources through the PRACE, GENCI, and MeteoFrance infrastructures allowed several facets of doubly-transcritical combustion to be explored:
\begin{itemize}
\item{The first LES of a doubly-transcritical LO\textsubscript{2}/ LCH\textsubscript{4} coaxial jet flame was used to investigate its structure as well as its intrinsic dynamics. These latter displayed several distinctive features never reported before. Most notably, break-up modes of the oxidizer dense core, as well as a vortex-shedding mode in the fuel stream were evidenced.}
\item{The first Flame Transfer Function of a doubly-transcritical LO\textsubscript{2}/LCH\textsubscript{4} coaxial jet flame was computed over a wide frequency range. This was achieved thanks to a series of LES where small-amplitude harmonic modulations were imposed at the fuel inlet to trigger linear flame fluctuations. Frequency-dependent preferential response regions were identified, and were observed to be strongly correlated to the generation of vortex-ring pairs in the methane annular jet. In addition, an analysis of the mechanisms contributing to the heat-release oscillations proved the prevalence of the species diffusivity variations.}
\item{The nonlinear response of a LO\textsubscript{2}/LCH\textsubscript{4} coaxial jet flame was assessed by performing another set of similar LES, at a larger modulation amplitude. The first Flame Describing Function of such flame was computed over a large frequency range. Saturation mechanisms were identified, as well as nonlinear production of higher harmonics.}
\item{Finally, the first DNS of a doubly-transcritical LO\textsubscript{2}/LCH\textsubscript{4} flame utilizing both complex chemistry and a conjugate heat transfer problem at the injector lip was conducted. This high-fidelity simulation aimed at assessing the effects of Flame-Wall Interactions on the flame root anchoring mechanism. Remarkably different stabilization modes were observed between this non-adiabatic flame and an adiabatic one. In particular, the strongly coupled unsteady conjugate heat transfer problem yielded coherent self-sustained  oscillations affecting both the flame root location and the temperature field within the injector lip. Such complex anchoring mechanism is expected to have first order effects on the overall flame topology and on its response to acoustic pulsations, which questions the use of adiabatic boundary conditions in simulations of LRE combustion.}
\end{itemize}
High-fidelity simulations' results discussed in this work are expected to provide a new and deeper insight into the dynamics of LRE cryogenic flames. The proposed analyzes of these results have the potential to guide the future development of theoretical models for the response of such flames to acoustic perturbations. In addition, quantitative transfer functions computed during this PhD thesis can be embedded into low-order acoustic networks for the prediction of thermoacoustic instabilities in multi-injectors LREs.

\newpage

\chapter*{Publications}

\paragraph{Ranked A papers} \mbox{}\\[-5mm]

This PhD thesis led to the following publications:
\begin{enumerate}[leftmargin=5.5mm]
\item{\textbf{Laurent, C.}, Bauerheim, M., Poinsot, T., \& Nicoud, F. (2019). A novel modal expansion method for low-order modeling of thermoacoustic instabilities in complex geometries. \textit{Combustion and Flame}, 206, 334-348.}
\item{\textbf{Laurent, C.}, Badhe, A., Nicoud, F. (2020). Including complex boundary conditions in low-order modeling of thermoacoustic instabilities. (Submitted to \textit{Combustion and Flame})}
\item{\textbf{Laurent, C.}, Esclapez, L., Maestro, D., Staffelbach, G., Cuenot, B., Selle, L., Schmitt, T., Duchaine, F., Poinsot, T. (2019). Flame wall interaction effects on the flame root stabilization mechanisms of a doubly-transcritical LO2/LCH4 cryogenic flame. \textit{Proceedings of the Combustion Institute}, 37(4), 5147-5154. }
\item{\textbf{Laurent, C.}, Staffelbach, G., Nicoud, F., Poinsot, T. (2020). Heat-release dynamics in a doubly-transcritical LO\textsubscript{2}/LCH\textsubscript{4} cryogenic coaxial jet flame subjected to fuel inflow acoustic modulation. (\textit{Proceedings of the Combustion Institute}, Accepted)}
\item{\textbf{Laurent, C.}, Poinsot, T. (2020). Nonlinear dynamics in the response of a doubly-transcritical  LO\textsubscript{2}/LCH\textsubscript{4} coaxial jet flame to fuel inflow acoustic perturbations. (\textit{Combustion and Flame}, Submitted)}
\end{enumerate}

\paragraph{Conferences} \mbox{}\\[-5mm]

\begin{itemize}
\item{Detailed Chemistry effects on the flame root stabilization mechanisms of a doubly-transcritical LO\textsubscript{2}/LCH\textsubscript{4} flame, 4\textsuperscript{eme} Colloque INCA, Paris, October 2017.}
\item{Flame wall interaction effects on the flame root stabilization mechanisms of a doubly-transcritical LO2/LCH4 cryogenic flame, Journ\'{e}e Fran\c{c}ois Lacas, Orl\'{e}ans,  January 2018.}
\item{Modal expansions for accurate thermoacoustic low-order models, CRCT, Toulouse, March 2018.}
\item{Flame wall interaction effects on the flame root stabilization mechanisms of a doubly-transcritical LO2/LCH4 cryogenic flame, 37\textsuperscript{th} International Symposium on Combustion, Dublin, August 2018.}
\end{itemize}

				
\begin{appendices}
\chapter{A library of state-space realizations} \label{Appendix:A}

\section{Acoustic subdomain} \label{sec:ss_realization_subdomain}

The state-space representation for a subdomain $\Omega_i$ belonging to an acoustic network is given in this section. It is assumed that the connection boundary $S_{ci}$ consists of $M_S$ sub-surfaces $(S_{ci}^{(m)})_{1 \leq m \leq M_S}$ of surface area $\Delta S_{ci}^{(m)}$, and located at the center-points $\vec{x}_{sm}$. It is also supposed that these sub-surfaces are small enough, such that acoustic variables can be approximated by their values at the points $\vec{x}_{sm}$. In addition, $\Omega_i$ contains $M_H$ heat-sources $(H_i^{(l)})_{1 \leq l \leq M_H}$. The block structure of the dynamics equation, derived from Eq.~\eqref{eq:press_final_equation_revisited}, writes:

{\small  
\setlength{\abovedisplayskip}{6pt}
\setlength{\belowdisplayskip}{\abovedisplayskip}
\setlength{\abovedisplayshortskip}{0pt}
\setlength{\belowdisplayshortskip}{3pt}
\begin{align}
\label{eq:statespace_subdomain_dyn_1}
\begin{aligned}
&\dfrac{d}{dt}
\underbrace{\begin{pmatrix}
\Gamma_1(t) \\
\dot{\Gamma}_1(t) \\
\vdots \\
\Gamma_{N}(t) \\
\dot{\Gamma}_N(t) \\
\end{pmatrix}}_{\mathbf{X}^{\Omega_i}(t)} 
=
\underbrace{\left(
\begin{array}{ccccc}
0 & 1 & & & \\
-\omega_1^2 & -\delta & & & \\
 & & \ddots &  & \\
 & & & 0 & 1 \\
  & & & -\omega_N^2 & -\delta \\
\end{array}
\right)}_{\mathbf{A}^{\Omega_i}}
\begin{pmatrix}
\Gamma_1(t) \\
\dot{\Gamma}_1(t) \\
\vdots \\
\Gamma_N(t) \\
\dot{\Gamma}_N(t)\\
\end{pmatrix} 
+ \\[5pt]
& 
\underbrace{\left(
\begin{array}{cccccc}
\mathbf{B}_{1}^{S_{ci}^{(1)}} & \hdots & \mathbf{B}_{1}^{S_{ci}^{(M_S)}} & \mathbf{B}_{1}^{H_{i}^{(1)}} & \hdots &  \mathbf{B}_{1}^{H_{i}^{(M_H)}}\\
\vdots & \ddots & \vdots &  \vdots & \ddots & \vdots\\
\mathbf{B}_{N}^{S_{ci}^{(1)}} & \hdots & \mathbf{B}_{N}^{S_{ci}^{(M_S)}} & \mathbf{B}_{N}^{H_{i}^{(1)}} & \hdots & \mathbf{B}_{N}^{H_{i}^{(M_H)}} \\
\end{array}
\right)}_{\mathbf{B}^{\Omega_i}}
\underbrace{\begin{pmatrix}
\mathbf{U}^{S_{ci}^{(1)}} \\
\vdots \\
\mathbf{U}^{S_{ci}^{(M_S)}} \\
Q_1 (t) \\
\vdots \\
Q_{M_H} (t) \\
\end{pmatrix}}_{\mathbf{U}^{\Omega_i}(t)}
\end{aligned}
\end{align}}%
where the blocks $\mathbf{B}_{n}^{S_{ci}^{(m)}}$, $\mathbf{U}^{S_{ci}^{(m)}}$ (used to compute the surface source terms from the boundaries $S_{ci}^{(m)}$), and $\mathbf{B}_{n}^{H_{i}^{(l)}}$ (used to compute the volume source terms from the heat-sources $H_i^{(l)}$) are given by:

{\small  
\setlength{\abovedisplayskip}{6pt}
\setlength{\belowdisplayskip}{\abovedisplayskip}
\setlength{\abovedisplayshortskip}{0pt}
\setlength{\belowdisplayshortskip}{3pt}
\begin{align}
\label{eq:statespace_subdomain_dyn_2}
\begin{aligned}
&  \mathbf{B}_{n}^{S_{ci}^{(m)}} = 
\rho_0 c_0^2 \Delta S_{ci}^{(m)}
\left(
\begin{array}{cc}
0 & 0 \\
-  \left[ \mathbf{\Lambda}^{-1}   \boldsymbol{ \phi} (\vec{x}_{sm})  \right]_n &  \left[ \mathbf{\Lambda}^{-1}    \boldsymbol{ \nabla_s \phi} (\vec{x}_{sm}) \right]_n \\
\end{array}
\right)
\\
& 
\mathbf{B}_{n}^{H_{i}^{(l)}} =
\begin{pmatrix}
0 \\
(\gamma -1) \left[ \mathbf{\Lambda}^{-1} \langle \boldsymbol{\phi} , \mathcal{H}_i^{(l)} \rangle  \right]_n\\
\end{pmatrix}
\ , \ 
\mathbf{U}^{S_{ci}^{(m)}} = 
\begin{pmatrix}
 u_s^{S_{ci}^{(m)}}  (\vec{x}_{sm},t) \\
\varphi^{S_{ci}^{(m)}}  (\vec{x}_{sm},t)\\
\end{pmatrix} 
\end{aligned}
\end{align}}

In this equation, the state vector $\mathbf{X}^{\Omega_i}(t)$ is of size $2N$, where $N$ is the number of eigenmodes used in the modal expansion. The dynamics matrix $\mathbf{A}^{\Omega_i}$ is block-diagonal of size $2N \times 2N$. The first $2 M_S$ columns of the input matrix $\mathbf{B}^{\Omega_i}$ and $2 M_S$ elements of the input vector $\mathbf{U}^{\Omega_i}(t)$ correspond to surface source terms imposed at adjacent boundaries $S_{ci}^{(m)}$. The $M_H$ last columns of $\mathbf{B}^{\Omega_i}$ and the $M_H$ last elements of $\mathbf{U}^{\Omega_i}(t)$ correspond to the volumetric heat release forcing.\par
In the state-space framework that is used here, any subsystem $\Omega_i$ outputs both the normal velocity $u_s(\vec{x}_{sm}) = \vec{u}(\vec{x}_{sm}).\vec{n}_s$ and the pressure $p(\vec{x}_{sm})$ at each one of the $M_S$ surface elements. In addition, it is also possible to incorporate in the output vector (not detailed here) pressure and velocity at any point within $\Omega_i$, such that those can then be passed as reference pressure/velocity to an active flame. Thus, the following equation is used to compute the output vector for the subdomain $\Omega_i$:

{\small  
\setlength{\abovedisplayskip}{6pt}
\setlength{\belowdisplayskip}{\abovedisplayskip}
\setlength{\abovedisplayshortskip}{0pt}
\setlength{\belowdisplayshortskip}{3pt}
\begin{align}
\label{eq:statespace_subdomain_out_1}
\underbrace{\begin{pmatrix}
\mathbf{Y}^{S_{ci}^{(1)}} \\
\vdots \\
\mathbf{Y}^{S_{ci}^{(M_S)}}\\
\end{pmatrix}}_{\mathbf{Y}^{\Omega_i}(t)} 
=
\underbrace{\left(
\begin{array}{ccc}
\mathbf{C}_ {1}^{S_{ci}^{(1)}} & \hdots & \mathbf{C}_{N}^{S_{ci}^{(1)}} \\
\vdots & \ddots & \vdots \\
\mathbf{C}_ {1}^{S_{ci}^{(M_S)}} & \hdots & \mathbf{C}_{N}^{S_{ci}^{(M_S)}} \\
\end{array}
\right)}_{\mathbf{C}^{\Omega_i}}
\underbrace{\begin{pmatrix}
\Gamma_1(t) \\
\dot{\Gamma}_1(t) \\
\vdots \\
\Gamma_N(t) \\
\dot{\Gamma}_N(t) \\
\end{pmatrix}}_{\mathbf{X}^{\Omega_i}(t)}
\end{align}}%
where the blocks $\mathbf{C}_{n}^{S_{ci}^{(m)}}$ and $\mathbf{Y}^{S_{ci}^{(m)}}$ are expressed as:

{\small  
\setlength{\abovedisplayskip}{6pt}
\setlength{\belowdisplayskip}{\abovedisplayskip}
\setlength{\abovedisplayshortskip}{0pt}
\setlength{\belowdisplayshortskip}{3pt}
\begin{align}
\label{eq:statespace_subdomain_out_2}
\mathbf{Y}^{S_{ci}^{(m)}} = 
\begin{pmatrix}
u_s  (\vec{x}_{sm}) \\
 p  (\vec{x}_{sm}) \\
\end{pmatrix}
\ , \ 
\mathbf{C}_{n}^{S_{ci}^{(m)}} = 
\left(
\begin{array}{cc}
-\dfrac{1}{\rho_0}  \nabla_s \phi_n   (\vec{x}_{sm})  & 0 \\
0 &   \phi_n  (\vec{x}_{sm}) \\
\end{array}
\right)
\end{align}}

Note that the feedthrough matrix $\mathbf{D}$ is zero. At the end, the state-space representation of the acoustics in the subdomain $\Omega_i$ is defined by Eq.~\eqref{eq:statespace_subdomain_dyn_1} to Eq.~\eqref{eq:statespace_subdomain_out_2}. It is a $2M_S$-input $2M_S$-output system (in the absence of heat release source terms), whose dynamics are described by a $2N \times 2N$ matrix.

\section{Acoustic subdomain with a complex connection boundary} \label{sec:ss_realization_subdomain_with_complex_boundary}

The state-space representation for the acoustics in a subdomain $\Omega_i$ needs to be adapted to account for complex impedance or conductivity on geometrically elaborate boundaries. For conciseness, $\Omega_i$ is assumed to be adjacent to only one complex boundary $S_{ci}^{(m)}$. Since other types of state-space interconnections (simple \textit{point-wise} boundaries, heat sources, etc.) do not require any adaptation, they are not considered here. The reformulated dynamics equation, formulated thanks to Eq.~\eqref{eq:subdomain_dynamical_system} and Eq.~\eqref{eq:surface_integral_projection}, has the following block structure:

{\small  
\setlength{\abovedisplayskip}{6pt}
\setlength{\belowdisplayskip}{\abovedisplayskip}
\setlength{\abovedisplayshortskip}{0pt}
\setlength{\belowdisplayshortskip}{3pt}
\begin{align}
\label{eq:statespace_subdomain_dyn_rev_1}
\begin{aligned}
&\dfrac{d}{dt}
\underbrace{\begin{pmatrix}
\Gamma_1(t) \\
\dot{\Gamma}_1(t) \\
\vdots \\
\Gamma_{N}(t) \\
\dot{\Gamma}_N(t) \\
\end{pmatrix}}_{\mathbf{X}^{\Omega_i}(t)} 
=
\underbrace{\left(
\begin{array}{ccccc}
0 & 1 & & & \\
-\omega_1^2 & -\alpha & & & \\
 & & \ddots &  & \\
 & & & 0 & 1 \\
  & & & -\omega_N^2 & -\alpha \\
\end{array}
\right)}_{\mathbf{A}^{\Omega_i}}
\begin{pmatrix}
\Gamma_1(t) \\
\dot{\Gamma}_1(t) \\
\vdots \\
\Gamma_N(t) \\
\dot{\Gamma}_N(t)\\
\end{pmatrix} 
+ \\[5pt]
& 
\underbrace{\left(
\begin{array}{ccc}
\mathbf{B}_ {1,1}^{S_{ci}^{(m)}} & \hdots & \mathbf{B}_{1,K_S}^{S_{ci}^{(m)}} \\
\vdots & \ddots & \vdots \\
\mathbf{B}_ {N,1}^{S_{ci}^{(m)}} & \hdots & \mathbf{B}_{N,K_S}^{S_{ci}^{(m)}} \\
\end{array}
\right)}_{\mathbf{B}^{\Omega_i}}
\underbrace{\begin{pmatrix}
\mathbf{U}_1^{S_{ci}^{(m)}} \\
\vdots \\
\mathbf{U}_{K_S}^{S_{ci}^{(m)}} \\
\end{pmatrix}}_{\mathbf{U}^{\Omega_i}(t)} \\
\end{aligned}
\end{align}}%
where the blocks $\mathbf{B}_{n,k}^{S_{ci}^{(m)}}$ and $\mathbf{U}_k^{S_{ci}^{(m)}}$, used to compute the surface forcing from the complex boundary $S_{ci}^{(m)}$, are given by:

{\small  
\setlength{\abovedisplayskip}{6pt}
\setlength{\belowdisplayskip}{\abovedisplayskip}
\setlength{\abovedisplayshortskip}{0pt}
\setlength{\belowdisplayshortskip}{3pt}
\begin{align}
\label{eq:statespace_subdomain_dyn_rev_2}
\begin{aligned}
&  \mathbf{B}_{n,k}^{S_{ci}^{(m)}} = 
\rho_0 c_0^2
\left(
\begin{array}{cc}
0 & 0 \\
- \mathpzc{s}_i  \left[ \mathbf{\Lambda}^{-1}   \left( \boldsymbol{ \phi} | \mathscr{K}_k^{(m)} \right) \right]_n &  \left[ \mathbf{\Lambda}^{-1}   \left( \boldsymbol{ \nabla_s \phi} | \mathscr{K}_k^{(m)} \right) \right]_n \\
\end{array}
\right)\\
& 
\mathbf{U}_k^{S_{ci}^{(m)}} = \dfrac{1}{\lambda_k} \begin{pmatrix}
\mathpzc{s}_i  \left( u_s^{S_{ci}^{(m)}}  | \mathscr{K}_k  \right) \\
\left( \varphi^{S_{ci}^{(m)}} |  \mathscr{K}_k \right)
\end{pmatrix}
= \begin{pmatrix}
\mu_k (t) \\
\nu_k (t)
\end{pmatrix} 
\end{aligned}
\end{align}}
where $\mathpzc{s}_i = \vec{u}_s^{\Omega_i} . \vec{e}_{\xi} = \pm 1$ defines the orientation of the subdomain surface normal with respect to the complex boundary $S_{ci}^{(m)}$. The state vector $\mathbf{X}^{\Omega_i}(t)$ is of size $2N$, with $N$ the number of eigenmodes used in the subdomain frame expansion. The input matrix $\mathbf{B}^{\Omega_i}$ ($2N \times 2 K_S$) serves to compute the surface source terms from the $2 K_S$ elements of the input vector $\mathbf{U}^{\Omega_i}$, which comprises the surface modal amplitudes of both the normal velocity and the acoustic potential from the complex boundary $S_{ci}^{(m)}$.\par

The subsystem $\Omega_i$ outputs both the projections of its normal velocity $u_s(\vec{x}_s)$ and  pressure $p (\vec{x}_s)$ onto each one of the $K_S$ surface modes of $S_{ci}^{(m)}$. This is achieved thanks to the output equation:

{\small  
\setlength{\abovedisplayskip}{6pt}
\setlength{\belowdisplayskip}{\abovedisplayskip}
\setlength{\abovedisplayshortskip}{0pt}
\setlength{\belowdisplayshortskip}{3pt}
\begin{align}
\label{eq:statespace_subdomain_out_rev_1}
\underbrace{\begin{pmatrix}
\mathbf{Y}_1^{S_{ci}^{(m)}} \\
\vdots \\
\mathbf{Y}_{K_S}^{S_{ci}^{(m)}}\\
\end{pmatrix}}_{\mathbf{Y}^{\Omega_i}(t)} 
=
\underbrace{\left(
\begin{array}{ccc}
\mathbf{C}_ {1,1}^{S_{ci}^{(m)}} & \hdots & \mathbf{C}_{1,N}^{S_{ci}^{(m)}} \\
\vdots & \ddots & \vdots \\
\mathbf{C}_ {K_S,1}^{S_{ci}^{(m)}} & \hdots & \mathbf{C}_{K_S,N}^{S_{ci}^{(m)}} \\
\end{array}
\right)}_{\mathbf{C}^{\Omega_i}}
\underbrace{\begin{pmatrix}
\Gamma_1(t) \\
\dot{\Gamma}_1(t) \\
\vdots \\
\Gamma_N(t) \\
\dot{\Gamma}_N(t) \\
\end{pmatrix}}_{\mathbf{X}^{\Omega_i}(t)}
\end{align}}%

where the blocks $\mathbf{C}_{k,n}^{S_{ci}^{(m)}}$ and $\mathbf{Y}_k^{S_{ci}^{(m)}}$ are expressed as:

{\small  
\setlength{\abovedisplayskip}{6pt}
\setlength{\belowdisplayskip}{\abovedisplayskip}
\setlength{\abovedisplayshortskip}{0pt}
\setlength{\belowdisplayshortskip}{3pt}
\begin{align}
\label{eq:statespace_subdomain_out_rev_2}
\mathbf{Y}_k^{S_{ci}^{(m)}} =  \mathpzc{s}_i \begin{pmatrix}
  \left( u_s  | \mathscr{K}_k  \right) \\
\left( p |  \mathscr{K}_k \right)
\end{pmatrix}
\ , \ 
\mathbf{C}_{k,n}^{S_{ci}^{(m)}} = 
\mathpzc{s}_i \left(
\begin{array}{cc}
-\dfrac{1}{\rho_0} \left( \nabla_s \phi_n  | \mathscr{K}_k  \right)  & 0 \\
0 &  \left( \phi_n  |  \mathscr{K}_k \right) \\
\end{array}
\right)
\end{align}}

Finally, the state-space realization of the subdomain $\Omega_i$ with one adjacent complex boundary $S_{ci}^{(m)}$ is a 2$K_S$-inputs-2$K_S$-outputs systems, whose intrinsic dynamics are governed by a $2N \times 2N$ matrix.

\section{Cross-section change between two long ducts} \label{sec:ss_realization_area_jump}

In this section, the state-space representation for an acoustically compact subdomain $\Omega_{sc}$ enclosing a cross-section change between two long ducts $\Omega_1$ and $\Omega_2$ of respective cross-section area $S_1$ and $S_2$ is derived. Although the acoustic field can be considered purely longitudinal in both ducts $\Omega_1$ and $\Omega_2$, it is locally multidimensional in the neighborhood of the section change. The subsystem $\Omega_{sc}$ is therefore defined such that it encloses the region where the acoustic flow is multidimensional. In practice, this translates into $L_{sc} \simeq \max(\sqrt{D_1},\sqrt{D_2})$. However, since the cross-section change is assumed acoustically compact (since $L_{sc} \ll L_1,L_2$), the detailed topological structure of the acoustic flow in $\Omega_{sc}$ has very little effect on the global eigenfrequencies and eigenmodes of the whole geometry. For this reason, the acoustic pressure in $\Omega_{sc}$ is not expanded onto an eigenmodes family, but instead a specific treatment relying on a more global description is applied. The volume-averaged pressure $\overline{p}(t)$ and velocity $\overline{u}(t)$ in $\Omega_{sc}$ are introduced. Volume-averaging the linearized Euler equations in $\Omega_{sc}$ yields (acoustic losses due to vorticity conversion being neglected):
\begin{align}
\label{eq:1Dduct_section_change_1}
\left\{
\begin{aligned}
&\dfrac{d \overline{u}}{dt} = \dfrac{1}{\rho_0 L_{sc}}(p^{\Omega_1}(t) - p^{\Omega_2}(t)) \\
& \dfrac{d \overline{p}}{dt} = \dfrac{2 \rho_0 c_0^2}{L_{sc}(S_1+S_2)} (S_1 u_s^{\Omega_1}(t) + S_2 u_s^{\Omega_2}(t))
\end{aligned}
\right.
\end{align}
where $p^{\Omega_1}(t)$ and $u_s^{\Omega_1}(t)$ (resp.~$p^{\Omega_2}(t)$ and $u_s^{\Omega_2}(t)$) are the pressure and the normal acoustic velocity imposed by the subdomain $\Omega_1$ (resp.~$\Omega_2$) onto $\Omega_{sc}$. The + sign in the second equation is due to the fact that the outer normals at $S_{c1}$ and $S_{c2}$ are defined with respect to $\Omega_1$ and $\Omega_2$ respectively, and are therefore pointing in opposite directions. For low frequencies, these conservation equations reduce to the classical quasi-static jump relations $p^{\Omega_1} = p^{\Omega_2}$ and $S_1 u_s^{\Omega_1} = - S_2 u_s^{\Omega_2}$. Pressures and normal velocities imposed by $\Omega_ {sc}$ onto the two ducts $\Omega_1$ and $\Omega_2$ can then be calculated thanks to 0\textsuperscript{th}-order approximations (\textit{i.e.} piece-wise constant functions):
\begin{align}
\label{eq:1Dduct_section_change_2}
\left\{
\begin{aligned}
& u_s(x_{sc}=0,t) = \dfrac{1}{2}(1+\dfrac{S_2}{S_1})\overline{u}(t) \ , \ p(x_{sc}=0,t) = \overline{p}(t) \\
& u_s(x_{sc}=L_{sc},t) = -\dfrac{1}{2}(1+\dfrac{S_1}{S_2})\overline{u}(t) \ , \ p(x_{sc}=L_{sc},t) = \overline{p}(t)
\end{aligned}
\right.
\end{align}

Utilizing these relations, the state-space realization of the subdomain $\Omega_{sc}$ then writes:

{\small  
\setlength{\abovedisplayskip}{6pt}
\setlength{\belowdisplayskip}{\abovedisplayskip}
\setlength{\abovedisplayshortskip}{0pt}
\setlength{\belowdisplayshortskip}{3pt}
\begin{align}
\label{eq:1Dduct_state_space_1}
\begin{aligned}
\dfrac{d}{dt} &
\underbrace{
\begin{pmatrix}
\overline{u}(t) \\
\overline{p}(t) \\
\overline{\varphi}(t)
\end{pmatrix}}_{\mathbf{X}^{(sc)}(t)}  =
\underbrace{
\begin{pmatrix}
0 & 0 & 0 \\
0 & 0 & 0 \\
0 & -1/ \rho_0 & 0
\end{pmatrix}}_{\mathbf{A}^{(sc)}}
\begin{pmatrix}
\overline{u}(t) \\
\overline{p}(t) \\
\overline{\varphi}(t)
\end{pmatrix} \\
&+ \underbrace{\begin{pmatrix}
0 & \dfrac{1}{\rho_0 L_{sc}} & 0 & -\dfrac{1}{\rho_0 L_{sc}}\\
\dfrac{2 S_1 c_0^2 \rho_0}{L_{sc}(S_1+S_2)} & 0 & \dfrac{2 S_2 c_0^2 \rho_0}{L_{sc}(S_1+S_2)} & 0 \\
0 & 0 & 0 & 0
\end{pmatrix}}_{\mathbf{B}^{(sc)}}
\underbrace{\begin{pmatrix}
u_s^{\Omega_1}(t) \\
p^{\Omega_1}(t) \\
u_s^{\Omega_2}(t) \\
p^{\Omega_2}(t)
\end{pmatrix}}_{\mathbf{U}^{(sc)}(t)}
\end{aligned}
\end{align}}%
The state-vector $\mathbf{X}^{(sc)}(t)$ contains the volume-averaged velocity $\overline{u}(t)$, pressure $\overline{p}(t)$, and acoustic potential $\overline{\varphi}(t) = - \int \overline{p} (t') dt' / \rho_0$. The first two components of the input vector $\mathbf{U}^{(sc)}(t)$ are imposed by the duct $\Omega_1$, while the last two are imposed by the duct $\Omega_2$. We then note $\mathbf{B}_1^{(sc)}$ (resp.~$\mathbf{B}_2^{(sc)}$) the matrix consisting of the first two columns (resp.~last two columns) of $\mathbf{B}^{(sc)}$. In Eq.~\eqref{eq:1Dduct_state_space_1}, the first two lines essentially impose acoustic momentum and acoustic volume flux conservation. For low frequencies, these conservation relations reduce to the classical quasi-static jump relations $p^{\Omega_1} = p^{\Omega_2}$ and $S_1 u_s^{\Omega_1} = - S_2 u_s^{\Omega_2}$. The third equation of the dynamical system is a time-integrator that facilitates the computation of the output vector, as the two ducts $\Omega_1$ and $\Omega_2$ require normal acoustic velocity and acoustic potential as inputs. Outputs for the state-space representation of $\Omega_{sc}$ are computed thanks to first-order approximations:
\begin{align}
\label{eq:1Dduct_state_space_2}
\underbrace{
\begin{pmatrix}
u_s(0,t)\\
\varphi(0,t) \\
u_s(L_{sc},t) \\
\varphi(L_{sc},t)
\end{pmatrix}}_{\mathbf{Y}^{(sc)}(t)} & =
\underbrace{
\begin{pmatrix}
\dfrac{1}{2}(1+\dfrac{S_2}{S_1}) & 0 & 0 \\
0 & 0 & 1 \\
-\dfrac{1}{2}(1+\dfrac{S_1}{S_2}) & 0 & 0 \\
0 & 0 & 1
\end{pmatrix}}_{\mathbf{C}^{(sc)}}
\begin{pmatrix}
\overline{u}(t) \\
\overline{p}(t) \\
\overline{\varphi}(t)
\end{pmatrix}
\end{align}
In Eq.~\eqref{eq:1Dduct_state_space_2}, the feedthrough matrix $\mathbf{D}$ is zero. The first two lines are outputs that are to be imposed to the first duct $\Omega_1$, while the last two lines are outputs that are to be imposed to the second duct $\Omega_2$. We then note $\mathbf{C}_1^{(sc)}$ (resp.~$\mathbf{C}_2^{(sc)}$) the matrix formed with the first two rows (resp.~last two rows) of $\mathbf{C}^{(sc)}$.\par

The specific treatment applied to the subdomain $\Omega_{sc}$ enclosing the neighboring region of the section change allows to derive a generic state-space representation for acoustically compact cross-section variations, independently of the exact geometry of this region. Indeed the use of volume-average acoustic variables and 0\textsuperscript{th}-order approximations, permits to avoid dealing with the topological complexity of the multidimensional acoustic flow in $\Omega_{sc}$. It is also worth noting that even though the derivation of the state-space representation for $\Omega_{sc}$ was here conducted in the absence of acoustic sources, and without conversion from acoustic to entropy or vorticity waves in the neighborhood of the section change, these assumptions could be omitted to obtain more general representations.\par

\section{Cross-section change between a long duct a large cavity} \label{sec:ss_realization_area_jump_1D3D}

In the case of a small domain of thickness $L_{sc}$ enclosing a cross-section change $\Omega_{sc}$ located between a long duct $\Omega_1$ giving on a large cavity $\Omega_2$, the flux conservation equation $S_1 u_s^{\Omega_1} = S_2 u_s^{\Omega_2}$ is degenerate (since $S_1 \ll S_2$ and $u_s^{\Omega_2} \ll u_s^{\Omega_1}$). The state-space realization of Appendix~\ref{sec:ss_realization_area_jump} therefore requires some adaptation:

{\small  
\setlength{\abovedisplayskip}{6pt}
\setlength{\belowdisplayskip}{\abovedisplayskip}
\setlength{\abovedisplayshortskip}{0pt}
\setlength{\belowdisplayshortskip}{3pt}
\begin{align}
\label{eq:area_jump_1D3D_dyn}
\dfrac{d}{dt} &
\underbrace{
\begin{pmatrix}
\overline{u}(t)
\end{pmatrix}}_{\mathbf{X}^{(sc)}(t)}  =
\underbrace{
\begin{pmatrix}
0 \\
\end{pmatrix}}_{\mathbf{A}^{(sc)}}
\begin{pmatrix}
\overline{u}(t)
\end{pmatrix}
+ \underbrace{\begin{pmatrix}
0 & \dfrac{1}{\rho_0 L_{sc}} & 0 & -\dfrac{1}{\rho_0 L_{sc}}
\end{pmatrix}}_{\mathbf{B}^{(sc)}}
\underbrace{\begin{pmatrix}
u_s^{\Omega_1}(t) \\
p^{\Omega_1}(t) \\
u_s^{\Omega_2}(t) \\
p^{\Omega_2}(t)
\end{pmatrix}}_{\mathbf{U}^{(sc)}(t)}
\end{align}}%
In the low-frequency limit, this dynamics equation only imposes the pressure continuity between both subdomains ($p^{\Omega_1}(t) ) = p^{\Omega_2}(t)$), without affecting the velocity. It is therefore necessary to account for the acoustic flux exchange in the output equation, such that the two subdomains are fully coupled. This is achieved by considering a non-zero feedthrough matrix:

\begin{align}
\label{eq:area_jump_1D3D_out}
\underbrace{
\begin{pmatrix}
u_s(0,t)\\
\varphi(0,t) \\
u_s(L_{sc},t) \\
\varphi(L_{sc},t)
\end{pmatrix}}_{\mathbf{Y}^{(sc)}(t)} & =
\underbrace{
\begin{pmatrix}
1 \\
0 \\
0 \\
0 
\end{pmatrix}}_{\mathbf{C}^{(sc)}}
\begin{pmatrix}
\overline{u}(t)
\end{pmatrix}
+ 
\underbrace{\begin{pmatrix}
0 & 0 & 0 & 0 \\
0 & 0 & 0 & 0 \\
-1 & 0 & 0 & 0 \\
0 & 0 & 0 & 0
\end{pmatrix}}_{\mathbf{D}^{(sc)}}
\begin{pmatrix}
u_s^{\Omega_1}(t) \\
p^{\Omega_1}(t) \\
u_s^{\Omega_2}(t) \\
p^{\Omega_2}(t)
\end{pmatrix}
\end{align}

\section{Active flame approximated by a Multi-Pole expansion} \label{sec:ss_realization_flame_PBF}

For a given active flame, fluctuations of heat release rate $Q(t)$  are governed by a FTF, that is approximated by a PBF expansion such as the one of Eq.~\eqref{eq:FTF_multi_pole}. In Chapter~\ref{chap:FRAME}, that is applied to a constant time-delay FTF $e^{-j \omega \tau}$, but this type of approximation is more generic, and its coefficients can be adjusted to fit FTF data points obtained through experiments or through pulsed LES. Recasting each Pole Base Function into the time-domain leads to the state-space realization of the flame:

{\scriptsize  
\setlength{\abovedisplayskip}{6pt}
\setlength{\belowdisplayskip}{\abovedisplayskip}
\setlength{\abovedisplayshortskip}{0pt}
\setlength{\belowdisplayshortskip}{3pt}
\begin{align}
\label{eq:statespace_FTF_1}
\begin{aligned}
&\dfrac{d}{dt}
\underbrace{\begin{pmatrix}
Z_1(t) \\
\dot{Z}_1(t) \\
\vdots \\
Z_{M_{PBF}}(t) \\
\dot{Z}_{M_{PBF}}(t) \\
\end{pmatrix}}_{\mathbf{X}^{(FTF)}(t)} 
=
\underbrace{\left(
\begin{array}{ccccc}
0 & 1 & & & \\
-\omega_{01}^2 & 2 c_1 & & & \\
 & & \ddots &  & \\
 & & & 0 & 1 \\
  & & & -\omega_{0M_{PBF}}^2 & 2 c_{M_{PBF}} \\
\end{array}
\right)}_{\mathbf{A}^{(FTF)}}
\begin{pmatrix}
Z_1(t) \\
\dot{Z}_1(t) \\
\vdots \\
Z_{M_{PBF}}(t) \\
\dot{Z}_{M_{PBF}}(t) \\
\end{pmatrix} 
+ \underbrace{\begin{pmatrix}
0 \\
-1/\overline{u} \\
\vdots \\
0 \\
-1/\overline{u} \\
\end{pmatrix}}_{\mathbf{B}^{(FTF)}} 
\underbrace{\begin{pmatrix}
u^{(ref)}(t) \\
\end{pmatrix}}_{\mathbf{U}^{(FTF)}(t)} \\
\end{aligned}
\end{align}}%

In Eq.~\eqref{eq:statespace_FTF_1}, the flame input vector $\mathbf{U}^{(FTF)}(t)$ has a single entry, the reference fluctuating velocity $u^{(ref)}(t)$. The input matrix $\mathbf{B}^{(FTF)}$ is simply used to normalize the reference velocity fluctuations by the mean flow at the reference point. Note that other formulations including a pressure value as reference are also possible. The state-vector $\mathbf{X}^{(FTF)}(t)$ contains abstract variables that serve as intermediates in the calculation of heat release rate. The state variable $\dot{Z}_k(t)$ can be interpreted as the proportion of normalized heat release fluctuating at frequencies contained in the band of width $2c_k$ centered around $\omega_{0k}$. This state-space realization is completed by the following output equation: 

{\scriptsize  
\setlength{\abovedisplayskip}{6pt}
\setlength{\belowdisplayskip}{\abovedisplayskip}
\setlength{\abovedisplayshortskip}{0pt}
\setlength{\belowdisplayshortskip}{3pt}
\begin{align}
\label{eq:statespace_FTF_2}
\begin{aligned}
\underbrace{\begin{pmatrix}
Q(t) \\
\end{pmatrix}}_{\mathbf{Y}^{(FTF)}(t)} = 
\underbrace{\left(
\begin{array}{ccccc}
0 & -2 \overline{Q} a_1 & \hdots & 0  &  -2 \overline{Q} a_{M_{PBF}} \\
\end{array}
\right)}_{\mathbf{C}^{(FTF)}}
\begin{pmatrix}
Z_1(t) \\
\dot{Z}_1(t) \\
\vdots \\
Z_{M_{PBF}}(t) \\
\dot{Z}_{M_{PBF}}(t) \\
\end{pmatrix} 
\\
\end{aligned}
\end{align}}%

The flame output vector $\mathbf{Y}^{(FTF)}(t)$ only comprises the heat release rate $Q(t)$, which is reconstructed from a linear combination of the individual components  $\dot{Z}_k(t)$ contained in the state-vector.

\paragraph{Remark} \mbox{}\\[-5mm]

\noindent Once a state-space realization (based on a PBF expansion) has been obtained for a FTF with a given time-delay $\tau$, deriving a new state-space representation for another FTF with a  time-delay $\tau'$ is straightforward and does not require to repeat the PBF expansion fitting procedure. Instead, the following relations can be used:
\begin{align}
\label{eq:PBF_rescale_1}
\exp \left(  -j \omega \tau' \right) =  \exp \left(  -j \left( \omega \dfrac{\tau'}{\tau} \right) \tau \right) \approx \sum_{k = 1}^{M_{PBF}} \dfrac{-2 a_k j \left( \omega  \dfrac{\tau'}{\tau} \right)}{ \left( \omega  \dfrac{\tau'}{\tau} \right)^2 + 2 c_k j \left(\omega  \dfrac{\tau'}{\tau} \right) - \omega_{0k}^2}
\end{align}
This directly yields a new PBF expansion for the FTF with a constant time-delay $\tau '$:
\begin{align}
\label{eq:PBF_rescale_2}
e^{  -j \omega \tau' } \approx \sum_{k = 1}^{M_{PBF}} \dfrac{-2 a_k ' j \omega}{\omega^2 + 2 c_k '  j \omega - \omega_{0k}'^2}
\end{align}
where the PBF coefficients are all rescaled in a similar fashion: $a_k ' = a_k \tau / \tau'$, $c_k ' = c_k \tau / \tau'$, and $ \omega_{0k}' = \omega_{0k} \tau / \tau'$. Note, however, that the FTF cutoff frequency and the width of the cutoff region are also rescaled by this same factor $\tau / \tau'$.

\section{Non-reflective boundary at the end of a long duct} \label{sec:ss_realization_NR}

Similarly to the case of a cross-section change between two long ducts (Appendix~\ref{sec:ss_realization_area_jump}), the state-space realization of an impedance $Z(j \omega) = 1/(r+ j \omega/ \omega_c)$ at the end of a duct $\Omega_1$ is derived by performing a volume averaging of the linearized Euler equations in a small control volume of thickness $L_b$ enclosing the boundary.  The resulting state-space representation writes:

{\small  
\setlength{\abovedisplayskip}{6pt}
\setlength{\belowdisplayskip}{\abovedisplayskip}
\setlength{\abovedisplayshortskip}{0pt}
\setlength{\belowdisplayshortskip}{3pt}
\begin{align}
\label{eq:NR_ss_1}
\dfrac{d}{dt}
\underbrace{
\begin{pmatrix}
\overline{p}(t) \\
\overline{\varphi}(t) \\
p^{\Omega_1} (t) \\
\dot{p}^{\Omega_1} (t)
\end{pmatrix}}_{\mathbf{X}^{(NR)}(t)}  =
\underbrace{
\begin{pmatrix}
0 & 0 & 0 & -\dfrac{c_0}{L_b \omega_c} \\
-1/ \rho_0 & 0 & 0 & 0 \\
0 & 0 & 0 & 1 \\
0 & 0 & -\omega_0^2 & -\omega_0
\end{pmatrix}}_{\mathbf{A}^{(NR)}}
\begin{pmatrix}
\overline{p}(t) \\
\overline{\varphi}(t) \\
p^{\Omega_1} (t) \\
\dot{p}^{\Omega_1} (t)
\end{pmatrix} 
+ \underbrace{\begin{pmatrix}
\dfrac{\rho_0 c_0^2}{L_b}  & -\dfrac{r c_0}{L_b}  \\
0 & 0 \\
0 & 0 \\
0 & \omega_0^2
\end{pmatrix}}_{\mathbf{B}^{(NR)}}
\underbrace{\begin{pmatrix}u_s^{\Omega_1}(t) \\
 p^{\Omega_1}(t)  
\end{pmatrix}}_{\mathbf{U}^{(NR)}(t)}
\end{align}}%
where $\omega_0$ is set to a very large value to build a time-derivator in the last two lines. The non-reflective condition appears as a particular case of this state-space equation: setting a large value for $\omega_c$ reduces the first line of Eq.~\eqref{eq:NR_ss_1} to the quasi-static equality $\rho_0 c_0 u_s^{\Omega_1}(t) = r p^{\Omega_1}(t)$, and choosing a value of $r$ very close to unity yields the zero-reflection relation. The state-space realization is completed by the following output equation:
\begin{align}
\label{eq:NR_ss_2}
\underbrace{
\begin{pmatrix}
u_s(x_{0s},t)\\\varphi(x_{0s},t)
\end{pmatrix}}_{\mathbf{Y}^{(NR)}(t)} & =
\underbrace{
\begin{pmatrix}
0 & 0 & 0 & 0\\
0 & 1 & 0 & 0
\end{pmatrix}}_{\mathbf{C}^{(NR)}}
\begin{pmatrix}
\overline{p}(t) \\
\overline{\varphi}(t)
\end{pmatrix}
\end{align}

\section{Complex connection boundary} \label{sec:ss_realization_complex_boundary}

The surface modal amplitudes $\nu_k(t)$ and $\mu_k (t)$ (Eq.~\eqref{eq:surface_modal_expansion}) entirely characterize the acoustics dynamics on the complex boundary $S_{ci}^{(m)}$, and are therefore used to build its state-space representation from Eq.~\eqref{eq:surface_dynamical_system}:

{\footnotesize  
\setlength{\abovedisplayskip}{6pt}
\setlength{\belowdisplayskip}{\abovedisplayskip}
\setlength{\abovedisplayshortskip}{0pt}
\setlength{\belowdisplayshortskip}{3pt}
\begin{align}
\label{eq:statespace_surface_dyn_1}
&\dfrac{d}{dt}
\underbrace{\begin{pmatrix}
\mathbf{X}^{S_{ci}^{(m)}}_1 \\
\vdots \\
\mathbf{X}^{S_{ci}^{(m)}}_{K_S} \\
\end{pmatrix}}_{\mathbf{X}^{S_{ci}^{(m)}}(t)} 
=
\underbrace{\left(
\begin{array}{ccc}
\mathbf{A}_{1}^{S_{ci}^{(m)}} &  & \\
 & \ddots & \\
 & & \mathbf{A}_{K_S}^{S_{ci}^{(m)}} \\
\end{array}
\right)}_{\mathbf{A}^{S_{ci}^{(m)}}}
\begin{pmatrix}
\mathbf{X}^{S_{ci}^{(m)}}_1 \\
\vdots \\
\mathbf{X}^{S_{ci}^{(m)}}_{K_S} \\
\end{pmatrix}
+
\underbrace{\left(
\begin{array}{ccc}
\mathbf{B}_{1}^{\Omega_{i,j}} &  &  \\
 & \ddots &  \\
 &  & \mathbf{B}_{K_S}^{\Omega_{i,j}} \\
\end{array}
\right)}_{\mathbf{B}^{S_{ci}^{(m)}}}
\underbrace{\begin{pmatrix}
\mathbf{U}_1^{\Omega_{i,j}} \\
\vdots \\
\mathbf{U}_{K_S}^{\Omega_{i,j}} \\
\end{pmatrix}}_{\mathbf{U}^{S_{ci}^{(m)}}(t)}
\end{align}}%
where the blocks $\mathbf{X}^{S_{ci}^{(m)}}_k$, $\mathbf{A}_k^{S_{ci}^{(m)}}$, $\mathbf{B}_{k}^{\Omega_{i,j}}$, and $\mathbf{U}_k^{\Omega_{i,j}}$ are defined as: 

{\footnotesize 
\setlength{\abovedisplayskip}{6pt}
\setlength{\belowdisplayskip}{\abovedisplayskip}
\setlength{\abovedisplayshortskip}{0pt}
\setlength{\belowdisplayshortskip}{3pt}
\begin{align}
\label{eq:statespace_surface_dyn_2}
\begin{aligned}
& \mathbf{X}^{S_{ci}^{(m)}}_k =
\begin{pmatrix}
\nu_k (t) \\
\dot{\nu}_k (t) \\
\mu_k (t) \\
\mathbf{X}_{\mathbf{K_R},k} (t)
\end{pmatrix} 
\ , \ 
\mathbf{A}_k^{S_{ci}^{(m)}} =
\left(
\begin{array}{cccc}
 0 & 1 & 0 & \mathbf{0} \\
- \omega_k^2 & 0 & 0 & \mathbf{0} \\
0 & 0 & 0 & \dfrac{1}{\lambda_k L_{\mathscr{D}} \overline{\rho_0}} \mathbf{C_{K_R}}\\
0 & 0 & 0 & \mathbf{A_{K_R}} 
\end{array}
\right)\\
&  \mathbf{B}_{k}^{\Omega_{i,j}} = 
\left(
\begin{array}{cccc}
0 & 0 & 0 & 0\\
-\dfrac{\overline{c_0}^2}{\lambda_k L_{\mathscr{D}}} & 0 & -\dfrac{\overline{c_0}^2}{\lambda_k L_{\mathscr{D}}} & 0 \\
0 & \dfrac{1}{\overline{\rho_0}\lambda_k L_{\mathscr{D}}} & 0 & \dfrac{1}{\overline{\rho_0} \lambda_k L_{\mathscr{D}}} \\
\mathbf{B_{K_R}} & 0 & 0 & 0 \\
\end{array}
\right)
 , 
\mathbf{U}_k^{\Omega_{i,j}} =  \begin{pmatrix}
\mathpzc{s}_i \left( u_s^{\Omega_i}  | \mathscr{K}_k  \right) \\
\mathpzc{s}_i \left( \varphi^{\Omega_i} |  \mathscr{K}_k \right) \\
\mathpzc{s}_j \left( u_s^{\Omega_j}  | \mathscr{K}_k  \right) \\
\mathpzc{s}_j \left( \varphi^{\Omega_j} |  \mathscr{K}_k \right) \\
\end{pmatrix}
\end{aligned}
\end{align}}%
In this equation, the subscripts ${}_{\mathbf{K_R}}$ refer to the SISO state-space realization associated to the inverse Fourier transform of the complex conductivity $K_R (j \omega)$. The blocks $\mathbf{B}_{k}^{\Omega_{i,j}}$ and $\mathbf{U}_k^{\Omega_{i,j}}$ are employed to compute the source terms from the two adjacent subdomains $\Omega_i$ and $\Omega_j$. To be consistent with the subdomains input vectors, $S_{ci}^{(m)}$ must output the modal amplitudes $\nu_k (t)$ and $\mu_k (t)$, which is achieved thanks to:

{\small  
\setlength{\abovedisplayskip}{6pt}
\setlength{\belowdisplayskip}{\abovedisplayskip}
\setlength{\abovedisplayshortskip}{0pt}
\setlength{\belowdisplayshortskip}{3pt}
\begin{align}
\label{eq:statespace_surface_out_1}
\underbrace{\begin{pmatrix}
\mathbf{Y}_1^{\Omega_{i,j}} \\
\vdots \\
\mathbf{Y}_{K_S}^{\Omega_{i,j}}\\
\end{pmatrix}}_{\mathbf{Y}^{S_{ci}^{(m)}}(t)} 
=
\underbrace{\left(
\begin{array}{ccc}
\mathbf{C}_ {1}^{\Omega_{i,j}} &  &  \\
 & \ddots &  \\
 &  & \mathbf{C}_ {K_S}^{\Omega_{i,j}} \\
\end{array}
\right)}_{\mathbf{C}^{S_{ci}^{(m)}}}
\underbrace{\begin{pmatrix}
\mathbf{X}^{S_{ci}^{(m)}}_1 \\
\vdots \\
\mathbf{X}^{S_{ci}^{(m)}}_{K_S} \\
\end{pmatrix}}_{\mathbf{X}^{S_{ci}^{(m)}}(t)}
\end{align}}%
with:

{\small  
\setlength{\abovedisplayskip}{6pt}
\setlength{\belowdisplayskip}{\abovedisplayskip}
\setlength{\abovedisplayshortskip}{0pt}
\setlength{\belowdisplayshortskip}{3pt}
\begin{align}
\label{eq:statespace_surface_out_2}
\mathbf{Y}_k^{\Omega_{i,j}} =  
\begin{pmatrix}
\mu_k (t) \\
\nu_k (t) \\
\mu_k (t) \\
\nu_k (t) \\
\end{pmatrix} 
\ , \ 
\mathbf{C}_{k}^{\Omega_{i,j}} = 
\left(
\begin{array}{cccc}
0  & 0 & 1 & \mathbf{0} \\
1  & 0 & 0 & \mathbf{0} \\
0  & 0 & 1 & \mathbf{0} \\
1  & 0 & 0 & \mathbf{0} \\
\end{array}
\right)
\end{align}}%

\section{Howe's model Rayleigh conductivity} \label{sec:ss_realization_conductivity}

For a Rayleigh conductivity $K_R$ expressed through the approximation to Howe's model given in Eq.~\eqref{eq:howe_approximation}, the inverse Fourier transform of the term $- j \omega \rho_u d^2 / K_R(j \omega)$ in Eq.~\eqref{eq:conductivity_relation} can be translated into a SISO state-space realization that is directly embedded into the complex boundary state-space realization provided in Eq.~\eqref{eq:statespace_surface_dyn_1}-\eqref{eq:statespace_surface_dyn_2}. Its expression writes:

\begin{align}
\label{eq:state_space_conductivity_dyn}
\dv{t} \underbrace{\left( z (t) \right)}_{\mathbf{X_{K_R}}(t)} = \underbrace{\left( - K_R^A/K_R^B \right)}_{\mathbf{A_{K_R}}} \left( z (t) \right) + \underbrace{\left( -1/ K_R^B \right)}_{\mathbf{B_{K_R}}} \left( \mathpzc{s}_i \left( u_s^{\Omega_i}  | \mathscr{K}_k  \right) \right)
\end{align}
and the output $Y_{K_R}$ is computed through:

\begin{align}
\label{eq:state_space_conductivity_out}
\left( Y_{K_R}(t) \right) = \underbrace{\left( - \rho_u d^2  \right)}_{\mathbf{C_{K_R}}}  \left( z (t) \right)
\end{align}

				
\chapter{Reference solution for a liner in a cylindrical domain} \label{Appendix:B}

Let us consider an annular and a cylindrical subdomains separated by a ribbon-like multi-perforated liner characterized by the Rayleigh conductivity $K_R (j \omega)$, such as the system presented in Sec.~\ref{sec:surface_modal_expansion_convergence}. In both the inner cylinder and the outer annulus, the pressure is solution of the classical homogeneous Helmholtz equation. In these two domains, the separation of variables $p (r,\theta) = R (r) \Theta (\theta)$ is performed, which yields a set of Bessel differential equations governing $R(r)$. Solutions of these equations write:
\begin{align}
\label{eq:theoretical_solution_radial_part}
R(r \leq R_1) = A J_n (k r) \ , \ R( R_1 < r \leq R_2) = B J_n (k r) + C Y_n (k r)
\end{align}
where $A,B,C$, and $k = \omega/c_0$ are complex constants to be determined, and $J_n$ and $Y_n$ are the Bessel functions of the 1\textsuperscript{st} and 2\textsuperscript{nd} kind, respectively. The azimuthal component $\Theta(\theta)$ is a simple linear combination of $\cos (n \theta)$ and $\sin (n \theta)$. The rigid-wall conditions at $r = R_2$ combined with the flux continuity at $r = R_1$ as well as pressure jump through the multi-perforated liner leads to a linear system $\mathbf{M} \mathbf{X} = \mathbf{0}$ where $\mathbf{X} = {}^t (A \  B \ C)$ and $\mathbf{M}$ is given by:

{\scriptsize  
\setlength{\abovedisplayskip}{6pt}
\setlength{\belowdisplayskip}{\abovedisplayskip}
\setlength{\abovedisplayshortskip}{0pt}
\setlength{\belowdisplayshortskip}{3pt}
\begin{align}
\label{eq:dispersion_cylinder_theoretical}
\left(
\begin{array}{ccc}
0 & J_n' (k R_2) & Y_n'(k R_2) \\
\dfrac{d^2}{K_R (j \omega)} k J_n'(k R_1) + J_n(k R_1) & -J_n(k R_1)  & - Y_n'(k R_1)  \\
J_n(k R_1) & -J_n(k R_1) + \dfrac{d^2}{K_R (j \omega)} k J_n'(k R_1) & - Y_n (k R_1) + \dfrac{d^2}{K_R (j \omega)} k Y_n' (k R_1) \\
\end{array}
\right)
\end{align}}

This linear system admits non-trivial solutions if and only if $det ( \mathbf{M}) = 0$, which results in a dispersion relation $F(\omega) = 0$ in the complex plane. The resolution of this equation yields the eigenfrequencies, their associated growth rates, and ultimately the mode shapes thanks to Eq.~\eqref{eq:theoretical_solution_radial_part}.

\end{appendices}



\def\rtn{\par \noindent }
\def\pskip{\rtn }

\def\aa{ Acta Astronautica  }
\def\aam{ Adv. Appl. Mech.  }
\def\aas{ Atomization and Sprays  }

\def\aiaap{ AIAA Paper  }
\def\aj{ AIAA J.  }
\def\ajou{ Aeronaut. J.  }
\def\annr{ Ann. Rev. Fluid Mech. }
\def\taj{ Astrophys. J. } 	
\def\as{ Astron. and Astrophys. }
\def\autcf{Reprinted by permission of Elsevier Science from }
\def\autcff{\copyright ~the Combustion Institute }
\def\autjfm{Reprinted with permission by Cambridge University Press }

\def\bbpc{ Ber. Bunsenges. Phys. Chem. }

\def\canm{ Comm. Appl. Num. Meth. }
\def\ces{ Chem. Eng. Sci. }
\def\cf{ Combust. Flame}
\def\cfl{ Comput. Fluids }
\def\cip{ Comput. Phys. }
\def\cmame{ Comput. Methods Appl. Mech. Eng. }
\def\cpam{ Commun. Pure Appl. Math. }
\def\cpc{ Computer Phys. Communications }
\def\cras{ C. R. Acad. Sci. }
\def\cst{ Combust. Sci. Technol. }
\def\ctm{ Combust. Theor. Model. }
\def\ctrsp{ Proc. of the Summer Program }
\def\ctrarb{ Annual Research Briefs }
\def\ptrsla{Phil. Trans. R. Soc. London A }
\def\zmp{Z. Math. Phys. }
\def\jsiam{J. Soc. Indust. Appl. Math. }
\def\pla{Physics Letters A }
\def\ptrsa{Phil. Trans. R. Soc. A }

\def\csat{ Composites Science and Technology }              	
\def\csat{ Composites Sci. and Tech. }              			
\def\csd{ Comput. Sci. Discov. }

\def\DHCRS{ Dynamics of Heterogeneous Combustion and Reacting Systems }

\def\ent{ Entropie }
\def\etme{ Eng. Turb. Modelling and Exp. } 
\def\exf{ Exp. Fluids }

\def\fd{ J. Fluid Dynamics }
\def\ftac{ Flow, Turb. and Combustion }

\def\icmf{ Int. Conf. Multiphase Flow }
\def\ijav{ Int. J. Acoust. Vib. }
\def\ijcfd{ Int. J. Comput. Fluid Dynamics }
\def\ijcm{ Int. J. Comput. Methods }
\def\ijhff{ Int. J. Heat Fluid Flow }
\def\ijmf{ Int. J. Multiphase Flow }
\def\ijmp{ Int. J. Modern Physics C }
\def\ijnme{ Int. J. Numer. Meth. Eng. }
\def\ijnmf{ Int. J.~Numer. Meth. Fluids }
\def\ijhmt{ Int. J.~Heat and Mass Transfer }
\def\ijts{ Int. J. of Therm. Sci. }
\def\iecf{Ind. Eng. Chem., Fundam.}
\def\ijck{ Int. J. Chem. Kinet. }                                      		
\def\ijck{ International Journal of Chemical Kinetics }         	
\def\ijhe{ Int. J. Hydrog. Energy }
\def\ijer{ Int. J. Engine Res. }
\def\iecr{ Ind. Eng. Chem. Res. }


\def\ja{ J. Aircraft }
\def\jacic{ J. Aeropace Comput. Inform. Comm. }
\def\jaes{ J. Aeronaut. Sci. }
\def\jam{ SIAM J. Appl. Math. }
\def\jame{ J. Appl. Mech. }
\def\jamp{ J. Appl. Phys. }
\def\jars{ J.~American Rocket Society }
\def\jas{ J. Atmos. Sci. }
\def\jasa{ J. Acous. Soc. Am. }
\def\jchp{ J. Chem. Phys. }
\def\jcht{ J. Chem. Thermodynamics }
\def\jcp{ J.~Comput. Phys. }
\def\je{ J. Energy }
\def\jep{ J. Eng. Gas Turb. and Power }
\def\jfe{ J. Fluids Eng. }
\def\jfm{ J.~Fluid Mech. }
\def\jfs{ J. Fluids Struct. }
\def\jht{ J. Heat Trans. }
\def\jie{ J. Inst. Energy }
\def\jmta{ J. M\'ec. Th\'eor. Appl. }
\def\jpp{ J.~Prop.~Power }
\def\jrnbs{ J. Res. Natl. Bur. Stand. }
\def\jsc{ J. Sci. Comput. }
\def\jsr{ J. Spacecrafts and Rockets }
\def\jsv{ J.~Sound Vib. }
\def\jssc{ SIAM J. Sci. Stat. Comput. }
\def\jt{ J.~Turb. }
\def\jtht{ J. Thermophysics and Heat Trans. }
\def\jtu{ J. Turbomach. }
\def\jpca{J. Phys. Chem. A}
\def\jpdap{ Journal of Physics D: Applied Physics }          	
\def\jpdap{ J. Phys. D: Appl. Phys. }          				
\def\jhm{ J. Hazard. Mater. }
\def\jlppi{ J. Loss Prev. Process Ind. }

\def\moc{ Math. Comp. }
\def\mwr{ Mon. Weather Rev. }

\def\ned{ Nuclear Eng. and Design }

\def\paa{ Prog. in Astronautics and Aeronautics }
\def\pas{ Prog. Aerospace Sci. }
\def\pcfd{ Prog. Comput. Fluid Dynamics }
\def\pci{ Proc. Combust. Inst. }
\def\pcp{ Prog. Comput. Phys. }
\def\pecs{ Prog. Energy Combust. Sci. }
\def\pf{ Phys. Fluids }
\def\pime{ Proc. Instn. Mech. Engrs. }
\def\plms{ Proc. London Math. Soc }
\def\ppsc{ Particle and Particle Systems Characterization }
\def\prl{ Phys. Rev. Lett. }
\def\prsl{ Proc. R. Soc. Lond. }
\def\prsla{ Proc. R. Soc. Lond. A }
\def\pt{ Powder Technology }
\def\pep{ Propellants, Explosives, Pyrotechnics }          	

\def\qjmam{ Q. J. Mech. Appl. Math. }

\def\ra{ La Rech. A\'{e }rospatiale }
\def\rpa{ Rev. Phys. Appl. }

\def\tcfd{ Theoret. Comput. Fluid Dynamics }
\def\tcsme{ ASME Trans. }
\def\ti{ Technique de l'Ing\'enieur }
\def\tsfp{ Turb. Shear Flow Phenomena }

\def\WSSCI{ WSS/CI }

\def\Litem{\par\noindent }
\font\smc=cmcsc10

\bibliographystyle{ieeetr}

\end{document}